%% file: main.tex
\documentclass[AMA,STIX1COL]{WileyNJD-v2}

\usepackage[utf8]{inputenc} 				% unicode formatted text

\usepackage[T1]{fontenc} 				% font encoding for umlauts
\usepackage{textcomp} 					% various symbols (e.g. ° and €)
\usepackage{amsmath}					% extended mathematics environment
\usepackage{amssymb} 					% extended mathematics symbol collection
\usepackage{mathtools}					% extended mathematics symbol collection
\usepackage{cases}                      % case in math environment (curly braces spanning multiple lines)
\usepackage[noabbrev]{cleveref}			% automatically print the type of reference
\usepackage[final]{microtype}			% improved justification
\usepackage{csquotes}					% quotes using \enquote{}
\usepackage{graphicx}                   % including and adjusting graphics of various formats
\usepackage{grffile}					% allows dots (prior to file-ending) in graphics files' names
\graphicspath{{figures/}}
\usepackage{subcaption}					% includes subfigure-environment
\usepackage{multirow}                   % tables with multiple rows
\usepackage{array}						% vertical centered tabular cells
\usepackage{makecell}					% line breaks within tabular cells
\usepackage{booktabs}					% thicker lines in tables
\usepackage{caption}
\usepackage{placeins}					% force floats to be print at \FloatBarrier before continuing
\usepackage[export]{adjustbox}			% vertical alignement of figures via \includegraphics[valign=c]

\usepackage{pgfplots}
\pgfplotsset{compat=newest}
\usepgfplotslibrary{groupplots,dateplot,fillbetween}
\usetikzlibrary{arrows,arrows.meta,patterns,shapes.arrows,calc,angles,quotes,babel}
\DeclareUnicodeCharacter{2212}{−}
\newlength\figureheight
\newlength\figurewidth
\newlength\mdist 	% distance of measure line and object
\newlength\radius	% radius of the Taylor bubble in the setup figure

\usetikzlibrary{external}
\usetikzlibrary{calc,angles,quotes}

\usepackage[colorinlistoftodos, textwidth=3.0cm]{todonotes}
\newcommand{\todoUR}[1][]{\todo[color=cyan!20, size=\footnotesize, author={Uli}, inline]}
\newcommand{\todoCS}[1][]{\todo[color=red!20, size=\footnotesize, author={Christoph}, inline]}

\definecolor{error-term-red}{RGB}{242,133,34}

\makeatletter
\def\mathcolor#1#{\@mathcolor{#1}}
\def\@mathcolor#1#2#3{%
	\protect\leavevmode
	\begingroup
	\color#1{#2}#3%
	\endgroup
}
\makeatother

\articletype{Research Article}%

\received{28 July 2022}
\revised{X YYYY ZZZZ}
\accepted{X YYYY ZZZZ}

\begin{document}

\title{Analysis and comparison of boundary condition variants in the free-surface lattice Boltzmann method}

\author[1]{Christoph Schwarzmeier*}
\author[1,2]{Ulrich Rüde}

\authormark{Schwarzmeier \textsc{et al.}}

\address[1]{\orgdiv{Chair for System Simulation}, \orgname{Friedrich-Alexander-Universität Erlangen-Nürnberg}, \orgaddress{Cauerstraße 11, 91058 Erlangen, \country{Germany}}}
\address[2]{\orgname{CERFACS}, \orgaddress{42 Avenue Gaspard Coriolis, 31057 Toulouse Cedex 1, \country{France}}}

\corres{*Christoph Schwarzmeier, Chair for System Simulation, Friedrich-Alexander-Universität Erlangen-Nürnberg, Cauerstraße 11, 91058 Erlangen, Germany.\\ \email{christoph.schwarzmeier@fau.de}}

\presentaddress{\orgdiv{Chair for System Simulation}, \orgname{Friedrich-Alexander-Universität Erlangen-Nürnberg}, \orgaddress{Cauerstraße 11, 91058 Erlangen, \country{Germany}}}

\abstract[Summary]{
The accuracy of the free-surface lattice Boltzmann method (FSLBM) depends significantly on the boundary condition employed at the free interface.
Ideally, the chosen boundary condition balances the forces exerted by the liquid and gas pressure.
Different variants of the same boundary condition are possible, depending on the number and choice of the particle distribution functions (PDFs) to which it is applied.	
This study analyzes and compares four variants, in which (i) the boundary condition is applied to all PDFs oriented in the opposite direction of the free interface's normal vector, including or (ii) excluding the central PDF.
While these variants overwrite existing information, the boundary condition can also be applied (iii) to only missing PDFs without dropping available data or (iv) to only missing PDFs but at least three PDFs as suggested in the literature.
It is shown that neither variant generally balances the forces exerted by the liquid and gas pressure at the free surface.
The four variants' accuracy was compared in five different numerical experiments covering various applications.
These include a standing gravity wave, a rectangular and cylindrical dam break, a rising Taylor bubble, and a droplet impacting a thin pool of liquid.
Overall, variant (iii) was substantially more accurate than the other variants in the numerical experiments performed in this study.
}


\jnlcitation{\cname{%
\author{Schwarzmeier C},
\author{Rüde U}}. 
\ctitle{\textbf{Analysis and comparison of boundary condition variants in the free surface lattice Boltzmann method}}. \cjournal{Int J Numer Meth Fluids}. \cyear{2022}, \cvol{XXX}.}

\maketitle

\input{src/introduction.tex}
\input{src/numerical-method.tex}
\input{src/boundary-condition.tex}
\input{src/numerical-experiments.tex}
\input{src/conclusion.tex}

\section*{Acknowledgments}
The authors thank the Deutsche Forschungsgemeinschaft (DFG, German Research Foundation) for funding project 408062554.\\
This work was supported by the SCALABLE project. This project has received funding from the European High-Performance Computing Joint Undertaking (JU) under grant agreement No 956000. The JU receives support from the European Union’s Horizon 2020 research and innovation programme and France, Germany, the Czech Republic.\\
The authors gratefully acknowledge the Gauss Centre for Supercomputing e.V. (www.gauss-centre.eu) for funding this project by providing computing time on the GCS Supercomputer SuperMUC at Leibniz Supercomputing Centre (www.lrz.de).\\
The authors gratefully acknowledge the scientific support and HPC resources provided by the Erlangen National High Performance Computing Center (NHR@FAU) of the Friedrich-Alexander-Universität Erlangen-Nürnberg (FAU). The hardware is funded by the German Research Foundation (DFG).\\
The authors appreciate the valuable discussions with Christoph Rettinger and Simon Bogner, and thank Sara Faghih-Naini and Jonas Plewinski for proofreading the manuscript.

%\subsection*{Author contributions}

\subsection*{Financial disclosure}

None reported.

\subsection*{Conflict of interest}

The authors declare no potential conflict of interests.

\section*{Supporting information}

The following supporting information is available as part of the online article:\\
An archive of the C++ source code used in this study.
It is part of the software framework waLBerla (version used here: \url{https://i10git.cs.fau.de/walberla/walberla/-/tree/01a28162ae1aacf7b96152c9f886ce54cc7f53ff}).
The ready-to-run simulation setups for all numerical experiments performed in this article are included in the directory \texttt{apps/showcases/FreeSurface}.

\appendix

\input{src/appendix.tex}

\bibliography{literature.bib}%

\end{document}

%% file: src/introduction.tex
%!TEX root = ../main.tex

\section{Introduction}\label{sec:int}
The free-surface lattice Boltzmann method (FSLBM)~\cite{korner2005LatticeBoltzmannModel} is a well-established approach for simulating free-surface flows with the lattice Boltzmann method (LBM).
In this context, \textit{free-surface flow} refers to an immiscible two-fluid flow problem, the flow dynamics of which are assumed to be entirely governed by the heavier fluid.
Consequently, the flow dynamics of the lighter fluid are neglected such that the problem reduces to a single-fluid flow with a free boundary~\cite{scardovelli1999DirectNumericalSimulation}.
The free boundary, that is, the interface, is tracked according to the volume-of-fluid (VOF) approach~\cite{hirt1981VolumeFluidVOF}.
There, an indicator denotes the affiliation to one of the fluids.
In this article, the lighter fluid is called gas phase, and the heavier fluid is referred to as the liquid phase.
The FSLBM has been successfully validated in simulations of different applications, including rising bubbles~\cite{donath2011VerificationSurfaceTension}, waves~\cite{zhao2013LatticeBoltzmannMethod}, dam break scenarios~\cite{janssen2011FreeSurfaceFlow}, drop impacts~\cite{lehmann2021EjectionMarineMicroplastics} and electron-beam melting~\cite{ammer2014SimulatingFastElectron}.
\par

There are other multiphase LBM models available in the literature.
Models such as the FSLBM, the level-set method~\cite{becker2009CombinedLatticeBGK}, the front-tracking approach~\cite{lallemand2007LatticeBoltzmannFronttracking}, and the color gradient model~\cite{gunstensen1991LatticeBoltzmannModel} represent the interface between the fluids in a sharp manner.
This is in contrast to models with a diffuse interface, such as phase-field models~\cite{inamuro2004LatticeBoltzmannMethod,zheng2005LatticeBoltzmannInterface,fakhari2017ImprovedLocalityPhasefield}, the free-energy model~\cite{swift1995LatticeBoltzmannSimulation}, and the pseudopotential model~\cite{shan1994SimulationNonidealGases}.
Sharp interface models generally require a lower computational resolution than models with a diffuse interface.
Despite this advantage in computational efficiency, the FSLBM is also inherently applicable to systems with (infinitely) large density and viscosity ratios.
However, the FSLBM can not be applied when the flow in the lighter fluid is relevant for the system's dynamics.
Additionally, the FSLBM's algorithm is relatively complicated when compared to phase-field models, for example.
More details on the FSLBM's advantages and disadvantages, and a comparison with an Allen--Cahn LBM phase-field model are presented in prior work~\cite{schwarzmeier2022ComparisonFreeSurface}.
\par

In the LBM, each cell of the computational grid contains particle distribution functions (PDFs), which represent the flow field information.
In every simulation time step, these PDFs stream to all cells in their direct surrounding.
Since the flow dynamics of the lighter phase are neglected in the FSLBM, gas cells do not carry valid PDF information.
Therefore, PDFs streaming from the gas towards the liquid phase are unavailable.
They must be reconstructed using a boundary condition for the free surface.
In the original FSLBM introduced by Körner et al.~\cite{korner2005LatticeBoltzmannModel}, the suggested boundary condition is not only applied to reconstruct missing PDFs.
Instead, it is applied so that existing PDFs from the liquid phase are also reconstructed.
Consequently, available information about the flow field is dropped.
Körner et al. argue that this is required to balance the forces exerted by the liquid and gas at the interface.
However, other authors have reported anisotropic artifacts~\cite{bogner2016CurvatureEstimationVolumeoffluid} or implausible simulation results~\cite{zhao2013LatticeBoltzmannMethod} when available information is overwritten by the boundary condition.
This article will show that the force-balance computation from Körner et al.~\cite{korner2005LatticeBoltzmannModel} must be corrected, as forces are only balanced in steady-state systems but not generally.
In the study presented here, other variants of applying the free boundary condition are analyzed, compared, and found to be more accurate than that from Körner et al.~\cite{korner2005LatticeBoltzmannModel}
\par

In the first section, the numerical foundations of the LBM and FSLBM are introduced.
Then, the balance of forces at the interface is computed for four different variants of applying the free-surface boundary condition.
These include the variant by Körner et al.~\cite{korner2005LatticeBoltzmannModel}, where PDFs are reconstructed based on the orientation of the interface-normal.
In this variant, existing PDFs are overwritten.
While not explicitly mentioned in the article~\cite{korner2005LatticeBoltzmannModel}, the central PDF must also be reconstructed in this variant.
The second variant under investigation is similar to the first one.
However, the central PDF is not overwritten.
In the third variant, only missing PDFs are reconstructed.
Therefore, no existing fluid flow information is dropped.
In the final variant, only missing but at least three PDFs are reconstructed~\cite{bogner2017DirectNumericalSimulation, thies2005LatticeBoltzmannModeling}.
The force-balance computations show that neither of these four variants generally balances the forces at the interface.
In five different numerical experiments, the boundary condition variants are then compared in different applications.
These include a standing gravity wave, a rectangular and cylindrical dam break, the rise of a Taylor bubble, and the formation of the splash crown when a drop impacts a pool of liquid.
Finally, it is concluded that it is preferable to avoid overwriting existing information.
Instead, only missing PDFs should be reconstructed with the free-surface boundary condition.
\par

The source code of the implementation used in this study is freely available as part of the open source C++ software framework \mbox{\textsc{waLBerla}}~\cite{bauer2021WaLBerlaBlockstructuredHighperformance} (\url{https://www.walberla.net}).
The version of waLBerla used in this article is provided in the supporting information.
\par

%% file: src/numerical-method.tex
%!TEX root = ../main.tex

\section{Numerical methods}\label{sec:nm}
This section introduces the numerical foundations of the lattice Boltzmann method and its extension to free-surface flows, the free-surface lattice Boltzmann method.
The section is based on Section 2 in articles~\cite{schwarzmeier2022ComparisonFreeSurface,schwarzmeier2022ComparisonRefillingSchemes} but is repeated here for completeness.
\par

\subsection{Lattice Boltzmann method}\label{sec:nm-lbm}
The lattice Boltzmann method is a relatively modern approach for simulating computational fluid dynamics.
A thorough introduction to the LBM is available in the literature~\cite{kruger2017LatticeBoltzmannMethod}.
Here, only its fundamental aspects are introduced.
\par

The LBM discretizes the Boltzmann equation from kinetic gas theory and describes the evolution of particle distribution functions on a uniformly discretized Cartesian lattice with spacing $\Delta x \in \mathbb{R^{+}}$.
In each lattice cell, the macroscopic fluid velocity is discretized with the D$d$Q$q$ velocity set, where $d \in \mathbb{N}$ refers to the lattice's spatial dimension and $q \in \mathbb{N}$ refers to the number of PDFs per cell.
A PDF $f_{i}(\boldsymbol{x}, t) \in \mathbb{R}$ with $i \in \{0, 1, \dots, q - 1\}$ represents the probability that there exists a population of virtual fluid particles at time $t \in \mathbb{R^{+}}$ and position $\boldsymbol{x} \in \mathbb{R}^{d}$ traveling with lattice velocity $\boldsymbol{c}_{i} \in \Delta x / \Delta t \, \{-1, 0, 1\}^{d}$.
The time is discretized by distinct time steps of length $\Delta t$.
The discrete lattice Boltzmann equation can be written in the subsequent steps of collision, also called relaxation,
\begin{equation}\label{eq:nm-lbm-collision}
f_{i}^{\star}(\boldsymbol{x}, t) = f_{i}(\boldsymbol{x}, t) + \Omega_{i}(\boldsymbol{x}, t) + F_{i}(\boldsymbol{x}, t)
\end{equation}
and streaming, also called propagation,
\begin{equation}\label{eq:nm-lbm-streaming}
f_{i}(\boldsymbol{x} + \boldsymbol{c}_{i}\Delta t, t+\Delta t) = f_{i}^{\star}(\boldsymbol{x}, t).
\end{equation}
In the collision step, the collision operator $\Omega_{i}(\boldsymbol{x}, t) \in \mathbb{R}$ relaxes the PDFs towards an equilibrium state $f_{i}^{\text{eq}}(\boldsymbol{x}, t)$ while being influenced by external forces $F_{i}(\boldsymbol{x}, t) \in \mathbb{R}$.
In the streaming step, the post-collision PDFs $f_{i}^{\star}(\boldsymbol{x}, t)$ stream to neighboring cells.
In the present article, the single relaxation time (SRT) collision operator
\begin{equation}\label{eq:nm-lbm-collision-operator}
\Omega_{i}(\boldsymbol{x}, t) = \frac{f_{i}(\boldsymbol{x}, t) - f_{i}^{\text{eq}}(\boldsymbol{x}, t)}{\tau} \Delta t
\end{equation}
is used with relaxation time $\tau > \Delta t / 2$.
The PDF's equilibrium can be derived from the continuous Maxwell--Boltzmann distribution~\cite{bauer2020TruncationErrorsD3Q19a} and is given by
\begin{equation}\label{eq:nm-lbm-equilibrium}
f_{i}^{\text{eq}}(\boldsymbol{x}, t) = w_{i}\rho\left(1 + \frac{\boldsymbol{u}\cdot\boldsymbol{c}_{i}}{c_{s}^{2}} + \frac{(\boldsymbol{u} \cdot \boldsymbol{c}_{i})^{2}}{2 c_{s}^{4}} - \frac{\boldsymbol{u} \cdot \boldsymbol{u}}{2 c_{s}^{2}} \right).
\end{equation}
It includes the lattice weights $w_{i} \in \mathbb{R}$, the lattice speed of sound $c_{s}^{2}$, the macroscopic fluid density $\rho \equiv \rho(\boldsymbol{x}, t) \in \mathbb{R^{+}}$, and the macroscopic fluid velocity $\boldsymbol{u} \equiv \boldsymbol{u}(\boldsymbol{x}, t) \in \mathbb{R}^{d}$.
In this study, the well-established D$2$Q$9$ and D$3$Q$19$ lattice models are used.
The corresponding lattice weights can be found in the literature~\cite{kruger2017LatticeBoltzmannMethod}. The lattice speed of sound $c_{s}^{2}=\sqrt{1/3} \, \Delta x / \Delta t$ defines the relation between the macroscopic fluid density $\rho(\boldsymbol{x}, t)$ and pressure $p(\boldsymbol{x}, t)=c_{s}^{2}\rho(\boldsymbol{x}, t)$.
The PDFs' zeroth-order moment is the density
\begin{equation}\label{eq:nm-lbm-density}
\rho(\boldsymbol{x}, t) = \sum_{i} f_{i}(\boldsymbol{x}, t)
\end{equation}
and the first-order moment reveals the macroscopic fluid velocity
\begin{equation}\label{eq:nm-lbm-velocity}
\boldsymbol{u}(\boldsymbol{x}, t) = \frac{\boldsymbol{F}(\boldsymbol{x}, t)\Delta t}{2 \rho(\boldsymbol{x}, t)} + \frac{1}{\rho(\boldsymbol{x}, t)}\sum_{i} \boldsymbol{c}_{i} f_{i}(\boldsymbol{x}, t),
\end{equation}
where $\boldsymbol{F}(\boldsymbol{x}, t) \in \mathbb{R}^{d}$ is an external force.
The fluid's kinematic viscosity
\begin{equation}\label{eq:nm-lbm-viscosity}
\nu = c_{s}^{2} \left(\tau - \frac{\Delta t}{2}\right)
\end{equation}
can be computed from the relaxation time $\tau$, that is, relaxation rate $\omega=1/\tau$.
In this article, the gravitational force, as part of $F_{i}$ in the LBM collision \eqref{eq:nm-lbm-collision}, was modeled according to Guo et al.~\cite{guo2002DiscreteLatticeEffects} with
\begin{equation}\label{eq:nm-lbm-force-guo}
F_{i}(\boldsymbol{x}, t) =
\left(1 - \frac{\Delta t}{2 \tau} \right)
w_{i}\left(\frac{\boldsymbol{c}_{i} - \boldsymbol{u}}{c_{s}^{2}} + \frac{(\boldsymbol{c}_{i} \cdot \boldsymbol{u}) \boldsymbol{c}_{i}}{c_{s}^4}\right) \cdot \boldsymbol{F}(\boldsymbol{x}, t),
\end{equation}
where $\boldsymbol{u} \equiv \boldsymbol{u}(\boldsymbol{x}, t)$ was used as before.
\par

For the simulations of the rectangular and cylindrical dam break in \Cref{sec:ne-rdb,sec:ne-cdb}, a Smagorinksy-type large eddy simulation turbulence model was employed~\cite{hou1996LatticeBoltzmannSubgrid,yu2005DNSDecayingIsotropic}.
With the user-chosen relaxation time $\tau_{0} > \Delta t / 2$, the model locally adjusts the collision operator's relaxation time $\tau(\boldsymbol{x}, t) = \tau_{0} + \tau_{t}(\boldsymbol{x}, t)$ with a contribution $\tau_{t}(\boldsymbol{x}, t) \in \mathbb{R}$ from the turbulence viscosity
\begin{equation}\label{eq:nm-lbm-turbulence-viscosity}
\nu_{t}(\boldsymbol{x}, t) \coloneqq \tau_t(\boldsymbol{x}, t) c_{s}^{2} = \left(C_{S}\Delta x_{\text{LES}}\right)^{2} \bar{S}(\boldsymbol{x}, t),
\end{equation}
where $\Delta x_{\text{LES}}$ is the filter length, $C_{S}$ is the Smagorinsky constant, and
\begin{equation}\label{eq:nm-lbm-strain-rate}
\bar{S}(\boldsymbol{x}, t) = \frac{\bar{Q}(\boldsymbol{x}, t)}{2\rho c_{s}^2 \tau_{0}}
\end{equation}
is the filtered strain rate tensor.
The filtered mean momentum flux
\begin{equation}\label{eq:nm-lbm-mean-momentum-flux}
\bar{Q}(\boldsymbol{x}, t) = \sqrt{2\sum_{\alpha, \beta}\bar{Q}_{\alpha, \beta}(\boldsymbol{x}, t)\bar{Q}_{\alpha, \beta}(\boldsymbol{x}, t)}
\end{equation}
is computed from the momentum fluxes
\begin{equation}\label{eq:nm-lbm-momentum-fluxes}
\bar{Q}_{\alpha, \beta}(\boldsymbol{x}, t) = \sum_{i} c_{i, \alpha} c_{i, \beta}\Bigl( f_{i}(\boldsymbol{x}, t) - f_{i}^{\text{eq}}(\boldsymbol{x}, t)\Bigr)
\end{equation}
as obtained from the second-order moments of the non-equilibrium parts of the PDFs.
The indices $\alpha$ and $\beta$ are used to refer to the components of a vector or tensor in index notation.
The turbulence relaxation time is then given by~\cite{hou1996LatticeBoltzmannSubgrid}
\begin{equation}\label{eq:nm-lbm-turublence-relaxation-time}
\tau_{t}(\boldsymbol{x}, t) = \frac{1}{2}\sqrt{\tau_{0}^{2} + 2\sqrt{2}(C_{S}\Delta x_{\text{LES}})^{2}(\rho c_{s}^4)^{-1}\bar{Q}(\boldsymbol{x}, t)}-\tau_{0}.
\end{equation}
In the simulations performed in this article, $\Delta x_{\text{LES}}=\Delta x$ and $C_{S}=0.1$~\cite{yu2005DNSDecayingIsotropic} were chosen.
\par

The bounce-back approach was used at solid obstacles with a no-slip boundary condition.
In this approach, PDFs streaming into solid obstacle cells are reflected reversely, that is, their original direction with index $i$ is reversed, denoted as $\bar{i}$, with lattice velocity $\boldsymbol{c}_{\bar{i}}=-\boldsymbol{c}_{i}$~\cite{kruger2017LatticeBoltzmannMethod}.
Free-slip boundary conditions are realized similarly, with the PDFs being reflected specularly.
Consequently, the normal velocity component of the incoming velocity $\boldsymbol{c}_{i}$ is reversed with $c_{j,n} = -c_{i,n}$, where  $\boldsymbol{c}_{j}$ is the resulting lattice velocity~\cite{kruger2017LatticeBoltzmannMethod}.
\par

As commonly used in the context of the LBM, $\Delta x=1$ and $\Delta t=1$ are chosen in the remainder of this article. 
Therefore, all quantities are denoted in the LBM unit system if not explicitly stated otherwise.
In all simulations, the LBM reference density $\rho_{0}=1$ and pressure $p_{0} = c_{s}^{2} \rho_{0} = 1/3$ were set.
The relaxation times $\tau$ or relaxation rates $\omega$ specified for the numerical experiments refer to the constant user-chosen values that are not yet adjusted by the Smagorinsky turbulence model.
\par

\subsection{Free-surface lattice Boltzmann method}\label{sec:nm-fslbm}
The free-surface lattice Boltzmann method used in this article is based on the approach from Körner et al.~\cite{korner2005LatticeBoltzmannModel}
It simulates a moving interface between two immiscible fluids, the heavier of which completely governs the flow dynamics of the system.
The immiscible two-fluid flow problem is therefore reduced to a single-fluid flow with a free boundary.
In practice, this simplification is valid if the densities and viscosities of the fluids differ substantially, such as in liquid--gas flow.
In the following, the heavier fluid is referred to as liquid, whereas the lighter fluid is referred to as gas.
\par

The interface between the liquid and the gas is treated as in the volume-of-fluid approach~\cite{hirt1981VolumeFluidVOF}, where each lattice cell gets assigned a fill level $\varphi (\boldsymbol{x}, t)$.
The fill level acts as an indicator describing the affiliation to one of the phases.
Cells can either be of liquid ($\varphi(\boldsymbol{x}, t)=1$), gas ($\varphi(\boldsymbol{x}, t)=0$), or interface type ($\varphi(\boldsymbol{x}, t) \in \left(0, 1\right)$).
The interface cells form a sharp and closed layer, which separates liquid and gas cells.
Interface and liquid cells are treated as regular LBM cells that contain PDFs and participate in the LBM collision~\eqref{eq:nm-lbm-collision} and streaming~\eqref{eq:nm-lbm-streaming}.
In contrast, agreeing with the free-surface assumption, gas cells neither contain PDFs nor participate in the LBM update.
\par

The liquid mass of each cell
\begin{equation}\label{eq:nm-fslbm-mass}
m\left(\boldsymbol{x}, t\right) = \varphi\left(\boldsymbol{x}, t\right) \rho\left(\boldsymbol{x}, t\right) \Delta x^{3}
\end{equation}
is determined by the cell's fill level $\varphi(\boldsymbol{x}, t)$, fluid density $\rho(\boldsymbol{x}, t)$, and volume $\Delta x^{3}$.
The mass flux between an interface cell and other cells is computed from the LBM streaming step via
\begin{equation}\label{eq:nm-fslbm-mass-flux}
\frac{\Delta m_{i}\left(\boldsymbol{x}, t\right)}{\Delta x^{3}} = 
\begin{cases}
0 & \boldsymbol{x} + \boldsymbol{c}_{i}\Delta t \in \text{gas} \\

f_{\overline{i}}^{\star}\left(\boldsymbol{x} + \boldsymbol{c}_{i}\Delta t, t\right) - f_{i}^{\star}\left(\boldsymbol{x}, t\right) & \boldsymbol{x} + \boldsymbol{c}_{i}\Delta t \in \text{liquid}\\

\frac{1}{2}\Bigl(\varphi\left(\boldsymbol{x}, t\right) + \varphi\left(\boldsymbol{x} + \boldsymbol{c}_{i}\Delta t, t\right) \Bigr)
\Bigl(f_{\overline{i}}^{\star}\left(\boldsymbol{x} + \boldsymbol{c}_{i}\Delta t, t\right) - 
f_{i}^{\star}\left(\boldsymbol{x}, t\right)\Bigr) & \boldsymbol{x} + \boldsymbol{c}_{i}\Delta t \in \text{interface},
\end{cases}
\end{equation}
where $\bar{i}$ denotes the inversion of the lattice direction, leading to $\boldsymbol{c}_{\overline{i}} = -\boldsymbol{c}_{i}$.
Note that the fluid density $\rho(\boldsymbol{x}, t)$ is computed by the PDFs' zeroth-order moment~\eqref{eq:nm-lbm-density}, that is, by the PDFs' sum.
Therefore, the PDFs' unit is the same as the macroscopic fluid density's unit, making \Cref{eq:nm-fslbm-mass,eq:nm-fslbm-mass-flux} consistent.
\par

In the implementation used here, interface cells are not immediately converted to liquid or gas cells when they become full ($\varphi(\boldsymbol{x}, t)=1$) or empty ($\varphi(\boldsymbol{x}, t)=0$).
Instead, the heuristically chosen threshold $\varepsilon_{\varphi}=10^{-2}$ is used to prevent oscillatory conversions~\cite{pohl2008HighPerformanceSimulation}.
Therefore, an interface cell converts to liquid or gas if $\varphi(\boldsymbol{x}, t)>1+\varepsilon_{\varphi}$ or $\varphi(\boldsymbol{x}, t)< 0-\varepsilon_{\varphi}$.
During such conversions, surrounding gas or liquid cells must be converted to interface cells to maintain a closed interface layer.
It is important to note that neither liquid nor gas cells can directly convert into one another but only to interface cells.
In case of conflicting conversions, the separation of liquid and gas is prioritized.
When converting an interface cell with fill level $\varphi^{\text{conv}}(\boldsymbol{x}, t)$ to gas or liquid, the fill level is forcefully set to $\varphi(\boldsymbol{x}, t)=0$ or $\varphi(\boldsymbol{x}, t)=1$ to ensure consistency with the cell type definitions.
This manual modification of the fill level may lead to small amounts of excessive mass $m_{\text{ex}}\left(\boldsymbol{x}, t\right)$ with
\begin{equation}\label{eq:nm-fslbm-excess-mass}
\frac{m_{\text{ex}}\left(\boldsymbol{x}, t\right)}{\rho\left(\boldsymbol{x}, t\right) \Delta x^{3}} = 
\begin{cases}
\varphi^{\text{conv}}\left(\boldsymbol{x}, t\right) - 1 & \text{if } \boldsymbol{x} \text{ is converted to liquid} \\
\varphi^{\text{conv}}\left(\boldsymbol{x}, t\right) & \text{if } \boldsymbol{x} \text{ is converted to gas}.
\end{cases}
\end{equation}
This excess mass is distributed evenly among all interface cells in the neighborhood of the converted cell to conserve the system's total mass.
\par

There may appear unnecessary interface cells without gas or liquid neighbors during a simulation.
In the implementation used in this study, these cells are forced to fill or empty by adjusting the mass flux~\eqref{eq:nm-fslbm-mass-flux}, as suggested by Thürey~\cite{thurey2007PhysicallyBasedAnimation}.
\par

When converting cells from interface to liquid or vice-versa, the PDFs of the cell are not modified.
In contrast, when converting interface cells to gas cells, the interface cells' PDFs are dropped.
However, no valid PDFs are available when converting gas cells to interface cells.
The PDFs of these cells are initialized with their equilibrium~\eqref{eq:nm-lbm-equilibrium} with $\rho(\boldsymbol{x}, t)$ and $\boldsymbol{u}(\boldsymbol{x}, t)$ averaged from all surrounding liquid and non-newly created interface cells.
\par

The LBM collision~\eqref{eq:nm-lbm-collision} and streaming~\eqref{eq:nm-lbm-streaming} is performed in all interface and liquid cells.
As opposed to Körner et al.~\cite{korner2005LatticeBoltzmannModel}, following other authors~\cite{bogner2017DirectNumericalSimulation,pohl2008HighPerformanceSimulation,donath2011WettingModelsParallel}, the gravitational force is not weighted with an interface cell's fill level in the LBM collision in this article.
\par

The macroscopic boundary condition at the free surface is given by~\cite{scardovelli1999DirectNumericalSimulation,bogner2017DirectNumericalSimulation}
\begin{equation}\label{eq:nm-fslbm-boundary-condition-macroscopic}
\begin{aligned}
p\left(\boldsymbol{x}, t\right) - p^{\text{G}}\left(\boldsymbol{x}, t\right) + p^{\text{L}}\left(\boldsymbol{x}, t\right) &=  2\mu\partial_{n}u_{n}\left(\boldsymbol{x}, t\right)\\
0 &= \partial_{t_{1}}u_{n}\left(\boldsymbol{x}, t\right) + \partial_{n}u_{t_{1}}\left(\boldsymbol{x}, t\right)\\
0 &= \partial_{t_{2}}u_{n}\left(\boldsymbol{x}, t\right) + \partial_{n}u_{t_{2}}\left(\boldsymbol{x}, t\right).
\end{aligned}
\end{equation}
It includes the gas pressure $p^{\text{G}}\left(\boldsymbol{x}, t\right)$, Laplace pressure $p^{\text{L}}\left(\boldsymbol{x}, t\right)$, tangent vectors $\boldsymbol{t}_{1}(\boldsymbol{x}, t) \in \mathbb{R}^{d}$ and $\boldsymbol{t}_{2}(\boldsymbol{x}, t) \in \mathbb{R}^{d}$, and normal vector $\boldsymbol{n}(\boldsymbol{x}, t)$.
Körner et al.~\cite{korner2005LatticeBoltzmannModel} suggested to use 
the LBM anti-bounce-back pressure boundary condition
\begin{equation}\label{eq:nm-fslbm-boundary-condition}
f_{i}^{\star}\left(\boldsymbol{x} - \boldsymbol{c}_{i}\Delta t, t\right)
= f_{i}^\text{eq}\left(\rho^{\text{G}}, \boldsymbol{u}\right)
+ f_{\overline{i}}^\text{eq}\left(\rho^{\text{G}}, \boldsymbol{u}\right)
- f_{\overline{i}}^{\star}\left(\boldsymbol{x}, t\right) % \quad \forall i: \boldsymbol{x} - \boldsymbol{c}_{i} \in \text{gas}
\end{equation} 
at the free interface, with the interface cell's velocity $\boldsymbol{u} \equiv \boldsymbol{u}\left(\boldsymbol{x}, t\right)$ and gas density $\rho^{\text{G}} \equiv \rho^{\text{G}}\left(\boldsymbol{x}, t\right)=p^{\text{G}}\left(\boldsymbol{x}, t\right)/c_{s}^{2}$.
Other formulations of the boundary condition have been investigated in the literature~\cite{thies2005LatticeBoltzmannModeling,bogner2015BoundaryConditionsFree}.
The free-surface boundary condition~\eqref{eq:nm-fslbm-boundary-condition} must be applied to all PDFs streaming from gas cells to interface cells as they are unavailable.
However, Körner et al.\cite{korner2005LatticeBoltzmannModel} proposed to reconstruct not only missing PDFs but also available PDFs.
Consequently, this approach drops existing flow-field information.
The theoretical justification and evaluation for this suggestion are discussed in close detail in \Cref{sec:fsbcv}.
Its implications are investigated in the numerical experiments in \Cref{sec:ne}.
At free-slip boundaries, the free-surface boundary condition~\eqref{eq:nm-fslbm-boundary-condition} must also be applied to specularly reflected PDFs originating from gas cells.
\par

The gas pressure
\begin{equation}\label{eq:nm-fslbm-pressure-gas}
p^{\text{G}}\left(\boldsymbol{x}, t\right) = p^{\text{V}}\left(t\right) - p^{\text{L}}\left(\boldsymbol{x}, t\right)
\end{equation}
incorporates the volume pressure $p^{\text{V}}(t)$ and the Laplace pressure $p^{\text{L}}(\boldsymbol{x}, t)$.
The volume pressure can be assumed constant in case of atmospheric pressure or result from changes in the volume of an enclosed gas volume, that is, bubble, according to
\begin{equation}\label{eq:nm-fslbm-pressure-volume}
p^{\text{V}}\left(t\right)=p^{\text{V}}\left(0\right) \frac{V\left(0\right)}{V\left(t\right)}.
\end{equation}
The Laplace pressure
\begin{equation}\label{eq:nm-fslbm-pressure-laplace}
p^{\text{L}}\left(\boldsymbol{x}, t\right)=2 \sigma \kappa\left(\boldsymbol{x}, t\right)
\end{equation}
is determined by the surface tension $\sigma \in \mathbb{R^{+}}$ and the interface curvature $\kappa(\boldsymbol{x}, t) \in \mathbb{R}$.
In the simulations shown in this article, the interface curvature
\begin{equation}
	\kappa(\boldsymbol{x}, t) =-\nabla \cdot \boldsymbol{\hat{n}}(\boldsymbol{x}, t)
\end{equation}
was computed using the finite difference method (FDM) following Bogner et al.~\cite{bogner2016CurvatureEstimationVolumeoffluid}
The normalized interface normal $\boldsymbol{\hat{n}}(\boldsymbol{x}, t) =  \boldsymbol{n}(\boldsymbol{x}, t) / |\boldsymbol{n}(\boldsymbol{x}, t)|$ was obtained with a weighted central FDM according to Parker and Youngs~\cite{parker1992TwoThreeDimensional} of
\begin{equation} \label{eq:nm-fslbm-normal}
	\boldsymbol{n}(\boldsymbol{x}, t) = \nabla \varphi(\boldsymbol{x}, t).
\end{equation}
The computation of $\boldsymbol{n}(\boldsymbol{x}, t) \in \mathbb{R}^{d}$ was modified near near solid obstacle cells according to Donath~\cite{donath2011WettingModelsParallel} so that the FDM's access pattern did not include obstacle cells.
The curvature $\kappa(\boldsymbol{x}, t)$ is effectively computed from a second-order derivative of the fill level $\varphi(\boldsymbol{x}, t)$.
Since $\varphi(\boldsymbol{x}, t)$ is a non-smooth indicator function, taking its second-order derivative introduces large errors.
To reduce these errors, the fill level as used in the normal computation~\eqref{eq:nm-fslbm-normal}, is smoothed using the \textit{K\textsubscript{8}}-Kernel from Reference~\cite{williams1999AccuracyConvergenceContinuum} with a support radius of 2.0.
A more detailed description of these steps, and a comparison with other curvature computation models is available in the work of Bogner et al.~\cite{bogner2016CurvatureEstimationVolumeoffluid}
A bubble model algorithm is used to track the volume pressure of bubbles during coalescence or segmentation~\cite{pohl2008HighPerformanceSimulation,anderl2014FreeSurfaceLattice}.
\par

%% file: src/boundary-condition.tex
%!TEX root = ../main.tex

\section{Free-surface boundary condition variants}\label{sec:fsbcv}
As mentioned in the preceding section, gas cells do not contain PDFs.
Therefore, PDFs propagating from gas to interface cells must be reconstructed in the LBM streaming step.
The reconstruction must satisfy the free-surface boundary condition~\eqref{eq:nm-fslbm-boundary-condition-macroscopic}, balancing the forces exerted by the liquid and gas pressure.
Following Körner et al.~\cite{korner2005LatticeBoltzmannModel}, the balance of the forces can be analyzed using an approach based on the momentum exchange method~\cite{ladd1994NumericalSimulationsParticulate,ladd1994NumericalSimulationsParticulatea}.
Assuming that the total force is determined by the PDFs streaming through the interface during one time step, the total force $\boldsymbol{F} \equiv \boldsymbol{F}(\boldsymbol{x}, t) \in \mathbb{R}^{d}$ exerted by the fluid on a surface element $\boldsymbol{n}(\boldsymbol{x}, t) \cdot A(\boldsymbol{x}, t)$ results from the momentum transported by the particles streaming through this element.
With the interface-normal $\boldsymbol{n}\equiv \boldsymbol{n}(\boldsymbol{x}, t) \in \mathbb{R}^{d}$ and surface area $A \equiv A(\boldsymbol{x}, t) \in \mathbb{R}$ the force is given by
\begin{equation}\label{eq:fsbcv-force-balance-general}
\frac{F_{\alpha}}{A}
=
-
n_{\beta} 
\left(
\sum_{i \in K}
f_{i}^{\star}(\boldsymbol{x},t)
(c_{i,\alpha} - u_{\alpha})(c_{i,\beta} - u_{\beta})
+
\sum_{i \in R}
f_{i}^{\star}(\boldsymbol{x} - \boldsymbol{c}_{i}\Delta t,t)
(c_{i,\alpha} - u_{\alpha})(c_{i,\beta} - u_{\beta})
\right).
\end{equation}
The macroscopic fluid velocity $\boldsymbol{u} \equiv \boldsymbol{u}(\boldsymbol{x}, t)$ is subtracted from the discrete lattice velocity $\boldsymbol{c}_{i}$ to satisfy Galilean invariance, making the analysis independent of the frame of reference.
Whereas PDFs contained in the first sum with $i \in K$ (\textit{keep}) are not modified, PDFs included in the second sum with $i \in R$ (\textit{reconstruct}) are reconstructed with the free-surface boundary condition~\eqref{eq:nm-fslbm-boundary-condition}.
\par

The forces at the interface are balanced if the force exerted from the liquid is equal to the force from the gas pressure $p^{\text{G}} \equiv p^{\text{G}}(\boldsymbol{x}, t) \in \mathbb{R}$, as denoted by
\begin{equation}
\frac{F_{\alpha}}{A}
\stackrel{!}{=}
-n_{\alpha} p^{\text{G}}.
\end{equation}
In the following four sections, the force balance for different definitions of $K$ and $R$ is computed.
The name of each section refers to the PDFs that are reconstructed in the respective variant.
\par

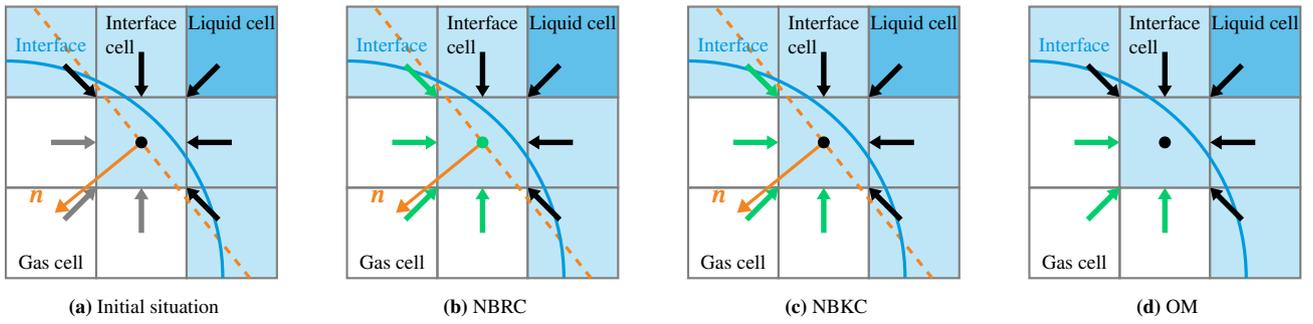
\begin{figure}[htbp]
	\centering
	\setlength{\figureheight}{0.2\textwidth}
	\setlength{\figurewidth}{0.2\textwidth}
	\begin{subfigure}[t]{0.245\textwidth}
		\centering
		\input{figures/reconstruction/initial.tex}%
		\caption{Initial situation}
		\label{fig:fsbcv-reconstruction-initial}
	\end{subfigure}
	\hfill
	\begin{subfigure}[t]{0.245\textwidth}
		\centering
		\input{figures/reconstruction/nbrc.tex}%
		\caption{NBRC}
		\label{fig:fsbcv-reconstruction-nbrc}
	\end{subfigure}
	\hfill
	\begin{subfigure}[t]{0.245\textwidth}
		\centering
		\input{figures/reconstruction/nbkc.tex}%
		\caption{NBKC}
		\label{fig:fsbcv-reconstruction-nbkc}
	\end{subfigure}
	\hfill
	\begin{subfigure}[t]{0.245\textwidth}
		\centering
		\input{figures/reconstruction/om.tex}%
		\caption{OM}
		\label{fig:fsbcv-reconstruction-om}
	\end{subfigure}
	\caption{\label{fig:fsbcv-reconstruction}
		PDFs, visualized as arrows, stream into neighboring lattice cells (\subref{fig:fsbcv-reconstruction-initial}) in the LBM streaming step.
		PDFs originating in liquid or interface cells are already available and displayed in black, whereas PDFs coming from gas cells must be reconstructed and are marked in gray.
		The PDFs reconstructed by the NBRC (\subref{fig:fsbcv-reconstruction-nbrc}), NBKC (\subref{fig:fsbcv-reconstruction-nbkc}) and OM (\subref{fig:fsbcv-reconstruction-om}) variants are colored in green.
		In the OM3 variant, at least three PDFs must be reconstructed.
		In the example here, this case is identical to the OM variant.
	}
\end{figure}

\subsection{Normal-based, reconstruct center (NBRC)}\label{sec:fsbcv-nbrc}
In the variant suggested in the original FSLBM model by Körner et al.~\cite{korner2005LatticeBoltzmannModel}, the PDFs are reconstructed based on the orientation of the interface-normal $\boldsymbol{n}$ with
\begin{equation}\label{eq:fsbcv-nbrc}
\begin{aligned}
K \coloneqq & \{i | \boldsymbol{n} \cdot \boldsymbol{c}_{i} < 0\}\\
R \coloneqq & \{i | \boldsymbol{n} \cdot \boldsymbol{c}_{i} \geq 0\}.
\end{aligned}
\end{equation}
The authors~\cite{korner2005LatticeBoltzmannModel} did not explicitly specify whether the central PDF $f_{0}$ must be reconstructed.
However, formally, $\boldsymbol{c}_{0} = 0$, so that it is included in $R$, the set of PDFs to be reconstructed.
Therefore, this variant is referred to as \textit{normal-based, reconstruct center} (NBRC) in this article.
\par

Note that although $\boldsymbol{c}_{0}=0$, the respective summand in the force-balance equation is not generally zero because of the subtraction with the velocity $\boldsymbol{u}$.
As illustrated in \Cref{fig:fsbcv-reconstruction-nbrc}, the NBRC variant overwrites existing PDFs, that is, it drops available information.
Körner et al.~\cite{korner2005LatticeBoltzmannModel} argue that this is required to maintain the balance of forces at the interface.
However, the central PDF is ignored in the force-balance computation in their article~\cite{korner2005LatticeBoltzmannModel}.
Therefore, the forces are not generally balanced. 
The corrected force balance is given by
\begin{equation}\label{eq:fsbcv-nbrc-force-balance}
\begin{split}
\frac{F_{\alpha}}{A}
=
&-
n_{\beta}
\left(
\sum_{i \in \{i | \boldsymbol{n} \cdot \boldsymbol{c}_{i} < 0\}}
f_{i}^{\star}
(\boldsymbol{x}, t)(c_{i,\alpha} - u_{\alpha})(c_{i,\beta} - u_{\beta})
+
\sum_{i \in \{i | \boldsymbol{n} \cdot \boldsymbol{c}_{i} \geq 0\}}
f_{i}^{\star}
(\boldsymbol{x} - \boldsymbol{c}_{i}\Delta t, t)(c_{i,\alpha} - u_{\alpha})(c_{i,\beta} - u_{\beta})
\right)
\\
=
&- n_{\alpha}
p^{\text{G}}
\\
& 
\mathcolor{error-term-red}{
+ n_{\beta}
\Bigl(
f_{0}^{\star}(\boldsymbol{x}, t)
-
f_{0}^{\text{eq}}(\rho^{\text{G}},\boldsymbol{u})
\Bigr)
u_{\alpha}u_{\beta}
+ 2 n_{\beta}
\sum_{i \in \{i | \boldsymbol{n} \cdot \boldsymbol{c}_{i} < 0\}}
\Bigl(
f_{i}^{\star}(\boldsymbol{x},t)
-
f_{i}^{\text{eq}}(\rho^{\text{G}},\boldsymbol{u})
\Bigr)
(
c_{i,\alpha}u_{\beta}
+
c_{i,\beta}u_{\alpha}
)
},
\end{split}
\end{equation}
where terms marked in orange are deviations from the desired balance of the forces.
This result shows that no general estimate of the error can be made.
The error depends on several non-constant quantities changing in time $t$ and location $\boldsymbol{x}$ in the simulation.
These quantities include the macroscopic velocity $\boldsymbol{u}(\boldsymbol{x}, t)$, the interface-normal $\boldsymbol{n}(\boldsymbol{x}, t)$, the gas density $\rho^{\text{G}}(\boldsymbol{x}, t)$, and the specific values of the post-collision PDFs $f_{i}^{\star}(\boldsymbol{x}, t)$.
The forces at the interface are only guaranteed to be balanced if the interface cell is at a steady-state with $\boldsymbol{u}(\boldsymbol{x}, t)=\boldsymbol{0}$.
\par

A step-by-step force-balance computation leading to the result in \Cref{eq:fsbcv-nbrc-force-balance} is available in \Cref{app:fbc-nbrc}.
\par

\subsection{Normal-based, keep center (NBKC)}\label{sec:fsbcv-nbkc}
In the \textit{normal-based, keep-center} (NBKC) variant, visualized in \Cref{fig:fsbcv-reconstruction-nbkc}, the central PDF $f_{0}$ is not modified.
The sets $K$ and $R$ are then defined by
\begin{equation}\label{eq:fsbcv-nbkc}
\begin{aligned}
K \coloneqq & \{i | \boldsymbol{n} \cdot \boldsymbol{c}_{i} < 0\}\\
R \coloneqq & \{i | \boldsymbol{n} \cdot \boldsymbol{c}_{i} > 0\}.
\end{aligned}
\end{equation}
The force-balance computation gives
\begin{equation}\label{eq:fsbcv-nbkc-force-balance}
\begin{split}
\frac{F_{\alpha}}{A}
=
&-
n_{\beta}
\left(
\sum_{i \in \{i | \boldsymbol{n} \cdot \boldsymbol{c}_{i} < 0\}}
f_{i}^{\star}
(\boldsymbol{x}, t)(c_{i,\alpha} - u_{\alpha})(c_{i,\beta} - u_{\beta})
+
\sum_{i \in \{i | \boldsymbol{n} \cdot \boldsymbol{c}_{i} > 0\}}
f_{i}^{\star}
(\boldsymbol{x} - \boldsymbol{c}_{i}\Delta t, t)(c_{i,\alpha} - u_{\alpha})(c_{i,\beta} - u_{\beta})
\right)
\\
=
&- n_{\alpha} p^{\text{G}}
\\
&\mathcolor{error-term-red}{
+
n_{\beta}
f_{0}^{\text{eq}}(\rho^{\text{G}},\boldsymbol{u})
u_{\alpha}u_{\beta}
+2 n_{\beta}
\sum_{i \in \{i | \boldsymbol{n} \cdot \boldsymbol{c}_{i} < 0\}}
\Bigl(
f_{i}^{\star}(\boldsymbol{x}, t)
-
f_{i}^{\text{eq}}(\rho^{\text{G}},\boldsymbol{u})
\Bigr)
(
c_{i,\alpha}u_{\beta}
+
c_{i,\beta}u_{\alpha}
)
}.
\end{split}
\end{equation}
where the terms marked in orange disturb the balance of the forces.
As for the NBRC in \Cref{sec:fsbcv-nbrc}, no general estimate about the error in the force balance can be made.
The sole exception are steady-state interface cells with $\boldsymbol{u}(\boldsymbol{x}, t)=\boldsymbol{0}$, where the forces are guaranteed to be balanced.
\par

\Cref{app:fbc-nbkc} presents the detailed force-balance computation that led to the result in \Cref{eq:fsbcv-nbkc-force-balance}.
\par

\subsection{Only missing (OM)}\label{sec:fsbcv-om}
In the literature, authors have noticed issues when reconstructing missing PDFs based on the orientation of the interface-normal as proposed by Körner et al~\cite{korner2005LatticeBoltzmannModel}.
These issues include anisotropic artifacts~\cite{bogner2016CurvatureEstimationVolumeoffluid} and the inability to model water wave propagation accurately~\cite{zhao2013LatticeBoltzmannMethod}.
However, neither reference explicitly specifies if the center PDF was reconstructed.
Therefore, it remains unclear whether these issues were observed with the NBRC or NBKC variant.
\par

The normal-based variants overwrite existing PDFs and do not generally balance the forces at the interface.
Therefore, an obvious alternative is to reconstruct only missing PDFs (OM), as in \Cref{fig:fsbcv-reconstruction-om}, without discarding any flow field information.
While this variant's balance of forces has not been analyzed theoretically in the literature, several authors~\cite{janssen2011FreeSurfaceFlow,bogner2016CurvatureEstimationVolumeoffluid,thorimbert2021ImplementationLatticeBoltzmann,biscarini2011ApplicationLatticeBoltzmann,cubeddu2017SimulationsBubbleGrowth,zhao2017LBMLESSimulationTransient,chiappini2018AnalysisFluidMotion,bublik2021ExperimentalValidationNumerical,huang2021ThreedimensionalSimulationReservoir} have used it without further reasoning.
\par

The sets $K$ and $R$ are defined by
\begin{equation}\label{eq:fsbcv-om-sets}
\begin{split}
K \equiv N^{-} &\coloneqq \{i|\boldsymbol{x} - \boldsymbol{c}_{i}\Delta t \in \text{non-gas}\}\\
N^{+} &\coloneqq \{i|\boldsymbol{x} + \boldsymbol{c}_{i}\Delta t \in \text{non-gas}\}\\
R \equiv G^{-} &\coloneqq \{i|\boldsymbol{x} - \boldsymbol{c}_{i}\Delta t \in \text{gas}\}\\
G^{+} &\coloneqq \{i|\boldsymbol{x} + \boldsymbol{c}_{i}\Delta t \in \text{gas}\}\\
T &\coloneqq \{0,1,\dots,q-1\}
\end{split}
\end{equation}
with the set $T$ (\textit{total}) containing all of a cell's PDFs.
The expression $\boldsymbol{x} - \boldsymbol{c}_{i}\Delta t$ denotes that a PDF streams from a neighboring cell to the current cell.
In contrast, a PDF with $\boldsymbol{x} + \boldsymbol{c}_{i}\Delta t$ streams from the current cell to a neighboring cell.
Therefore, conforming with the free-surface boundary condition~\eqref{eq:nm-fslbm-boundary-condition} and general force-balance computation~\eqref{eq:fsbcv-force-balance-general}, only PDFs streaming from a neighboring gas cell to this cell with $i \in G^{-}$ are reconstructed.
Note that the sets are related by
\begin{equation}\label{eq:fsbcv-om-set-relations}
\begin{split}
i \in N^{-} \quad &\Leftrightarrow \quad \bar{i} \in N^{+}\\
i \in G^{-} \quad &\Leftrightarrow \quad \bar{i} \in G^{+}.
\end{split}
\end{equation}
An index $i$ is not exclusively in only one set but can be part of $N^{-}$ and $N^{+}$, or $G^{-}$ and $G^{+}$.
The central PDF $f_{0}$ with $i=0$ belongs to the interface cell itself and will always be a non-gas cell with $i=0 \in N^{-}$ and $i=0 \in N^{+}$ such that
\begin{align}
G^{-} \cup N^{-} = T\\
G^{+} \cup N^{+} = T.
\end{align}
The force balance is then given by
\begin{equation}\label{eq:fsbcv-om-force-balance}
\begin{split}
\frac{F_{\alpha}}{A}
=
&-
n_{\beta}
\left(
\sum_{i \in N^{-}}
f_{i}^{\star}
(\boldsymbol{x}, t)(c_{i,\alpha} - u_{\alpha})(c_{i,\beta} - u_{\beta})
+
\sum_{i \in G^{-}}
f_{i}^{\star}
(\boldsymbol{x} - \boldsymbol{c}_{i}\Delta t, t)(c_{i,\alpha} - u_{\alpha})(c_{i,\beta} - u_{\beta})
\right)
\\
=
& -n_{\alpha} p^{\text{G}}
\\
&
\mathcolor{error-term-red}{
- n_{\beta}
\sum_{i \in N^{-}}
f_{i}^{\star}(\boldsymbol{x}, t)(c_{i,\alpha}c_{i,\beta} - c_{i,\alpha}u_{\beta} - c_{i,\beta}u_{\alpha} + u_{\alpha}u_{\beta})
+ n_{\beta}
\sum_{i \in G^{+}}
f_{i}^{\star}(\boldsymbol{x}, t)(c_{i,\alpha}c_{i,\beta} + c_{i,\alpha}u_{\beta} + c_{i,\beta}u_{\alpha} + u_{\alpha}u_{\beta})
}
\\
&
\mathcolor{error-term-red}{
+ n_{\beta}
\sum_{i \in T \setminus G^{-} \cup G^{+}} f_{i}^{\text{eq}}(\rho^{\text{G}},\boldsymbol{u})
(c_{i,\alpha}c_{i,\beta} - c_{i,\alpha}u_{\beta} - c_{i,\beta}u_{\alpha} + u_{\alpha}u_{\beta} )
- 2 n_{\beta}
\sum_{i \in G^{+}}
f_{i}^{\text{eq}}(\rho^{\text{G}},\boldsymbol{u})(c_{i,\alpha}u_{\beta} + c_{i,\beta}u_{\alpha})
},
\end{split}
\end{equation}
where the terms disturbing the force balance are marked in orange.
As for the NBRC and NBKC variant, it is impossible to predict the error made in the force-balance computation generally.
However, in contrast to the normal-based variants, the OM variant does not guarantee the balance of the forces at a steady-state interface.
\par

The detailed step-by-step force-balance computation is available in \Cref{app:fbc-om}.

\subsection{Only missing but at least three (OM3)}\label{sec:fsbcv-om3}
Bogner~\cite{bogner2017DirectNumericalSimulation} and Thies~\cite{thies2005LatticeBoltzmannModeling} have argued that it is not sufficient to only reconstruct missing PDFs, but it has to be ensured that at least three PDFs are reconstructed (OM3) with the free-surface boundary condition~\eqref{eq:nm-fslbm-boundary-condition}.
Otherwise, the macroscopic boundary condition at the free interface~\eqref{eq:nm-fslbm-boundary-condition-macroscopic} would be underdetermined.
However, neither of these references provides a rigorous mathematical proof for this statement.
The authors suggest using the variant from Körner et al.~\cite{korner2005LatticeBoltzmannModel} as a fallback in case less than three PDFs are missing in a cell.
In this study, this fourth variant will also be investigated numerically using the NBKC variant as fallback for such cases.
Formally, the sets $K$ and $R$ are then defined as
\begin{equation}\label{eq:fsbcv-om3}
\begin{aligned}
K \coloneqq &
\begin{cases}
\{i | \boldsymbol{x} - \boldsymbol{c}_{i}\Delta t \in \text{non-gas}\} & \text{ if} \quad |\{i | \boldsymbol{x} - \boldsymbol{c}_{i}\Delta t \in \text{gas}\}| \geq 3 \\
\{i | \boldsymbol{n} \cdot \boldsymbol{c}_{i} < 0\} & \text{ else}
\end{cases}\\
R \coloneqq &
\begin{cases}
\mathrlap{\{i | \boldsymbol{x} - \boldsymbol{c}_{i}\Delta t \in \text{gas}\}}\hphantom{\{i | \boldsymbol{x} - \boldsymbol{c}_{i}\Delta t \in \text{non-gas}\}} & \text{ if} \quad |\{i | \boldsymbol{x} - \boldsymbol{c}_{i}\Delta t \in \text{gas}\}| \geq 3 \\
\{i | \boldsymbol{n} \cdot \boldsymbol{c}_{i} \geq 0\} & \text{ else}.
\end{cases}
\end{aligned}
\end{equation}
Depending on the number of missing PDFs in the individual interface cell, the resulting force balance will be either similar to the NBKC variant with \Cref{eq:fsbcv-nbkc-force-balance} or OM variant in \Cref{eq:fsbcv-om-force-balance}.
Consequently, no general error estimate can be made here.
\par

%% file: figures/reconstruction/initial.tex
\begin{tikzpicture}
\definecolor{dodgerblue0154222}{RGB}{0,154,222}
\definecolor{myblue0}{RGB}{96,192,234}
\definecolor{myblue1}{RGB}{191,230,247}

\definecolor{darkorange24213334}{RGB}{242,133,34}

% gas cells
\node[
rectangle,
anchor=center,
minimum width=1/3*\figurewidth,
minimum height=1/3*\figureheight,
draw=black!50,
line width=0.8pt
] (r) at (1/6*\figurewidth,1/6*\figureheight) {};

\node[
rectangle,
anchor=center,
minimum width=1/3*\figurewidth,
minimum height=1/3*\figureheight,
draw=black!50,
line width=0.8pt,
] (r) at (1/6*\figurewidth,3/6*\figureheight) {};

\node[
rectangle,
anchor=center,
minimum width=1/3*\figurewidth,
minimum height=1/3*\figureheight,
draw=black!50,
line width=0.8pt,
] (r) at (3/6*\figurewidth,1/6*\figureheight) {};

\node[
rectangle,
anchor=south,
font=\footnotesize
] (r) at (1/6*\figurewidth,0) {Gas cell};

% interface cells
\node[
rectangle,
anchor=center,
minimum width=1/3*\figurewidth,
minimum height=1/3*\figureheight,
draw=black!50,
line width=0.8pt,
fill=myblue1
] (r) at (3/6*\figurewidth,3/6*\figureheight) {};

\node[
rectangle,
anchor=center,
minimum width=1/3*\figurewidth,
minimum height=1/3*\figureheight,
draw=black!50,
line width=0.8pt,
fill=myblue1
] (r) at (3/6*\figurewidth,5/6*\figureheight) {};

\node[
rectangle,
anchor=center,
minimum width=1/3*\figurewidth,
minimum height=1/3*\figureheight,
draw=black!50,
line width=0.8pt,
fill=myblue1
] (r) at (5/6*\figurewidth,3/6*\figureheight) {};

\node[
rectangle,
anchor=center,
minimum width=1/3*\figurewidth,
minimum height=1/3*\figureheight,
draw=black!50,
line width=0.8pt,
fill=myblue1
] (r) at (5/6*\figurewidth,1/6*\figureheight) {};

\node[
rectangle,
anchor=center,
minimum width=1/3*\figurewidth,
minimum height=1/3*\figureheight,
draw=black!50,
line width=0.8pt,
fill=myblue1
] (r) at (1/6*\figurewidth,5/6*\figureheight) {};

\node[
rectangle,
anchor=north,
align=left,
font=\footnotesize
] (r) at (0.5*\figurewidth,\figureheight) {Interface\\ cell};

% liquid cell
\node[
rectangle,
anchor=center,
minimum width=1/3*\figurewidth,
minimum height=1/3*\figureheight,
draw=black!50,
line width=0.8pt,
fill=myblue0
] (r) at (5/6*\figurewidth,5/6*\figureheight) {};

\node[
rectangle,
anchor=north,
font=\footnotesize
] (r) at (5/6*\figurewidth,\figureheight) {Liquid cell};

%interface normal
\draw[very thick, dashed, darkorange24213334] (0.9\figurewidth,0)--(0.1\figurewidth,\figureheight);
\draw[very thick, darkorange24213334, ->, -{Triangle[scale=0.8]}] (0.5\figurewidth,0.5\figureheight)--(0.18\figurewidth,0.24\figureheight) node [pos=1,above left] {$\boldsymbol{n}$};

% interface
\node[
dodgerblue0154222,
rectangle,
anchor=south west,
font=\footnotesize
] (r) at (0, 0.8*\figureheight) {Interface};

\draw [very thick,dodgerblue0154222](0.8*\figurewidth,0) arc [start angle=0, delta angle=90, radius=0.8*\figureheight];

% PDFs
\draw[black,fill=black] (0.5*\figurewidth,0.5*\figureheight) circle [radius=0.02\figurewidth];
\draw[line width=0.02*\figurewidth,->, -{Triangle[scale=0.5]}] (5/6*\figurewidth,0.5*\figureheight)--(4/6*\figurewidth,0.5*\figureheight) node {};
\draw[line width=0.02*\figurewidth,->, -{Triangle[scale=0.5]}] (4/6*\figurewidth+0.11785*\figurewidth,4/6*\figureheight+0.11785*\figureheight)--(4/6*\figurewidth,4/6*\figureheight) node {};
\draw[line width=0.02*\figurewidth,->, -{Triangle[scale=0.5]}] (0.5*\figurewidth,5/6*\figureheight)--(0.5*\figurewidth,4/6*\figureheight) node {};
\draw[line width=0.02*\figurewidth,->, -{Triangle[scale=0.5]}] (2/6*\figurewidth-0.11785*\figurewidth,4/6*\figureheight+0.11785*\figureheight)--(2/6*\figurewidth,4/6*\figureheight) node {};
\draw[black!50, line width=0.02*\figurewidth,->, -{Triangle[scale=0.5]}] (1/6*\figurewidth,0.5*\figureheight) -- (2/6*\figurewidth,0.5*\figureheight) node {};
\draw[black!50, line width=0.02*\figurewidth,->, -{Triangle[scale=0.5]}] (2/6*\figurewidth-0.11785*\figurewidth,2/6*\figureheight-0.11785*\figureheight)--(2/6*\figurewidth,2/6*\figureheight) node {};
\draw[black!50,line width=0.02*\figurewidth,->, -{Triangle[scale=0.5]}] (0.5*\figurewidth,1/6*\figureheight)--(0.5*\figurewidth,2/6*\figureheight) node {};
\draw[line width=0.02*\figurewidth,->, -{Triangle[scale=0.5]}] (4/6*\figurewidth+0.11785*\figurewidth,2/6*\figureheight-0.11785*\figureheight)--(4/6*\figurewidth,2/6*\figureheight) node {};
\end{tikzpicture}

%% file: figures/reconstruction/nbrc.tex
\begin{tikzpicture}
\definecolor{dodgerblue0154222}{RGB}{0,154,222}
\definecolor{myblue0}{RGB}{96,192,234}
\definecolor{myblue1}{RGB}{191,230,247}

\definecolor{darkorange24213334}{RGB}{242,133,34}

\definecolor{springgreen0205108}{RGB}{0,205,108}

% gas cells
\node[
rectangle,
anchor=center,
minimum width=1/3*\figurewidth,
minimum height=1/3*\figureheight,
draw=black!50,
line width=0.8pt
] (r) at (1/6*\figurewidth,1/6*\figureheight) {};

\node[
rectangle,
anchor=center,
minimum width=1/3*\figurewidth,
minimum height=1/3*\figureheight,
draw=black!50,
line width=0.8pt,
] (r) at (1/6*\figurewidth,3/6*\figureheight) {};

\node[
rectangle,
anchor=center,
minimum width=1/3*\figurewidth,
minimum height=1/3*\figureheight,
draw=black!50,
line width=0.8pt,
] (r) at (3/6*\figurewidth,1/6*\figureheight) {};

\node[
rectangle,
anchor=south,
font=\footnotesize
] (r) at (1/6*\figurewidth,0) {Gas cell};

% interface cells
\node[
rectangle,
anchor=center,
minimum width=1/3*\figurewidth,
minimum height=1/3*\figureheight,
draw=black!50,
line width=0.8pt,
fill=myblue1
] (r) at (3/6*\figurewidth,3/6*\figureheight) {};

\node[
rectangle,
anchor=center,
minimum width=1/3*\figurewidth,
minimum height=1/3*\figureheight,
draw=black!50,
line width=0.8pt,
fill=myblue1
] (r) at (3/6*\figurewidth,5/6*\figureheight) {};

\node[
rectangle,
anchor=center,
minimum width=1/3*\figurewidth,
minimum height=1/3*\figureheight,
draw=black!50,
line width=0.8pt,
fill=myblue1
] (r) at (5/6*\figurewidth,3/6*\figureheight) {};

\node[
rectangle,
anchor=center,
minimum width=1/3*\figurewidth,
minimum height=1/3*\figureheight,
draw=black!50,
line width=0.8pt,
fill=myblue1
] (r) at (5/6*\figurewidth,1/6*\figureheight) {};

\node[
rectangle,
anchor=center,
minimum width=1/3*\figurewidth,
minimum height=1/3*\figureheight,
draw=black!50,
line width=0.8pt,
fill=myblue1
] (r) at (1/6*\figurewidth,5/6*\figureheight) {};

\node[
rectangle,
anchor=north,
align=left,
font=\footnotesize
] (r) at (0.5*\figurewidth,\figureheight) {Interface\\ cell};

% liquid cell
\node[
rectangle,
anchor=center,
minimum width=1/3*\figurewidth,
minimum height=1/3*\figureheight,
draw=black!50,
line width=0.8pt,
fill=myblue0
] (r) at (5/6*\figurewidth,5/6*\figureheight) {};

\node[
rectangle,
anchor=north,
font=\footnotesize
] (r) at (5/6*\figurewidth,\figureheight) {Liquid cell};

%interface normal
\draw[very thick, dashed, darkorange24213334] (0.9\figurewidth,0)--(0.1\figurewidth,\figureheight);
\draw[very thick, darkorange24213334, ->, -{Triangle[scale=0.8]}] (0.5\figurewidth,0.5\figureheight)--(0.18\figurewidth,0.24\figureheight) node [pos=1,above left] {$\boldsymbol{n}$};

% interface
\node[
dodgerblue0154222,
rectangle,
anchor=south west,
font=\footnotesize
] (r) at (0, 0.8*\figureheight) {Interface};

\draw [very thick,dodgerblue0154222](0.8*\figurewidth,0) arc [start angle=0, delta angle=90, radius=0.8*\figureheight];

% PDFs
\draw[springgreen0205108,fill=springgreen0205108] (0.5*\figurewidth,0.5*\figureheight) circle [radius=0.02\figurewidth];
\draw[line width=0.02*\figurewidth,->, -{Triangle[scale=0.5]}] (5/6*\figurewidth,0.5*\figureheight)--(4/6*\figurewidth,0.5*\figureheight) node {};
\draw[line width=0.02*\figurewidth,->, -{Triangle[scale=0.5]}] (4/6*\figurewidth+0.11785*\figurewidth,4/6*\figureheight+0.11785*\figureheight)--(4/6*\figurewidth,4/6*\figureheight) node {};
\draw[line width=0.02*\figurewidth,->, -{Triangle[scale=0.5]}] (0.5*\figurewidth,5/6*\figureheight)--(0.5*\figurewidth,4/6*\figureheight) node {};
\draw[springgreen0205108, line width=0.02*\figurewidth,->, -{Triangle[scale=0.5]}] (2/6*\figurewidth-0.11785*\figurewidth,4/6*\figureheight+0.11785*\figureheight)--(2/6*\figurewidth,4/6*\figureheight) node {};
\draw[springgreen0205108, line width=0.02*\figurewidth,->, -{Triangle[scale=0.5]}] (1/6*\figurewidth,0.5*\figureheight) -- (2/6*\figurewidth,0.5*\figureheight) node {};
\draw[springgreen0205108, line width=0.02*\figurewidth,->, -{Triangle[scale=0.5]}] (2/6*\figurewidth-0.11785*\figurewidth,2/6*\figureheight-0.11785*\figureheight)--(2/6*\figurewidth,2/6*\figureheight) node {};
\draw[springgreen0205108, line width=0.02*\figurewidth,->, -{Triangle[scale=0.5]}] (0.5*\figurewidth,1/6*\figureheight)--(0.5*\figurewidth,2/6*\figureheight) node {};
\draw[line width=0.02*\figurewidth,->, -{Triangle[scale=0.5]}] (4/6*\figurewidth+0.11785*\figurewidth,2/6*\figureheight-0.11785*\figureheight)--(4/6*\figurewidth,2/6*\figureheight) node {};
\end{tikzpicture}

%% file: figures/reconstruction/nbkc.tex
\begin{tikzpicture}
\definecolor{dodgerblue0154222}{RGB}{0,154,222}
\definecolor{myblue0}{RGB}{96,192,234}
\definecolor{myblue1}{RGB}{191,230,247}

\definecolor{darkorange24213334}{RGB}{242,133,34}

\definecolor{springgreen0205108}{RGB}{0,205,108}

% gas cells
\node[
rectangle,
anchor=center,
minimum width=1/3*\figurewidth,
minimum height=1/3*\figureheight,
draw=black!50,
line width=0.8pt
] (r) at (1/6*\figurewidth,1/6*\figureheight) {};

\node[
rectangle,
anchor=center,
minimum width=1/3*\figurewidth,
minimum height=1/3*\figureheight,
draw=black!50,
line width=0.8pt,
] (r) at (1/6*\figurewidth,3/6*\figureheight) {};

\node[
rectangle,
anchor=center,
minimum width=1/3*\figurewidth,
minimum height=1/3*\figureheight,
draw=black!50,
line width=0.8pt,
] (r) at (3/6*\figurewidth,1/6*\figureheight) {};

\node[
rectangle,
anchor=south,
font=\footnotesize
] (r) at (1/6*\figurewidth,0) {Gas cell};

% interface cells
\node[
rectangle,
anchor=center,
minimum width=1/3*\figurewidth,
minimum height=1/3*\figureheight,
draw=black!50,
line width=0.8pt,
fill=myblue1
] (r) at (3/6*\figurewidth,3/6*\figureheight) {};

\node[
rectangle,
anchor=center,
minimum width=1/3*\figurewidth,
minimum height=1/3*\figureheight,
draw=black!50,
line width=0.8pt,
fill=myblue1
] (r) at (3/6*\figurewidth,5/6*\figureheight) {};

\node[
rectangle,
anchor=center,
minimum width=1/3*\figurewidth,
minimum height=1/3*\figureheight,
draw=black!50,
line width=0.8pt,
fill=myblue1
] (r) at (5/6*\figurewidth,3/6*\figureheight) {};

\node[
rectangle,
anchor=center,
minimum width=1/3*\figurewidth,
minimum height=1/3*\figureheight,
draw=black!50,
line width=0.8pt,
fill=myblue1
] (r) at (5/6*\figurewidth,1/6*\figureheight) {};

\node[
rectangle,
anchor=center,
minimum width=1/3*\figurewidth,
minimum height=1/3*\figureheight,
draw=black!50,
line width=0.8pt,
fill=myblue1
] (r) at (1/6*\figurewidth,5/6*\figureheight) {};

\node[
rectangle,
anchor=north,
align=left,
font=\footnotesize
] (r) at (0.5*\figurewidth,\figureheight) {Interface\\ cell};

% liquid cell
\node[
rectangle,
anchor=center,
minimum width=1/3*\figurewidth,
minimum height=1/3*\figureheight,
draw=black!50,
line width=0.8pt,
fill=myblue0
] (r) at (5/6*\figurewidth,5/6*\figureheight) {};

\node[
rectangle,
anchor=north,
font=\footnotesize
] (r) at (5/6*\figurewidth,\figureheight) {Liquid cell};

%interface normal
\draw[very thick, dashed, darkorange24213334] (0.9\figurewidth,0)--(0.1\figurewidth,\figureheight);
\draw[very thick, darkorange24213334, ->, -{Triangle[scale=0.8]}] (0.5\figurewidth,0.5\figureheight)--(0.18\figurewidth,0.24\figureheight) node [pos=1,above left] {$\boldsymbol{n}$};

% interface
\node[
dodgerblue0154222,
rectangle,
anchor=south west,
font=\footnotesize
] (r) at (0, 0.8*\figureheight) {Interface};

\draw [very thick,dodgerblue0154222](0.8*\figurewidth,0) arc [start angle=0, delta angle=90, radius=0.8*\figureheight];

% PDFs
\draw[black,fill=black] (0.5*\figurewidth,0.5*\figureheight) circle [radius=0.02\figurewidth];
\draw[line width=0.02*\figurewidth,->, -{Triangle[scale=0.5]}] (5/6*\figurewidth,0.5*\figureheight)--(4/6*\figurewidth,0.5*\figureheight) node {};
\draw[line width=0.02*\figurewidth,->, -{Triangle[scale=0.5]}] (4/6*\figurewidth+0.11785*\figurewidth,4/6*\figureheight+0.11785*\figureheight)--(4/6*\figurewidth,4/6*\figureheight) node {};
\draw[line width=0.02*\figurewidth,->, -{Triangle[scale=0.5]}] (0.5*\figurewidth,5/6*\figureheight)--(0.5*\figurewidth,4/6*\figureheight) node {};
\draw[springgreen0205108, line width=0.02*\figurewidth,->, -{Triangle[scale=0.5]}] (2/6*\figurewidth-0.11785*\figurewidth,4/6*\figureheight+0.11785*\figureheight)--(2/6*\figurewidth,4/6*\figureheight) node {};
\draw[springgreen0205108, line width=0.02*\figurewidth,->, -{Triangle[scale=0.5]}] (1/6*\figurewidth,0.5*\figureheight) -- (2/6*\figurewidth,0.5*\figureheight) node {};
\draw[springgreen0205108, line width=0.02*\figurewidth,->, -{Triangle[scale=0.5]}] (2/6*\figurewidth-0.11785*\figurewidth,2/6*\figureheight-0.11785*\figureheight)--(2/6*\figurewidth,2/6*\figureheight) node {};
\draw[springgreen0205108, line width=0.02*\figurewidth,->, -{Triangle[scale=0.5]}] (0.5*\figurewidth,1/6*\figureheight)--(0.5*\figurewidth,2/6*\figureheight) node {};
\draw[line width=0.02*\figurewidth,->, -{Triangle[scale=0.5]}] (4/6*\figurewidth+0.11785*\figurewidth,2/6*\figureheight-0.11785*\figureheight)--(4/6*\figurewidth,2/6*\figureheight) node {};
\end{tikzpicture}

%% file: figures/reconstruction/om.tex
\begin{tikzpicture}
\definecolor{dodgerblue0154222}{RGB}{0,154,222}
\definecolor{myblue0}{RGB}{96,192,234}
\definecolor{myblue1}{RGB}{191,230,247}

\definecolor{darkorange24213334}{RGB}{242,133,34}

\definecolor{springgreen0205108}{RGB}{0,205,108}

% gas cells
\node[
rectangle,
anchor=center,
minimum width=1/3*\figurewidth,
minimum height=1/3*\figureheight,
draw=black!50,
line width=0.8pt
] (r) at (1/6*\figurewidth,1/6*\figureheight) {};

\node[
rectangle,
anchor=center,
minimum width=1/3*\figurewidth,
minimum height=1/3*\figureheight,
draw=black!50,
line width=0.8pt,
] (r) at (1/6*\figurewidth,3/6*\figureheight) {};

\node[
rectangle,
anchor=center,
minimum width=1/3*\figurewidth,
minimum height=1/3*\figureheight,
draw=black!50,
line width=0.8pt,
] (r) at (3/6*\figurewidth,1/6*\figureheight) {};

\node[
rectangle,
anchor=south,
font=\footnotesize
] (r) at (1/6*\figurewidth,0) {Gas cell};

% interface cells
\node[
rectangle,
anchor=center,
minimum width=1/3*\figurewidth,
minimum height=1/3*\figureheight,
draw=black!50,
line width=0.8pt,
fill=myblue1
] (r) at (3/6*\figurewidth,3/6*\figureheight) {};

\node[
rectangle,
anchor=center,
minimum width=1/3*\figurewidth,
minimum height=1/3*\figureheight,
draw=black!50,
line width=0.8pt,
fill=myblue1
] (r) at (3/6*\figurewidth,5/6*\figureheight) {};

\node[
rectangle,
anchor=center,
minimum width=1/3*\figurewidth,
minimum height=1/3*\figureheight,
draw=black!50,
line width=0.8pt,
fill=myblue1
] (r) at (5/6*\figurewidth,3/6*\figureheight) {};

\node[
rectangle,
anchor=center,
minimum width=1/3*\figurewidth,
minimum height=1/3*\figureheight,
draw=black!50,
line width=0.8pt,
fill=myblue1
] (r) at (5/6*\figurewidth,1/6*\figureheight) {};

\node[
rectangle,
anchor=center,
minimum width=1/3*\figurewidth,
minimum height=1/3*\figureheight,
draw=black!50,
line width=0.8pt,
fill=myblue1
] (r) at (1/6*\figurewidth,5/6*\figureheight) {};

\node[
rectangle,
anchor=north,
align=left,
font=\footnotesize
] (r) at (0.5*\figurewidth,\figureheight) {Interface\\ cell};

% liquid cell
\node[
rectangle,
anchor=center,
minimum width=1/3*\figurewidth,
minimum height=1/3*\figureheight,
draw=black!50,
line width=0.8pt,
fill=myblue0
] (r) at (5/6*\figurewidth,5/6*\figureheight) {};

\node[
rectangle,
anchor=north,
font=\footnotesize
] (r) at (5/6*\figurewidth,\figureheight) {Liquid cell};

%%interface normal
%\draw[very thick, dashed, darkorange24213334] (0.9\figurewidth,0)--(0.1\figurewidth,\figureheight);
%\draw[very thick, darkorange24213334, ->, -{Triangle[scale=0.8]}] (0.5\figurewidth,0.5\figureheight)--(0.18\figurewidth,0.24\figureheight) node [pos=1,above left] {$\boldsymbol{n}$};

% interface
\node[
dodgerblue0154222,
rectangle,
anchor=south west,
font=\footnotesize
] (r) at (0, 0.8*\figureheight) {Interface};

\draw [very thick,dodgerblue0154222](0.8*\figurewidth,0) arc [start angle=0, delta angle=90, radius=0.8*\figureheight];

% PDFs
\draw[black,fill=black] (0.5*\figurewidth,0.5*\figureheight) circle [radius=0.02\figurewidth];
\draw[line width=0.02*\figurewidth,->, -{Triangle[scale=0.5]}] (5/6*\figurewidth,0.5*\figureheight)--(4/6*\figurewidth,0.5*\figureheight) node {};
\draw[line width=0.02*\figurewidth,->, -{Triangle[scale=0.5]}] (4/6*\figurewidth+0.11785*\figurewidth,4/6*\figureheight+0.11785*\figureheight)--(4/6*\figurewidth,4/6*\figureheight) node {};
\draw[line width=0.02*\figurewidth,->, -{Triangle[scale=0.5]}] (0.5*\figurewidth,5/6*\figureheight)--(0.5*\figurewidth,4/6*\figureheight) node {};
\draw[black, line width=0.02*\figurewidth,->, -{Triangle[scale=0.5]}] (2/6*\figurewidth-0.11785*\figurewidth,4/6*\figureheight+0.11785*\figureheight)--(2/6*\figurewidth,4/6*\figureheight) node {};
\draw[springgreen0205108, line width=0.02*\figurewidth,->, -{Triangle[scale=0.5]}] (1/6*\figurewidth,0.5*\figureheight) -- (2/6*\figurewidth,0.5*\figureheight) node {};
\draw[springgreen0205108, line width=0.02*\figurewidth,->, -{Triangle[scale=0.5]}] (2/6*\figurewidth-0.11785*\figurewidth,2/6*\figureheight-0.11785*\figureheight)--(2/6*\figurewidth,2/6*\figureheight) node {};
\draw[springgreen0205108, line width=0.02*\figurewidth,->, -{Triangle[scale=0.5]}] (0.5*\figurewidth,1/6*\figureheight)--(0.5*\figurewidth,2/6*\figureheight) node {};
\draw[line width=0.02*\figurewidth,->, -{Triangle[scale=0.5]}] (4/6*\figurewidth+0.11785*\figurewidth,2/6*\figureheight-0.11785*\figureheight)--(4/6*\figurewidth,2/6*\figureheight) node {};
\end{tikzpicture}

%% file: src/numerical-experiments.tex
%!TEX root = ../main.tex

\section{Numerical experiments}\label{sec:ne}
This section compares the boundary condition variants introduced and analyzed in \Cref{sec:fsbcv} using five numerical experiments.
The chosen test cases are partly identical to the ones suggested in prior articles~\cite{schwarzmeier2022ComparisonFreeSurface,schwarzmeier2022ComparisonRefillingSchemes}.
They include the simulation of a standing gravity wave, the collapse of a rectangular and cylindrical liquid column, the rise of a Taylor bubble, and the impact of a drop into a thin film of liquid.
The description of the test cases, simulation setups, and figures are similar to those from References\cite{schwarzmeier2022ComparisonFreeSurface,schwarzmeier2022ComparisonRefillingSchemes} but are repeated here for completeness.
All simulations were performed with double-precision floating-point arithmetic.
\par

\subsection{Gravity wave}\label{sec:ne-gw}
A gravity wave is a standing wave with a phase boundary between two immiscible fluids.
The wave's flow dynamics are entirely governed by gravitational forces, whereas surface tension forces are neglected.
The simulation results were compared to the analytical model~\cite{dingemans1997WaterWavePropagation,lamb1975Hydrodynamics}.
\par

\subsubsection{Simulation setup}\label{sec:ne-gw-ss}
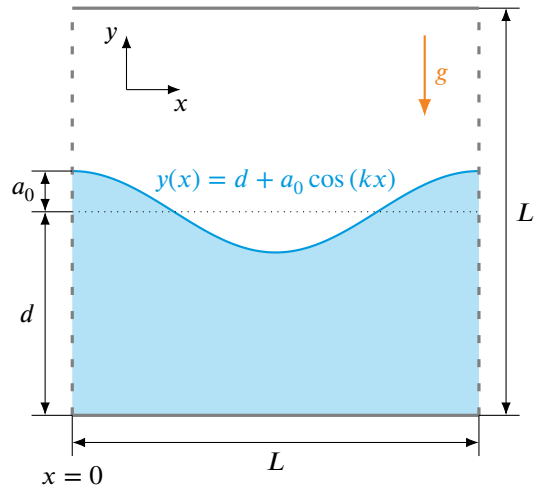
\begin{figure}[htbp]
	\centering
	\setlength{\figureheight}{0.3\textwidth}
	\setlength{\figurewidth}{0.3\textwidth}
	\setlength\mdist{0.02\textwidth}
	\input{figures/gravity-wave/setup.tex}%
	\caption{\label{fig:ne-gw-setup}
		Simulation setup of the two-dimensional gravity wave test case with wavelength $L$, liquid depth $d$, initial wave amplitude $a_{0}$, wavenumber $k=2\pi/L$, and gravitational acceleration $g$.
		No-slip boundary conditions were used at the domain walls in $y$-direction.
		The domain's side walls in the $x$-direction were periodic.
		C.\ Schwarzmeier, M.\ Holzer, T.\ Mitchell, M.\ Lehmann, F.\ Häusl, U.\ Rüde, Comparison of free surface and conservative Allen--Cahn phase field lattice Boltzmann method, arXiv preprint\cite{schwarzmeier2022ComparisonFreeSurface}, 2022; licensed under a Creative Commons Attribution (CC BY) license; the colors were changed from the original.
	}
\end{figure}

As illustrated in \Cref{fig:ne-gw-setup}, a gravity wave of wavelength $L$ was simulated in a two-dimensional quadratic domain of size $L\times L \times 1$ ($x$-, $y$-, $z$-direction) with $L \in \{200, 400, 800\}$ lattice cells.
The interface at the phase boundary was initialized with the profile given by $y(x) = d + a_{0} \cos\left(kx \right)$ with liquid depth $d=0.5L$, initial amplitude $a_{0}=0.01L$, and wavenumber $k=2\pi/L$.
In the $y$-direction, the domain was confined by walls with no-slip boundary conditions, whereas it was periodic in the $x$-direction.
The liquid was initialized with hydrostatic pressure according to the gravitational acceleration $g$, so the LBM pressure at $y=d$ was equal to the constant atmospheric volume pressure $p^{\text{V}}(t) = p_{0}$.
The relaxation rate $\omega=1.8$ was chosen and kept constant for any computational domain resolution to conform with diffusive scaling~\cite{kruger2017LatticeBoltzmannMethod}.
The system is characterized by the Reynolds number
\begin{equation} \label{eq:re-wave}
\mathrm{Re} \coloneqq \frac{a_{0}\omega_{0}L}{\nu} = 10,
\end{equation}
which is defined with the angular frequency of the wave
\begin{equation}
\omega_{0} = \sqrt{g k \, \mathrm{tanh} \left(k d \right)},
\end{equation}
and kinematic fluid viscosity $\nu$.
Because of the gravitational acceleration $g$, the initial profile evolved into a standing wave that oscillated around $d$.
It was dampened by viscous forces.
The non-dimensionalized surface elevation $a^{*}(x, t) = a(x, t)/a_{0}$ and non-dimensionalized time $t^{*} = t\omega_{0}$ were monitored at the left domain border, that is, at $x=0$ every $t^{*} = 0.01$.
The simulations were performed until $t^{*} = 40$, which was found to be sufficient for the wave's motion to be fully decayed.
\par

\subsubsection{Analytical model}\label{sec:ne-gw-am}
An analytical model for the gravity wave's motion is derived by linearizing the continuity and Euler equations with a free-surface boundary condition~\cite{dingemans1997WaterWavePropagation}.
The standing wave's amplitude
\begin{equation}
a(x,t) = a_{D}(t) \cos \left( kx - \omega_{0} t \right) + d,
\end{equation}
is obtained under the assumption of an inviscid fluid with zero damping $a_{D}(t)=a_{0}$.
Viscous damping is considered by~\cite{lamb1975Hydrodynamics}
\begin{equation}
a_{D}(t) = a_{0} \mathrm{e}^{-2 \nu k^{2} t}.
\end{equation}
The analytical model is applicable if $k |a_{0}| \ll 1$ and $k |a_{0}| \ll kd$~\cite{dingemans1997WaterWavePropagation}, which is true in this study with $k |a_{0}| = 0.02\pi \ll 1 < kd = \pi$.
\par

\subsubsection{Results and discussion}\label{sec:ne-gw-rad}
\begin{figure}[htbp]
	\centering
	\setlength{\figureheight}{0.4\textwidth}
	\setlength{\figurewidth}{1\textwidth}
	\input{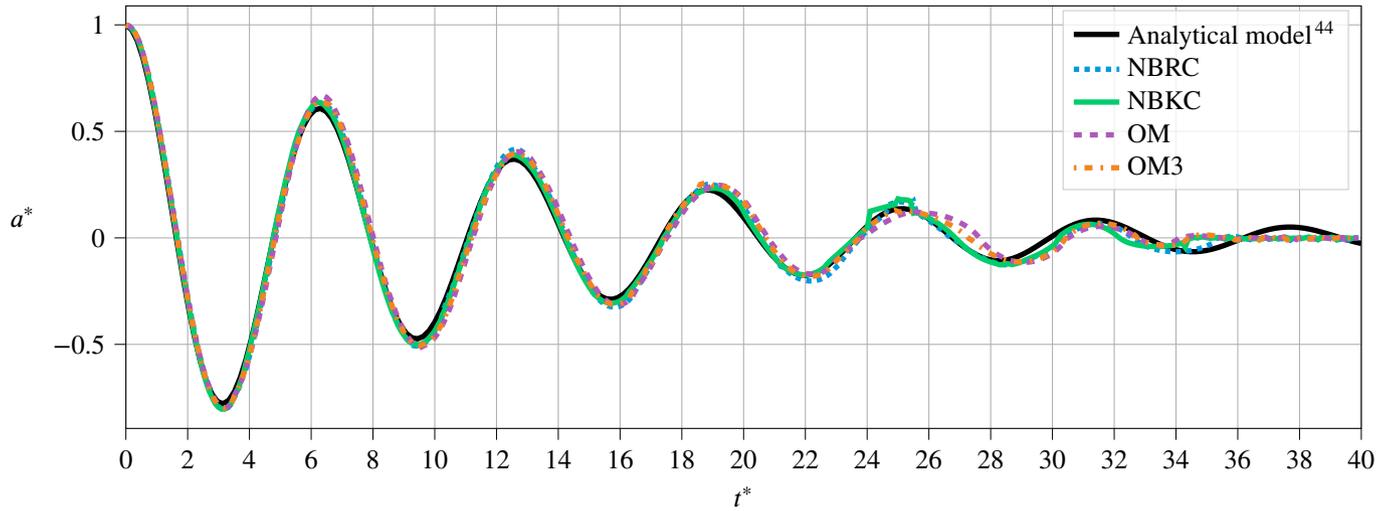}%
	\caption{\label{fig:gravity-wave-radults}
		Simulated surface elevation of the gravity wave in terms of non-dimensional amplitude $a^{*}(0,t^{*})$ and time $t^{*}$.
		The simulations were performed with a computational domain resolution, that is, wavelength of $L=800$ lattice cells.
		The different boundary condition variants show only minor differences and agree well with the analytical model~\cite{dingemans1997WaterWavePropagation}.
	}
\end{figure}

\Cref{fig:gravity-wave-radults} shows the gravity wave simulated with the boundary condition variants presented in \Cref{sec:fsbcv} at a wavelength of $L=800$ lattice cells.
The simulation results with all variants agreed well with the analytical model before $t^{*} \approx 24$.
More noticeable differences are visible in the later course of the simulation.
However, it must be pointed out that the FSLBM requires the wave's amplitude to range over at least one, but preferably multiple interface cells to capture the interface's motion significantly well~\cite{schwarzmeier2022ComparisonFreeSurface}.
This deficiency of the FSLBM is also visible in the grid convergence study provided in \Cref{fig:app-ne-gravity-wave-convergence} in \Cref{app:ne-gw}.
There, it is apparent that the number of meaningfully simulated wave periods decreased when decreasing the computational domain resolution, that is, $L$.
Therefore, the assessment in this test case should be made on the first periods of the simulated wave, where all variants are of similar accuracy.
\par

In summary, the gravity wave test case does not allow a clear conclusion regarding the boundary condition variant to be selected.
\par

\subsection{Rectangular dam break}\label{sec:ne-rdb}
In a rectangular dam break test case, a rectangular liquid column collapses and spreads at the bottom surface.
The test case is regularly used as a numerical benchmark to validate free-surface flow simulations~\cite{janssen2011FreeSurfaceFlow,sato2022ComparativeStudyCumulant,moraga2015VOFFVMPrediction}.
The experiments from Martin and Moyce~\cite{martin1952PartIVExperimental} were used as reference data for the simulations in this section.
\par

\subsubsection{Simulation setup}\label{sec:ne-rdb-ss}
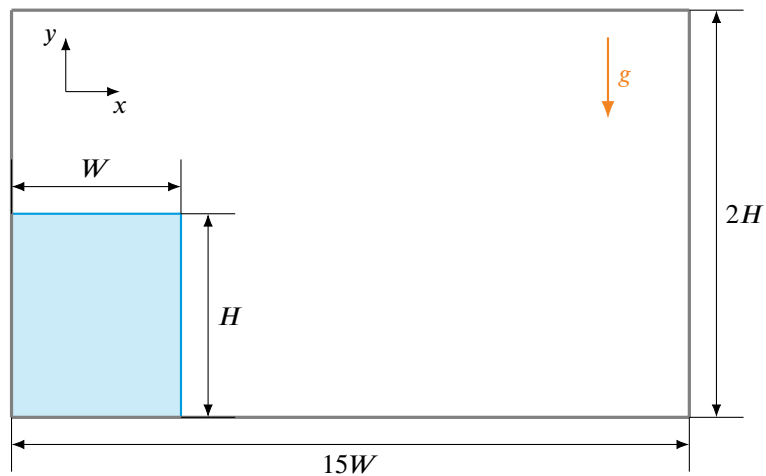
\begin{figure}[htbp]
	\centering
	\setlength{\figureheight}{0.3\textwidth}
	\setlength{\figurewidth}{0.5\textwidth}
	\setlength\mdist{0.02\textwidth}
	\input{figures/dam-break-rectangular/setup.tex}%
	\caption{\label{fig:dam-break-rectangular-setup}
		Simulation setup of the two-dimensional rectangular dam break test case with the liquid column's initial width $W$ and height $H$.
		The gravitational acceleration $g$ acted in negative $y$-direction and led to the liquid column's collapse.
		Free-slip boundary conditions were set at all domain walls.
		C.\ Schwarzmeier, U.\ Rüde, Comparison of refilling schemes in the free-surface lattice Boltzmann method, arXiv preprint\cite{schwarzmeier2022ComparisonRefillingSchemes}, 2022; licensed under a Creative Commons Attribution (CC BY) license.
	}
\end{figure}

The setup was chosen to resemble the reference experiments~\cite{martin1952PartIVExperimental} and is shown in \Cref{fig:dam-break-rectangular-setup}.
In a two-dimensional domain of size $15W \times 2H \times 1$ ($x$-, $y$-, $z$-direction), a rectangular liquid column of width $W \in \{50, 100, 200\}$ lattice cells and height $H=2W$ was positioned at the domain's left wall in the $x$-direction.
The gravitational acceleration $g$ acted in the negative $y$-direction.
Accordingly, the liquid was initialized with hydrostatic pressure, so the LBM pressure at $y=H$ was equal to the constant atmospheric gas pressure $p^{\text{V}}(t) = p_{0}$.
Free-slip boundary conditions were set at all domain borders, and wetting effects were not considered.
The chosen relaxation rate $\omega=1.9995$ was kept constant for all computational domain resolutions as specified by $W$, conforming with diffusive scaling.
The simulations were performed using the turbulence model presented in \Cref{sec:nm-lbm} with Smagorinsky constant $C_{S}=0.1$~\cite{yu2005DNSDecayingIsotropic}.
The Galilei number
\begin{equation}\label{eq:ne-rdb-ss-ga}
\text{Ga} \coloneqq \frac{g W^{3}}{\nu^{2}} = 1.83\cdot 10^{9}
\end{equation}
relates the gravitational to viscous forces. The Bond number
\begin{equation}\label{eq:ne-rdb-ss-bo}
\text{Bo} \coloneqq \frac{\Delta \rho g W^{2}}{\sigma} = 445
\end{equation}
defines the relation between gravitational and surface tension forces.
In these dimensionless numbers, $\nu$ is the kinematic viscosity,  $\sigma$ is the surface tension, and $\Delta \rho$ is the density difference between the liquid and the gas phase.
Note that $\Delta \rho = \rho$ in a free-surface system, as the gas phase density is assumed to be zero.
While the reference experiments~\cite{martin1952PartIVExperimental} were performed with water, the authors did not provide fluid properties.
With given initial column width $W=0.05715$\,m, Ga and Bo as specified above were computed assuming water~\cite{rumble2021CRCHandbookChemistry} at 25\,\textdegree C with the fluid density $\rho \approx 1000$\,kg/m\textsuperscript{3}, kinematic viscosity $\nu \approx 10^{-6}$\,m/s\textsuperscript{2}, surface tension $\sigma \approx 7.2\cdot 10^{-2}$\,kg/s\textsuperscript{2}, and gravitational acceleration $g=9.81$\,m/s\textsuperscript{2}.
\par

The liquid column's residual height $h(t)$ and width $w(t)$ were monitored during the simulation, where
$h(t)$ was obtained by finding the uppermost interface cell at the left domain wall, that is, at $x=0$.
The width $w(t)$ was obtained by searching for the rightmost interface cell at the bottom domain wall, that is, at $y=0$.
Following Martin and Moyce~\cite{martin1952PartIVExperimental}, the height $h^{*}(t) \coloneqq h(t) / H$, width $w^{*}(t) \coloneqq w(t) / W$, and time $t^{*} \coloneqq t \sqrt{2g/W}$ were non-dimensionalized, with $h(t^{*})$ and $w(t^{*})$ being monitored every $t^{*}=0.01$.
In agreement with the experimental data, the simulations were stopped at $w^{*}(t^{*}) \geq 14$.
\par

\subsubsection{Results and discussion}\label{sec:ne-rdb-rad}
\begin{figure}[htbp]
	\centering
	\setlength{\figureheight}{0.425\textwidth}
	\setlength{\figurewidth}{0.475\textwidth}
	\begin{subfigure}[t]{0.49\textwidth}
		\centering
		\input{figures/dam-break-rectangular/w-200-y.tex}%
	\end{subfigure}
	\hfill
	\begin{subfigure}[t]{0.49\textwidth}
		\centering
		\input{figures/dam-break-rectangular/w-200-x.tex}%
	\end{subfigure}
	\caption{\label{fig:dam-break-rectangular-200}
		Simulated rectangular dam break with non-dimensionalized residual dam height $h^{*}(t^{*})$, width $w^{*}(t^{*})$, and time $t^{*}$.
		The simulations were performed with a computational domain resolution, that is, initial dam width of $W=200$ lattice cells.
		The sudden jumps in width observed with the NBKC and OM3 variant were caused by droplets moving faster than the liquid front spread.
		These droplets separated in the early phase of the dam break, as visualized in \Cref{fig:dam-break-rectangular-droplet}. Similarly, the droplets shown in \Cref{fig:dam-break-rectangular-mesh} for the NBRC variant disturbed the evaluation algorithm and led to the increase of $h^{*}(t^{*})$ rather than to its decrease as expected.
	}
\end{figure}
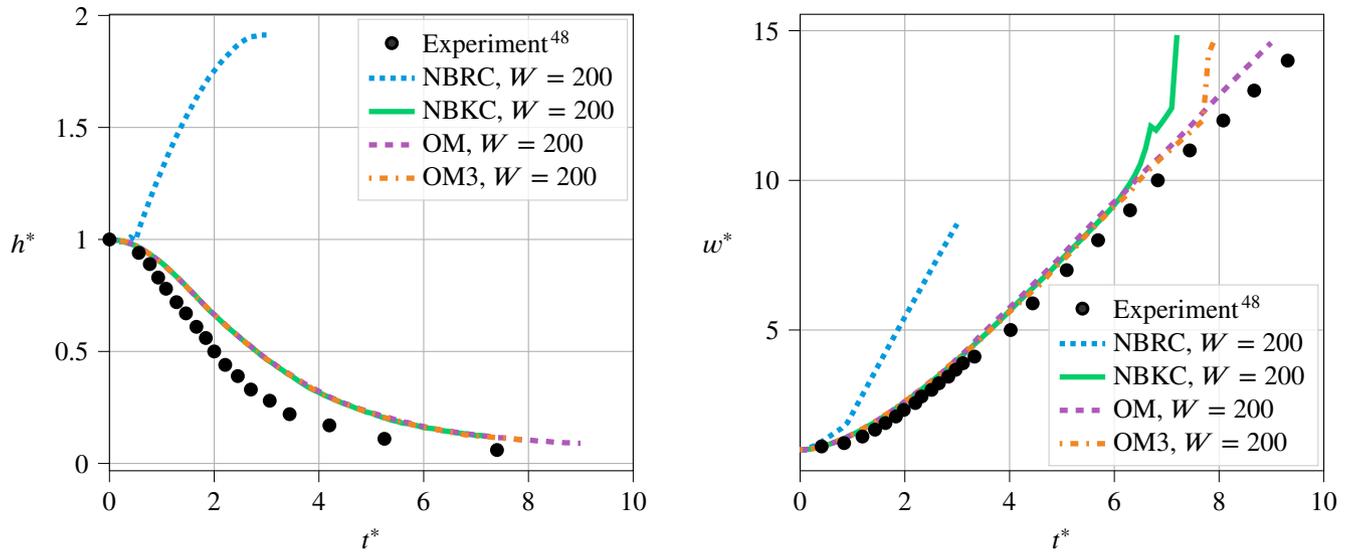

\begin{figure}[htbp]
	\centering
	\begin{tabular}{
			>{\centering\arraybackslash}m{0.05\textwidth}
			>{\centering\arraybackslash}m{0.5\textwidth}
		}	
		NBRC &
		\includegraphics[width=0.5\textwidth]{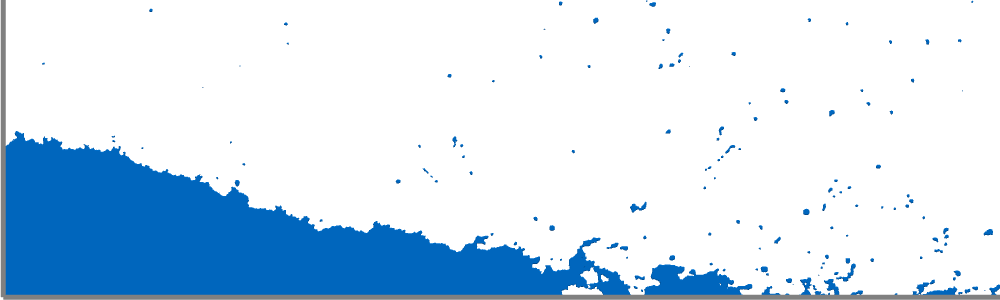}\\
		
		NBKC &
		\includegraphics[width=0.5\textwidth]{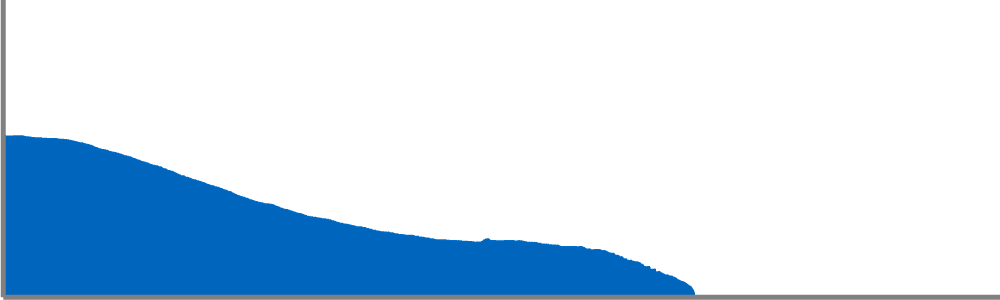}\\
		
		OM & 
		\includegraphics[width=0.5\textwidth]{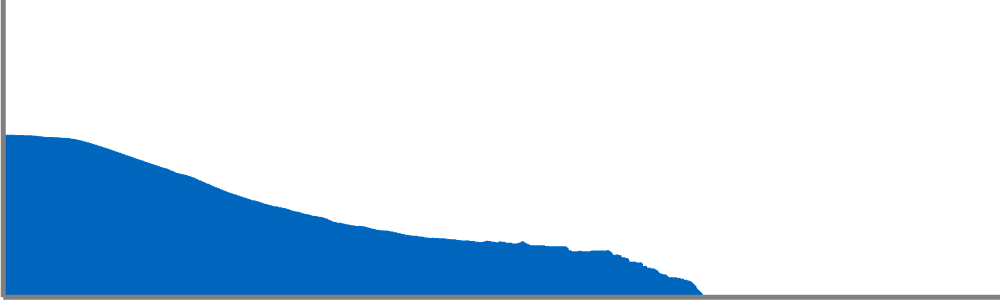}\\
		
		OM3 &
		\includegraphics[width=0.5\textwidth]{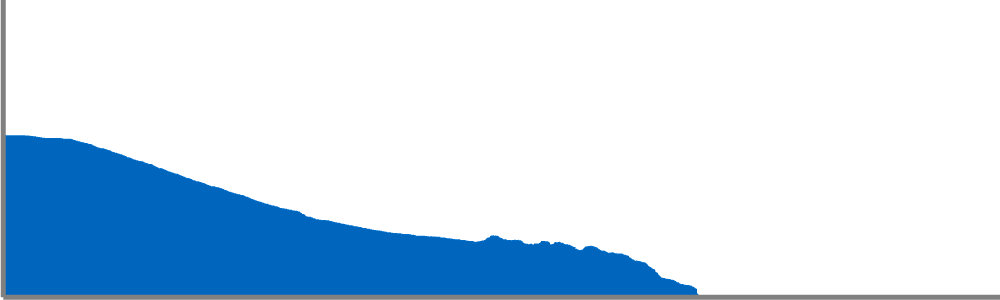}\\
	\end{tabular}
	\caption{\label{fig:dam-break-rectangular-mesh}
		Contour of the simulated rectangular dam break at $t^{*}=3$ with an initial dam width of $W=200$ lattice cells.
		The NBRC variant led to non-physical effects, as splashing was not reported in the reference experiments~\cite{martin1952PartIVExperimental}.
	}
\end{figure}

\begin{figure}[htbp]
	\centering
	\includegraphics[width=0.7\textwidth]{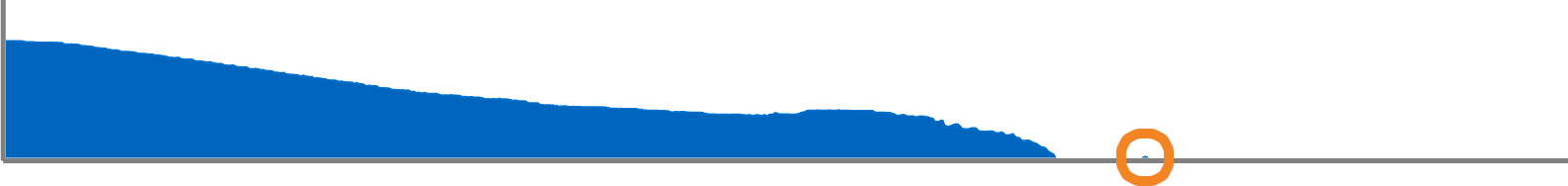}
	\caption{
		Contour of the simulated rectangular dam break at $t^{*}=4$ with the NBKC variant and an initial column width of $W=100$ lattice cells.
		The droplet marked by the orange circle has detached from the liquid column.
	}
	\label{fig:dam-break-rectangular-droplet}
\end{figure}

\Cref{fig:dam-break-rectangular-200} shows the spread of the liquid column with an initial width of $W=200$ lattice cells.
All but the NBRC variant produced similarly accurate results and moderately agreed with the experimental data~\cite{martin1952PartIVExperimental}.
In contrast, with the NBRC variant, the liquid column's collapse was characterized by the detachment of many droplets, as visualized in \Cref{fig:dam-break-rectangular-mesh}.
Splashing such as this was neither present in the other variants nor reported to be observed in the reference experiments.
It was less pronounced at lower computational domain resolutions.
However, single droplets were also present in the NBKC and OM3 variants, leading to the sudden jumps in $w^{*}(t^{*})$ in \Cref{fig:dam-break-rectangular-200}.
As shown in \Cref{fig:dam-break-rectangular-droplet}, single droplets separated in the early phase of the dam collapse, moving faster than the liquid front spread.
In the case of $W=100$ with the NBKC variant, the droplet shown in \Cref{fig:dam-break-rectangular-droplet} even led to a numerically unstable simulation.
There, the droplet's velocity exceeded the lattice speed of sound $c_{s}^{2}$, which generally is a result of numerical instabilities in the LBM~\cite{kruger2017LatticeBoltzmannMethod}.\par

A grid refinement study of this test case is presented in \Cref{fig:dam-break-rectangular-convergence} in \Cref{app:ne-rdb}, showing that only the OM variant converged reasonably well.
All other variants were subject to the detachment of droplets, as mentioned earlier.\par

Considering the above observations, the OM variant could be identified as the most accurate for the test case shown here.
\par

\FloatBarrier

\subsection{Cylindrical dam break}\label{sec:ne-cdb}
The rectangular dam break test case in \Cref{sec:ne-rdb} is extended to a cylindrical dam break.
The reference experiments are again taken from Martin and Moyce~\cite{martin1952PartIVExperimental}.
This test case was chosen to evaluate the effect of the boundary condition variant on the rotational symmetry, as
Bogner~\cite{bogner2016CurvatureEstimationVolumeoffluid} reported anisotropic artifacts when using a normal-based variant such as the NBKC or NBRC variant.
\par

\subsubsection{Simulation setup}\label{sec:ne-cdb-ss}
\begin{figure}[htbp]
	\centering
	\setlength{\figureheight}{0.3\textwidth}
	\setlength{\figurewidth}{0.5\textwidth}
	\setlength\mdist{0.02\textwidth}
	\input{figures/dam-break-cylindrical/setup.tex}%
	\caption{
		\label{fig:dam-break-cylindrical-setup}
		Simulation setup of the three-dimensional cylindrical dam break test case.
		A cylindrical liquid column of diameter $D$ and height $H$ was initialized in the domain's center.
		It collapsed due to the gravitational acceleration $g$ acting in negative $z$-direction.
		Free-slip boundary conditions were set at all domain borders.
		C.\ Schwarzmeier, U.\ Rüde, Comparison of refilling schemes in the free-surface lattice Boltzmann method, arXiv preprint\cite{schwarzmeier2022ComparisonRefillingSchemes}, 2022; licensed under a Creative Commons Attribution (CC BY) license.
	}
\end{figure}
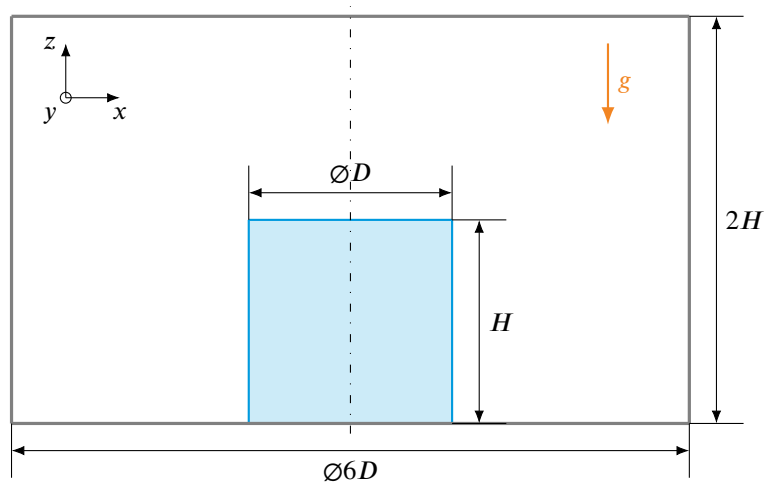

As visualized in \Cref{fig:dam-break-cylindrical-setup}, a cylindrical liquid column of diameter $D \in \{50, 100, 200\}$ lattice cells and height $H=D$ was placed at the center of the three-dimensional domain of size $6D \times 6D \times 2H$ ($x$-, $y$-, $z$-direction).
In other aspects, the setup was similar to the one of the rectangular dam break in \Cref{sec:ne-rdb-ss}.
However, in the definitions of the Galilei~\eqref{eq:ne-rdb-ss-ga} and Bond number~\eqref{eq:ne-rdb-ss-bo}, the characteristic length $0.5D$ was used.
\par

During the simulation, the liquid column's radius $r(t)$ was monitored.
It is defined as the distance of the liquid front to the column's initial center of symmetry such that $r(0) = D/2$.
As the liquid column's collapse was observed not to be symmetric in the numerical experiments, $r(t)$ was computed for every interface cell detected by a \textit{seed-fill} algorithm~\cite{pavlidis1982AlgorithmsGraphicsImage} starting at an arbitrary domain boundary.
In practice, this implied that only the outermost interface cells were detected, that is, the interface cells at the spreading liquid's front.
A statistical sample was then used to evaluate $r(t)$ by computing the maximum, minimum, and mean values of $r(t)$ at every $t^{*}=0.01$.
The radius $r^{*}(t) \coloneqq 2r(t)/D$ and time $t^{*} \coloneqq t \sqrt{4g/D}$ were non-dimensionalized as in the reference data from the literature~\cite{martin1952PartIVExperimental}.
In agreement with the reference experiments, the simulations were performed until $r_{\text{max}}^{*}(t^{*}) \geq 4.33$, where $r_{\text{max}}^{*}(t^{*})$ is the non-dimensionalized maximum liquid front radius.
\par

\subsubsection{Results and discussion}\label{sec:ne-cdb-rad}
\begin{figure}[htbp]
	\centering
	\setlength{\figureheight}{0.5\textwidth}
	\setlength{\figurewidth}{1\textwidth}
	\input{figures/dam-break-cylindrical/d-200.tex}%
	\caption{\label{fig:dam-break-cylindrical-200}
		Simulated cylindrical dam break with non-dimensionalized liquid column radius $r^{*}(t^{*})$ and time $t^{*}$.
		The simulations were performed with a computational domain resolution, that is, initial column diameter of $D=200$ lattice cells.
		The markers represent the mean value of $r^{*}(t^{*})$, and the error bars indicate its maximum and minimum.
		The OM variant agreed best with the experimental data~\cite{martin1952PartIVExperimental} and had the smallest error bars.
		It maintained the column's rotational symmetry better than the other boundary condition variants.
	}
\end{figure}
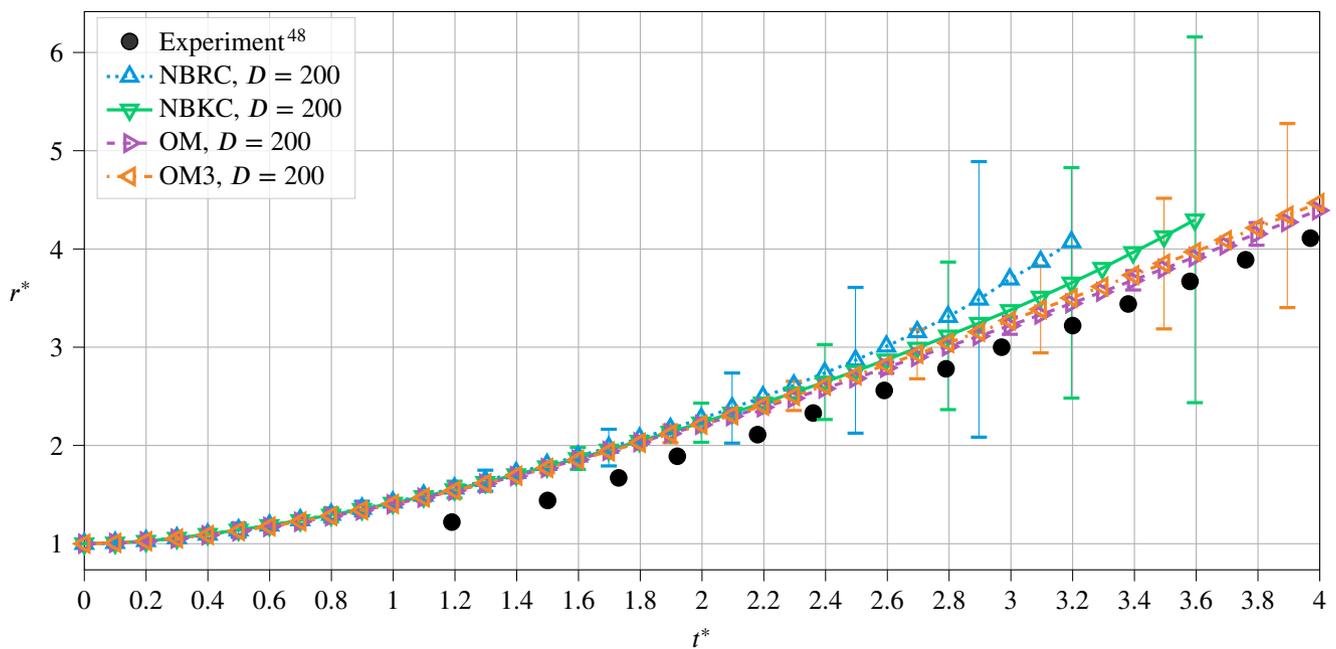

\begin{figure}[htbp]
	\centering
	\begin{subfigure}[t]{0.49\textwidth}
		\centering
		\includegraphics[width=\textwidth]{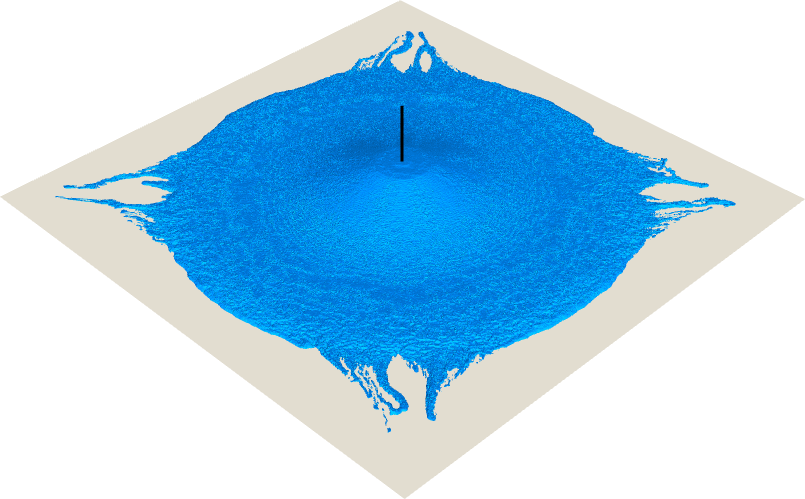}%
		\caption{\label{fig:dam-break-cylindrical-mesh-nbrc}NBRC}
	\end{subfigure}
	\hfill
	\begin{subfigure}[t]{0.49\textwidth}
		\centering
		\includegraphics[width=\textwidth]{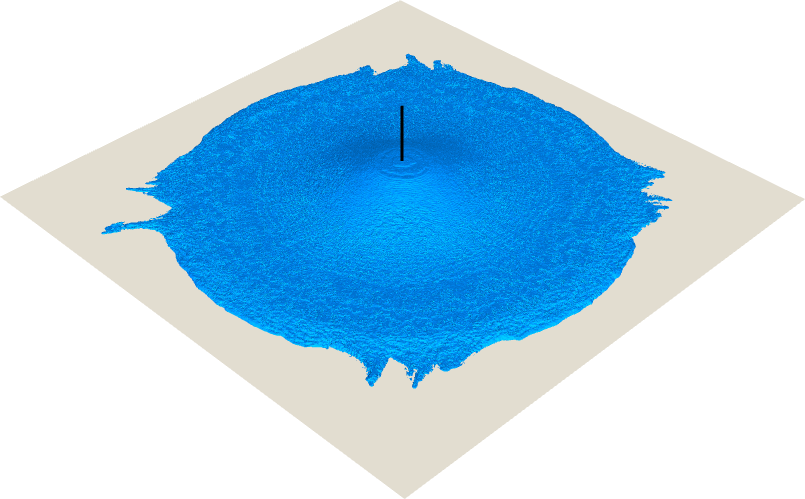}%
		\caption{\label{fig:dam-break-cylindrical-mesh-nbkc}NBKC}
	\end{subfigure}
	\hfill
	\begin{subfigure}[t]{0.49\textwidth}
		\centering
		\includegraphics[width=\textwidth]{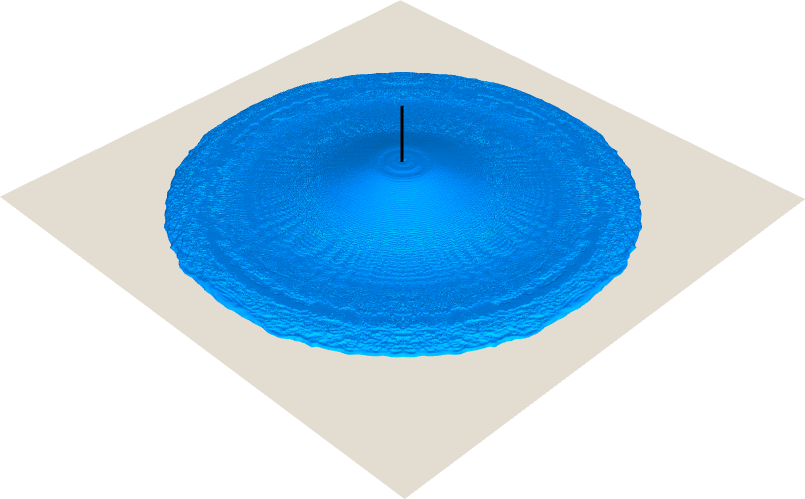}%
		\caption{\label{fig:dam-break-cylindrical-mesh-om}OM}
	\end{subfigure}
	\hfill
	\begin{subfigure}[t]{0.49\textwidth}
		\centering
		\includegraphics[width=\textwidth]{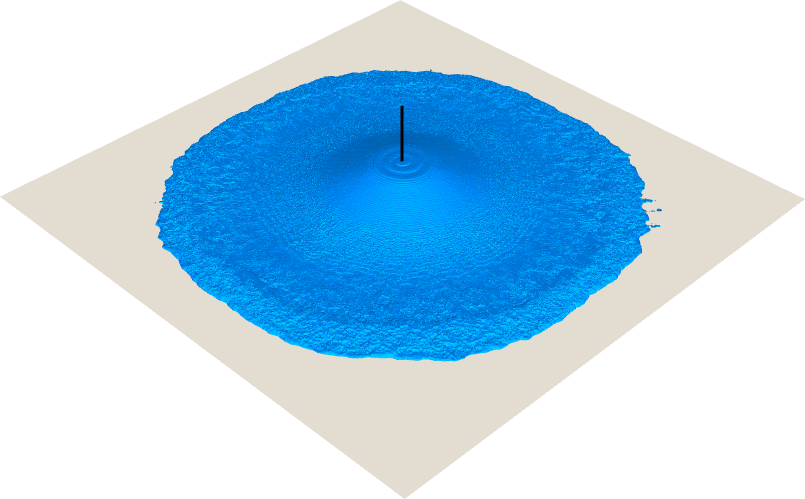}%
		\caption{\label{fig:dam-break-cylindrical-mesh-om3}OM3}
	\end{subfigure}
	\hfill
	\caption{\label{fig:dam-break-cylindrical-mesh}
		Shape of the simulated cylindrical dam break at non-dimensionalized time $t^{*}=3$.
		The simulations were performed with an initial column diameter of $D=200$ lattice cells.
		The black line indicates the column's initial center of symmetry.
		The OM variant maintained the rotational symmetry well, whereas the other approaches showed more deviations.
	}
\end{figure}

\Cref{fig:dam-break-cylindrical-200} compares the simulation results for a computational domain resolution equivalent to $D=200$ lattice cells with the experimental data~\cite{martin1952PartIVExperimental}.
The markers show the mean value of the non-dimensionalized radius $r^{*}(t^{*})$.
The error bars indicate the maximum and minimum values of $r^{*}(t^{*})$.
It is immediately apparent that the OM variant agreed best with the measurements from the literature.
It has the smallest error bars in \Cref{fig:dam-break-cylindrical-200}, and therefore the lowest standard deviation in $r^{*}(t^{*})$ when compared to the other boundary condition variants.
All other variants have significantly larger error bars, indicating that they did not maintain the rotational symmetric nature of the liquid column during its collapse.
This observation agrees with the one reported by Bogner~\cite{bogner2016CurvatureEstimationVolumeoffluid}.
Qualitatively, the rotationally symmetry during the collapse is shown in \Cref{fig:dam-break-cylindrical-mesh} at $t^{*} = 3$.
The solid black line indicates the liquid column's initial center of origin.
It can be seen that the NBRC and NBKC variants significantly deviated from rotational symmetry.
\par

The grid refinement study is presented in \Cref{fig:dam-break-cylindrical-convergence} in \Cref{app:ne-cdb} and shows that all of the presented boundary condition variants converged well.
\par

As in the rectangular dam break test case, OM variant was most accurate in this benchmark.
\par

\FloatBarrier

\subsection{Taylor bubble}\label{sec:ne-tb}
A Taylor bubble is a gas bubble rising in a cylindrical tube through stagnant liquid due to buoyancy forces.
Its length is multiple times its diameter.
It has an elongated shape and its leading edge becomes round.
The simulation results were compared to the experimental data from Bugg and Saad~\cite{bugg2002VelocityFieldTaylor}.
\par

\subsubsection{Simulation setup}\label{sec:ne-tb-ss}
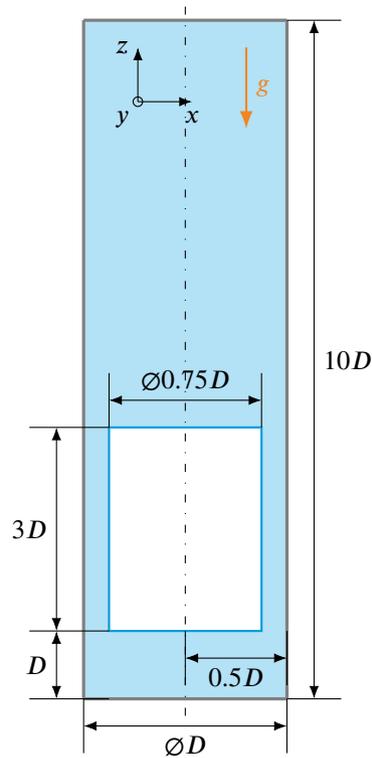
\begin{figure}[htbp]
	\centering
	\setlength{\figureheight}{0.5\textwidth}
	\setlength{\figurewidth}{0.15\textwidth}
	\setlength\mdist{0.02\textwidth}
	\input{figures/taylor-bubble/setup.tex}%
	\caption{\label{fig:taylor-bubble-setup}
		Simulation setup of the three-dimensional Taylor bubble test case with an initially cylindrical gas bubble in a cylindrical tube of diameter $D$.
		The gravitational acceleration $g$ acted in the negative $z$-direction.
		No-slip boundary conditions were applied at the tube and all domain walls.
		C.\ Schwarzmeier, M.\ Holzer, T.\ Mitchell, M.\ Lehmann, F.\ Häusl, U.\ Rüde, Comparison of free surface and conservative Allen-Cahn phase field lattice Boltzmann method, arXiv preprint\cite{schwarzmeier2022ComparisonFreeSurface}, 2022; licensed under a Creative Commons Attribution (CC BY) license; the colors were changed from the original.
	}
\end{figure}

The simulation setup resembled that of the reference experiments~\cite{bugg2002VelocityFieldTaylor} and is illustrated in \Cref{fig:taylor-bubble-setup}.
The no-slip domain walls formed a cylindrical tube of diameter $D = \{32, 64, 128\}$ lattice cells, pointing in the $z$-direction in a three-dimensional computational domain of size $1D \times 1D \times 10D$ ($x \times y \times z$).
The gas bubble was initialized as a cylinder oriented in the $z$-direction with a diameter of $0.75D$ and a length of $3D$.
It was initially located $D$ above the domain's bottom wall with the volumetric gas pressure $p^{\text{V}}(t) = p_{0}$.
The remainder of the domain was filled with a resting liquid that was initialized with hydrostatic pressure according to the gravitational acceleration $g$.
Therefore, the pressure was initially equivalent to $p_{0}$ at $5D$ in the $z$-direction.
All simulations were performed with the relaxation rate $\omega = 1.8$, conforming with diffusive scaling.
The fluid mechanics of the setup are characterized by the Morton number
\begin{equation}
\text{Mo} \coloneqq \frac{g \mu^{4}}{\rho \sigma^{3}} = 0.015
\end{equation}
that describes the ratio of viscous to surface tension forces.
It includes the surface tension $\sigma$, the dynamic fluid viscosity $\mu$, and the liquid density $\rho$.
The Bond number $\text{Bo}=100$~\eqref{eq:ne-rdb-ss-bo}, is used with characteristic length $D$.
The evaluations were performed in terms of the non-dimensionalized bubble radius $r^{*}(t) \coloneqq r(t)/(0.5D)$, axial location $z^{*}(t) \coloneqq z(t)/D$, and time $t^{*} \coloneqq t\sqrt{g/D}$.
\par

\subsubsection{Results and discussion}\label{sec:ne-tb-rad}
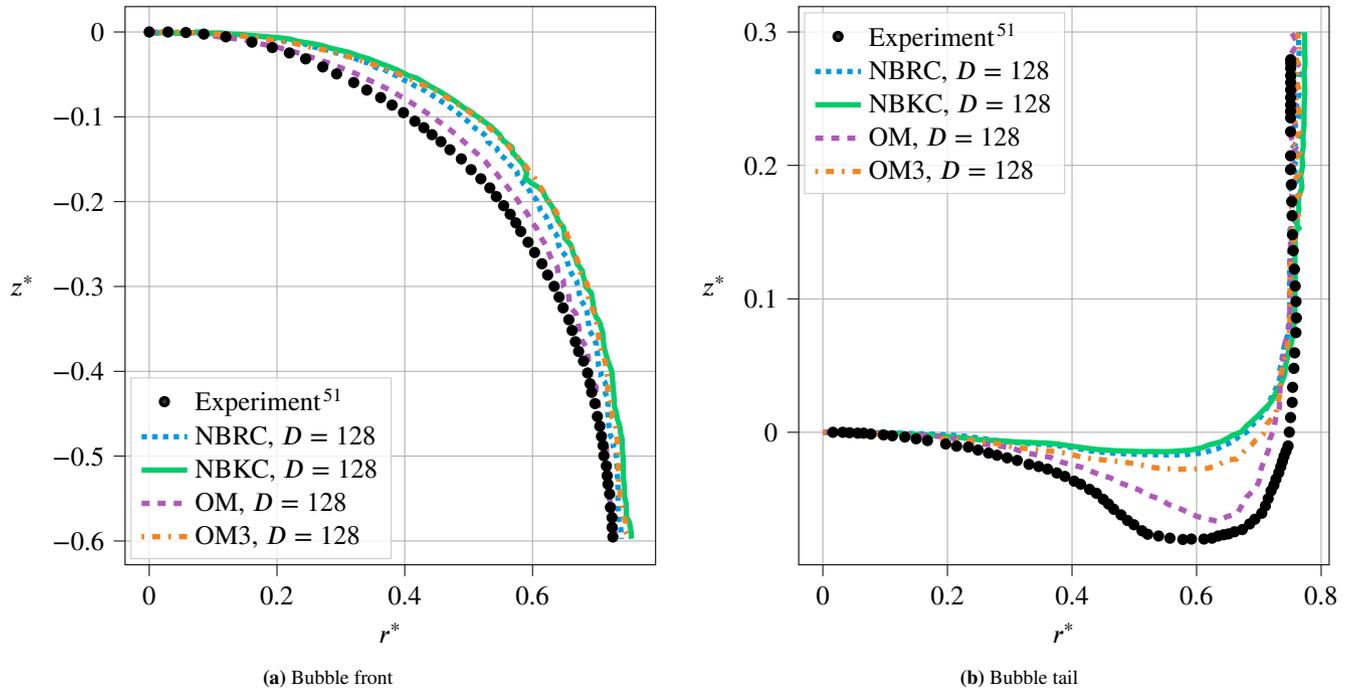
\begin{figure}[htbp]
	\centering
	\setlength{\figureheight}{0.5\textwidth}
	\setlength{\figurewidth}{0.48\textwidth}
	\begin{subfigure}[t]{0.49\textwidth}
		\centering
		\input{figures/taylor-bubble/d-128-shape-front.tex}%
		\caption{\label{fig:taylor-bubble-shape-front}Bubble front}
	\end{subfigure}
	\hfill
	\begin{subfigure}[t]{0.49\textwidth}
		\centering
		\input{figures/taylor-bubble/d-128-shape-tail.tex}%
		\caption{\label{fig:taylor-bubble-shape-tail}Bubble tail}
	\end{subfigure}
	\caption{\label{fig:taylor-bubble-shape}
		Simulated shape of the Taylor bubble's front and tail.
		The simulations were performed with a computational domain resolution, that is, tube diameter of $D=128$ lattice cells.
		The comparison with experimental data~\cite{bugg2002VelocityFieldTaylor} was drawn in terms of the non-dimensionalized axial location $z^{*}$ and radial location $r^{*}$ at time $t^{*}=15$.
		The OM variant was most accurate.
	}
\end{figure}

The simulated Reynolds number
\begin{equation}
\text{Re} \coloneqq \frac{D u}{\nu}
\end{equation}
is listed in \Cref{tab:taylor-bubble-re} for different tube diameters $D$.
Re is computed with the kinematic viscosity $\nu$ and the bubble's rise velocity $u$.
The latter was obtained from the bubble's center of mass in the $z$-direction at times $t^{*}=10$ and $t^{*}=15$.
All boundary condition variants agreed reasonably well with the experimental data.
The OM variant was the most accurate with an error of approximately $4$\,\% at $D=128$.
\Cref{fig:taylor-bubble-shape} compares the bubble's shape at its front and tail at time $t^{*}=15$ with the experimental observations.
All variants generally produced plausible results.
The OM variant most closely resembled the bubble's shape from the experimental measurements.
For the other variants, no clear trend is visible.
\par

The grid refinement study in \Cref{tab:taylor-bubble-re} and \Cref{fig:taylor-bubble-shape-convergence} in \Cref{app:ne-tb} shows that all boundary condition variants converged well.
\par

As in both dam break test cases, the OM variant was the most accurate in this benchmark.
\par

\begin{table}[htbp]
	\centering
	\begin{tabular}{
			>{\raggedright}m{0.2\textwidth}
			>{\centering\arraybackslash}m{0.1\textwidth}
			>{\centering\arraybackslash}m{0.1\textwidth}
			>{\centering\arraybackslash}m{0.1\textwidth}
		}
		
		\toprule
		$D$ & $32$ & $64$ & $128$ \\
		\midrule
		
		Re\textsubscript{NBRC} & $21.92$ & $23.77$ & $25.12$ \\
		
		Re\textsubscript{NBKC} & $20.98$ & $22.86$ & $24.27$ \\
		
		Re\textsubscript{OM} & $24.12$ & $25.35$ & $25.89$ \\
		
		Re\textsubscript{OM3} & $22.68$ & $23.96$ & $24.74$ \\
		
		\midrule
		Re\textsubscript{Experiment}~\cite{bugg2002VelocityFieldTaylor} & \multicolumn{3}{c}{$27$} \\
		\bottomrule
	\end{tabular}
	\caption{
		\label{tab:taylor-bubble-re}
		Reynolds number Re of the simulated Taylor bubble for different computational domain resolutions as specified by the tube diameter $D$.
		The bubble's rise velocity, as used to compute Re, was obtained from the Taylor bubble's location in axial direction at time $t^{*}=10$ and $t^{*}=15$.
	}
\end{table}

\FloatBarrier

\subsection{Drop impact}\label{sec:ne-di}
In the final test case, the vertical impact of a drop into a pool of liquid was simulated.
Due to the lack of quantitative experimental data in the reference experiments from Wang and Chen~\cite{wang2000SplashingImpactSingle}, only a qualitative comparison with a photograph could be made here.
\par

\subsubsection{Simulation setup}\label{sec:ne-di-ss}
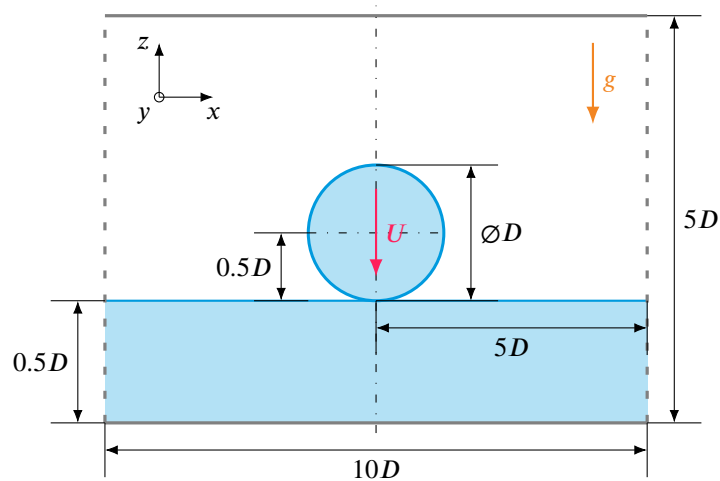
\begin{figure}[htbp]
	\centering
	\setlength{\figureheight}{0.3\textwidth}
	\setlength{\figurewidth}{0.4\textwidth}
	\setlength\mdist{0.02\textwidth}
	\input{figures/drop-impact/setup.tex}%
	\caption{\label{fig:drop-impact-setup}
		Simulation setup of the drop impact test case.
		A spherical drop of liquid with diameter $D$ was initialized right above the surface of a liquid pool of height $0.5D$ in a domain of size $10D \times 10D \times 5D$.
		The gravitational acceleration $g$ acted in the negative $z$-direction and the droplet was initialized with impact velocity $U$ in the same direction.
		The domain's side walls in $x$- and $y$-direction were periodic, whereas the domain's top and bottom walls in $z$-direction were set to no-slip.
		C.\ Schwarzmeier, M.\ Holzer, T.\ Mitchell, M.\ Lehmann, F.\ Häusl, U.\ Rüde, Comparison of free surface and conservative Allen--Cahn phase field lattice Boltzmann method, arXiv preprint\cite{schwarzmeier2022ComparisonFreeSurface}, 2022; licensed under a Creative Commons Attribution (CC BY) license; the colors were changed from the original.
	}
\end{figure}

The simulation setup was chosen to conform with the reference experiments~\cite{wang2000SplashingImpactSingle}.
As illustrated in \Cref{fig:drop-impact-setup}, a spherical droplet with a diameter of $D=80$ lattice cells was initialized in a three-dimensional computational domain of size $10D \times 10D \times 5D$ ($x$-, $y$-, $z$-direction) lattice cells.
The droplet was located at the surface of a thin liquid film of height $0.5D$ and had an initialized impact velocity $U$ in the negative $z$-direction.
The gravitational acceleration $g$ also acted in the negative $z$-direction.
In the drop and in the liquid film, hydrostatic pressure according to $g$ was initialized.
Accordingly, the pressure at the pool's surface was equal to the constant atmospheric volumetric gas pressure $p^{\text{V}}(t) = p_{0}$.
There were no-slip boundary conditions at the top and bottom domain walls in the $z$-direction. The domain walls in the $x$- and $y$-direction were periodic.
The relaxation rate was chosen $\omega=1.989$.
The droplet's impact is described by the Weber number
\begin{equation}
\text{We} \coloneqq \frac{\rho U^{2}D}{\sigma} = 2010
\end{equation}
that relates inertial and surface tension forces, and the Ohnesorge number
\begin{equation}
\text{Oh} \coloneqq \frac{\mu}{\sqrt{\sigma \rho D}} = 0.0384
\end{equation}
that relates viscous to inertial and surface tension forces.
These dimensionless numbers include the surface tension $\sigma$, dynamic viscosity $\mu$, and liquid density $\rho$.

A fluid with density $\rho = 1200$\, kg/m\textsuperscript{3} and dynamic viscosity $\mu = 0.022$\,kg/(m$\cdot$s) was used in the experiments~\cite{wang2000SplashingImpactSingle}.
Assuming $g=9.81$\,m/s\textsuperscript{2}, the definition of the system is closed by $\text{Bo}=3.18$~\eqref{eq:ne-rdb-ss-bo} with characteristic length $D$.
As observed by Lehmann et al.~\cite{lehmann2021EjectionMarineMicroplastics}, the non-dimensionalized time $t^{*} \coloneqq t U/D$ must be offset by $t^{*}=0.16$ for comparison with the numerical simulations as set up in the study here.
\par

\subsubsection{Results and discussion}\label{sec:ne-di-rad}
\begin{figure}[htbp]
	\centering
	\begin{tabular}{
			>{\centering\arraybackslash}m{0.025\textwidth}
			>{\centering\arraybackslash}m{0.45\textwidth}
			>{\centering\arraybackslash}m{0.45\textwidth}
		}
		
		Experiment~\cite{wang2000SplashingImpactSingle} & \multicolumn{2}{c}{\includegraphics[width=0.24\textwidth,valign=c]{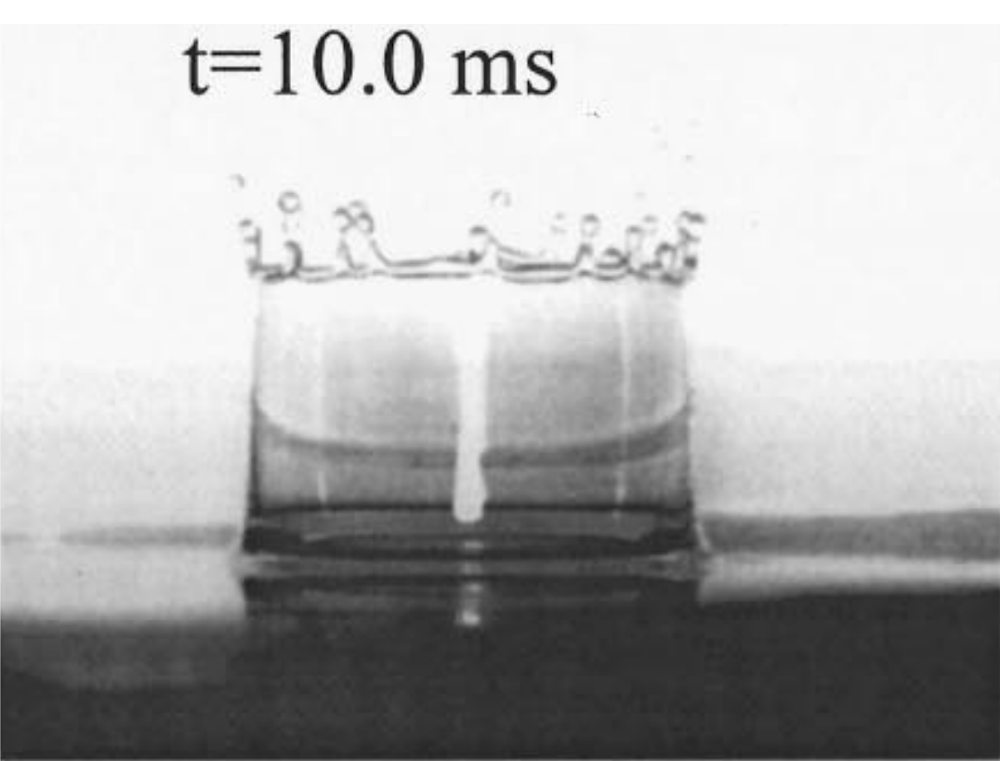}} \\
		
		NBRC & \includegraphics[width=0.29\textwidth]{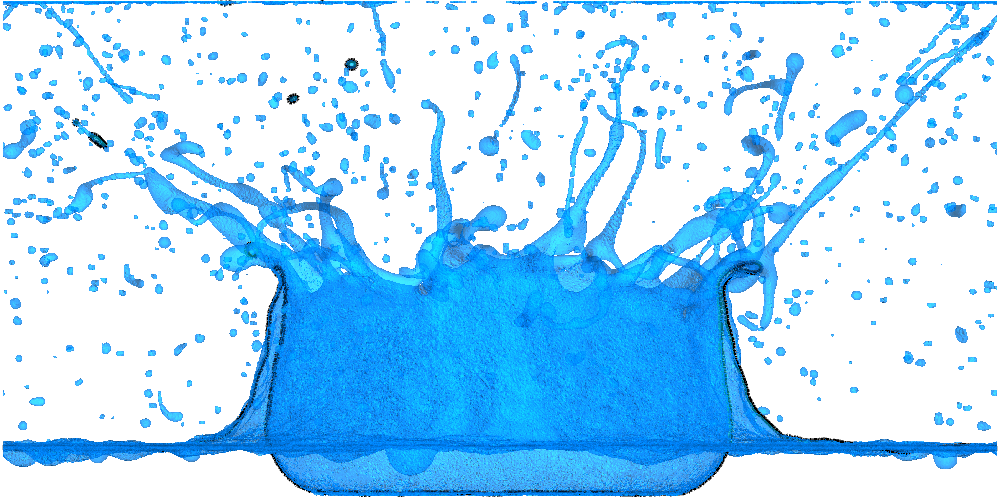} & \includegraphics[width=0.29\textwidth]{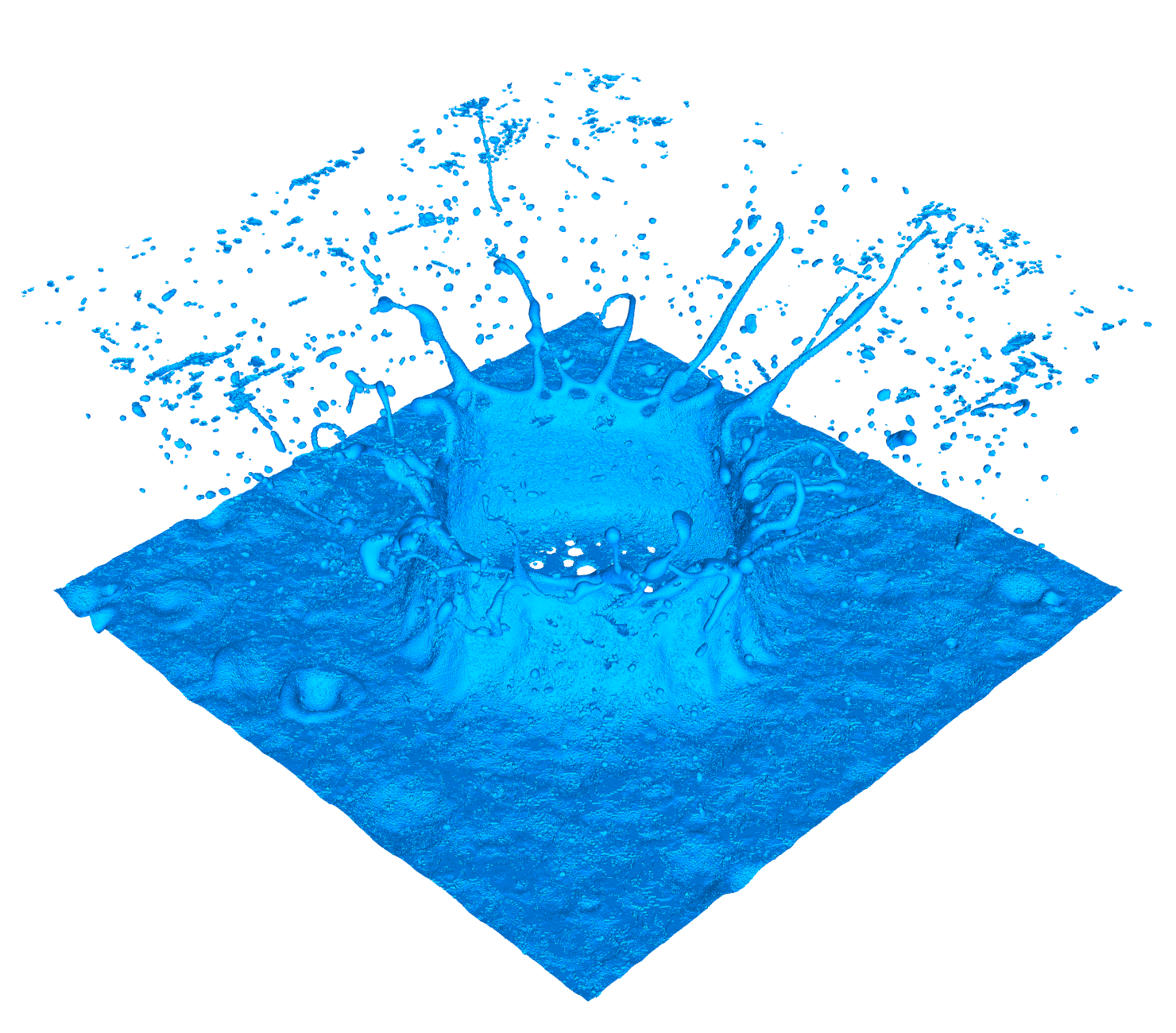} \\
		
		NBKC & \includegraphics[width=0.29\textwidth]{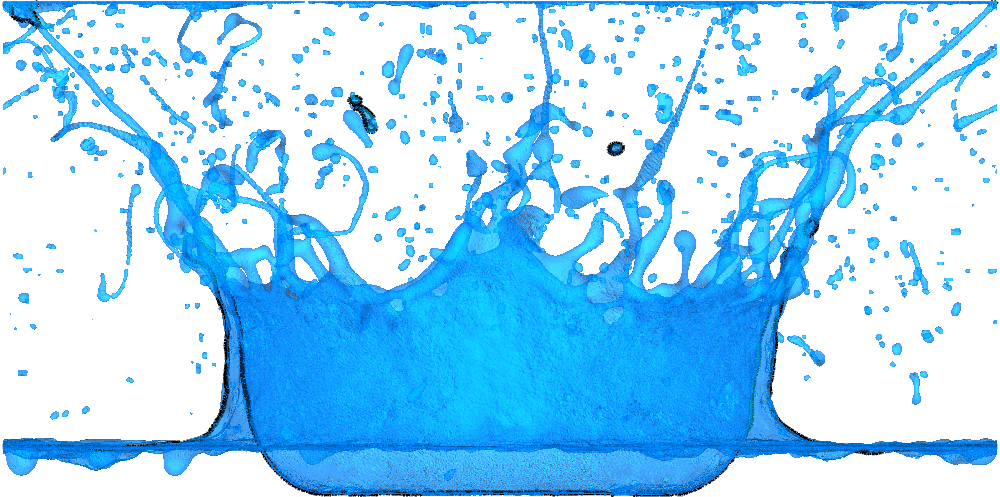} & \includegraphics[width=0.29\textwidth]{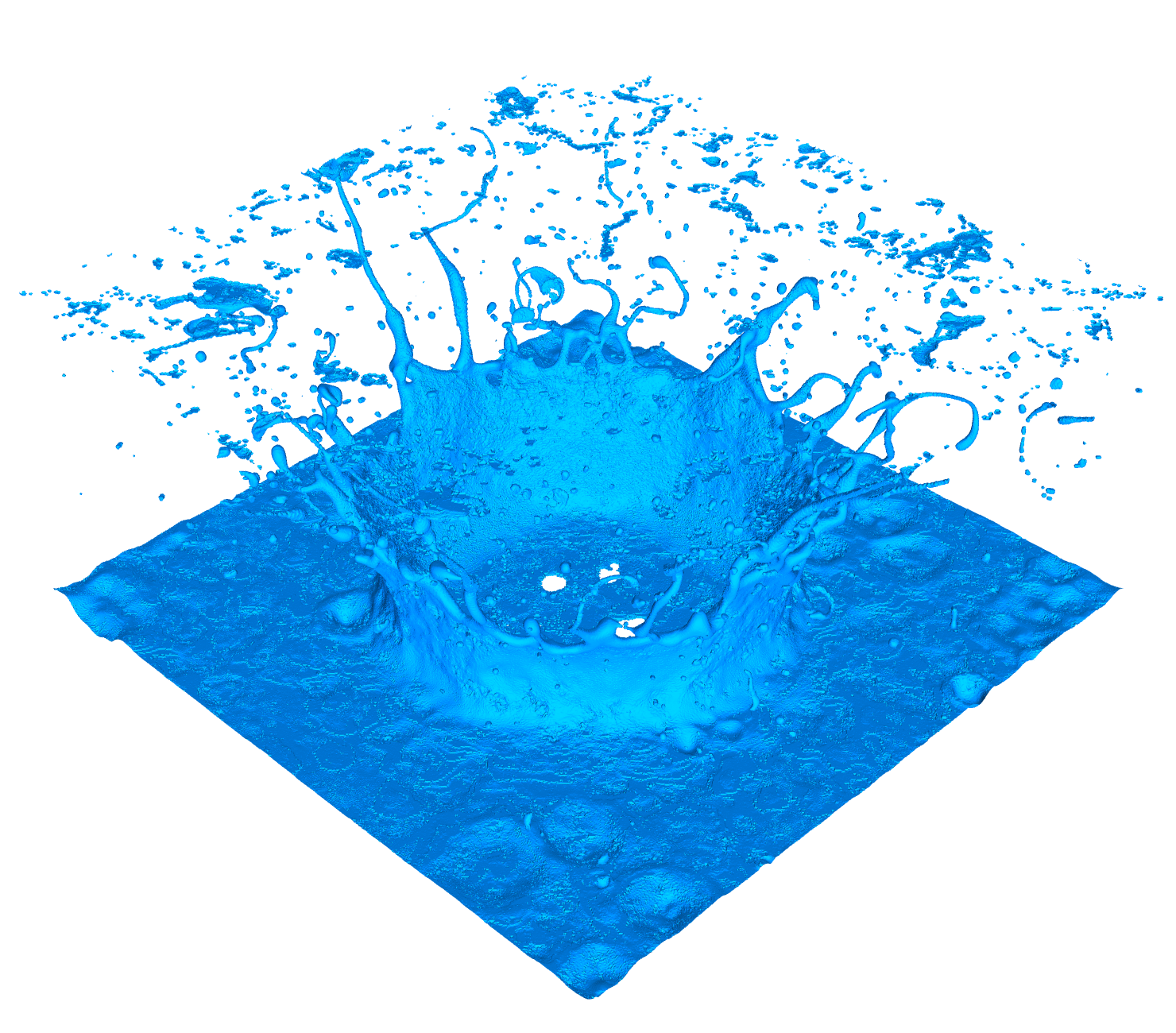} \\
				
		OM & \includegraphics[width=0.29\textwidth]{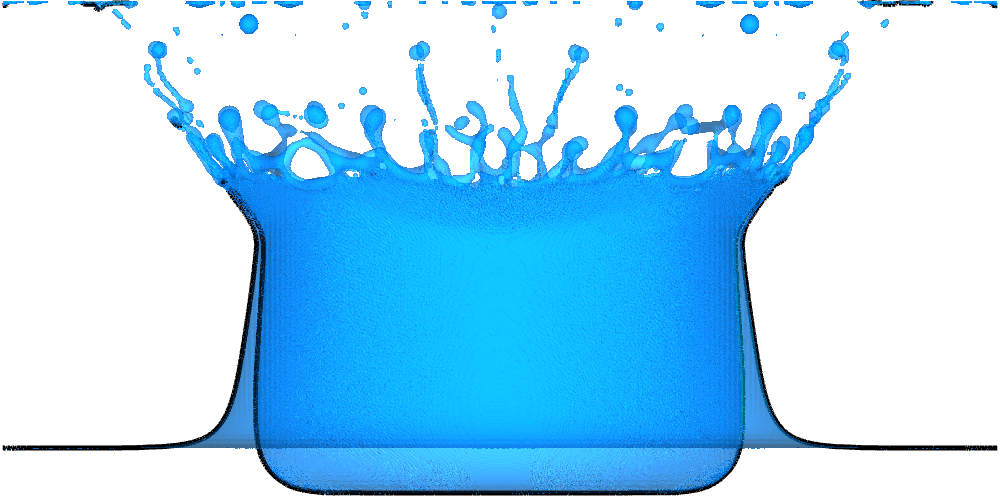} & \includegraphics[width=0.29\textwidth]{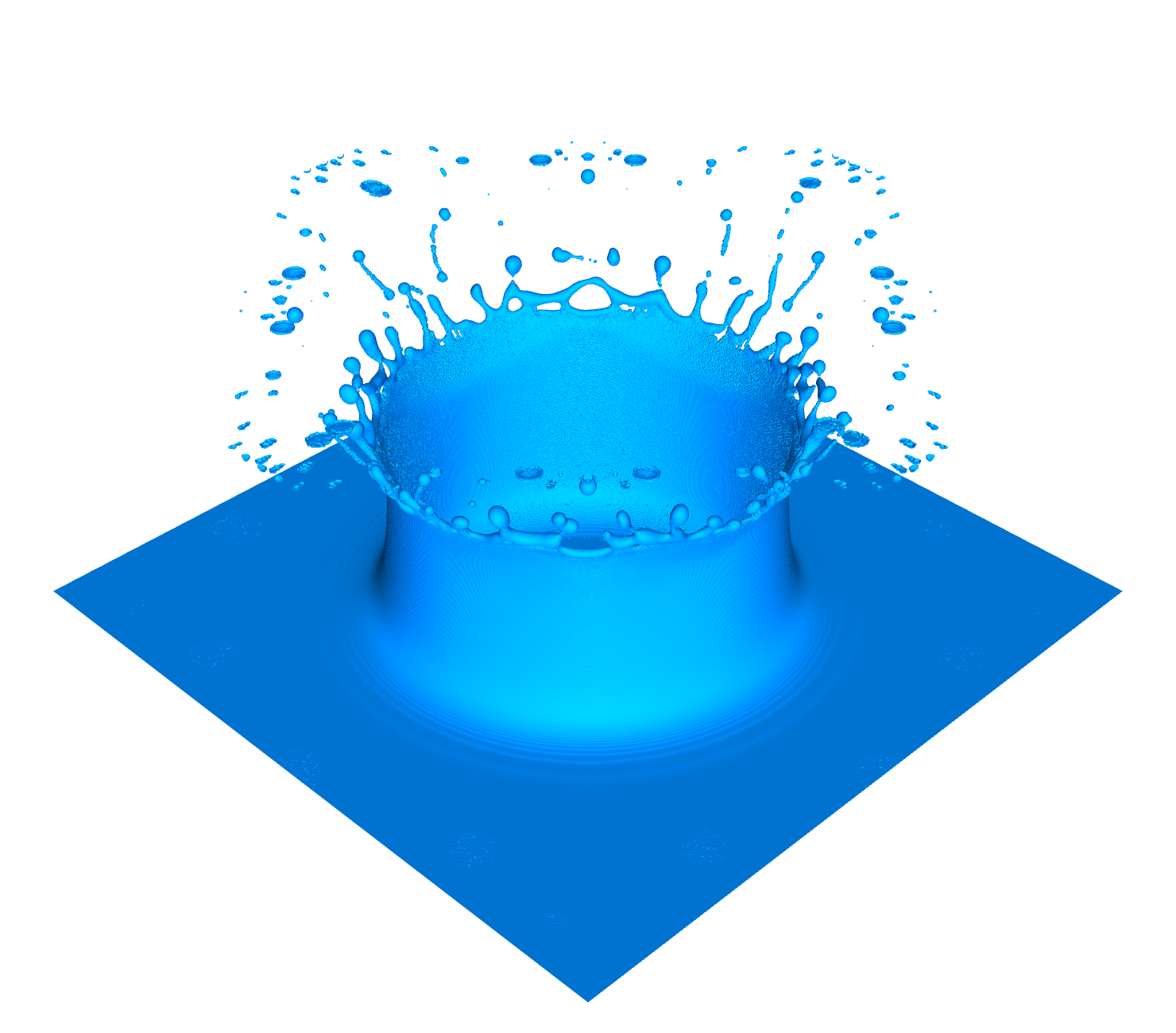} \\
						
		OM3 & \includegraphics[width=0.29\textwidth]{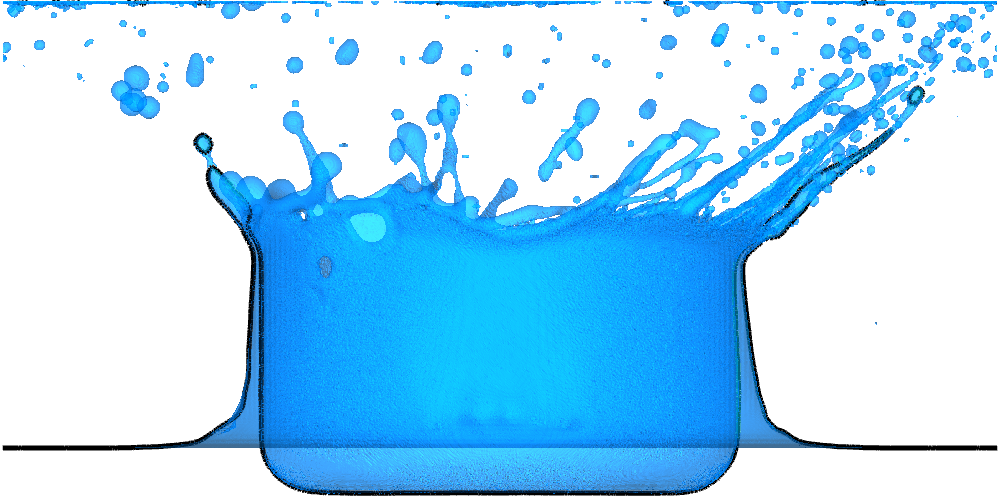} & \includegraphics[width=0.29\textwidth]{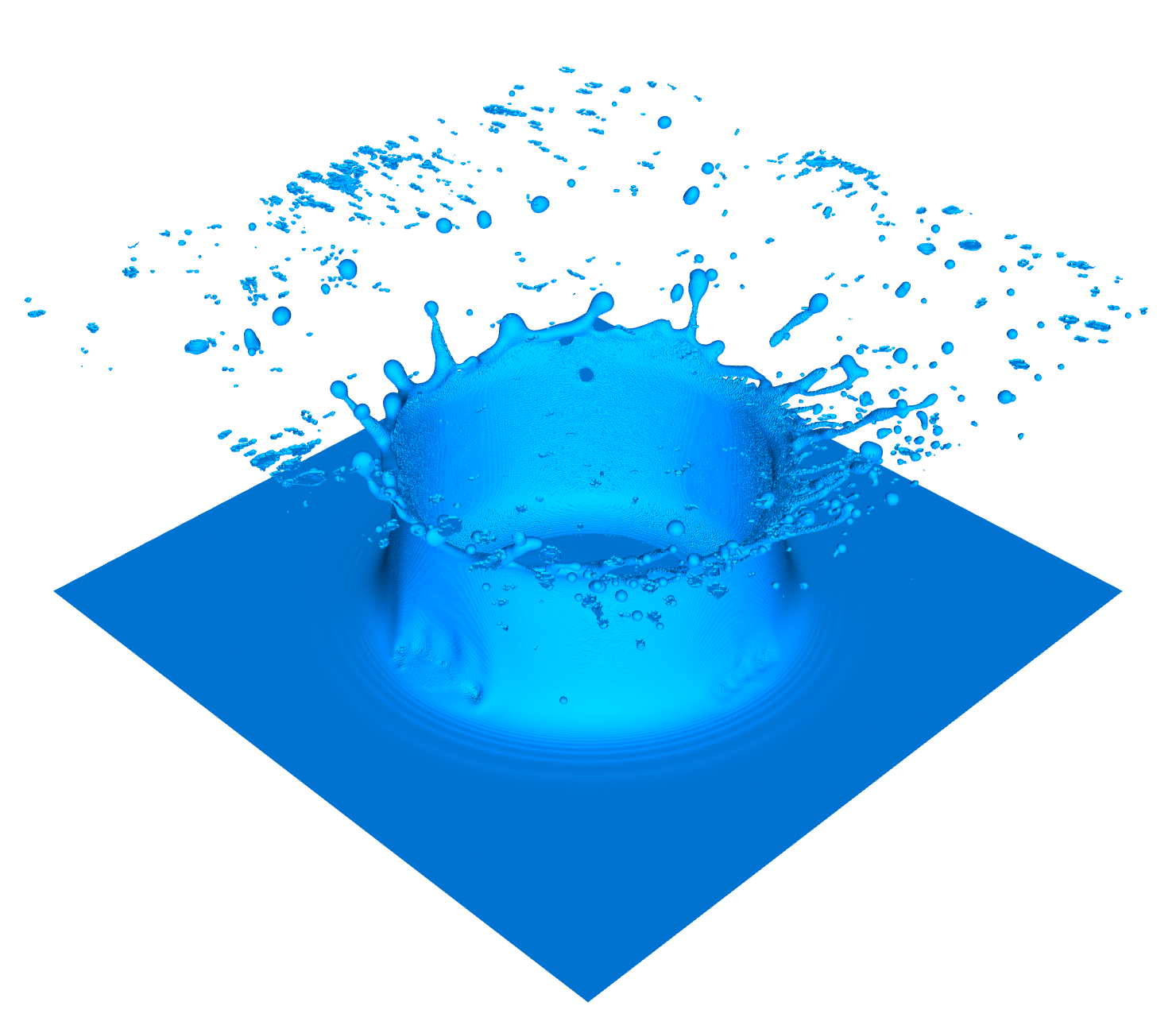} \\
	\end{tabular}
	\caption{
		\label{fig:drop-mesh}
		Simulated splash crown of the drop impact at non-dimensionalized time $t^{*}=12$.
		The simulations were performed with a computational domain resolution, that is, initial drop diameter of $D=80$ lattice cells.
		The solid black lines illustrate the crown's contour in a centrally located cross-section with normal in the $x$-direction.
		All but the OM variant showed significant anisotropic artifacts.
		In the NBRC and NBKC variant, droplets that have detached from the splash crown in an earlier phase, have already hit the liquid pool's surface.
		The photograph of the laboratory experiment is reproduced from A.-B.\ Wang, C.-C.\ Chen, Splashing impact of a single drop onto very thin liquid films~\cite{wang2000SplashingImpactSingle}, Physics of Fluids, 12, 2000, with the permission of AIP Publishing.
	}
\end{figure}

\Cref{fig:drop-mesh} shows the drop impact, that is, splash crown formation at $t^{*}=12$ with the solid black line indicating the contour in a central cross-section with normal in the $x$-direction.
Since no scale bars are provided in the photographs of the experiment~\cite{wang2000SplashingImpactSingle}, the simulations could only be validated and compared qualitatively.
Visually, the OM variant produced the most realistic results.
More specifically, in agreement with the observations for the cylindrical dam break in \Cref{sec:ne-cdb-rad}, it showed the least anisotropic behavior of all tested variants.
In contrast, the NBRC and NBKC variant overestimated splashing, that is, the detachment of smaller droplets.
However, these droplets had the shape of a thread rather than of a sphere as in the reference experiments.
Eventually, the droplets fell due to the influence of gravity and reached the liquid film's surface, as can be seen by the impacts there.
The OM3 variant was more accurate than the normal-based variants but was also subject to anisotropy, as clearly visible in the side view in \Cref{fig:drop-mesh}.
\par

Again, as in the numerical experiments from the preceding sections, the OM variant was the most accurate in this test case.
\par

\FloatBarrier

%% file: figures/gravity-wave/setup.tex
\begin{tikzpicture}
\definecolor{darkorange24213334}{RGB}{242,133,34}
\definecolor{dodgerblue0154222}{RGB}{0,154,222}

\begin{axis}%
[width=\figurewidth,
height=\figureheight,
xmin=0,
xmax=1,
ymin=0,
ymax=1,
ticks=none,
axis lines=none,
clip=false,
scale only axis
]
\addplot[thick, name path=f, domain=0:1,samples=50,smooth,dodgerblue0154222] {0.5+0.1*cos(deg(pi*x*2))};

\path[name path=axis] (axis cs:0,0) -- (axis cs:1,0);

\addplot [thick, color=dodgerblue0154222, fill=dodgerblue0154222!30] fill between[of=f and axis,	soft clip={domain=0:1}];
\end{axis}

% borders left and right
\draw[very thick, loosely dashed, black!50] (0,0)--(0,\figureheight);
\draw[very thick, loosely dashed, black!50] (\figurewidth,0)--(\figurewidth,\figureheight);

% borders top and bottom
\draw[very thick,, black!50] (0,\figureheight)--(\figurewidth,\figureheight);
\draw[very thick,, black!50] (0,0)--(\figurewidth,0);

% domain height
\draw[<->, >=Latex] (\figurewidth+\mdist,0)--(\figurewidth+\mdist,\figureheight) node [pos=0.5,right] {$L$};
\draw[-] (\figurewidth,0)--(\figurewidth+1.5\mdist,0);
\draw[-] (\figurewidth,\figureheight)--(\figurewidth+1.5\mdist,\figureheight);

% domain width
\draw[<->, >=Latex] (0,-\mdist)--(\figurewidth,-\mdist) node [pos=0.5,below] {$L$};
\draw[-] (0,0)--(0,-1.5\mdist) node [below] {$x=0$};
\draw[-] (\figurewidth,0)--(\figurewidth,-1.5\mdist);

% liquid height
\draw[dotted] (0,0.5\figureheight)--(\figurewidth,0.5\figureheight);
\draw[<->, >=Latex] (-\mdist,0)--(-\mdist,0.5\figureheight) node [pos=0.5,left] {$d$};
\draw[-] (-1.5\mdist,0)--(0,0);
\draw[-] (-1.5\mdist,0.5\figureheight)--(0,0.5\figureheight);

% initial amplitude
\draw[<->, >=Latex] (-\mdist,0.5\figureheight)--(-\mdist,0.6\figureheight) node [pos=0.5,left] {$a_{0}$};
\draw[-] (-1.5\mdist,0.6\figureheight)--(0,0.6\figureheight);

% initial profile label
\node[
rectangle,
anchor=south,
dodgerblue0154222] at (0.5\figurewidth,0.52\figureheight) {$y(x) = d + a_{0} \cos\left(kx \right)$};

% vertical line of symmetry
%\draw[loosely dashdotted] (0.5\figurewidth,-0.025\figureheight)--(0.5\figurewidth,1.025\figureheight);

% gravity
\draw[thick, ->, >=Latex, darkorange24213334] (\figurewidth-2\mdist,\figureheight-\mdist)--(\figurewidth-2\mdist,\figureheight-4\mdist) node [pos=0.5,right] {$g$};

% coordinate system
\draw[->, >=Latex] (2\mdist,\figureheight-3\mdist)--(4\mdist,\figureheight-3\mdist) node [below] {$x$};
\draw[->, >=Latex] (2\mdist,\figureheight-3\mdist)--(2\mdist,\figureheight-1\mdist) node [left] {$y$};
\end{tikzpicture}

%% file: figures/dam-break-rectangular/setup.tex
\begin{tikzpicture}
\definecolor{darkorange24213334}{RGB}{242,133,34}
\definecolor{dodgerblue0154222}{RGB}{0,154,222}

% liquid column
\draw[thick, fill=dodgerblue0154222!20, draw=none] (0,0) rectangle (0.25\figurewidth,0.5\figureheight);
\draw[thick, dodgerblue0154222] (0,0.5\figureheight)--(0.25\figurewidth,0.5\figureheight);
\draw[thick, dodgerblue0154222] (0.25\figurewidth,0)--(0.25\figurewidth,0.5\figureheight);

% borders left and right
\draw[very thick, black!50] (0,0)--(0,\figureheight);
\draw[very thick, black!50] (\figurewidth,0)--(\figurewidth,\figureheight);

% borders top and bottom
\draw[very thick, black!50] (0,\figureheight)--(\figurewidth,\figureheight);
\draw[very thick, black!50] (0,0)--(\figurewidth,0);

% domain height
\draw[<->, >=Latex] (\figurewidth+\mdist,0)--(\figurewidth+\mdist,\figureheight) node [pos=0.5,right] {$2H$};
\draw[-] (\figurewidth,0)--(\figurewidth+2\mdist,0);
\draw[-] (\figurewidth,\figureheight)--(\figurewidth+2\mdist,\figureheight);

% domain width
\draw[<->, >=Latex] (0,-\mdist)--(\figurewidth,-\mdist) node [pos=0.5,below] {$15 W$};
\draw[-] (0,0)--(0,-2\mdist);
\draw[-] (\figurewidth,0)--(\figurewidth,-2\mdist);

% liquid column height
\draw[<->, >=Latex] (0.25\figurewidth+\mdist,0)--(0.25\figurewidth+\mdist,0.5\figureheight) node [pos=0.5,right] {$H$};
\draw[-] (0.25\figurewidth,0.5\figureheight)--(0.25\figurewidth+2\mdist,0.5\figureheight);
\draw[-] (0.25\figurewidth,0)--(0.25\figurewidth+2\mdist,0);

% liquid column diameter
\draw[<->, >=Latex] (0,0.5\figureheight+\mdist)--(0.25\figurewidth,0.5\figureheight+\mdist) node [pos=0.5,above] {$W$};
\draw[-] (0,0.5\figureheight)--(0,0.5\figureheight+2\mdist);
\draw[-] (0.25\figurewidth,0.5\figureheight)--(0.25\figurewidth,0.5\figureheight+2\mdist);

% gravity
\draw[thick, ->, >=Latex, darkorange24213334] (\figurewidth-3\mdist,\figureheight-\mdist)--(\figurewidth-3\mdist,\figureheight-4\mdist) node [pos=0.5,right] {$g$};

% coordinate system
\draw[->, >=Latex] (2\mdist,\figureheight-3\mdist)--(4\mdist,\figureheight-3\mdist) node [below] {$x$};
\draw[->, >=Latex] (2\mdist,\figureheight-3\mdist)--(2\mdist,\figureheight-1\mdist) node [left] {$y$};
\end{tikzpicture}

%% file: figures/dam-break-rectangular/w-200-y.tex
% This file was created with tikzplotlib v0.10.1.
\begin{tikzpicture}

\definecolor{darkgray176}{RGB}{176,176,176}
\definecolor{darkorange24213334}{RGB}{242,133,34}
\definecolor{dodgerblue0154222}{RGB}{0,154,222}
\definecolor{lightgray204}{RGB}{204,204,204}
\definecolor{mediumorchid17588186}{RGB}{175,88,186}
\definecolor{springgreen0205108}{RGB}{0,205,108}

\begin{axis}[
height=\figureheight,
legend cell align={left},
legend style={
  fill opacity=0.8,
  draw opacity=1,
  text opacity=1,
  at={(0.99,0.99)},
  anchor=north east,
  draw=lightgray204
},
tick align=outside,
tick pos=left,
width=\figurewidth,
x grid style={darkgray176},
xlabel={\(\displaystyle t^{*}\)},
xmajorgrids,
xmin=0, xmax=10,
xtick style={color=black},
y grid style={darkgray176},
ylabel style={rotate=-90.0},
ylabel={\(\displaystyle h^{*}\)},
ymajorgrids,
ymin=-0.032625, ymax=2.005125,
ytick style={color=black}
]
\addplot [very thick, black, mark=*, mark size=2, mark options={solid}, only marks]
table {%
0 1
0.559999942779541 0.940000057220459
0.769999980926514 0.889999985694885
0.930000066757202 0.829999923706055
1.08000004291534 0.779999971389771
1.27999997138977 0.720000028610229
1.46000003814697 0.670000076293945
1.6599999666214 0.610000014305115
1.8400000333786 0.559999942779541
2 0.5
2.21000003814697 0.440000057220459
2.45000004768372 0.389999985694885
2.70000004768372 0.330000042915344
3.05999994277954 0.279999971389771
3.44000005722046 0.220000028610229
4.19999980926514 0.169999957084656
5.25 0.110000014305115
7.40000009536743 0.059999942779541
};
\addlegendentry{Experiment~\cite{martin1952PartIVExperimental}}
\addplot [line width=2pt, dodgerblue0154222, dotted]
table {%
0 1
0.099940299987793 0.997499942779541
0.199880957603455 0.990000009536743
0.299821019172668 0.995000004768372
0.399760961532593 1.00999999046326
0.499701023101807 0.995000004768372
0.599642038345337 1.07000005245209
0.799521923065186 1.19500005245209
0.899461984634399 1.25499999523163
0.99940299987793 1.30999994277954
1.09933996200562 1.36749994754791
1.19928002357483 1.41999995708466
1.29921996593475 1.47000002861023
1.39916002750397 1.51750004291534
1.49909996986389 1.5625
1.69897997379303 1.64750003814697
1.89885997772217 1.7224999666214
2.09875011444092 1.78250002861023
2.19868993759155 1.80999994277954
2.29862999916077 1.83249998092651
2.39857006072998 1.85749995708466
2.49850988388062 1.87750005722046
2.59844994544983 1.89250004291534
2.69839000701904 1.90499997138977
2.79833006858826 1.90750002861023
2.89826989173889 1.91250002384186
2.99820995330811 1.91250002384186
};
\addlegendentry{NBRC, $W=200$}
\addplot [line width=2pt, springgreen0205108]
table {%
0 1
0.199880957603455 0.995000004768372
0.499701023101807 0.972500085830688
0.599642038345337 0.960000038146973
0.799521923065186 0.930000066757202
0.99940299987793 0.894999980926514
1.09933996200562 0.875
1.39916002750397 0.807500004768372
1.49909996986389 0.782500028610229
1.59904003143311 0.759999990463257
1.89885997772217 0.684999942779541
1.9988100528717 0.664999961853027
2.19868993759155 0.620000004768372
2.59844994544983 0.539999961853027
2.69839000701904 0.522500038146973
2.89826989173889 0.482499957084656
2.99820995330811 0.465000033378601
3.09815001487732 0.450000047683716
3.19809007644653 0.432500004768372
3.29802989959717 0.417500019073486
3.39796996116638 0.399999976158142
3.4979100227356 0.387500047683716
3.59785008430481 0.370000004768372
3.79772996902466 0.340000033378601
3.89767003059387 0.327499985694885
3.99761009216309 0.317499995231628
4.09754991531372 0.305000066757202
4.49731016159058 0.264999985694885
4.59724998474121 0.257499933242798
4.69718980789185 0.247499942779541
4.79713010787964 0.240000009536743
4.89706993103027 0.230000019073486
4.99701023101807 0.225000023841858
5.0969500541687 0.215000033378601
5.39677000045776 0.192499995231628
5.4967098236084 0.1875
5.59665012359619 0.180000066757202
5.69658994674683 0.177500009536743
5.89648008346558 0.167500019073486
5.99641990661621 0.159999966621399
6.19630002975464 0.154999971389771
6.39618015289307 0.144999980926514
6.4961199760437 0.142500042915344
6.59605979919434 0.137500047683716
6.89588022232056 0.129999995231628
6.99582004547119 0.125
7.09575986862183 0.125
7.19570016860962 0.120000004768372
};
\addlegendentry{NBKC, $W=200$}
\addplot [line width=2pt, mediumorchid17588186, dashed]
table {%
0 1
0.199880957603455 0.995000004768372
0.299821019172668 0.990000009536743
0.399760961532593 0.982500076293945
0.499701023101807 0.972500085830688
0.599642038345337 0.960000038146973
0.799521923065186 0.930000066757202
0.99940299987793 0.894999980926514
1.09933996200562 0.872499942779541
1.19928002357483 0.852499961853027
1.29921996593475 0.829999923706055
1.39916002750397 0.805000066757202
1.49909996986389 0.782500028610229
1.59904003143311 0.757499933242798
1.69897997379303 0.735000014305115
1.79892003536224 0.710000038146973
2.09875011444092 0.642499923706055
2.19868993759155 0.622499942779541
2.29862999916077 0.600000023841858
2.49850988388062 0.559999942779541
2.59844994544983 0.542500019073486
2.79833006858826 0.502500057220459
2.99820995330811 0.467499971389771
3.09815001487732 0.452499985694885
3.19809007644653 0.434999942779541
3.69778990745544 0.360000014305115
4.19749021530151 0.297500014305115
4.49731016159058 0.267500042915344
4.59724998474121 0.259999990463257
4.69718980789185 0.25
5.19688987731934 0.212499976158142
5.29683017730713 0.207499980926514
5.59665012359619 0.184999942779541
5.69658994674683 0.182500004768372
5.79653978347778 0.174999952316284
6.39618015289307 0.144999980926514
6.4961199760437 0.144999980926514
6.59605979919434 0.139999985694885
6.69600009918213 0.137500047683716
6.79593992233276 0.132499933242798
6.99582004547119 0.127500057220459
7.09575986862183 0.122499942779541
7.19570016860962 0.122499942779541
7.29563999176025 0.120000004768372
7.39557981491089 0.115000009536743
7.49552011489868 0.115000009536743
7.89527988433838 0.105000019073486
7.99522018432617 0.105000019073486
8.19509983062744 0.100000023841858
8.29504013061523 0.100000023841858
8.4949197769165 0.0950000286102295
8.5948600769043 0.0950000286102295
8.69480037689209 0.0924999713897705
8.79473972320557 0.0924999713897705
8.89468002319336 0.0900000333786011
8.99462032318115 0.0900000333786011
};
\addlegendentry{OM, $W=200$}
\addplot [line width=2pt, darkorange24213334, dash pattern=on 1pt off 3pt on 3pt off 3pt]
table {%
0 1
0.199880957603455 0.995000004768372
0.299821019172668 0.990000009536743
0.399760961532593 0.980000019073486
0.499701023101807 0.972500085830688
0.599642038345337 0.960000038146973
0.799521923065186 0.930000066757202
0.99940299987793 0.894999980926514
1.09933996200562 0.872499942779541
1.19928002357483 0.852499961853027
1.29921996593475 0.829999923706055
1.39916002750397 0.805000066757202
1.49909996986389 0.782500028610229
1.69897997379303 0.732500076293945
2.19868993759155 0.620000004768372
2.69839000701904 0.519999980926514
2.89826989173889 0.485000014305115
2.99820995330811 0.465000033378601
3.09815001487732 0.450000047683716
3.19809007644653 0.432500004768372
3.4979100227356 0.387500047683716
3.59785008430481 0.375
3.89767003059387 0.330000042915344
4.09754991531372 0.305000066757202
4.49731016159058 0.264999985694885
4.59724998474121 0.257499933242798
4.69718980789185 0.247499942779541
5.0969500541687 0.217499971389771
5.19688987731934 0.207499980926514
5.39677000045776 0.197499990463257
5.4967098236084 0.190000057220459
5.59665012359619 0.184999942779541
5.69658994674683 0.177500009536743
6.096360206604 0.157500028610229
6.19630002975464 0.154999971389771
6.4961199760437 0.139999985694885
6.59605979919434 0.137500047683716
6.79593992233276 0.127500057220459
6.89588022232056 0.127500057220459
6.99582004547119 0.122499942779541
7.09575986862183 0.122499942779541
7.49552011489868 0.112499952316284
7.59545993804932 0.112499952316284
7.89527988433838 0.105000019073486
};
\addlegendentry{OM3, $W=200$}
\end{axis}

\end{tikzpicture}

%% file: figures/dam-break-rectangular/w-200-x.tex
% This file was created with tikzplotlib v0.10.1.
\begin{tikzpicture}

\definecolor{darkgray176}{RGB}{176,176,176}
\definecolor{darkorange24213334}{RGB}{242,133,34}
\definecolor{dodgerblue0154222}{RGB}{0,154,222}
\definecolor{lightgray204}{RGB}{204,204,204}
\definecolor{mediumorchid17588186}{RGB}{175,88,186}
\definecolor{springgreen0205108}{RGB}{0,205,108}

\begin{axis}[
height=\figureheight,
legend cell align={left},
legend style={
  fill opacity=0.8,
  draw opacity=1,
  text opacity=1,
  at={(0.99,0.01)},
  anchor=south east,
  draw=lightgray204
},
tick align=outside,
tick pos=left,
width=\figurewidth,
x grid style={darkgray176},
xlabel={\(\displaystyle t^{*}\)},
xmajorgrids,
xmin=0, xmax=10,
xtick style={color=black},
y grid style={darkgray176},
ylabel style={rotate=-90.0},
ylabel={\(\displaystyle w^{*}\)},
ymajorgrids,
ymin=0.29675, ymax=15.54825,
ytick style={color=black}
]
\addplot [very thick, black, mark=*, mark size=2, mark options={solid}, only marks]
table {%
0.409999966621399 1.11000001430511
0.839999914169312 1.22000002861023
1.19000005722046 1.44000005722046
1.42999994754791 1.66999995708466
1.62999999523163 1.88999998569489
1.83000004291534 2.10999989509583
1.98000001907349 2.32999992370605
2.20000004768372 2.55999994277954
2.3199999332428 2.77999997138977
2.50999999046326 3
2.65000009536743 3.22000002861023
2.82999992370605 3.44000005722046
2.97000002861023 3.67000007629395
3.10999989509583 3.89000010490417
3.32999992370605 4.1100001335144
4.01999998092651 5
4.44000005722046 5.8899998664856
5.09000015258789 7
5.69000005722046 8
6.30000019073486 9
6.82999992370605 10
7.44000005722046 11
8.07999992370605 12
8.67000007629395 13
9.3100004196167 14
};
\addlegendentry{Experiment~\cite{martin1952PartIVExperimental}}
\addplot [line width=2pt, dodgerblue0154222, dotted]
table {%
0 1
0.099940299987793 0.990000009536743
0.199880957603455 1.08000004291534
0.299821019172668 1.15499997138977
0.399760961532593 1.25
0.499701023101807 1.375
0.599642038345337 1.49000000953674
0.699581980705261 1.61000001430511
0.799521923065186 1.72500002384186
0.899461984634399 1.89499998092651
0.99940299987793 2.21000003814697
1.59904003143311 4.15500020980835
1.9988100528717 5.42999982833862
2.69839000701904 7.63000011444092
2.99820995330811 8.54500007629395
};
\addlegendentry{NBRC, $W=200$}
\addplot [line width=2pt, springgreen0205108]
table {%
0 1
0.099940299987793 1.00499999523163
0.199880957603455 1.02499997615814
0.299821019172668 1.05499994754791
0.499701023101807 1.14499998092651
0.599642038345337 1.21000003814697
0.699581980705261 1.26999998092651
0.799521923065186 1.3400000333786
0.899461984634399 1.41999995708466
0.99940299987793 1.50499999523163
1.09933996200562 1.59500002861023
1.19928002357483 1.69000005722046
1.39916002750397 1.88999998569489
1.49909996986389 1.99500000476837
1.79892003536224 2.33999991416931
1.9988100528717 2.58999991416931
2.19868993759155 2.84999990463257
2.39857006072998 3.11999988555908
2.49850988388062 3.26500010490417
2.69839000701904 3.54500007629395
3.09815001487732 4.15000009536743
3.59785008430481 4.95499992370605
3.79772996902466 5.28999996185303
3.89767003059387 5.46000003814697
4.09754991531372 5.81500005722046
4.29743003845215 6.15999984741211
4.79713010787964 7.00500011444092
4.99701023101807 7.36999988555908
5.39677000045776 8.07499980926514
5.89648008346558 8.97500038146973
5.99641990661621 9.17500019073486
6.096360206604 9.39000034332275
6.19630002975464 9.63500022888184
6.29623985290527 9.89000034332275
6.39618015289307 10.1850004196167
6.4961199760437 10.5450000762939
6.59605979919434 11.0600004196167
6.69600009918213 11.8050003051758
6.79593992233276 11.6750001907349
6.89588022232056 11.9049997329712
6.99582004547119 12.1499996185303
7.09575986862183 12.414999961853
7.19570016860962 14.8549995422363
};
\addlegendentry{NBKC, $W=200$}
\addplot [line width=2pt, mediumorchid17588186, dashed]
table {%
0 1
0.099940299987793 1.00499999523163
0.199880957603455 1.02499997615814
0.299821019172668 1.05999994277954
0.399760961532593 1.10000002384186
0.499701023101807 1.14999997615814
0.599642038345337 1.21000003814697
0.699581980705261 1.27499997615814
0.799521923065186 1.35000002384186
0.99940299987793 1.50999999046326
1.09933996200562 1.60000002384186
1.19928002357483 1.69500005245209
1.39916002750397 1.89499998092651
1.59904003143311 2.11500000953674
1.69897997379303 2.23000001907349
1.89885997772217 2.48000001907349
1.9988100528717 2.60999989509583
2.09875011444092 2.74499988555908
2.19868993759155 2.875
2.29862999916077 3.00999999046326
2.39857006072998 3.15000009536743
2.59844994544983 3.42000007629395
2.69839000701904 3.55999994277954
2.79833006858826 3.71000003814697
2.89826989173889 3.86999988555908
2.99820995330811 4.03999996185303
3.09815001487732 4.19500017166138
3.19809007644653 4.3600001335144
3.39796996116638 4.7350001335144
3.79772996902466 5.40999984741211
3.89767003059387 5.57000017166138
4.09754991531372 5.91499996185303
4.19749021530151 6.07999992370605
4.39736986160278 6.43499994277954
4.49731016159058 6.61499977111816
4.79713010787964 7.13000011444092
4.89706993103027 7.30499982833862
5.0969500541687 7.67500019073486
5.19688987731934 7.86999988555908
5.29683017730713 8.05500030517578
5.4967098236084 8.40999984741211
5.79653978347778 8.94499969482422
5.99641990661621 9.27999973297119
6.096360206604 9.4350004196167
6.19630002975464 9.60499954223633
6.29623985290527 9.78499984741211
6.39618015289307 9.97999954223633
6.4961199760437 10.164999961853
6.69600009918213 10.5100002288818
6.89588022232056 10.8299999237061
7.29563999176025 11.5349998474121
7.49552011489868 11.8950004577637
7.69540023803711 12.2799997329712
7.99522018432617 12.8249998092651
8.09515953063965 13.0150003433228
8.4949197769165 13.7299995422363
8.5948600769043 13.8950004577637
8.69480037689209 14.0699996948242
8.79473972320557 14.2550001144409
8.89468002319336 14.4300003051758
8.99462032318115 14.5950002670288
};
\addlegendentry{OM, $W=200$}
\addplot [line width=2pt, darkorange24213334, dash pattern=on 1pt off 3pt on 3pt off 3pt]
table {%
0 1
0.099940299987793 1.00499999523163
0.199880957603455 1.02499997615814
0.299821019172668 1.05999994277954
0.399760961532593 1.10000002384186
0.499701023101807 1.14999997615814
0.599642038345337 1.21000003814697
0.699581980705261 1.27999997138977
0.899461984634399 1.42999994754791
0.99940299987793 1.51499998569489
1.09933996200562 1.60500001907349
1.29921996593475 1.80499994754791
1.39916002750397 1.9099999666214
1.49909996986389 2.01999998092651
1.59904003143311 2.13499999046326
1.69897997379303 2.23499989509583
1.79892003536224 2.35500001907349
1.89885997772217 2.49000000953674
1.9988100528717 2.61500000953674
2.09875011444092 2.73499989509583
2.19868993759155 2.875
2.29862999916077 2.99499988555908
2.39857006072998 3.11999988555908
2.49850988388062 3.26500010490417
2.59844994544983 3.43499994277954
2.69839000701904 3.58500003814697
2.79833006858826 3.74499988555908
2.89826989173889 3.875
2.99820995330811 4.01000022888184
3.19809007644653 4.32000017166138
3.29802989959717 4.45499992370605
3.39796996116638 4.60500001907349
3.69778990745544 5.1100001335144
3.79772996902466 5.25500011444092
3.89767003059387 5.44500017166138
3.99761009216309 5.59999990463257
4.09754991531372 5.7649998664856
4.29743003845215 6.07999992370605
4.49731016159058 6.39499998092651
4.59724998474121 6.54500007629395
4.69718980789185 6.7350001335144
4.79713010787964 6.94999980926514
4.89706993103027 7.13500022888184
4.99701023101807 7.30499982833862
5.0969500541687 7.46500015258789
5.19688987731934 7.6399998664856
5.39677000045776 8.04500007629395
5.59665012359619 8.40999984741211
5.69658994674683 8.60000038146973
5.79653978347778 8.77000045776367
5.89648008346558 8.96500015258789
6.29623985290527 9.60000038146973
6.39618015289307 9.83500003814697
6.4961199760437 10.0450000762939
6.79593992233276 10.5799999237061
6.89588022232056 10.7449998855591
7.49552011489868 11.6800003051758
7.59545993804932 11.8500003814697
7.69540023803711 12.0100002288818
7.79534006118774 14.2049999237061
7.89527988433838 14.6350002288818
};
\addlegendentry{OM3, $W=200$}
\end{axis}

\end{tikzpicture}

%% file: figures/dam-break-cylindrical/setup.tex
\begin{tikzpicture}
\definecolor{darkorange24213334}{RGB}{242,133,34}
\definecolor{dodgerblue0154222}{RGB}{0,154,222}

% liquid column
\draw[thick, fill=dodgerblue0154222!20, draw=none] (0.5\figurewidth-0.25\figureheight,0) rectangle (0.5\figurewidth+0.25\figureheight,0.5\figureheight);
\draw[thick, dodgerblue0154222] (0.5\figurewidth-0.25\figureheight,0.5\figureheight)--(0.5\figurewidth+0.25\figureheight,0.5\figureheight);
\draw[thick, dodgerblue0154222] (0.5\figurewidth-0.25\figureheight,0)--(0.5\figurewidth-0.25\figureheight,0.5\figureheight);
\draw[thick, dodgerblue0154222] (0.5\figurewidth+0.25\figureheight,0)--(0.5\figurewidth+0.25\figureheight,0.5\figureheight);

% borders left and right
\draw[very thick, black!50] (0,0)--(0,\figureheight);
\draw[very thick, black!50] (\figurewidth,0)--(\figurewidth,\figureheight);

% borders top and bottom
\draw[very thick, black!50] (0,\figureheight)--(\figurewidth,\figureheight);
\draw[very thick, black!50] (0,0)--(\figurewidth,0);

% domain height
\draw[<->, >=Latex] (\figurewidth+\mdist,0)--(\figurewidth+\mdist,\figureheight) node [pos=0.5,right] {$2H$};
\draw[-] (\figurewidth,0)--(\figurewidth+2\mdist,0);
\draw[-] (\figurewidth,\figureheight)--(\figurewidth+2\mdist,\figureheight);

% domain width
\draw[<->, >=Latex] (0,-\mdist)--(\figurewidth,-\mdist) node [pos=0.5,below] {$\varnothing 6D$};
\draw[-] (0,0)--(0,-2\mdist);
\draw[-] (\figurewidth,0)--(\figurewidth,-2\mdist);

% liquid column height
\draw[<->, >=Latex] (0.5\figurewidth+0.25\figureheight+\mdist,0)--(0.5\figurewidth+0.25\figureheight+\mdist,0.5\figureheight) node [pos=0.5,right] {$H$};
\draw[-] (0.5\figurewidth+0.25\figureheight,0.5\figureheight)--(0.5\figurewidth+0.25\figureheight+2\mdist,0.5\figureheight);
\draw[-] (0.5\figurewidth+0.25\figureheight,0)--(0.5\figurewidth+0.25\figureheight+2\mdist,0);

% liquid column diameter
\draw[<->, >=Latex] (0.5\figurewidth-0.25\figureheight,0.5\figureheight+\mdist)--(0.5\figurewidth+0.25\figureheight,0.5\figureheight+\mdist) node [pos=0.5,above] {$\varnothing D$};
\draw[-] (0.5\figurewidth-0.25\figureheight,0.5\figureheight)--(0.5\figurewidth-0.25\figureheight,0.5\figureheight+2\mdist);
\draw[-] (0.5\figurewidth+0.25\figureheight,0.5\figureheight)--(0.5\figurewidth+0.25\figureheight,0.5\figureheight+2\mdist);

% vertical line of symmetry
\draw[loosely dashdotted] (0.5\figurewidth,-0.025\figureheight)--(0.5\figurewidth,1.025\figureheight);

% gravity
\draw[thick, ->, >=Latex, darkorange24213334] (\figurewidth-3\mdist,\figureheight-\mdist)--(\figurewidth-3\mdist,\figureheight-4\mdist) node [pos=0.5,right] {$g$};

% coordinate system
\draw[->, >=Latex] (2\mdist,\figureheight-3\mdist)--(4\mdist,\figureheight-3\mdist) node [below] {$x$};
\draw[->, >=Latex] (2\mdist,\figureheight-3\mdist)--(2\mdist,\figureheight-1\mdist) node [left] {$z$};
\draw[draw=black] (2\mdist,\figureheight-3\mdist) circle [radius=0.2\mdist] node[opacity=1, below left] {$y$};
\end{tikzpicture}

%% file: figures/dam-break-cylindrical/d-200.tex
% This file was created with tikzplotlib v0.10.1.
\begin{tikzpicture}

\definecolor{darkgray176}{RGB}{176,176,176}
\definecolor{darkorange24213334}{RGB}{242,133,34}
\definecolor{dodgerblue0154222}{RGB}{0,154,222}
\definecolor{lightgray204}{RGB}{204,204,204}
\definecolor{mediumorchid17588186}{RGB}{175,88,186}
\definecolor{springgreen0205108}{RGB}{0,205,108}

\begin{axis}[
height=\figureheight,
legend cell align={left},
legend style={
  fill opacity=0.8,
  draw opacity=1,
  text opacity=1,
  at={(0.01,0.99)},
  anchor=north west,
  draw=lightgray204
},
tick align=outside,
tick pos=left,
width=\figurewidth,
x grid style={darkgray176},
xlabel={\(\displaystyle t^{*}\)},
xmajorgrids,
xmin=0, xmax=4,
xtick style={color=black},
y grid style={darkgray176},
ylabel style={rotate=-90.0},
ylabel={\(\displaystyle r^{*}\)},
ymajorgrids,
ymin=0.7340387, ymax=6.4160953,
ytick style={color=black}
]
\addplot [semithick, black, mark=*, mark size=3, mark options={solid}, only marks]
table {%
1.19000005722046 1.22000002861023
1.5 1.44000005722046
1.73000001907349 1.66999995708466
1.91999995708466 1.88999998569489
2.1800000667572 2.10999989509583
2.35999989509583 2.32999992370605
2.58999991416931 2.55999994277954
2.78999996185303 2.77999997138977
2.97000002861023 3
3.20000004768372 3.22000002861023
3.38000011444092 3.44000005722046
3.57999992370605 3.67000007629395
3.75999999046326 3.89000010490417
3.97000002861023 4.1100001335144
};
\addlegendentry{Experiment~\cite{martin1952PartIVExperimental}}
\path [draw=dodgerblue0154222]
(axis cs:0.0998772,0.992314)
--(axis cs:0.0998772,1.026866);

\path [draw=dodgerblue0154222]
(axis cs:0.499386,1.06634)
--(axis cs:0.499386,1.20894);

\path [draw=dodgerblue0154222]
(axis cs:0.898895,1.28295)
--(axis cs:0.898895,1.43203);

\path [draw=dodgerblue0154222]
(axis cs:1.2984,1.53411)
--(axis cs:1.2984,1.74831);

\path [draw=dodgerblue0154222]
(axis cs:1.69791,1.7917)
--(axis cs:1.69791,2.16474);

\path [draw=dodgerblue0154222]
(axis cs:2.09742,2.02401)
--(axis cs:2.09742,2.73717);

\path [draw=dodgerblue0154222]
(axis cs:2.49693,2.12507)
--(axis cs:2.49693,3.60855);

\path [draw=dodgerblue0154222]
(axis cs:2.89644,2.08431)
--(axis cs:2.89644,4.88923);

\addplot [very thick, dodgerblue0154222, mark=-, mark size=3, mark options={solid}, only marks, forget plot]
table {%
0.0998772382736206 0.992313981056213
0.499386072158813 1.06633996963501
0.898895025253296 1.28295004367828
1.29840004444122 1.53410995006561
1.69790995121002 1.79170000553131
2.09741997718811 2.02400994300842
2.49692988395691 2.12507009506226
2.89644002914429 2.08431005477905
};
\addplot [very thick, dodgerblue0154222, mark=-, mark size=3, mark options={solid}, only marks, forget plot]
table {%
0.0998772382736206 1.02686595916748
0.499386072158813 1.20894002914429
0.898895025253296 1.43202996253967
1.29840004444122 1.74830996990204
1.69790995121002 2.16474008560181
2.09741997718811 2.73716998100281
2.49692988395691 3.60855007171631
2.89644002914429 4.8892297744751
};
\path [draw=springgreen0205108]
(axis cs:0,1)
--(axis cs:0,1);

\path [draw=springgreen0205108]
(axis cs:0.399509,1.05195)
--(axis cs:0.399509,1.14179);

\path [draw=springgreen0205108]
(axis cs:0.799018,1.22012)
--(axis cs:0.799018,1.36388);

\path [draw=springgreen0205108]
(axis cs:1.19853,1.4655)
--(axis cs:1.19853,1.63636);

\path [draw=springgreen0205108]
(axis cs:1.59804,1.75655)
--(axis cs:1.59804,1.97943);

\path [draw=springgreen0205108]
(axis cs:1.99754,2.03211)
--(axis cs:1.99754,2.43051);

\path [draw=springgreen0205108]
(axis cs:2.39705,2.2642)
--(axis cs:2.39705,3.02662);

\path [draw=springgreen0205108]
(axis cs:2.79656,2.36487)
--(axis cs:2.79656,3.86677);

\path [draw=springgreen0205108]
(axis cs:3.19607,2.48213)
--(axis cs:3.19607,4.82811);

\path [draw=springgreen0205108]
(axis cs:3.59558,2.43576)
--(axis cs:3.59558,6.15782);

\addplot [very thick, springgreen0205108, mark=-, mark size=3, mark options={solid}, only marks, forget plot]
table {%
0 1
0.399508953094482 1.05194997787476
0.799018025398254 1.22011995315552
1.19852995872498 1.46549999713898
1.59803998470306 1.75654995441437
1.99753999710083 2.03210997581482
2.39704990386963 2.26419997215271
2.79656004905701 2.36487007141113
3.19606995582581 2.48213005065918
3.59558010101318 2.43576002120972
};
\addplot [very thick, springgreen0205108, mark=-, mark size=3, mark options={solid}, only marks, forget plot]
table {%
0 1
0.399508953094482 1.14179003238678
0.799018025398254 1.36388003826141
1.19852995872498 1.63636004924774
1.59803998470306 1.97942996025085
1.99753999710083 2.4305100440979
2.39704990386963 3.02661991119385
2.79656004905701 3.86677002906799
3.19606995582581 4.8281102180481
3.59558010101318 6.15782022476196
};
\path [draw=mediumorchid17588186]
(axis cs:0.199754,1.01106)
--(axis cs:0.199754,1.03902);

\path [draw=mediumorchid17588186]
(axis cs:0.599263,1.15637)
--(axis cs:0.599263,1.18661);

\path [draw=mediumorchid17588186]
(axis cs:0.998772,1.38493)
--(axis cs:0.998772,1.41763);

\path [draw=mediumorchid17588186]
(axis cs:1.39828,1.6719)
--(axis cs:1.39828,1.70488);

\path [draw=mediumorchid17588186]
(axis cs:1.79779,2.00306)
--(axis cs:1.79779,2.06358);

\path [draw=mediumorchid17588186]
(axis cs:2.1973,2.34277)
--(axis cs:2.1973,2.43321);

\path [draw=mediumorchid17588186]
(axis cs:2.59681,2.7344)
--(axis cs:2.59681,2.8474);

\path [draw=mediumorchid17588186]
(axis cs:2.99632,3.13147)
--(axis cs:2.99632,3.30867);

\path [draw=mediumorchid17588186]
(axis cs:3.39583,3.58253)
--(axis cs:3.39583,3.77999);

\path [draw=mediumorchid17588186]
(axis cs:3.79533,4.03773)
--(axis cs:3.79533,4.26837);

\path [draw=mediumorchid17588186]
(axis cs:4.19484,4.50736)
--(axis cs:4.19484,4.75158);

\path [draw=mediumorchid17588186]
(axis cs:4.59435,4.93004)
--(axis cs:4.59435,5.29408);

\addplot [very thick, mediumorchid17588186, mark=-, mark size=3, mark options={solid}, only marks, forget plot]
table {%
0.199753999710083 1.01105999946594
0.599262952804565 1.15637004375458
0.998772025108337 1.38493001461029
1.3982800245285 1.67190003395081
1.79779005050659 2.00306010246277
2.19729995727539 2.34277009963989
2.59681010246277 2.73440003395081
2.99632000923157 3.13146996498108
3.39582991600037 3.58253002166748
3.79533004760742 4.03773021697998
4.19483995437622 4.50735998153687
};
\addplot [very thick, mediumorchid17588186, mark=-, mark size=3, mark options={solid}, only marks, forget plot]
table {%
0.199753999710083 1.03901994228363
0.599262952804565 1.18660998344421
0.998772025108337 1.4176299571991
1.3982800245285 1.70487999916077
1.79779005050659 2.06358003616333
2.19729995727539 2.43320989608765
2.59681010246277 2.84739995002747
2.99632000923157 3.30867004394531
3.39582991600037 3.77998995780945
3.79533004760742 4.26837015151978
4.19483995437622 4.75158023834229
};
\path [draw=darkorange24213334]
(axis cs:0.299632,1.02549)
--(axis cs:0.299632,1.08149);

\path [draw=darkorange24213334]
(axis cs:0.69914,1.19877)
--(axis cs:0.69914,1.25449);

\path [draw=darkorange24213334]
(axis cs:1.09865,1.43287)
--(axis cs:1.09865,1.50967);

\path [draw=darkorange24213334]
(axis cs:1.49816,1.72225)
--(axis cs:1.49816,1.82013);

\path [draw=darkorange24213334]
(axis cs:1.89767,2.02801)
--(axis cs:1.89767,2.20865);

\path [draw=darkorange24213334]
(axis cs:2.29718,2.35582)
--(axis cs:2.29718,2.65278);

\path [draw=darkorange24213334]
(axis cs:2.69668,2.67884)
--(axis cs:2.69668,3.18308);

\path [draw=darkorange24213334]
(axis cs:3.09619,2.94247)
--(axis cs:3.09619,3.82919);

\path [draw=darkorange24213334]
(axis cs:3.4957,3.18632)
--(axis cs:3.4957,4.51592);

\path [draw=darkorange24213334]
(axis cs:3.89521,3.40316)
--(axis cs:3.89521,5.27662);

\addplot [very thick, darkorange24213334, mark=-, mark size=3, mark options={solid}, only marks, forget plot]
table {%
0.29963207244873 1.02549004554749
0.699140071868896 1.19877004623413
1.0986499786377 1.43287003040314
1.49816000461578 1.72224998474121
1.89767003059387 2.02800989151001
2.29717993736267 2.35581994056702
2.69668006896973 2.67883992195129
3.09618997573853 2.94247007369995
3.49569988250732 3.18632006645203
3.8952100276947 3.40316009521484
};
\addplot [very thick, darkorange24213334, mark=-, mark size=3, mark options={solid}, only marks, forget plot]
table {%
0.29963207244873 1.08149003982544
0.699140071868896 1.25449001789093
1.0986499786377 1.50967001914978
1.49816000461578 1.8201299905777
1.89767003059387 2.2086501121521
2.29717993736267 2.65278005599976
2.69668006896973 3.18307995796204
3.09618997573853 3.82919001579285
3.49569988250732 4.51592016220093
3.8952100276947 5.27661991119385
};
\addplot [very thick, dodgerblue0154222, dotted, mark=triangle, mark size=3.5, mark options={solid,fill opacity=0}]
table {%
0 1
0.0998772382736206 1.00959002971649
0.199753999710083 1.02849996089935
0.29963207244873 1.05751001834869
0.399508953094482 1.09397995471954
0.499386072158813 1.13763999938965
0.599262952804565 1.18703997135162
0.699140071868896 1.2406200170517
0.799018025398254 1.29674005508423
0.898895025253296 1.35748994350433
0.998772025108337 1.42174994945526
1.0986499786377 1.49075996875763
1.19852995872498 1.56394994258881
1.29840004444122 1.64120995998383
1.3982800245285 1.72121000289917
1.49816000461578 1.80343997478485
1.59803998470306 1.88923001289368
1.69790995121002 1.97821998596191
1.79779005050659 2.07202005386353
1.89767003059387 2.16972994804382
1.99753999710083 2.27239990234375
2.09741997718811 2.38058996200562
2.19729995727539 2.49265003204346
2.29717993736267 2.61149001121521
2.39704990386963 2.73761010169983
2.49692988395691 2.86681008338928
2.59681010246277 3.00845003128052
2.69668006896973 3.1548900604248
2.79656004905701 3.31029009819031
2.89644002914429 3.48676991462708
2.99632000923157 3.69159007072449
3.09618997573853 3.87050008773804
3.19606995582581 4.0701699256897
};
\addlegendentry{NBRC, $D=200$}
\addplot [very thick, springgreen0205108, mark=triangle, mark size=3.5, mark options={solid,rotate=180,fill opacity=0}]
table {%
0 1
0.0998772382736206 1.00853002071381
0.199753999710083 1.02604997158051
0.29963207244873 1.05704998970032
0.399508953094482 1.09686994552612
0.499386072158813 1.14031994342804
0.599262952804565 1.18720996379852
0.699140071868896 1.23769998550415
0.799018025398254 1.29200005531311
0.898895025253296 1.350909948349
0.998772025108337 1.41391003131866
1.0986499786377 1.48038995265961
1.19852995872498 1.55093002319336
1.29840004444122 1.62583994865417
1.3982800245285 1.70369005203247
1.49816000461578 1.78443002700806
1.59803998470306 1.86799001693726
1.69790995121002 1.95476996898651
1.79779005050659 2.04407000541687
1.89767003059387 2.13595008850098
1.99753999710083 2.23130989074707
2.09741997718811 2.33071994781494
2.19729995727539 2.43276000022888
2.29717993736267 2.53732991218567
2.39704990386963 2.64541006088257
2.49692988395691 2.75644993782043
2.59681010246277 2.87204003334045
2.69668006896973 2.99218988418579
2.79656004905701 3.11581993103027
2.89644002914429 3.24502992630005
2.99632000923157 3.37477993965149
3.09618997573853 3.51165008544922
3.19606995582581 3.65511989593506
3.29594993591309 3.80441999435425
3.39582991600037 3.96540999412537
3.49569988250732 4.1263599395752
3.59558010101318 4.29679012298584
};
\addlegendentry{NBKC, $D=200$}
\addplot [very thick, mediumorchid17588186, dashed, mark=triangle, mark size=3.5, mark options={solid,rotate=270,fill opacity=0}]
table {%
0 1
0.0998772382736206 1.00802004337311
0.199753999710083 1.02504003047943
0.29963207244873 1.05154001712799
0.399508953094482 1.08580994606018
0.499386072158813 1.1258499622345
0.599262952804565 1.17148995399475
0.699140071868896 1.22225999832153
0.799018025398254 1.27874004840851
0.898895025253296 1.33844995498657
0.998772025108337 1.40128004550934
1.0986499786377 1.46823000907898
1.19852995872498 1.5382000207901
1.29840004444122 1.61097002029419
1.3982800245285 1.68839001655579
1.49816000461578 1.77014994621277
1.59803998470306 1.85602998733521
1.69790995121002 1.94463002681732
1.79779005050659 2.03331995010376
1.89767003059387 2.12198996543884
1.99753999710083 2.20986008644104
2.09741997718811 2.29793000221252
2.19729995727539 2.38798999786377
2.29717993736267 2.48127007484436
2.39704990386963 2.5792601108551
2.49692988395691 2.68318009376526
2.59681010246277 2.79089999198914
2.69668006896973 2.8999400138855
2.79656004905701 3.00723004341125
2.89644002914429 3.11297988891602
2.99632000923157 3.22006988525391
3.09618997573853 3.33053994178772
3.19606995582581 3.44525003433228
3.29594993591309 3.56321001052856
3.39582991600037 3.68126010894775
3.49569988250732 3.79921007156372
3.59558010101318 3.91667008399963
3.69546008110046 4.03455018997192
3.79533004760742 4.1530499458313
3.8952100276947 4.27276992797852
3.99509000778198 4.39237022399902
4.09497022628784 4.51060009002686
};
\addlegendentry{OM, $D=200$}
\addplot [very thick, darkorange24213334, dash pattern=on 1pt off 3pt on 3pt off 3pt, mark=triangle, mark size=3.5, mark options={solid,rotate=90,fill opacity=0}]
table {%
0 1
0.0998772382736206 1.00839996337891
0.199753999710083 1.02583003044128
0.29963207244873 1.05349004268646
0.399508953094482 1.08822000026703
0.499386072158813 1.12920999526978
0.599262952804565 1.17536997795105
0.699140071868896 1.22662997245789
0.799018025398254 1.28251004219055
0.898895025253296 1.34174001216888
0.998772025108337 1.40468001365662
1.0986499786377 1.47126996517181
1.19852995872498 1.54129004478455
1.29840004444122 1.61483001708984
1.3982800245285 1.69135999679565
1.49816000461578 1.7711900472641
1.59803998470306 1.853639960289
1.69790995121002 1.93918001651764
1.79779005050659 2.02765989303589
1.89767003059387 2.11833000183105
1.99753999710083 2.21150994300842
2.09741997718811 2.306960105896
2.19729995727539 2.40439009666443
2.29717993736267 2.50430011749268
2.39704990386963 2.60747003555298
2.49692988395691 2.71319007873535
2.59681010246277 2.82088994979858
2.69668006896973 2.93095993995667
2.79656004905701 3.0432300567627
2.89644002914429 3.15610003471375
2.99632000923157 3.27114009857178
3.09618997573853 3.38582992553711
3.19606995582581 3.50124001502991
3.29594993591309 3.6172399520874
3.39582991600037 3.73386001586914
3.49569988250732 3.85111999511719
3.59558010101318 3.97030997276306
3.69546008110046 4.09104013442993
3.79533004760742 4.21481990814209
3.8952100276947 4.33989000320435
3.99509000778198 4.46775007247925
4.09497022628784 4.59554004669189
};
\addlegendentry{OM3, $D=200$}
\end{axis}

\end{tikzpicture}

%% file: figures/taylor-bubble/setup.tex
\begin{tikzpicture}
\definecolor{dodgerblue0154222}{RGB}{0,154,222}
\definecolor{darkorange24213334}{RGB}{242,133,34}
\setlength\radius{0.375\figurewidth}

% background filling
\draw [fill=dodgerblue0154222!30, draw=none] (0,0) rectangle (\figurewidth,\figureheight);

% borders left and right
\draw[very thick, black!50] (0,0)--(0,\figureheight);
\draw[very thick, black!50] (\figurewidth,0)--(\figurewidth,\figureheight);

% borders top and bottom
\draw[very thick, black!50] (0,\figureheight)--(\figurewidth,\figureheight);
\draw[very thick, black!50] (0,0)--(\figurewidth,0);

% domain height
\draw[<->, >=Latex] (\figurewidth+\mdist,0)--(\figurewidth+\mdist,\figureheight) node [pos=0.5,right] {$10D$};
\draw[-] (\figurewidth,0)--(\figurewidth+2\mdist,0);
\draw[-] (\figurewidth,\figureheight)--(\figurewidth+2\mdist,\figureheight);

% domain width
\draw[<->, >=Latex] (0,-\mdist)--(\figurewidth,-\mdist) node [pos=0.5,below] {$\varnothing D$};
\draw[-] (0,0)--(0,-2\mdist);
\draw[-] (\figurewidth,0)--(\figurewidth,-2\mdist);

% cylindrical bubble
\draw [fill=white, draw=dodgerblue0154222, thick] (0.5\figurewidth-\radius,0.1\figureheight) rectangle (0.5\figurewidth+\radius,0.4\figureheight);

% cylinder diameter
\draw[<->, >=Latex] (0.5\figurewidth-\radius,0.4\figureheight+\mdist)--(0.5\figurewidth+\radius,0.4\figureheight+\mdist) node [pos=0.5,above] {$\varnothing 0.75 D$};
\draw[-] (0.5\figurewidth-\radius,0.4\figureheight)--(0.5\figurewidth-\radius,0.4\figureheight+2\mdist);
\draw[-] (0.5\figurewidth+\radius,0.4\figureheight)--(0.5\figurewidth+\radius,0.4\figureheight+2\mdist);

% cylinder position x
\draw[<->, >=Latex] (0.5\figurewidth,0.1\figureheight-\mdist)--(\figurewidth,0.1\figureheight-\mdist) node [pos=0.5,below] {$0.5 D$};
\draw[-] (0.5\figurewidth,0.1\figureheight)--(0.5\figurewidth,0.1\figureheight-2\mdist);
\draw[-] (\figurewidth,0.1\figureheight)--(\figurewidth,0.1\figureheight-2\mdist);

% cylinder position y
\draw[<->, >=Latex] (-\mdist,0)--(-\mdist,0.1\figureheight) node [pos=0.5,left] {$D$};
\draw[-] (-2\mdist,0.1\figureheight)--(0.5\figurewidth-\radius,0.1\figureheight);
\draw[-] (-2\mdist,0)--(0.5\figurewidth-\radius,0);

% cylinder length
\draw[<->, >=Latex] (-\mdist,0.1\figureheight)--(-\mdist,0.4\figureheight) node [pos=0.5,left] {$3D$};
\draw[-] (-2\mdist,0.4\figureheight)--(0.5\figurewidth-\radius,0.4\figureheight);

% vertical line of symmetry
\draw[loosely dashdotted] (0.5\figurewidth,-0.025\figureheight)--(0.5\figurewidth,1.025\figureheight);

% gravity
\draw[thick, ->, >=Latex, darkorange24213334] (\figurewidth-1.5\mdist,\figureheight-\mdist)--(\figurewidth-1.5\mdist,\figureheight-4\mdist) node [pos=0.5,right] {$g$};

% coordinate system
\draw[->, >=Latex] (2\mdist,\figureheight-3\mdist)--(4\mdist,\figureheight-3\mdist) node [below] {$x$};
\draw[->, >=Latex] (2\mdist,\figureheight-3\mdist)--(2\mdist,\figureheight-1\mdist) node [left] {$z$};
\draw[draw=black] (2\mdist,\figureheight-3\mdist) circle [radius=0.175\mdist] node[opacity=1, below left] {$y$};
\end{tikzpicture}

%% file: figures/taylor-bubble/d-128-shape-front.tex
% This file was created with tikzplotlib v0.10.1.
\begin{tikzpicture}

\definecolor{darkgray176}{RGB}{176,176,176}
\definecolor{darkorange24213334}{RGB}{242,133,34}
\definecolor{dodgerblue0154222}{RGB}{0,154,222}
\definecolor{lightgray204}{RGB}{204,204,204}
\definecolor{mediumorchid17588186}{RGB}{175,88,186}
\definecolor{springgreen0205108}{RGB}{0,205,108}

\begin{axis}[
height=\figureheight,
legend cell align={left},
legend style={
  fill opacity=0.8,
  draw opacity=1,
  text opacity=1,
  at={(0.01,0.01)},
  anchor=south west,
  draw=lightgray204
},
tick align=outside,
tick pos=left,
width=\figurewidth,
x grid style={darkgray176},
xlabel={\(\displaystyle r^{*}\)},
xmajorgrids,
xmin=-0.037747681140625, xmax=0.792701303953125,
xtick style={color=black},
y grid style={darkgray176},
ylabel style={rotate=-90.0},
ylabel={\(\displaystyle z^{*}\)},
ymajorgrids,
ymin=-0.628112339982421, ymax=0.0299101114277343,
ytick style={color=black}
]
\addplot [very thick, black, mark=*, mark size=1.5, mark options={solid}, only marks]
table {%
0 0
0.0293359756469727 0
0.0573979616165161 -0.000633955001831055
0.0854599475860596 -0.00253605842590332
0.119897961616516 -0.00570595264434814
0.160714030265808 -0.0120450258255005
0.192602038383484 -0.0183850526809692
0.219388008117676 -0.0247249603271484
0.244897961616516 -0.0316979885101318
0.271683931350708 -0.0412089824676514
0.293367981910706 -0.0494489669799805
0.320153951644897 -0.0595920085906982
0.340561985969543 -0.0684679746627808
0.362244009971619 -0.0773429870605469
0.380102038383484 -0.0862189531326294
0.39795994758606 -0.0950939655303955
0.415816068649292 -0.10523796081543
0.429846048355103 -0.112844944000244
0.442602038383484 -0.121086955070496
0.456632018089294 -0.129328012466431
0.470664024353027 -0.138838052749634
0.48852002620697 -0.149615049362183
0.503826022148132 -0.162294030189514
0.519132018089294 -0.173071980476379
0.533164024353027 -0.183848977088928
0.543367981910706 -0.193992018699646
0.554846048355103 -0.204769968986511
0.565052032470703 -0.214913964271545
0.57397997379303 -0.225057005882263
0.581632018089294 -0.235200047492981
0.593111991882324 -0.247879028320312
0.603316068649292 -0.259925007820129
0.614795923233032 -0.273237943649292
0.623723983764648 -0.286550998687744
0.633928060531616 -0.299864053726196
0.641582012176514 -0.312543034553528
0.64795994758606 -0.325222969055176
0.656888008117676 -0.339169979095459
0.66198992729187 -0.351848959922791
0.667092084884644 -0.365162014961243
0.672194004058838 -0.376574039459229
0.679846048355103 -0.387984991073608
0.686223983764648 -0.401932001113892
0.690052032470703 -0.414610981941223
0.692601919174194 -0.424754977226257
0.697704076766968 -0.438068032264709
0.701529979705811 -0.453283071517944
0.705358028411865 -0.464694023132324
0.707906007766724 -0.476740002632141
0.71045994758606 -0.48751699924469
0.711734056472778 -0.49956202507019
0.715561985969543 -0.511608004570007
0.716835975646973 -0.523019075393677
0.718111991882324 -0.533161997795105
0.720664024353027 -0.544573068618774
0.721935987472534 -0.560423016548157
0.72448992729187 -0.572468042373657
0.72448992729187 -0.583878993988037
0.725765943527222 -0.595291018486023
};
\addlegendentry{Experiment~\cite{bugg2002VelocityFieldTaylor}}
\addplot [line width=2pt, dodgerblue0154222, dotted]
table {%
0.00718438625335693 -0.000845909118652344
0.0143687725067139 -0.00139141082763672
0.046875 -0.00121593475341797
0.0540593862533569 -0.00100040435791016
0.0696843862533569 -0.00146961212158203
0.078125 -0.0013427734375
0.09375 -0.00200939178466797
0.100934386253357 -0.00190162658691406
0.116559386253357 -0.00247287750244141
0.179059386253357 -0.00548362731933594
0.195866942405701 -0.00776767730712891
0.196861147880554 -0.00798320770263672
0.210309386253357 -0.0103273391723633
0.234375 -0.0128173828125
0.241559386253357 -0.0137519836425781
0.250243306159973 -0.0157957077026367
0.257184386253357 -0.0175189971923828
0.265625 -0.0189542770385742
0.272809386253357 -0.0198879241943359
0.3125 -0.0289936065673828
0.320938587188721 -0.0308094024658203
0.32243800163269 -0.0314207077026367
0.335309386253357 -0.035212516784668
0.34375 -0.0367317199707031
0.34846019744873 -0.0392332077026367
0.352316975593567 -0.0408134460449219
0.359375 -0.0435142517089844
0.366559386253357 -0.0438385009765625
0.370067358016968 -0.0453577041625977
0.382184386253357 -0.0500583648681641
0.390049338340759 -0.0523796081542969
0.397233724594116 -0.0556488037109375
0.397809386253357 -0.0562810897827148
0.40625 -0.0592355728149414
0.414773106575012 -0.0631380081176758
0.421875 -0.0658807754516602
0.429059386253357 -0.0682163238525391
0.432668566703796 -0.0696926116943359
0.433956027030945 -0.0704832077026367
0.439852952957153 -0.0730695724487305
0.450053930282593 -0.0782957077026367
0.463809728622437 -0.0849227905273438
0.486507058143616 -0.0976772308349609
0.491994023323059 -0.101733207702637
0.498781204223633 -0.105325698852539
0.504284262657166 -0.109725475311279
0.507184386253357 -0.111666679382324
0.511088967323303 -0.113945484161377
0.522809386253357 -0.119936943054199
0.525839567184448 -0.1217942237854
0.529499292373657 -0.125170707702637
0.530149221420288 -0.129391193389893
0.538434386253357 -0.134541511535645
0.539688229560852 -0.134677410125732
0.544450759887695 -0.137203693389893
0.545729517936707 -0.140795707702637
0.550062894821167 -0.145016193389893
0.554059386253357 -0.147465705871582
0.555920481681824 -0.148608207702637
0.561533212661743 -0.152828693389893
0.564598083496094 -0.156420707702637
0.566142678260803 -0.157570362091064
0.569684386253357 -0.159511566162109
0.572773575782776 -0.161790370941162
0.575428247451782 -0.164233207702637
0.576350092887878 -0.167825222015381
0.579346537590027 -0.172045707702637
0.582954406738281 -0.175098896026611
0.586212515830994 -0.179319381713867
0.587304592132568 -0.179858207702637
0.592197179794312 -0.183450222015381
0.59589409828186 -0.187670707702637
0.596835255622864 -0.191262722015381
0.597355842590332 -0.191550731658936
0.600934386253357 -0.194067478179932
0.603610038757324 -0.195626735687256
0.607202291488647 -0.199075222015381
0.608983278274536 -0.203295707702637
0.611418724060059 -0.206887722015381
0.612608790397644 -0.211108207702637
0.613765478134155 -0.212473392486572
0.616559386253357 -0.214203357696533
0.620062232017517 -0.21669340133667
0.62200927734375 -0.218920707702637
0.627128601074219 -0.226733207702637
0.628400206565857 -0.230325222015381
0.630802750587463 -0.234545707702637
0.632547378540039 -0.235372066497803
0.636462807655334 -0.238137722015381
0.639839053153992 -0.242358207702637
0.640837669372559 -0.245950222015381
0.640576004981995 -0.250170707702637
0.642084717750549 -0.253762722015381
0.644861698150635 -0.257983207702637
0.645623207092285 -0.258378505706787
0.647809386253357 -0.260150909423828
0.650324702262878 -0.262598514556885
0.654074907302856 -0.265795707702637
0.654024600982666 -0.269387722015381
0.655476450920105 -0.273608207702637
0.657028317451477 -0.277200222015381
0.659357905387878 -0.281420707702637
0.661445021629333 -0.285317420959473
0.665713310241699 -0.289376735687256
0.668895959854126 -0.292825222015381
0.670094609260559 -0.297045707702637
0.671589016914368 -0.300637722015381
0.672677755355835 -0.304858207702637
0.673116087913513 -0.308450222015381
0.675124883651733 -0.312670707702637
0.676433205604553 -0.316891193389893
0.683257699012756 -0.324703693389893
0.685060977935791 -0.328295707702637
0.686607599258423 -0.336108207702637
0.686629056930542 -0.339700222015381
0.687751770019531 -0.343920707702637
0.688580513000488 -0.351733207702637
0.691662669181824 -0.355016708374023
0.695064187049866 -0.359236717224121
0.695538401603699 -0.359545707702637
0.69703996181488 -0.363137722015381
0.699538350105286 -0.367358207702637
0.699215054512024 -0.370950222015381
0.699924111366272 -0.375170707702637
0.701619625091553 -0.378762722015381
0.703484654426575 -0.394387722015381
0.705299377441406 -0.398608207702637
0.707375764846802 -0.402200222015381
0.706995964050293 -0.406420707702637
0.708030462265015 -0.408256053924561
0.712900638580322 -0.412476539611816
0.713899254798889 -0.414233207702637
0.714739799499512 -0.417825222015381
0.713026404380798 -0.422045707702637
0.713960409164429 -0.425637722015381
0.716577053070068 -0.429858207702637
0.717926979064941 -0.437670707702637
0.718185663223267 -0.441262722015381
0.717712879180908 -0.445483207702637
0.718509078025818 -0.456887722015381
0.718981862068176 -0.461108207702637
0.719657182693481 -0.464700222015381
0.721117377281189 -0.468920707702637
0.72104549407959 -0.472512722015381
0.722337007522583 -0.476733207702637
0.724837183952332 -0.47892427444458
0.726069450378418 -0.483144760131836
0.727894186973572 -0.484545707702637
0.726191520690918 -0.488137722015381
0.728225827217102 -0.492358207702637
0.730474472045898 -0.495950222015381
0.731310129165649 -0.500170707702637
0.731539964675903 -0.503762722015381
0.731421828269958 -0.507983207702637
0.732370257377625 -0.511575222015381
0.732640266418457 -0.515795707702637
0.732000827789307 -0.519387722015381
0.733501791954041 -0.527200222015381
0.733608245849609 -0.535012722015381
0.733912110328674 -0.539233207702637
0.733792185783386 -0.550637722015381
0.734832406044006 -0.558450222015381
0.734418749809265 -0.562670707702637
0.735434770584106 -0.570483207702637
0.7350252866745 -0.574075222015381
0.735776543617249 -0.578295707702637
0.737069725990295 -0.581887722015381
0.738217711448669 -0.586108207702637
0.739514589309692 -0.597512722015381
};
\addlegendentry{NBRC, $D=128$}
\addplot [line width=2pt, springgreen0205108]
table {%
0 -0.000968456268310547
0.0124937295913696 -0.00109338760375977
0.015625 -0.000874519348144531
0.03125 -0.000796318054199219
0.046875 -0.00156164169311523
0.0625 -0.00143718719482422
0.0749937295913696 0
0.078125 0
0.09375 -0.00128078460693359
0.11250627040863 -0.00109386444091797
0.12813127040863 -0.00256252288818359
0.140625 -0.00287485122680664
0.15625 -0.00399923324584961
0.203125 -0.00679636001586914
0.21875 -0.00815582275390625
0.234375 -0.0113430023193359
0.25 -0.0130777359008789
0.265625 -0.0152177810668945
0.278953671455383 -0.0179996490478516
0.28125 -0.0186243057250977
0.296875 -0.0213742256164551
0.3125 -0.024014949798584
0.319997787475586 -0.0261569023132324
0.328125 -0.0282807350158691
0.34375 -0.0316557884216309
0.351272821426392 -0.0340476036071777
0.359375 -0.0364212989807129
0.376500010490417 -0.0412650108337402
0.390625 -0.0456085205078125
0.40625 -0.0516395568847656
0.41874372959137 -0.054450511932373
0.4375 -0.0631241798400879
0.453125 -0.0694522857666016
0.461733460426331 -0.073422908782959
0.476136207580566 -0.0803275108337402
0.477930665016174 -0.0810947418212891
0.490574598312378 -0.0881400108337402
0.492131471633911 -0.0892047882080078
0.5 -0.0923905372619629
0.507194638252258 -0.0959525108337402
0.508575558662415 -0.0975179672241211
0.515625 -0.100738048553467
0.521156549453735 -0.10376501083374
0.522778987884521 -0.105330467224121
0.52647864818573 -0.106781959533691
0.534939527511597 -0.11157751083374
0.536903977394104 -0.113142967224121
0.548475861549377 -0.120955467224121
0.55476975440979 -0.128767967224121
0.557235836982727 -0.12973690032959
0.5625 -0.132851123809814
0.566823601722717 -0.13501501083374
0.568242788314819 -0.136580467224121
0.569164752960205 -0.14282751083374
0.571670651435852 -0.144252300262451
0.578125 -0.149999141693115
0.579318523406982 -0.15064001083374
0.581097364425659 -0.152205467224121
0.587587833404541 -0.15845251083374
0.587669253349304 -0.160017967224121
0.591080069541931 -0.16626501083374
0.588781237602234 -0.172512054443359
0.59375 -0.174985885620117
0.59688127040863 -0.177257061004639
0.598902583122253 -0.178313732147217
0.609375 -0.183172702789307
0.616554021835327 -0.188137054443359
0.617646813392639 -0.18970251083374
0.618294954299927 -0.191267967224121
0.623100996017456 -0.19851541519165
0.625 -0.199889659881592
0.630389332771301 -0.20532751083374
0.631457090377808 -0.206892967224121
0.634543061256409 -0.21314001083374
0.635826826095581 -0.214705467224121
0.639812350273132 -0.22095251083374
0.640048503875732 -0.222630977630615
0.640625 -0.22295618057251
0.64850652217865 -0.230330467224121
0.649718403816223 -0.23657751083374
0.650580406188965 -0.242824554443359
0.652050018310547 -0.24439001083374
0.659538984298706 -0.250637054443359
0.661258101463318 -0.253767967224121
0.664070725440979 -0.261580467224121
0.667943835258484 -0.26782751083374
0.670255064964294 -0.274074554443359
0.67245352268219 -0.27564001083374
0.678243398666382 -0.281887054443359
0.678619146347046 -0.285017967224121
0.680630683898926 -0.29126501083374
0.681563854217529 -0.292830467224121
0.683050513267517 -0.29907751083374
0.683726906776428 -0.300570964813232
0.691438674926758 -0.30689001083374
0.69228732585907 -0.308455467224121
0.696905732154846 -0.33032751083374
0.699779391288757 -0.336574554443359
0.701897144317627 -0.33814001083374
0.706337571144104 -0.344387054443359
0.706243634223938 -0.34595251083374
0.706857323646545 -0.347517967224121
0.709780931472778 -0.35376501083374
0.710719347000122 -0.363142967224121
0.711838006973267 -0.370955467224121
0.714481711387634 -0.378767967224121
0.716893076896667 -0.38501501083374
0.717246890068054 -0.386580467224121
0.720561504364014 -0.39315938949585
0.721428871154785 -0.394392967224121
0.724152565002441 -0.40064001083374
0.724484443664551 -0.402205467224121
0.726063013076782 -0.417830467224121
0.726978421211243 -0.425642967224121
0.727759718894958 -0.433455467224121
0.727947115898132 -0.43970251083374
0.72828209400177 -0.441267967224121
0.731244444847107 -0.449080467224121
0.732763290405273 -0.456047534942627
0.736466646194458 -0.464705467224121
0.737753510475159 -0.47095251083374
0.737584471702576 -0.472517967224121
0.740543723106384 -0.480330467224121
0.740247368812561 -0.488142967224121
0.741063117980957 -0.495955467224121
0.741912603378296 -0.50220251083374
0.741815567016602 -0.503767967224121
0.742087364196777 -0.511580467224121
0.741971969604492 -0.519392967224121
0.742497205734253 -0.527205467224121
0.743478298187256 -0.535017967224121
0.744359135627747 -0.54907751083374
0.745048046112061 -0.550642967224121
0.746809601783752 -0.55689001083374
0.746408820152283 -0.558455467224121
0.746896147727966 -0.56470251083374
0.746608734130859 -0.570949554443359
0.75 -0.575780391693115
0.752124786376953 -0.578762054443359
0.752453446388245 -0.581892967224121
0.754406332969666 -0.589705467224121
0.754631161689758 -0.59595251083374
0.754953622817993 -0.597517967224121
};
\addlegendentry{NBKC, $D=128$}
\addplot [line width=2pt, mediumorchid17588186, dashed]
table {%
0 0
0.03125 -0.000117301940917969
0.0546867847442627 -0.000624656677246094
0.0703117847442627 -0.00140666961669922
0.0859367847442627 -0.00265598297119141
0.109375 -0.00574207305908203
0.117186784744263 -0.00656223297119141
0.140625 -0.00847721099853516
0.148436784744263 -0.00921916961669922
0.171875 -0.0136327743530273
0.179686784744263 -0.0148439407348633
0.195311784744263 -0.0167970657348633
0.21875 -0.0219535827636719
0.226561784744263 -0.0233592987060547
0.243249416351318 -0.0269927978515625
0.244311809539795 -0.02734375
0.28237509727478 -0.0370311737060547
0.302778959274292 -0.04296875
0.304686784744263 -0.0437498092651367
0.320311784744263 -0.0481252670288086
0.370997667312622 -0.06640625
0.386256694793701 -0.0726957321166992
0.410544157028198 -0.0836715698242188
0.423317909240723 -0.08984375
0.445311784744263 -0.101093292236328
0.467406988143921 -0.11328125
0.480708599090576 -0.12109375
0.497566223144531 -0.132851600646973
0.505377769470215 -0.138280868530273
0.514861822128296 -0.14453125
0.525735139846802 -0.15234375
0.533960103988647 -0.16015625
0.544907093048096 -0.16796875
0.552094459533691 -0.17578125
0.554686784744263 -0.177499771118164
0.561839818954468 -0.18359375
0.5696702003479 -0.19179630279541
0.570311784744263 -0.192187309265137
0.577710151672363 -0.19921875
0.583280563354492 -0.20703125
0.585936784744263 -0.208906173706055
0.592608690261841 -0.21484375
0.596686840057373 -0.22265625
0.605717897415161 -0.23046875
0.610264778137207 -0.23828125
0.616608619689941 -0.246562480926514
0.617186784744263 -0.247031211853027
0.623249292373657 -0.25390625
0.62621808052063 -0.26171875
0.630749225616455 -0.26953125
0.632811784744263 -0.271015644073486
0.638952255249023 -0.27734375
0.640139818191528 -0.28515625
0.648139953613281 -0.29296875
0.649077415466309 -0.30078125
0.656139850616455 -0.30859375
0.657389879226685 -0.31640625
0.659405469894409 -0.32421875
0.665991544723511 -0.330351829528809
0.66792106628418 -0.33203125
0.671366214752197 -0.343749523162842
0.67238974571228 -0.34765625
0.672999143600464 -0.35546875
0.677014827728271 -0.36328125
0.681780576705933 -0.368594169616699
0.683874130249023 -0.37109375
0.685921192169189 -0.37890625
0.687374114990234 -0.38671875
0.687843084335327 -0.39453125
0.689014911651611 -0.40234375
0.692155361175537 -0.41015625
0.699264764785767 -0.41796875
0.699671030044556 -0.42578125
0.701749324798584 -0.43359375
0.703264951705933 -0.44140625
0.703999280929565 -0.44921875
0.703686714172363 -0.45703125
0.705296039581299 -0.46484375
0.708788156509399 -0.476562023162842
0.709811687469482 -0.48046875
0.710936784744263 -0.481468677520752
0.716311693191528 -0.48828125
0.715546131134033 -0.49609375
0.717741250991821 -0.507812023162842
0.718389749526978 -0.51171875
0.71870231628418 -0.523437023162842
0.718733549118042 -0.53515625
0.720389842987061 -0.55078125
0.723702430725098 -0.55859375
0.723889827728271 -0.56640625
0.728842973709106 -0.58203125
0.729233503341675 -0.58984375
0.732483625411987 -0.59765625
};
\addlegendentry{OM, $D=128$}
\addplot [line width=2pt, darkorange24213334, dash pattern=on 1pt off 3pt on 3pt off 3pt]
table {%
0 -0.000233650207519531
0.015625 -3.814697265625e-05
0.0390616655349731 -7.72476196289062e-05
0.058336615562439 -0.000546455383300781
0.140625 -0.00449180603027344
0.164061665534973 -0.00593662261962891
0.181364178657532 -0.00804710388183594
0.189175844192505 -0.00933551788330078
0.195311665534973 -0.0103120803833008
0.226561665534973 -0.0133590698242188
0.250316381454468 -0.0175771713256836
0.257811665534973 -0.0189838409423828
0.273436665534973 -0.0210151672363281
0.304686665534973 -0.0275774002075195
0.322160363197327 -0.0312108993530273
0.324008584022522 -0.0317964553833008
0.335936665534973 -0.0348434448242188
0.353542327880859 -0.038945198059082
0.355522632598877 -0.0396089553833008
0.390833497047424 -0.049687385559082
0.403174638748169 -0.053593635559082
0.437865614891052 -0.0657806396484375
0.448040843009949 -0.0696873664855957
0.47269880771637 -0.0804290771484375
0.485555410385132 -0.0864839553833008
0.496659755706787 -0.0919919013977051
0.507811665534973 -0.0978903770446777
0.523436665534973 -0.106952667236328
0.528367877006531 -0.109921455383301
0.544518113136292 -0.121639728546143
0.552131652832031 -0.127264976501465
0.56100070476532 -0.133358955383301
0.570311665534973 -0.141561985015869
0.587077379226685 -0.156015396118164
0.588217973709106 -0.156796455383301
0.594952344894409 -0.164608955383301
0.602647662162781 -0.171640396118164
0.603733539581299 -0.172421455383301
0.609905362129211 -0.180233955383301
0.617186665534973 -0.188358783721924
0.624030470848083 -0.195858955383301
0.627905368804932 -0.203671455383301
0.636795997619629 -0.211483955383301
0.64071786403656 -0.219296455383301
0.643889784812927 -0.227108955383301
0.652671098709106 -0.234921455383301
0.655874133110046 -0.242733955383301
0.658030390739441 -0.250546455383301
0.664764881134033 -0.257804393768311
0.665467858314514 -0.258358955383301
0.670592784881592 -0.270077228546143
0.672436714172363 -0.273983955383301
0.673405408859253 -0.281796455383301
0.685811638832092 -0.297421455383301
0.68759286403656 -0.313046455383301
0.694905400276184 -0.328671455383301
0.695311665534973 -0.329311847686768
0.698905467987061 -0.336483955383301
0.702577233314514 -0.352108955383301
0.703928709030151 -0.363827228546143
0.704530358314514 -0.367733955383301
0.709374070167542 -0.378674983978271
0.711842894554138 -0.383358955383301
0.714171051979065 -0.391171455383301
0.717545986175537 -0.406796455383301
0.718436717987061 -0.414608955383301
0.717421054840088 -0.422421455383301
0.721936702728271 -0.438046455383301
0.724905371665955 -0.445858955383301
0.724077343940735 -0.453671455383301
0.729561686515808 -0.461483955383301
0.732545971870422 -0.481014728546143
0.733264803886414 -0.484921455383301
0.733546018600464 -0.492733955383301
0.734436631202698 -0.500546455383301
0.734256982803345 -0.512264728546143
0.734092950820923 -0.516171455383301
0.735233545303345 -0.523983955383301
0.734733581542969 -0.531796455383301
0.736780405044556 -0.539608955383301
0.736967921257019 -0.547421455383301
0.739577293395996 -0.555233955383301
0.742624163627625 -0.568714618682861
0.743061661720276 -0.570858955383301
0.745796084403992 -0.578671455383301
0.744046092033386 -0.586483955383301
0.748952269554138 -0.594296455383301
0.747069716453552 -0.598202228546143
};
\addlegendentry{OM3, $D=128$}
\end{axis}

\end{tikzpicture}

%% file: figures/taylor-bubble/d-128-shape-tail.tex
% This file was created with tikzplotlib v0.10.1.
\begin{tikzpicture}

\definecolor{darkgray176}{RGB}{176,176,176}
\definecolor{darkorange24213334}{RGB}{242,133,34}
\definecolor{dodgerblue0154222}{RGB}{0,154,222}
\definecolor{lightgray204}{RGB}{204,204,204}
\definecolor{mediumorchid17588186}{RGB}{175,88,186}
\definecolor{springgreen0205108}{RGB}{0,205,108}

\begin{axis}[
height=\figureheight,
legend cell align={left},
legend style={
  fill opacity=0.8,
  draw opacity=1,
  text opacity=1,
  at={(0.01,0.99)},
  anchor=north west,
  draw=lightgray204
},
tick align=outside,
tick pos=left,
width=\figurewidth,
x grid style={darkgray176},
xlabel={\(\displaystyle r^{*}\)},
xmajorgrids,
xmin=-0.0387082636359375, xmax=0.812873536354688,
xtick style={color=black},
y grid style={darkgray176},
ylabel style={rotate=-90.0},
ylabel={\(\displaystyle z^{*}\)},
ymajorgrids,
ymin=-0.0993263449601563, ymax=0.319011244163282,
ytick style={color=black}
]
\addplot [very thick, black, mark=*, mark size=1.5, mark options={solid}, only marks]
table {%
0.751070022583008 0.279245018959045
0.751060009002686 0.276661038398743
0.75104808807373 0.272783041000366
0.751029968261719 0.267291069030762
0.751013994216919 0.262120962142944
0.750993967056274 0.256628036499023
0.750977993011475 0.250813007354736
0.750959992408752 0.245319962501526
0.750944018363953 0.240473031997681
0.750927925109863 0.235626935958862
0.750891923904419 0.225288033485413
0.75083601474762 0.207193970680237
0.750802040100098 0.196854948997498
0.752063989639282 0.185546040534973
0.752671957015991 0.172943949699402
0.753286004066467 0.16228199005127
0.753888010978699 0.148064970970154
0.755147933959961 0.136109948158264
0.757050037384033 0.122215986251831
0.75830602645874 0.109614968299866
0.759566068649292 0.0979830026626587
0.760175943374634 0.085705041885376
0.759490013122559 0.0747189521789551
0.75749397277832 0.059533953666687
0.756160020828247 0.0479030609130859
0.754166007041931 0.0336869955062866
0.753479957580566 0.0223790407180786
0.750848054885864 0.0110709667205811
0.748866081237793 8.59498977661133e-05
0.746886014938354 -0.0102519989013672
0.743628025054932 -0.0141290426254272
0.740370035171509 -0.0180050134658813
0.737761974334717 -0.0218809843063354
0.735153913497925 -0.0257569551467896
0.731894016265869 -0.0302799940109253
0.727335929870605 -0.0351250171661377
0.724077939987183 -0.0396469831466675
0.719521999359131 -0.0438460111618042
0.714959979057312 -0.0496599674224854
0.710399985313416 -0.0551519393920898
0.704541921615601 -0.0606429576873779
0.693494081497192 -0.0651619434356689
0.68504798412323 -0.0687140226364136
0.673354029655457 -0.072909951210022
0.661018013954163 -0.0745220184326172
0.651278018951416 -0.0761339664459229
0.643490076065063 -0.0767780542373657
0.63440203666687 -0.0777440071105957
0.624661922454834 -0.0793570280075073
0.611681938171387 -0.0799989700317383
0.591563940048218 -0.0799920558929443
0.577935934066772 -0.0803110599517822
0.55717396736145 -0.0793349742889404
0.539010047912598 -0.0777130126953125
0.521492004394531 -0.0764150619506836
0.512416005134583 -0.0738279819488525
0.499451994895935 -0.0693000555038452
0.48907995223999 -0.0660660266876221
0.478709936141968 -0.0618619918823242
0.467043995857239 -0.0576579570770264
0.458620071411133 -0.0537780523300171
0.450191974639893 -0.0502209663391113
0.439177989959717 -0.0456939935684204
0.428156018257141 -0.0427830219268799
0.413892030715942 -0.0389009714126587
0.40351402759552 -0.0366359949111938
0.388602018356323 -0.0330770015716553
0.375632047653198 -0.0304880142211914
0.360715985298157 -0.0275750160217285
0.34644603729248 -0.0256320238113403
0.330878019332886 -0.0240110158920288
0.312067985534668 -0.02109694480896
0.295850038528442 -0.0191529989242554
0.27963399887085 -0.0172100067138672
0.266011953353882 -0.0155899524688721
0.249145984649658 -0.0133229494094849
0.233579993247986 -0.0110559463500977
0.214761972427368 -0.0104039907455444
0.197247982025146 -0.0087820291519165
0.168053984642029 -0.0061880350112915
0.149888038635254 -0.00488996505737305
0.131721973419189 -0.00359201431274414
0.111608028411865 -0.00261604785919189
0.0973340272903442 -0.00196504592895508
0.0772199630737305 -0.0013120174407959
0.0655399560928345 -0.000661969184875488
0.055806040763855 -0.000658988952636719
0.0428299903869629 -0.000331997871398926
0.0317980051040649 -5.00679016113281e-06
0.0162240266799927 0
};
\addlegendentry{Experiment~\cite{bugg2002VelocityFieldTaylor}}
\addplot [line width=2pt, dodgerblue0154222, dotted]
table {%
0 0
0.0625 -0.000128746032714844
0.116559386253357 -0.000597953796386719
0.163434386253357 -0.00116825103759766
0.203125 -0.00206565856933594
0.225934386253357 -0.00287437438964844
0.24935507774353 -0.00423908233642578
0.28125 -0.00640106201171875
0.304059386253357 -0.00749540328979492
0.34375 -0.00874233245849609
0.375 -0.00987672805786133
0.390625 -0.0107383728027344
0.397809386253357 -0.0111908912658691
0.429059386253357 -0.0139174461364746
0.453125 -0.0151605606079102
0.46875 -0.0157217979431152
0.491559386253357 -0.0162811279296875
0.522809386253357 -0.0167636871337891
0.554059386253357 -0.0169081687927246
0.569684386253357 -0.016873836517334
0.578125 -0.0168862342834473
0.585309386253357 -0.0166134834289551
0.59375 -0.0159635543823242
0.609375 -0.0157699584960938
0.625 -0.0132536888122559
0.63247013092041 -0.0125041007995605
0.647809386253357 -0.00908851623535156
0.654993772506714 -0.0075836181640625
0.664114356040955 -0.00469160079956055
0.664705991744995 -0.00440883636474609
0.670618772506714 -0.00196599960327148
0.679059386253357 -0.000290870666503906
0.681954741477966 0.000858783721923828
0.686740517616272 0.00312089920043945
0.690395355224609 0.00604963302612305
0.694684386253357 0.00768041610717773
0.698176026344299 0.00964164733886719
0.701417207717896 0.0109333992004395
0.706616640090942 0.0137395858764648
0.710309386253357 0.0166959762573242
0.711408615112305 0.0173320770263672
0.71329927444458 0.0187458992004395
0.716928720474243 0.0229663848876953
0.720621585845947 0.0265583992004395
0.723541975021362 0.0296435356140137
0.727730512619019 0.0332355499267578
0.729106307029724 0.0343708992004395
0.732659816741943 0.0385913848876953
0.734024882316589 0.0421833992004395
0.733611345291138 0.0464038848876953
0.73504102230072 0.0499958992004395
0.737066745758057 0.0542163848876953
0.737928867340088 0.0578083992004395
0.73950719833374 0.0620288848876953
0.741573691368103 0.0659499168395996
0.744266271591187 0.0698413848876953
0.746328234672546 0.0734333992004395
0.747037291526794 0.0776538848876953
0.748064637184143 0.0812458992004395
0.749098658561707 0.0890583992004395
0.749712944030762 0.101091384887695
0.74963390827179 0.104683399200439
0.749330043792725 0.108903884887695
0.749554753303528 0.120308399200439
0.749538779258728 0.128120899200439
0.749567866325378 0.135933399200439
0.750200867652893 0.140153884887695
0.753103375434875 0.143745899200439
0.752672910690308 0.147966384887695
0.749914050102234 0.151558399200439
0.753822088241577 0.15489673614502
0.761951327323914 0.159370899200439
0.757184386253357 0.161758899688721
0.752972483634949 0.165189743041992
0.749380350112915 0.167183399200439
0.757184386253357 0.171984672546387
0.763198375701904 0.174995899200439
0.757184386253357 0.178383350372314
0.750148296356201 0.182808399200439
0.760582566261292 0.188244819641113
0.764743804931641 0.190620899200439
0.761345624923706 0.193142890930176
0.757184386253357 0.195889949798584
0.752306342124939 0.198433399200439
0.757184386253357 0.201093673706055
0.760625720024109 0.203633308410645
0.764508366584778 0.206245899200439
0.761067032814026 0.209356784820557
0.757184386253357 0.21312141418457
0.755508542060852 0.214058399200439
0.757184386253357 0.214925765991211
0.760769367218018 0.218072414398193
0.764829397201538 0.221870899200439
0.761368632316589 0.226091384887695
0.757520794868469 0.229683399200439
0.761224985122681 0.233275413513184
0.764820694923401 0.237495899200439
0.76028573513031 0.245308399200439
0.764573693275452 0.253120899200439
0.762598633766174 0.257341384887695
0.761650323867798 0.260933399200439
0.763870239257812 0.265153884887695
0.764983773231506 0.268745899200439
0.762471079826355 0.276558399200439
0.764811396598816 0.284370899200439
0.763788461685181 0.292183399200439
0.764109253883362 0.296403884887695
0.764590620994568 0.299995899200439
};
\addlegendentry{NBRC, $D=128$}
\addplot [line width=2pt, springgreen0205108]
table {%
0 0
0.03125 -0.000116825103759766
0.125 -0.000846385955810547
0.206410884857178 -0.00332498550415039
0.234375 -0.00458765029907227
0.25 -0.00553560256958008
0.29374372959137 -0.00668573379516602
0.3125 -0.00732612609863281
0.359375 -0.00809526443481445
0.375 -0.00866842269897461
0.40311872959137 -0.0106415748596191
0.421875 -0.011782169342041
0.4375 -0.0127663612365723
0.46875 -0.0138339996337891
0.515625 -0.014493465423584
0.53125 -0.0143780708312988
0.546875 -0.0145244598388672
0.5625 -0.0143089294433594
0.578125 -0.0137500762939453
0.59061872959137 -0.0134000778198242
0.59375 -0.0130524635314941
0.60787832736969 -0.0117135047912598
0.62186872959137 -0.00911855697631836
0.625 -0.00896215438842773
0.640625 -0.00688266754150391
0.643258333206177 -0.00604391098022461
0.654477834701538 -0.00327634811401367
0.65625 -0.00255584716796875
0.671875 0.000343799591064453
0.679831385612488 0.00453615188598633
0.698524832725525 0.0123486518859863
0.705291867256165 0.0157256126403809
0.715111970901489 0.0201611518859863
0.720384478569031 0.023406982421875
0.725706934928894 0.0279736518859863
0.726924896240234 0.0295391082763672
0.734375 0.0354657173156738
0.735016584396362 0.0357861518859863
0.736508131027222 0.0373516082763672
0.741668105125427 0.0435986518859863
0.742717623710632 0.0498456954956055
0.74673318862915 0.0576581954956055
0.746974349021912 0.0592236518859863
0.749110698699951 0.0630092620849609
0.750720143318176 0.0645751953125
0.752207040786743 0.0670361518859863
0.755005598068237 0.0732831954956055
0.754989862442017 0.0748486518859863
0.756451606750488 0.0810956954956055
0.757159352302551 0.0826611518859863
0.758240342140198 0.0982861518859863
0.757765650749207 0.104533195495605
0.758490681648254 0.112345695495605
0.758318543434143 0.113911151885986
0.75830614566803 0.121723651885986
0.757793784141541 0.129536151885986
0.758350014686584 0.137348651885986
0.758015632629395 0.145161151885986
0.760464429855347 0.151408195495605
0.764200806617737 0.152973651885986
0.761431932449341 0.154539108276367
0.761506915092468 0.160786151885986
0.764268040657043 0.167033195495605
0.764108300209045 0.168598651885986
0.762933969497681 0.174845695495605
0.763839483261108 0.176411151885986
0.765834808349609 0.178809642791748
0.768733382225037 0.182658195495605
0.769171714782715 0.184223651885986
0.767125010490417 0.192036151885986
0.770443558692932 0.199848651885986
0.769181609153748 0.206095695495605
0.769206643104553 0.207661151885986
0.771193265914917 0.213908195495605
0.771374821662903 0.215473651885986
0.770812630653381 0.221720695495605
0.770975470542908 0.223286151885986
0.772412180900574 0.229533195495605
0.772556185722351 0.231098651885986
0.772553324699402 0.238911151885986
0.772937536239624 0.246723651885986
0.773756146430969 0.254536151885986
0.773415565490723 0.262348651885986
0.77267849445343 0.268595695495605
0.772716045379639 0.270161151885986
0.774165272712708 0.276408195495605
0.774149656295776 0.277973651885986
0.773599982261658 0.284220695495605
0.773759603500366 0.285786151885986
0.773553133010864 0.293598651885986
0.77336573600769 0.299845695495605
};
\addlegendentry{NBKC, $D=128$}
\addplot [line width=2pt, mediumorchid17588186, dashed]
table {%
0 0
0.0390617847442627 -0.000109195709228516
0.0859367847442627 -0.000726699829101562
0.148436784744263 -0.00183582305908203
0.171875 -0.00248813629150391
0.1875 -0.00316047668457031
0.195311784744263 -0.0035548210144043
0.214087009429932 -0.00511741638183594
0.226561784744263 -0.00649213790893555
0.242186784744263 -0.0078125
0.296875 -0.0111522674560547
0.304686784744263 -0.0119137763977051
0.320311784744263 -0.0145468711853027
0.335936784744263 -0.0162811279296875
0.359375 -0.0182380676269531
0.367186784744263 -0.0189919471740723
0.382811784744263 -0.0222654342651367
0.398436784744263 -0.0243358612060547
0.414061784744263 -0.0259842872619629
0.445311784744263 -0.0319609642028809
0.460936784744263 -0.0339765548706055
0.474828720092773 -0.0372495651245117
0.476561784744263 -0.0377731323242188
0.507811784744263 -0.0431637763977051
0.512266397476196 -0.0445389747619629
0.523436784744263 -0.0472731590270996
0.539061784744263 -0.0496950149536133
0.551988124847412 -0.053187370300293
0.554686784744263 -0.054023265838623
0.570311784744263 -0.056617259979248
0.601561784744263 -0.0631952285766602
0.632811784744263 -0.0666637420654297
0.648436784744263 -0.0632500648498535
0.664061784744263 -0.0610857009887695
0.66624927520752 -0.0601639747619629
0.679686784744263 -0.0487656593322754
0.692257404327393 -0.0430507659912109
0.695311784744263 -0.041562557220459
0.70015549659729 -0.0367264747619629
0.702608585357666 -0.0289139747619629
0.70810866355896 -0.0211014747619629
0.710936784744263 -0.0183749198913574
0.715405464172363 -0.0132889747619629
0.717296123504639 -0.00547647476196289
0.724202394485474 0.00233602523803711
0.726717948913574 0.0101485252380371
0.732717990875244 0.0179610252380371
0.733999252319336 0.0335860252380371
0.735202312469482 0.0413985252380371
0.738030433654785 0.0492110252380371
0.740499258041382 0.0648360252380371
0.742186784744263 0.0669922828674316
0.745967864990234 0.0726485252380371
0.74898362159729 0.0804610252380371
0.749733686447144 0.0882735252380371
0.74967098236084 0.0960860252380371
0.750335216522217 0.115617752075195
0.750702381134033 0.139055252075195
0.750921010971069 0.142961025238037
0.752202272415161 0.154680252075195
0.753389835357666 0.166398525238037
0.751514911651611 0.174211025238037
0.754358530044556 0.182023525238037
0.756639957427979 0.189836025238037
0.7538743019104 0.197648525238037
0.752811670303345 0.205461025238037
0.758655309677124 0.212472438812256
0.759499311447144 0.213273525238037
0.757811784744263 0.214515686035156
0.751718044281006 0.221086025238037
0.757811784744263 0.227351665496826
0.759936809539795 0.228898525238037
0.757811784744263 0.23118782043457
0.753499269485474 0.236711025238037
0.757811784744263 0.24469518661499
0.75373363494873 0.252336025238037
0.761311769485474 0.260148525238037
0.757811784744263 0.262726783752441
0.75204610824585 0.267961025238037
0.763202428817749 0.275773525238037
0.754725456237793 0.281425952911377
0.751639842987061 0.283586025238037
0.760358333587646 0.289734363555908
0.762905359268188 0.291398525238037
0.752858638763428 0.299211025238037
};
\addlegendentry{OM, $D=128$}
\addplot [line width=2pt, darkorange24213334, dash pattern=on 1pt off 3pt on 3pt off 3pt]
table {%
0 0
0.03125 -0.000191211700439453
0.0546866655349731 -0.000562667846679688
0.0859366655349731 -0.00145292282104492
0.15625 -0.00392961502075195
0.195311665534973 -0.00499200820922852
0.234375 -0.00600385665893555
0.25 -0.00660181045532227
0.265625 -0.00746488571166992
0.273436665534973 -0.00797653198242188
0.304686665534973 -0.0109376907348633
0.328125 -0.0123085975646973
0.382811665534973 -0.0149297714233398
0.400661110877991 -0.016754150390625
0.414061665534973 -0.0184845924377441
0.429686665534973 -0.0197267532348633
0.460936665534973 -0.0212111473083496
0.484375 -0.0224299430847168
0.492186665534973 -0.0229296684265137
0.53125 -0.026425838470459
0.539061665534973 -0.026890754699707
0.570311665534973 -0.0277185440063477
0.59375 -0.0275077819824219
0.617186665534973 -0.0270156860351562
0.63304603099823 -0.0250000953674316
0.633280396461487 -0.0248827934265137
0.648436665534973 -0.0207657814025879
0.664061665534973 -0.020078182220459
0.674882531166077 -0.0148358345031738
0.679686665534973 -0.012601375579834
0.688702344894409 -0.00925779342651367
0.695311665534973 -0.00560951232910156
0.706827282905579 -0.00144529342651367
0.710936665534973 0.00305461883544922
0.713139772415161 0.00636720657348633
0.72763979434967 0.0141797065734863
0.729967951774597 0.0219922065734863
0.734483599662781 0.0298047065734863
0.735710144042969 0.0415239334106445
0.736046075820923 0.0454297065734863
0.741327404975891 0.0550742149353027
0.744108557701111 0.0610547065734863
0.747421026229858 0.0688672065734863
0.749171018600464 0.0766797065734863
0.749889731407166 0.0844922065734863
0.750108480453491 0.0962114334106445
0.750139832496643 0.104023933410645
0.750342965126038 0.127461433410645
0.750421047210693 0.139179706573486
0.757811665534973 0.145406246185303
0.760717868804932 0.146992206573486
0.757811665534973 0.148531436920166
0.750280380249023 0.154804706573486
0.757811665534973 0.160062313079834
0.762670993804932 0.162617206573486
0.757811665534973 0.165195465087891
0.750327348709106 0.170429706573486
0.764905452728271 0.178242206573486
0.750967860221863 0.186054706573486
0.761584758758545 0.191734313964844
0.765358567237854 0.193867206573486
0.757811665534973 0.198671817779541
0.75204598903656 0.201679706573486
0.757811665534973 0.204601764678955
0.765546083450317 0.209492206573486
0.757811665534973 0.214797019958496
0.752874135971069 0.217304706573486
0.757811665534973 0.219758033752441
0.765655398368835 0.225117206573486
0.757811665534973 0.232211112976074
0.756358504295349 0.232929706573486
0.757811665534973 0.233804702758789
0.764811635017395 0.240742206573486
0.758374214172363 0.248554706573486
0.765264749526978 0.256367206573486
0.75860857963562 0.264179706573486
0.765389800071716 0.271992206573486
0.760217905044556 0.279804706573486
0.764499187469482 0.287617206573486
0.762670993804932 0.295429706573486
0.763553977012634 0.299336433410645
};
\addlegendentry{OM3, $D=128$}
\end{axis}

\end{tikzpicture}

%% file: figures/drop-impact/setup.tex
\begin{tikzpicture}
\definecolor{dodgerblue0154222}{RGB}{0,154,222}
\definecolor{darkorange24213334}{RGB}{242,133,34}
\definecolor{mediumorchid17588186}{RGB}{175,88,186}
\definecolor{crimson2553191}{RGB}{255,31,91}
\setlength\radius{0.125\figurewidth}

% coordinates used for impact velocity angle
\coordinate (intersectPoint) at (0.5\figurewidth,0.3\figureheight+4\radius);
\coordinate (obliquePoint) at (0.6\figurewidth,0.3\figureheight+2.5\radius);
\coordinate (verticalPoint) at (0.5\figurewidth,0.3\figureheight);

% liquid pool
\draw [thick, fill=dodgerblue0154222!30, draw=none] (0,0) rectangle (\figurewidth,0.3\figureheight);
\draw[thick, dodgerblue0154222] (0,0.3\figureheight)--(\figurewidth,0.3\figureheight);

% borders left and right
\draw[very thick, loosely dashed, black!50] (0,0)--(0,\figureheight);
\draw[very thick, loosely dashed, black!50] (\figurewidth,0)--(\figurewidth,\figureheight);

% borders top and bottom
\draw[very thick, black!50] (0,\figureheight)--(\figurewidth,\figureheight);
\draw[very thick, black!50] (0,0)--(\figurewidth,0);

% domain height
\draw[<->, >=Latex] (\figurewidth+\mdist,0)--(\figurewidth+\mdist,\figureheight) node [pos=0.5,right] {$5D$};
\draw[-] (\figurewidth,0)--(\figurewidth+2\mdist,0);
\draw[-] (\figurewidth,\figureheight)--(\figurewidth+2\mdist,\figureheight);

% domain width
\draw[<->, >=Latex] (0,-\mdist)--(\figurewidth,-\mdist) node [pos=0.5,below] {$10D$};
\draw[-] (0,0)--(0,-2\mdist);
\draw[-] (\figurewidth,0)--(\figurewidth,-2\mdist);

% liquid pool height
\draw[<->, >=Latex] (-\mdist,0)--(-\mdist,0.3\figureheight) node [pos=0.5,left] {$0.5D$};
\draw[-] (-2\mdist,0.3\figureheight)--(0,0.3\figureheight);
\draw[-] (-2\mdist,0)--(0,0);

% drop
\draw[thick, draw=dodgerblue0154222, fill=dodgerblue0154222!30, fill opacity=1, line width=0.4mm] (0.5\figurewidth,0.3\figureheight+\radius) circle [radius=\radius] node {};
\draw[loosely dashdotted] (0.5\figurewidth-\radius,0.3\figureheight+\radius)--(0.5\figurewidth+\radius,0.3\figureheight+\radius);
%\draw[loosely dashdotted] (intersectPoint) -- (verticalPoint);

% drop position y
\draw[<->, >=Latex] (0.5\figurewidth-\radius-\mdist,0.3\figureheight)--(0.5\figurewidth-\radius-\mdist,0.3\figureheight+\radius) node [pos=0.5,left] {$0.5D$};
\draw[-] (0.5\figurewidth-\radius-2\mdist,0.3\figureheight+\radius)--(0.5\figurewidth-\radius,0.3\figureheight+\radius);
\draw[-] (0.5\figurewidth-\radius-2\mdist,0.3\figureheight)--(0.5\figurewidth-\radius,0.3\figureheight);

% drop position x
\draw[<->, >=Latex] (0.5\figurewidth,0.3\figureheight-\mdist)--(\figurewidth,0.3\figureheight-\mdist) node [pos=0.5,below] {$5D$};
\draw[-] (0.5\figurewidth,0.3\figureheight)--(0.5\figurewidth,0.3\figureheight-2\mdist);
\draw[-] (\figurewidth,0.3\figureheight)--(\figurewidth,0.3\figureheight-2\mdist);

% drop diameter
\draw[<->, >=Latex] (0.5\figurewidth+\radius+\mdist,0.3\figureheight)--(0.5\figurewidth+\radius+\mdist,0.3\figureheight+2\radius) node [pos=0.5,right] {$\varnothing D$};
\draw[-] (0.5\figurewidth,0.3\figureheight+2\radius)--(0.5\figurewidth+\radius+2\mdist,0.3\figureheight+2\radius);
\draw[-] (0.5\figurewidth,0.3\figureheight)--(0.5\figurewidth+\radius+2\mdist,0.3\figureheight);

% vertical line of symmetry
\draw[loosely dashdotted] (0.5\figurewidth,-0.025\figureheight)--(0.5\figurewidth,1.025\figureheight);

% impact velocity
\draw[thick, ->, >=Latex, crimson2553191] (0.5\figurewidth,0.3\figureheight+1.65\radius)--(0.5\figurewidth,0.3\figureheight+0.35\radius) node [pos=0.5, right] {$U$};

% gravity
\draw[thick, ->, >=Latex, darkorange24213334] (\figurewidth-2\mdist,\figureheight-\mdist)--(\figurewidth-2\mdist,\figureheight-4\mdist) node [pos=0.5,right] {$g$};

% coordinate system
\draw[->, >=Latex] (2\mdist,\figureheight-3\mdist)--(4\mdist,\figureheight-3\mdist) node [below] {$x$};
\draw[->, >=Latex] (2\mdist,\figureheight-3\mdist)--(2\mdist,\figureheight-1\mdist) node [left] {$z$};
\draw[draw=black] (2\mdist,\figureheight-3\mdist) circle [radius=0.175\mdist] node[opacity=1, below left] {$y$};
\end{tikzpicture}

%% file: src/conclusion.tex
%!TEX root = ../main.tex

\section{Conclusions}\label{sec:conclusion}
In this study, different variants for free-surface boundary conditions in the FSLBM~\cite{korner2005LatticeBoltzmannModel} were compared.
The FSLBM assumes a free surface and neglects the fluid dynamics in the gas phase of a liquid--gas system.
Accordingly, no PDFs are stored in the gas phase, and PDFs streaming from gas cells to interface cells must be reconstructed with a free-surface boundary condition.
In the original formulation of the FSLBM, these missing PDFs are reconstructed based on the orientation of the interface-normal~\cite{korner2005LatticeBoltzmannModel}.
However, with this approach, existing information about the flow field is overwritten.
The authors argued that this would be required to balance the forces exerted by the liquid and gas pressure.
\par

In this article, four different variants for reconstructing missing PDFs were under investigation.
These include normal-based variants, where the central PDF is reconstructed (NBRC)~\cite{korner2005LatticeBoltzmannModel} or kept (NBKC).
As opposed to these, only missing PDFs are reconstructed in the OM variant.
In the OM3 variant, only missing but at least three PDFs are reconstructed~\cite{bogner2017DirectNumericalSimulation,thies2005LatticeBoltzmannModeling}, falling back to the NBKC variant otherwise.
\par

It was mathematically shown that neither of the variants generally balances the forces at a free interface in motion.
However, the OM variant was found to be the most accurate in five numerical experiments, whereas the other variants were subject to anisotropic artifacts and numerical instabilities.
It can be concluded that for the FSLBM~\cite{korner2005LatticeBoltzmannModel} considered in this article, only missing PDFs should be reconstructed, and no information about the flow field should be dropped.
\par

%% file: src/appendix.tex
%!TEX root = ../main.tex

\section{Force-balance computation}\label{app:fbc}

This section provides the step-by-step computation of the force balance at the free interface for the boundary condition variants presented in \Cref{sec:fsbcv}.

\subsection{Normal-based, reconstruct center (NBRC)}\label{app:fbc-nbrc}
Using the definitions of $K$, $R$ \eqref{eq:fsbcv-nbrc} and the free-surface boundary condition~\eqref{eq:nm-fslbm-boundary-condition}, the general force balance~\eqref{eq:fsbcv-force-balance-general} becomes
\begin{equation}\label{eq:fbc-nbrc-1}
\begin{split}
\frac{F_{\alpha}}{A} = 
&- n_{\beta}
\sum_{i \in \{i | \boldsymbol{n} \cdot \boldsymbol{c}_{i} < 0\}}
f_{i}^{\star}(\boldsymbol{x}, t)(c_{i,\alpha} - u_{\alpha})(c_{i,\beta} - u_{\beta})
\\
&- n_{\beta}
\sum_{i \in \{i | \boldsymbol{n} \cdot \boldsymbol{c}_{i} \geq 0\}}
\Bigl(
f_{i}^{\text{eq}}(\rho^{\text{G}},\boldsymbol{u})
+
f_{\bar{i}}^{\text{eq}}(\rho^{\text{G}},\boldsymbol{u})
-
f_{\bar{i}}^{\star}(\boldsymbol{x},t)\Bigr)
(c_{i,\alpha} - u_{\alpha})(c_{i,\beta} - u_{\beta})
\\
=
&- n_{\beta}
\sum_{i \in \{i | \boldsymbol{n} \cdot \boldsymbol{c}_{i} < 0\}}
f_{i}^{\star}(\boldsymbol{x}, t)(c_{i,\alpha} - u_{\alpha})(c_{i,\beta} - u_{\beta})
+ n_{\beta}
\sum_{i \in \{i | \boldsymbol{n} \cdot \boldsymbol{c}_{i} \geq 0\}}
f_{\bar{i}}^{\star}(\boldsymbol{x},t)
(c_{i,\alpha} - u_{\alpha})(c_{i,\beta} - u_{\beta})
\\
&- n_{\beta}
\sum_{i \in \{i | \boldsymbol{n} \cdot \boldsymbol{c}_{i} \geq 0\}}
f_{i}^{\text{eq}}(\rho^{\text{G}},\boldsymbol{u})(c_{i,\alpha} - u_{\alpha})(c_{i,\beta} - u_{\beta})
- n_{\beta}
\sum_{i \in \{i | \boldsymbol{n} \cdot \boldsymbol{c}_{i} \geq 0\}}
f_{\bar{i}}^{\text{eq}}(\rho^{\text{G}},\boldsymbol{u})(c_{i,\alpha} - u_{\alpha})(c_{i,\beta} - u_{\beta}).
\end{split}
\end{equation}
Each PDF in a cell can either be left unmodified or reconstructed but not both at the same time.
Consequently, a PDF can only be exclusively in either $\{i|\boldsymbol{n} \cdot \boldsymbol{c}_{i} < 0\}$ or $\{i|\boldsymbol{n} \cdot \boldsymbol{c}_{i} \geq 0\}$.
The union of these sets must contain all $q$ PDFs of the cell.
These conditions are formally denoted as
\begin{equation}\label{eq:fbc-nbrc-sets-union}
\{i|\boldsymbol{n} \cdot \boldsymbol{c}_{i} \geq 0\} \cap \{i|\boldsymbol{n} \cdot \boldsymbol{c}_{i} < 0\} \stackrel{!}{=} \emptyset
\quad\land\quad
\{i|\boldsymbol{n} \cdot \boldsymbol{c}_{i} \geq 0\} \cup \{i|\boldsymbol{n} \cdot \boldsymbol{c}_{i} < 0\} \stackrel{!}{=} \{0, 1, \dots, q-1\}
\end{equation}
where $q$ is defined by the chosen velocity set D$d$Q$q$.
In the velocity sets generally employed for simulating hydrodynamics with the LBM, the central lattice velocity is zero, $\boldsymbol{c}_{0}=\boldsymbol{0}$~\cite{kruger2017LatticeBoltzmannMethod}.
Accordingly, the corresponding dot product with the interface-normal $\boldsymbol{n}$ is also zero, $\boldsymbol{n} \cdot \boldsymbol{c}_{0}=0$.
This implies that with $K=\{i|\boldsymbol{n} \cdot \boldsymbol{c}_{i} < 0\}$ and $R=\{i|\boldsymbol{n} \cdot \boldsymbol{c}_{i} \geq 0\}$ chosen as by Körner et al.\cite{korner2005LatticeBoltzmannModel}, the central post-collision PDF $f_{0}^{\star}$ must be reconstructed.
Without loss of generality, it is assumed now that the normal is non-zero $\boldsymbol{n} \neq \boldsymbol{0}$ to simplify the analysis.
Then, $\boldsymbol{n} \cdot \boldsymbol{c}_{i} = 0$ if and only if $i=0=\bar{i}$.
\par

The central PDF $f_{0}$ is extracted from the second sum in the force balance~\eqref{eq:fbc-nbrc-1}
\begin{equation}\label{eq:fbc-nbrc-extract-zero}
\begin{split}
\sum_{i \in \{i | \boldsymbol{n} \cdot \boldsymbol{c}_{i} \geq 0\}}
&f_{\bar{i}}^{\star}(\boldsymbol{x},t)
(c_{i,\alpha} - u_{\alpha})(c_{i,\beta} - u_{\beta})
\\
&=
f_{0}^{\star}(\boldsymbol{x},t)
u_{\alpha}u_{\beta}
+
\sum_{i \in \{i | \boldsymbol{n} \cdot \boldsymbol{c}_{i} > 0\}}
f_{\bar{i}}^{\star}(\boldsymbol{x},t)
(c_{i,\alpha} - u_{\alpha})(c_{i,\beta} - u_{\beta})
\\
&=
f_{0}^{\star}(\boldsymbol{x},t)
u_{\alpha}u_{\beta}
+
\sum_{i \in \{i | -\boldsymbol{n} \cdot \boldsymbol{c_{\bar{i}}} > 0\}}
f_{\bar{i}}^{\star}(\boldsymbol{x},t)
(-c_{\bar{i},\alpha} - u_{\alpha})(-c_{\bar{i},\beta} - u_{\beta})
\\
&=
f_{0}^{\star}(\boldsymbol{x},t)
u_{\alpha}u_{\beta}
+
\sum_{i \in \{i | \boldsymbol{n} \cdot \boldsymbol{c_{\bar{i}}} < 0\}}
f_{\bar{i}}^{\star}(\boldsymbol{x},t)
(c_{\bar{i},\alpha} + u_{\alpha})(c_{\bar{i},\beta} + u_{\beta})
\\
&=
f_{0}^{\star}(\boldsymbol{x},t)
u_{\alpha}u_{\beta}
+
\sum_{i \in \{i | \boldsymbol{n} \cdot \boldsymbol{c}_{i} < 0\}}
f_{i}^{\star}(\boldsymbol{x},t)
(c_{i,\alpha} + u_{\alpha})(c_{i,\beta} + u_{\beta}),
\end{split}
\end{equation}
where the range of the sum $i \in \{i | \boldsymbol{n} \cdot \boldsymbol{c}_{i} > 0\}$ was reverted to $i \in \{i | \boldsymbol{n} \cdot \boldsymbol{c}_{i} < 0\}$, with $f_{\bar{i}}$ referring to the PDF for the direction $\boldsymbol{c_{\bar{i}}} = -\boldsymbol{c}_{i}$.
In the last step, $\bar{i}$ was substituted with $i$.
It is important to note that the substitution changes the effective indices of the sum, as the corresponding set defining the indices is also changed.
Analogously, $f_{0}^{\text{eq}}$ is extracted from the fourth sum in the force balance~\eqref{eq:fbc-nbrc-1} giving
\begin{equation}\label{eq:fbc-nbrc-extract-zero-fq}
\begin{split}
\sum_{i \in \{i | \boldsymbol{n} \cdot \boldsymbol{c}_{i} \geq 0\}}
&f_{\bar{i}}^{\text{eq}}(\rho^{\text{G}},\boldsymbol{u})
(c_{i,\alpha} - u_{\alpha})(c_{i,\beta} - u_{\beta})
\\
&=
f_{0}^{\text{eq}}(\rho^{\text{G}},\boldsymbol{u})
u_{\alpha}u_{\beta}
+
\sum_{i \in \{i | \boldsymbol{n} \cdot \boldsymbol{c}_{i} < 0\}}
f_{i}^{\text{eq}}(\rho^{\text{G}},\boldsymbol{u})
(c_{i,\alpha} + u_{\alpha})(c_{i,\beta} + u_{\beta}).
\end{split}
\end{equation}
\par

Inserting the transformations~\eqref{eq:fbc-nbrc-extract-zero} and~\eqref{eq:fbc-nbrc-extract-zero-fq} in the force balance~\eqref{eq:fbc-nbrc-1} gives
\begin{equation}\label{eq:fbc-nbrc-2}
\begin{split}
\frac{F_{\alpha}}{A} =
&- n_{\beta}
\sum_{i \in \{i | \boldsymbol{n} \cdot \boldsymbol{c}_{i} < 0\}}
f_{i}^{\star}(\boldsymbol{x}, t)(c_{i,\alpha} - u_{\alpha})(c_{i,\beta} - u_{\beta})
\\
&+ n_{\beta}
\Biggr(
f_{0}^{\star}(\boldsymbol{x}, t)u_{\alpha}u_{\beta}
+
\sum_{i \in \{i | \boldsymbol{n} \cdot \boldsymbol{c}_{i} < 0\}}
f_{i}^{\star}(\boldsymbol{x}, t)(c_{i,\alpha} + u_{\alpha})(c_{i,\beta} + u_{\beta})
\Biggl)
\\
&- n_{\beta}
\sum_{i \in \{i | \boldsymbol{n} \cdot \boldsymbol{c}_{i} \geq 0\}}
f_{i}^{\text{eq}}(\rho^{\text{G}},\boldsymbol{u})(c_{i,\alpha} - u_{\alpha})(c_{i,\beta} - u_{\beta})
\\
&- n_{\beta}
\Biggl(
f_{0}^{\text{eq}}(\rho^{\text{G}},\boldsymbol{u})u_{\alpha}u_{\beta}
+
\sum_{i \in \{i | \boldsymbol{n} \cdot \boldsymbol{c}_{i} < 0\}}
f_{i}^{\text{eq}}(\rho^{\text{G}},\boldsymbol{u})(c_{i,\alpha} + u_{\alpha})(c_{i,\beta} + u_{\beta})
\Biggr)
\\
=
&- n_{\beta}
\sum_{i \in \{i | \boldsymbol{n} \cdot \boldsymbol{c}_{i} < 0\}}
f_{i}^{\star}(\boldsymbol{x}, t)
(
c_{i,\alpha}c_{i,\beta}
-
c_{i,\alpha}u_{\beta}
-
c_{i,\beta}u_{\alpha}
+
u_{\alpha}u_{\beta}
)
\\
&+ n_{\beta}
f_{0}^{\star}(\boldsymbol{x}, t)u_{\alpha}u_{\beta}
+ n_{\beta}
\sum_{i \in \{i | \boldsymbol{n} \cdot \boldsymbol{c}_{i} < 0\}}
f_{i}^{\star}(\boldsymbol{x}, t)
(
c_{i,\alpha}c_{i,\beta}
+
c_{i,\alpha}u_{\beta}
+
c_{i,\beta}u_{\alpha}
+
u_{\alpha}u_{\beta}
)
\\
&- n_{\beta}
\sum_{i \in \{i | \boldsymbol{n} \cdot \boldsymbol{c}_{i} \geq 0\}}
f_{i}^{\text{eq}}(\rho^{\text{G}},\boldsymbol{u})
(
c_{i,\alpha}c_{i,\beta}
-
c_{i,\alpha}u_{\beta}
-
c_{i,\beta}u_{\alpha}
+
u_{\alpha}u_{\beta}
)
\\
&- n_{\beta}
f_{0}^{\text{eq}}(\rho^{\text{G}},\boldsymbol{u})u_{\alpha}u_{\beta}
- n_{\beta}
\sum_{i \in \{i | \boldsymbol{n} \cdot \boldsymbol{c}_{i} < 0\}}
f_{i}^{\text{eq}}(\rho^{\text{G}},\boldsymbol{u})
(
c_{i,\alpha}c_{i,\beta}
+
c_{i,\alpha}u_{\beta}
+
c_{i,\beta}u_{\alpha}
+
u_{\alpha}u_{\beta}
)
\\
=
& n_{\beta}
\Bigl(
f_{0}^{\star}(\boldsymbol{x}, t)
-
f_{0}^{\text{eq}}(\rho^{\text{G}},\boldsymbol{u})
\Bigr)
u_{\alpha}u_{\beta}
\\
&+ 2 n_{\beta}
\sum_{i \in \{i | \boldsymbol{n} \cdot \boldsymbol{c}_{i} < 0\}}
f_{i}^{\star}(\boldsymbol{x},t)
(
c_{i,\alpha}u_{\beta}
+
c_{i,\beta}u_{\alpha}
)
\\
&- n_{\beta}
\sum_{i}
f_{i}^{\text{eq}}(\rho^{\text{G}},\boldsymbol{u})
(
c_{i,\alpha}c_{i,\beta}
+
u_{\alpha}u_{\beta}
)
\\
&+ n_{\beta}
\sum_{i \in \{i | \boldsymbol{n} \cdot \boldsymbol{c}_{i} \geq 0\}}
f_{i}^{\text{eq}}(\rho^{\text{G}},\boldsymbol{u})
(
c_{i,\alpha}u_{\beta}
+
c_{i,\beta}u_{\alpha}
)
- n_{\beta}
\sum_{i \in \{i | \boldsymbol{n} \cdot \boldsymbol{c}_{i} < 0\}}
f_{i}^{\text{eq}}(\rho^{\text{G}},\boldsymbol{u})
(
c_{i,\alpha}u_{\beta}
+
c_{i,\beta}u_{\alpha}
),
\end{split}
\end{equation}
where in the last step, the sets' property~\eqref{eq:fbc-nbrc-sets-union} was used to combine the summands containing $f_{i}^{\text{eq}}(\rho^{\text{G}},\boldsymbol{u})$ from different sums.
The final two sums of the force balance~\eqref{eq:fbc-nbrc-2} are extended with zero by adding and subtracting $n_{\beta}
\sum_{i \in \{i | \boldsymbol{n} \cdot \boldsymbol{c}_{i} < 0\}}f_{i}^{\text{eq}}(\rho^{\text{G}},\boldsymbol{u})(c_{i,\alpha}u_{\beta}+c_{i,\beta}u_{\alpha})$, which leads to
\begin{equation}\label{eq:fbc-nbrc-extend-zero}
\begin{split}
n_{\beta}
&\sum_{i \in \{i | \boldsymbol{n} \cdot \boldsymbol{c}_{i} \geq 0\}}
f_{i}^{\text{eq}}(\rho^{\text{G}},\boldsymbol{u})
(
c_{i,\alpha}u_{\beta}
+
c_{i,\beta}u_{\alpha}
)
- n_{\beta}
\sum_{i \in \{i | \boldsymbol{n} \cdot \boldsymbol{c}_{i} < 0\}}
f_{i}^{\text{eq}}(\rho^{\text{G}},\boldsymbol{u})
(
c_{i,\alpha}u_{\beta}
+
c_{i,\beta}u_{\alpha}
)
\\
=
&n_{\beta}
\sum_{i \in \{i | \boldsymbol{n} \cdot \boldsymbol{c}_{i} \geq 0\}}
f_{i}^{\text{eq}}(\rho^{\text{G}},\boldsymbol{u})
(
c_{i,\alpha}u_{\beta}
+
c_{i,\beta}u_{\alpha}
)
+ n_{\beta}
\sum_{i \in \{i | \boldsymbol{n} \cdot \boldsymbol{c}_{i} < 0\}}
f_{i}^{\text{eq}}(\rho^{\text{G}},\boldsymbol{u})
(
c_{i,\alpha}u_{\beta}
+
c_{i,\beta}u_{\alpha}
)
\\
&- n_{\beta}
\sum_{i \in \{i | \boldsymbol{n} \cdot \boldsymbol{c}_{i} < 0\}}
f_{i}^{\text{eq}}(\rho^{\text{G}},\boldsymbol{u})
(
c_{i,\alpha}u_{\beta}
+
c_{i,\beta}u_{\alpha}
)
- n_{\beta}
\sum_{i \in \{i | \boldsymbol{n} \cdot \boldsymbol{c}_{i} < 0\}}
f_{i}^{\text{eq}}(\rho^{\text{G}},\boldsymbol{u})
(
c_{i,\alpha}u_{\beta}
+
c_{i,\beta}u_{\alpha}
)
\\
=
&n_{\beta}
\sum_{i}
f_{i}^{\text{eq}}(\rho^{\text{G}},\boldsymbol{u})
(
c_{i,\alpha}u_{\beta}
+
c_{i,\beta}u_{\alpha}
)
-
2 n_{\beta}
\sum_{i \in \{i | \boldsymbol{n} \cdot \boldsymbol{c}_{i} < 0\}}
f_{i}^{\text{eq}}(\rho^{\text{G}},\boldsymbol{u})
(
c_{i,\alpha}u_{\beta}
+
c_{i,\beta}u_{\alpha}
),
\end{split}
\end{equation}
where again the sets' property~\eqref{eq:fbc-nbrc-sets-union} was used to combine the sums ranging over different sets.
\par

The final expression for the balance of the forces is obtained when inserting the transformation \eqref{eq:fbc-nbrc-extend-zero} into the force balance \eqref{eq:fbc-nbrc-2}
\begin{equation}
\begin{split}
\frac{F_{\alpha}}{A} =
& n_{\beta}
\Bigl(
f_{0}^{\star}(\boldsymbol{x}, t)
-
f_{0}^{\text{eq}}(\rho^{\text{G}},\boldsymbol{u})
\Bigr)
u_{\alpha}u_{\beta}
\\
&+ 2 n_{\beta}
\sum_{i \in \{i | \boldsymbol{n} \cdot \boldsymbol{c}_{i} < 0\}}
f_{i}^{\star}(\boldsymbol{x},t)
(
c_{i,\alpha}u_{\beta}
+
c_{i,\beta}u_{\alpha}
)
\\
&- n_{\beta}
\sum_{i}
f_{i}^{\text{eq}}(\rho^{\text{G}},\boldsymbol{u})
(
c_{i,\alpha}c_{i,\beta}
+
u_{\alpha}u_{\beta}
-
c_{i,\alpha}u_{\beta}
-
c_{i,\beta}u_{\alpha}
)
\\
& - 2 n_{\beta}
\sum_{i \in \{i | \boldsymbol{n} \cdot \boldsymbol{c}_{i} < 0\}}
f_{i}^{\text{eq}}(\rho^{\text{G}},\boldsymbol{u})
(
c_{i,\alpha}u_{\beta}
+
c_{i,\beta}u_{\alpha}
)
\\
=
& n_{\beta}
\Bigl(
f_{0}^{\star}(\boldsymbol{x}, t)
-
f_{0}^{\text{eq}}(\rho^{\text{G}},\boldsymbol{u})
\Bigr)
u_{\alpha}u_{\beta}
\\
&+ 2 n_{\beta}
\Biggl(
\sum_{i \in \{i | \boldsymbol{n} \cdot \boldsymbol{c}_{i} < 0\}}
f_{i}^{\star}(\boldsymbol{x},t)
(
c_{i,\alpha}u_{\beta}
+
c_{i,\beta}u_{\alpha}
)
-
\sum_{i \in \{i | \boldsymbol{n} \cdot \boldsymbol{c}_{i} < 0\}}
f_{i}^{\text{eq}}(\rho^{\text{G}},\boldsymbol{u})
(
c_{i,\alpha}u_{\beta}
+
c_{i,\beta}u_{\alpha}
)
\Biggr)
\\
&- n_{\beta}
\bigl(
\rho^{\text{G}} c_{s}^{2}\delta_{\alpha\beta} + \rho^{\text{G}} u_{\alpha}u_{\beta}
+
\rho^{\text{G}} u_{\alpha}u_{\beta}
-
\rho^{\text{G}} u_{\alpha} u_{\beta}
-
\rho^{\text{G}} u_{\alpha} u_{\beta}
\bigr)
\\
=
& n_{\beta}
\Bigl(
f_{0}^{\star}(\boldsymbol{x}, t)
-
f_{0}^{\text{eq}}(\rho^{\text{G}},\boldsymbol{u})
\Bigr)
u_{\alpha}u_{\beta}
\\
&+ 2 n_{\beta}
\sum_{i \in \{i | \boldsymbol{n} \cdot \boldsymbol{c}_{i} < 0\}}
\Bigl(
f_{i}^{\star}(\boldsymbol{x},t)
-
f_{i}^{\text{eq}}(\rho^{\text{G}},\boldsymbol{u})
\Bigr)
(
c_{i,\alpha}u_{\beta}
+
c_{i,\beta}u_{\alpha}
)
\\
&- n_{\alpha}
p^{\text{G}},
\end{split}
\end{equation}
where $p^{\text{G}} = \rho c_{s}^{2}$ was used.
This result has been obtained using the equilibrium moments~\cite{kruger2017LatticeBoltzmannMethod}
\begin{equation}\label{eq:fbc-eq-moments}
\begin{split}
&\Pi^{\text{eq}}(\rho^{\text{G}},\boldsymbol{u}) = \sum_{i} f_{i}^{\text{eq}}(\rho^{\text{G}},\boldsymbol{u}) = \rho^{\text{G}}\\
&\Pi^{\text{eq}}_{\alpha}(\rho^{\text{G}},\boldsymbol{u}) = \sum_{i} f_{i}^{\text{eq}}(\rho^{\text{G}},\boldsymbol{u})c_{i,\alpha} = \rho^{\text{G}} u_{\alpha}\\
&\Pi^{\text{eq}}_{\alpha\beta}(\rho^{\text{G}},\boldsymbol{u}) = 
\sum_{i} f_{i}^{\text{eq}}(\rho^{\text{G}},\boldsymbol{u}) c_{i,\alpha}c_{i,\beta} = 
\rho^{\text{G}} c_{s}^{2}\delta_{\alpha\beta} + \rho^{\text{G}} u_{\alpha}u_{\beta},
\end{split}
\end{equation}
where $\delta_{\alpha \beta}$ is the Kronecker delta.
The balance of the forces at the interface is disturbed by the expression
\begin{equation}
n_{\beta}
\Bigl(
f_{0}^{\star}(\boldsymbol{x}, t)
-
f_{0}^{\text{eq}}(\rho^{\text{G}},\boldsymbol{u})
\Bigr)
u_{\alpha}u_{\beta}
+ 2 n_{\beta}
\sum_{i \in \{i | \boldsymbol{n} \cdot \boldsymbol{c}_{i} < 0\}}
\Bigl(
f_{i}^{\star}(\boldsymbol{x},t)
-
f_{i}^{\text{eq}}(\rho^{\text{G}},\boldsymbol{u})
\Bigr)
(
c_{i,\alpha}u_{\beta}
+
c_{i,\beta}u_{\alpha}
).
\end{equation}

\subsection{Normal-based, keep center (NBKC)}\label{app:fbc-nbkc}
Using the same procedure as for the NBRC in \Cref{app:fbc-nbrc} but replacing $\boldsymbol{n} \cdot \boldsymbol{c}_{i} \geq 0$ with $\boldsymbol{n} \cdot \boldsymbol{c}_{i} > 0$, the additional terms for $i=0$ vanish in the transformations~\eqref{eq:fbc-nbrc-extract-zero} and~\eqref{eq:fbc-nbrc-extract-zero-fq}, so that the force balance~\eqref{eq:fbc-nbrc-2} becomes
\begin{equation}\label{eq:fbc-nbkc-1}
\begin{split}
\frac{F_{\alpha}}{A} =
&- n_{\beta}
\sum_{i \in \{i | \boldsymbol{n} \cdot \boldsymbol{c}_{i} < 0\}}
f_{i}^{\star}(\boldsymbol{x}, t)
(c_{i,\alpha} - u_{\alpha})(c_{i,\beta} - u_{\beta})
+ n_{\beta}
\sum_{i \in \{i | \boldsymbol{n} \cdot \boldsymbol{c}_{i} < 0\}}
f_{i}^{\star}(\boldsymbol{x}, t)
(c_{i,\alpha} + u_{\alpha})(c_{i,\beta} + u_{\beta})
\\
&- n_{\beta}
\sum_{i \in \{i | \boldsymbol{n} \cdot \boldsymbol{c}_{i} > 0\}}
f_{i}^{\text{eq}}(\rho^{\text{G}},\boldsymbol{u})
(c_{i,\alpha} - u_{\alpha})(c_{i,\beta} - u_{\beta})
- n_{\beta}
\sum_{i \in \{i | \boldsymbol{n} \cdot \boldsymbol{c}_{i} < 0\}}
f_{i}^{\text{eq}}(\rho^{\text{G}},\boldsymbol{u})
(c_{i,\alpha} + u_{\alpha})(c_{i,\beta} + u_{\beta})
\\
=
&- n_{\beta}
\sum_{i \in \{i | \boldsymbol{n} \cdot \boldsymbol{c}_{i} < 0\}}
f_{i}^{\star}(\boldsymbol{x}, t)
(
c_{i,\alpha}c_{i,\beta}
-
c_{i,\alpha}u_{\beta}
-
c_{i,\beta}u_{\alpha}
+
u_{\alpha}u_{\beta}
)
\\
&+ n_{\beta}
\sum_{i \in \{i | \boldsymbol{n} \cdot \boldsymbol{c}_{i} < 0\}}
f_{i}^{\star}(\boldsymbol{x}, t)
(
c_{i,\alpha}c_{i,\beta}
+
c_{i,\alpha}u_{\beta}
+
c_{i,\beta}u_{\alpha}
+
u_{\alpha}u_{\beta}
)
\\
&- n_{\beta}
\sum_{i \in \{i | \boldsymbol{n} \cdot \boldsymbol{c}_{i} > 0\}}
f_{i}^{\text{eq}}(\rho^{\text{G}},\boldsymbol{u})
(
c_{i,\alpha}c_{i,\beta}
-
c_{i,\alpha}u_{\beta}
-
c_{i,\beta}u_{\alpha}
+
u_{\alpha}u_{\beta}
)
\\
&- n_{\beta}
\sum_{i \in \{i | \boldsymbol{n} \cdot \boldsymbol{c}_{i} < 0\}}
f_{i}^{\text{eq}}(\rho^{\text{G}},\boldsymbol{u})
(
c_{i,\alpha}c_{i,\beta}
+
c_{i,\alpha}u_{\beta}
+
c_{i,\beta}u_{\alpha}
+
u_{\alpha}u_{\beta}
)
\\
=
&2 n_{\beta}
\sum_{i \in \{i | \boldsymbol{n} \cdot \boldsymbol{c}_{i} < 0\}}
f_{i}^{\star}(\boldsymbol{x}, t)
(
c_{i,\alpha}u_{\beta}
+
c_{i,\beta}u_{\alpha}
)
\\
&- n_{\beta}
\sum_{i \in \{i | \boldsymbol{n} \cdot \boldsymbol{c}_{i} > 0\}}
f_{i}^{\text{eq}}(\rho^{\text{G}},\boldsymbol{u})
(
c_{i,\alpha}c_{i,\beta}
-
c_{i,\alpha}u_{\beta}
-
c_{i,\beta}u_{\alpha}
+
u_{\alpha}u_{\beta}
)
\\
&- n_{\beta}
\sum_{i \in \{i | \boldsymbol{n} \cdot \boldsymbol{c}_{i} < 0\}}
f_{i}^{\text{eq}}(\rho^{\text{G}},\boldsymbol{u})
(
c_{i,\alpha}c_{i,\beta}
+
c_{i,\alpha}u_{\beta}
+
c_{i,\beta}u_{\alpha}
+
u_{\alpha}u_{\beta}
),
\end{split}
\end{equation}
The last two sums of the force balance~\eqref{eq:fbc-nbkc-1} can be combined by extending with zero, that is, by adding and subtracting $n_{\beta} f_{0}^{\text{eq}}(\rho^{\text{G}},\boldsymbol{u})(c_{0,\alpha}c_{0,\beta}-c_{0,\alpha}u_{\beta}-c_{0,\beta}u_{\alpha}+u_{\alpha}u_{\beta}) = n_{\beta} f_{0}^{\text{eq}}(\rho^{\text{G}},\boldsymbol{u})u_{\alpha}u_{\beta}$ so that
\begin{equation}\label{eq:fbc-extend-zero-1}
\begin{split}
- n_{\beta}
&\sum_{i \in \{i | \boldsymbol{n} \cdot \boldsymbol{c}_{i} > 0\}}
f_{i}^{\text{eq}}(\rho^{\text{G}},\boldsymbol{u})
(
c_{i,\alpha}c_{i,\beta}
-
c_{i,\alpha}u_{\beta}
-
c_{i,\beta}u_{\alpha}
+
u_{\alpha}u_{\beta}
)
- n_{\beta}
\sum_{i \in \{i | \boldsymbol{n} \cdot \boldsymbol{c}_{i} < 0\}}
f_{i}^{\text{eq}}(\rho^{\text{G}},\boldsymbol{u})
(
c_{i,\alpha}c_{i,\beta}
+
c_{i,\alpha}u_{\beta}
+
c_{i,\beta}u_{\alpha}
+
u_{\alpha}u_{\beta}
)
\\
=
&- n_{\beta}
\sum_{i \in \{i | \boldsymbol{n} \cdot \boldsymbol{c}_{i} \geq 0\}}
f_{i}^{\text{eq}}(\rho^{\text{G}},\boldsymbol{u})
(
c_{i,\alpha}c_{i,\beta}
-
c_{i,\alpha}u_{\beta}
-
c_{i,\beta}u_{\alpha}
+
u_{\alpha}u_{\beta}
)
+
n_{\beta}
f_{0}^{\text{eq}}(\rho^{\text{G}},\boldsymbol{u})
(
c_{0,\alpha}c_{0,\beta}
-
c_{0,\alpha}u_{\beta}
-
c_{0,\beta}u_{\alpha}
+
u_{\alpha}u_{\beta}
)
\\
&- n_{\beta}
\sum_{i \in \{i | \boldsymbol{n} \cdot \boldsymbol{c}_{i} < 0\}}
f_{i}^{\text{eq}}(\rho^{\text{G}},\boldsymbol{u})
(
c_{i,\alpha}c_{i,\beta}
+
c_{i,\alpha}u_{\beta}
+
c_{i,\beta}u_{\alpha}
+
u_{\alpha}u_{\beta}
)
\\
=
&- n_{\beta}
\sum_{i}
f_{i}^{\text{eq}}(\rho^{\text{G}},\boldsymbol{u})
(
c_{i,\alpha}c_{i,\beta}
+
u_{\alpha}u_{\beta}
)
+
n_{\beta}
f_{0}^{\text{eq}}(\rho^{\text{G}},\boldsymbol{u})u_{\alpha}u_{\beta}
\\
&+ n_{\beta}
\sum_{i \in \{i | \boldsymbol{n} \cdot \boldsymbol{c}_{i} \geq 0\}}
f_{i}^{\text{eq}}(\rho^{\text{G}},\boldsymbol{u})
(
c_{i,\alpha}u_{\beta}
+
c_{i,\beta}u_{\alpha}
)
- n_{\beta}
\sum_{i \in \{i | \boldsymbol{n} \cdot \boldsymbol{c}_{i} < 0\}}
f_{i}^{\text{eq}}(\rho^{\text{G}},\boldsymbol{u})
(
c_{i,\alpha}u_{\beta}
+
c_{i,\beta}u_{\alpha}
).
\end{split}
\end{equation}
Similar as before, the last two sums of the intermediate result~\eqref{eq:fbc-extend-zero-1} are combined by adding and subtracting $n_{\beta}\sum_{i \in \{i | \boldsymbol{n} \cdot \boldsymbol{c}_{i} < 0\}} f_{i}^{\text{eq}}(\rho^{\text{G}},\boldsymbol{u})(c_{i,\alpha}u_{\beta}+c_{i,\beta}u_{\alpha})$, which leads to
\begin{equation}\label{eq:fbc-extend-zero-2}
\begin{split}
n_{\beta}
&\sum_{i \in \{i | \boldsymbol{n} \cdot \boldsymbol{c}_{i} \geq 0\}}
f_{i}^{\text{eq}}(\rho^{\text{G}},\boldsymbol{u})
(
c_{i,\alpha}u_{\beta}
+
c_{i,\beta}u_{\alpha}
)
- n_{\beta}
\sum_{i \in \{i | \boldsymbol{n} \cdot \boldsymbol{c}_{i} < 0\}}
f_{i}^{\text{eq}}(\rho^{\text{G}},\boldsymbol{u})
(
c_{i,\alpha}u_{\beta}
+
c_{i,\beta}u_{\alpha}
)
\\
=
&n_{\beta}
\sum_{i \in \{i | \boldsymbol{n} \cdot \boldsymbol{c}_{i} \geq 0\}}
f_{i}^{\text{eq}}(\rho^{\text{G}},\boldsymbol{u})
(
c_{i,\alpha}u_{\beta}
+
c_{i,\beta}u_{\alpha}
)
+ n_{\beta}
\sum_{i \in \{i | \boldsymbol{n} \cdot \boldsymbol{c}_{i} < 0\}}
f_{i}^{\text{eq}}(\rho^{\text{G}},\boldsymbol{u})
(
c_{i,\alpha}u_{\beta}
+
c_{i,\beta}u_{\alpha}
)
\\
&- 2n_{\beta}
\sum_{i \in \{i | \boldsymbol{n} \cdot \boldsymbol{c}_{i} < 0\}}
f_{i}^{\text{eq}}(\rho^{\text{G}},\boldsymbol{u})
(
c_{i,\alpha}u_{\beta}
+
c_{i,\beta}u_{\alpha}
)
\\
=
&n_{\beta}
\sum_{i}
f_{i}^{\text{eq}}(\rho^{\text{G}},\boldsymbol{u})
(
c_{i,\alpha}u_{\beta}
+
c_{i,\beta}u_{\alpha}
)
- 2n_{\beta}
\sum_{i \in \{i | \boldsymbol{n} \cdot \boldsymbol{c}_{i} < 0\}}
f_{i}^{\text{eq}}(\rho^{\text{G}},\boldsymbol{u})
(
c_{i,\alpha}u_{\beta}
+
c_{i,\beta}u_{\alpha}
).
\end{split}
\end{equation}

Combining the transformations~\eqref{eq:fbc-extend-zero-1} and~\eqref{eq:fbc-extend-zero-2}, and inserting the result in the force balance~\eqref{eq:fbc-nbkc-1} gives
\begin{equation}
\begin{split}
\frac{F_{\alpha}}{A}
=
&2 n_{\beta}
\sum_{i \in \{i | \boldsymbol{n} \cdot \boldsymbol{c}_{i} < 0\}}
\Bigl(
f_{i}^{\star}(\boldsymbol{x}, t)
(
c_{i,\alpha}u_{\beta}
+
c_{i,\beta}u_{\alpha}
)
-
f_{i}^{\text{eq}}(\rho^{\text{G}},\boldsymbol{u})
(
c_{i,\alpha}u_{\beta}
+
c_{i,\beta}u_{\alpha}
)
\Bigr)
+
n_{\beta}
f_{0}^{\text{eq}}(\rho^{\text{G}},\boldsymbol{u})
(
u_{\alpha}u_{\beta}
)
\\
&- n_{\beta}
\sum_{i}
f_{i}^{\text{eq}}(\rho^{\text{G}},\boldsymbol{u})
(
c_{i,\alpha}c_{i,\beta}
-
c_{i,\alpha}u_{\beta}
-
c_{i,\beta}u_{\alpha}
+
u_{\alpha}u_{\beta}
)
\\
=
&2 n_{\beta}
\sum_{i \in \{i | \boldsymbol{n} \cdot \boldsymbol{c}_{i} < 0\}}
\Bigl(
f_{i}^{\star}(\boldsymbol{x}, t)
-
f_{i}^{\text{eq}}(\rho^{\text{G}},\boldsymbol{u})
\Bigr)
(
c_{i,\alpha}u_{\beta}
+
c_{i,\beta}u_{\alpha}
)
+
n_{\beta}
f_{0}^{\text{eq}}(\rho^{\text{G}},\boldsymbol{u})
(
u_{\alpha}u_{\beta}
)
\\
&- n_{\beta}
\bigl(
\rho^{\text{G}} c_{s}^{2} \delta_{\alpha\beta}
+
\rho^{\text{G}} u_{\alpha}u_{\beta}
-
\rho^{\text{G}} u_{\alpha}u_{\beta}
-
\rho^{\text{G}} u_{\alpha}u_{\beta}
+
\rho^{\text{G}} u_{\alpha}u_{\beta}
\bigr)
\\
=
&2 n_{\beta}
\sum_{i \in \{i | \boldsymbol{n} \cdot \boldsymbol{c}_{i} < 0\}}
\Bigl(
f_{i}^{\star}(\boldsymbol{x}, t)
-
f_{i}^{\text{eq}}(\rho^{\text{G}},\boldsymbol{u})
\Bigr)
(
c_{i,\alpha}u_{\beta}
+
c_{i,\beta}u_{\alpha}
)
+
n_{\beta}
f_{0}^{\text{eq}}(\rho^{\text{G}},\boldsymbol{u})
(
u_{\alpha}u_{\beta}
)
\\
&
- n_{\alpha} p^{\text{G}},
\end{split}
\end{equation}
where the equilibrium moments~\eqref{eq:fbc-eq-moments} were used.
The expression
\begin{equation}
2 n_{\beta}
\sum_{i \in \{i | \boldsymbol{n} \cdot \boldsymbol{c}_{i} < 0\}}
\Bigl(
f_{i}^{\star}(\boldsymbol{x}, t)
-
f_{i}^{\text{eq}}(\rho^{\text{G}},\boldsymbol{u})
\Bigr)
(
c_{i,\alpha}u_{\beta}
+
c_{i,\beta}u_{\alpha}
)
+
n_{\beta}
f_{0}^{\text{eq}}(\rho^{\text{G}},\boldsymbol{u})
(
u_{\alpha}u_{\beta}
)
\end{equation}
disturbs the force balance at the interface.
\par

\subsection{Only missing (OM)}\label{app:fbc-om}
With $K$ and $R$~\eqref{eq:fsbcv-om-sets}, and the free-surface boundary condition~\eqref{eq:nm-fslbm-boundary-condition}, the force balance \eqref{eq:fsbcv-force-balance-general} becomes
\begin{equation}\label{eq:fbc-om-1}
\begin{split}
\frac{F_{\alpha}}{A} = 
&- n_{\beta}
\sum_{i \in N^{-}}
f_{i}^{\star}(\boldsymbol{x}, t)(c_{i,\alpha} - u_{\alpha})(c_{i,\beta} - u_{\beta})
\\
&- n_{\beta}
\sum_{i \in G^{-}}
\Bigl(
f_{i}^{\text{eq}}(\rho^{\text{G}},\boldsymbol{u})
+
f_{\bar{i}}^{\text{eq}}(\rho^{\text{G}},\boldsymbol{u})
-
f_{\bar{i}}^{\star}(\boldsymbol{x},t)\Bigr)
(c_{i,\alpha} - u_{\alpha})(c_{i,\beta} - u_{\beta}).
\end{split}
\end{equation}
Using the set relations~\eqref{eq:fsbcv-om-set-relations} gives
\begin{equation}\label{eq:fbc-om-revert}
\begin{split}
\sum_{i \in G^{-}}
&f_{\bar{i}}^{\star}(\boldsymbol{x}, t)(c_{i,\alpha} - u_{\alpha})(c_{i,\beta} - u_{\beta})
=
\sum_{i \in G^{-}}
f_{\bar{i}}^{\star}(\boldsymbol{x}, t)(- c_{\bar{i},\alpha} - u_{\alpha})(- c_{\bar{i},\beta} - u_{\beta})
=
\\
&\sum_{i \in G^{-}}
f_{\bar{i}}^{\star}(\boldsymbol{x}, t)(c_{\bar{i},\alpha} + u_{\alpha})(c_{\bar{i},\beta} + u_{\beta})
=
\sum_{i \in G^{+}}
f_{i}^{\star}(\boldsymbol{x}, t)(c_{i,\alpha} + u_{\alpha})(c_{i,\beta} + u_{\beta}),
\end{split}
\end{equation}
and analogously
\begin{equation}\label{eq:fbc-om-revert-eq}
\begin{split}
\sum_{i \in G^{-}}
&f_{\bar{i}}^{\text{eq}}(\rho^{\text{G}},\boldsymbol{u})(c_{i,\alpha} - u_{\alpha})(c_{i,\beta} - u_{\beta})
\\
=
&\sum_{i \in G^{+}}
f_{i}^{\text{eq}}(\rho^{\text{G}},\boldsymbol{u})(c_{i,\alpha} + u_{\alpha})(c_{i,\beta} + u_{\beta}).
\end{split}
\end{equation}
Inserting the transformations~\eqref{eq:fbc-om-revert} and~\eqref{eq:fbc-om-revert-eq} into the force balance~\eqref{eq:fbc-om-1} leads to
\begin{equation}\label{eq:fbc-om-2}
\begin{split}
\frac{F_{\alpha}}{A} = 
&- n_{\beta}
\sum_{i \in N^{-}}
f_{i}^{\star}(\boldsymbol{x}, t)(c_{i,\alpha} - u_{\alpha})(c_{i,\beta} - u_{\beta})
+ n_{\beta}
\sum_{i \in G^{+}}
f_{i}^{\star}(\boldsymbol{x}, t)(c_{i,\alpha} + u_{\alpha})(c_{i,\beta} + u_{\beta})
\\
&- n_{\beta}
\sum_{i \in G^{-}}
f_{i}^{\text{eq}}(\rho^{\text{G}},\boldsymbol{u})(c_{i,\alpha} - u_{\alpha})(c_{i,\beta} - u_{\beta})
- n_{\beta}
\sum_{i \in G^{+}}
f_{i}^{\text{eq}}(\rho^{\text{G}},\boldsymbol{u})(c_{i,\alpha} + u_{\alpha})(c_{i,\beta} + u_{\beta}).
\end{split}
\end{equation}
The last two sums of the force balance~\eqref{eq:fbc-om-2} are extended with zero by adding and subtracting the term 
$n_{\beta} \sum_{i \in G^{+}} f_{i}^{\text{eq}}(\rho^{\text{G}},\boldsymbol{u})(c_{i,\alpha}u_{\beta} + c_{i,\beta}u_{\alpha})$
giving
\begin{equation}\label{eq:fbc-om-extend-zero-1}
\begin{split}
- n_{\beta}
\sum_{i \in G^{-}}
&f_{i}^{\text{eq}}(\rho^{\text{G}},\boldsymbol{u})(c_{i,\alpha} - u_{\alpha})(c_{i,\beta} - u_{\beta})
- n_{\beta}
\sum_{i \in G^{+}}
f_{i}^{\text{eq}}(\rho^{\text{G}},\boldsymbol{u})(c_{i,\alpha} + u_{\alpha})(c_{i,\beta} + u_{\beta})
\\
=
&
- n_{\beta}
\sum_{i \in G^{-}}
f_{i}^{\text{eq}}(\rho^{\text{G}},\boldsymbol{u})
(c_{i,\alpha}c_{i,\beta} - c_{i,\alpha}u_{\beta} - c_{i,\beta}u_{\alpha} + u_{\alpha}u_{\beta})
- n_{\beta}
\sum_{i \in G^{+}}
f_{i}^{\text{eq}}(\rho^{\text{G}},\boldsymbol{u})
(c_{i,\alpha}c_{i,\beta} + c_{i,\alpha}u_{\beta} + c_{i,\beta}u_{\alpha} + u_{\alpha}u_{\beta})
\\
=
&- n_{\beta}
\sum_{i \in G^{-} \cup G^{+}}
f_{i}^{\text{eq}}(\rho^{\text{G}},\boldsymbol{u})
(c_{i,\alpha}c_{i,\beta} + u_{\alpha}u_{\beta})
+ n_{\beta}
\sum_{i \in G^{-}}
f_{i}^{\text{eq}}(\rho^{\text{G}},\boldsymbol{u})(c_{i,\alpha}u_{\beta} + c_{i,\beta}u_{\alpha})
- n_{\beta}
\sum_{i \in G^{+}}
f_{i}^{\text{eq}}(\rho^{\text{G}},\boldsymbol{u})(c_{i,\alpha}u_{\beta} + c_{i,\beta}u_{\alpha})
\\
=
&- n_{\beta}
\sum_{i \in G^{-} \cup G^{+}}
f_{i}^{\text{eq}}(\rho^{\text{G}},\boldsymbol{u})
(c_{i,\alpha}c_{i,\beta} + u_{\alpha}u_{\beta})
+ n_{\beta}
\sum_{i \in G^{-} \cup G^{+}}
f_{i}^{\text{eq}}(\rho^{\text{G}},\boldsymbol{u})(c_{i,\alpha}u_{\beta} + c_{i,\beta}u_{\alpha})
- 2 n_{\beta}
\sum_{i \in G^{+}}
f_{i}^{\text{eq}}(\rho^{\text{G}},\boldsymbol{u})(c_{i,\alpha}u_{\beta} + c_{i,\beta}u_{\alpha})
\\
=
&- n_{\beta}
\sum_{i \in G^{-} \cup G^{+}}
f_{i}^{\text{eq}}(\rho^{\text{G}},\boldsymbol{u})
(c_{i,\alpha}c_{i,\beta} + u_{\alpha}u_{\beta} - c_{i,\alpha}u_{\beta} - c_{i,\beta}u_{\alpha})
- 2 n_{\beta}
\sum_{i \in G^{+}}
f_{i}^{\text{eq}}(\rho^{\text{G}},\boldsymbol{u})(c_{i,\alpha}u_{\beta} + c_{i,\beta}u_{\alpha}).
\end{split}
\end{equation}
The intermediate result~\eqref{eq:fbc-om-extend-zero-1} is again extended by subtracting and adding the term
$n_{\beta}\sum_{i \in T \setminus (G^{-} \cup G^{+})}f_{i}^{\text{eq}}(\rho^{\text{G}},\boldsymbol{u})(c_{i,\alpha}c_{i,\beta} + u_{\alpha}u_{\beta} - c_{i,\alpha}u_{\beta} - c_{i,\beta}u_{\alpha})$ to obtain
\begin{equation}\label{eq:fbc-om-extend-zero-2}
\begin{split}
- n_{\beta}
&\sum_{i \in G^{-} \cup G^{+}}
f_{i}^{\text{eq}}(\rho^{\text{G}},\boldsymbol{u})
(c_{i,\alpha}c_{i,\beta} + u_{\alpha}u_{\beta} - c_{i,\alpha}u_{\beta} - c_{i,\beta}u_{\alpha})
- 2 n_{\beta}
\sum_{i \in G^{+}}
f_{i}^{\text{eq}}(\rho^{\text{G}},\boldsymbol{u})(c_{i,\alpha}u_{\beta} + c_{i,\beta}u_{\alpha})
\\
=
& - n_{\beta}
\sum_{i \in T} f_{i}^{\text{eq}}(\rho^{\text{G}},\boldsymbol{u})
(c_{i,\alpha}c_{i,\beta} + u_{\alpha}u_{\beta} - c_{i,\alpha}u_{\beta} - c_{i,\beta}u_{\alpha})
+
n_{\beta}
\sum_{i \in T \setminus (G^{-} \cup G^{+})} f_{i}^{\text{eq}}(\rho^{\text{G}},\boldsymbol{u})
(c_{i,\alpha}c_{i,\beta} + u_{\alpha}u_{\beta} - c_{i,\alpha}u_{\beta} - c_{i,\beta}u_{\alpha})
\\
& - 2 n_{\beta}
\sum_{i \in G^{+}}
f_{i}^{\text{eq}}(\rho^{\text{G}},\boldsymbol{u})(c_{i,\alpha}u_{\beta} + c_{i,\beta}u_{\alpha}),
\end{split}
\end{equation}
which can be rewritten using the the equilibrium moments~\eqref{eq:fbc-eq-moments} as
\begin{equation}\label{eq:fbc-om-extend-zero-3}
\begin{split}
- n_{\beta}
&\sum_{i \in T} f_{i}^{\text{eq}}(\rho^{\text{G}},\boldsymbol{u})
(c_{i,\alpha}c_{i,\beta} + u_{\alpha}u_{\beta} - c_{i,\alpha}u_{\beta} - c_{i,\beta}u_{\alpha})
+
n_{\beta}
\sum_{i \in T \setminus (G^{-} \cup G^{+})} f_{i}^{\text{eq}}(\rho^{\text{G}},\boldsymbol{u})
(c_{i,\alpha}c_{i,\beta} + u_{\alpha}u_{\beta} - c_{i,\alpha}u_{\beta} - c_{i,\beta}u_{\alpha})
\\
& - 2 n_{\beta}
\sum_{i \in G^{+}}
f_{i}^{\text{eq}}(\rho^{\text{G}},\boldsymbol{u})(c_{i,\alpha}u_{\beta} + c_{i,\beta}u_{\alpha})
\\
=
& - n_{\beta}
\bigl(
\rho^{\text{G}} c_{s}^{2} \delta_{\alpha\beta}
+
\rho^{\text{G}} u_{\alpha}u_{\beta}
+
\rho^{\text{G}} u_{\alpha}u_{\beta}
-
\rho^{\text{G}} u_{\alpha}u_{\beta}
-
\rho^{\text{G}} u_{\alpha}u_{\beta}
\bigr)
+
n_{\beta}
\sum_{i \in T \setminus (G^{-} \cup G^{+})} f_{i}^{\text{eq}}(\rho^{\text{G}},\boldsymbol{u})
(c_{i,\alpha}c_{i,\beta} + u_{\alpha}u_{\beta} - c_{i,\alpha}u_{\beta} - c_{i,\beta}u_{\alpha})
\\
& - 2 n_{\beta}
\sum_{i \in G^{+}}
f_{i}^{\text{eq}}(\rho^{\text{G}},\boldsymbol{u})(c_{i,\alpha}u_{\beta} + c_{i,\beta}u_{\alpha}).
\end{split}
\end{equation}
Inserting the transformation~\eqref{eq:fbc-om-extend-zero-3} into the force balance~\eqref{eq:fbc-om-2} gives
\begin{equation}\label{eq:fbc-om-3}
\begin{split}
\frac{F_{\alpha}}{A} = 
&- n_{\beta}
\sum_{i \in N^{-}}
f_{i}^{\star}(\boldsymbol{x}, t)(c_{i,\alpha}c_{i,\beta} - c_{i,\alpha}u_{\beta} - c_{i,\beta}u_{\alpha} + u_{\alpha}u_{\beta})
+ n_{\beta}
\sum_{i \in G^{+}}
f_{i}^{\star}(\boldsymbol{x}, t)(c_{i,\alpha}c_{i,\beta} + c_{i,\alpha}u_{\beta} + c_{i,\beta}u_{\alpha} + u_{\alpha}u_{\beta})
\\
& + n_{\beta}
\sum_{i \in T \setminus (G^{-} \cup G^{+})} f_{i}^{\text{eq}}(\rho^{\text{G}},\boldsymbol{u})
(c_{i,\alpha}c_{i,\beta} + u_{\alpha}u_{\beta} - c_{i,\alpha}u_{\beta} - c_{i,\beta}u_{\alpha})
- 2 n_{\beta}
\sum_{i \in G^{+}}
f_{i}^{\text{eq}}(\rho^{\text{G}},\boldsymbol{u})(c_{i,\alpha}u_{\beta} + c_{i,\beta}u_{\alpha})
\\
& - n_{\beta}
\rho^{\text{G}} c_{s}^{2} \delta_{\alpha\beta}
\\
=
&- n_{\beta}
\sum_{i \in N^{-}}
f_{i}^{\star}(\boldsymbol{x}, t)(c_{i,\alpha}c_{i,\beta} - c_{i,\alpha}u_{\beta} - c_{i,\beta}u_{\alpha} + u_{\alpha}u_{\beta})
+ n_{\beta}
\sum_{i \in G^{+}}
f_{i}^{\star}(\boldsymbol{x}, t)(c_{i,\alpha}c_{i,\beta} + c_{i,\alpha}u_{\beta} + c_{i,\beta}u_{\alpha} + u_{\alpha}u_{\beta})
\\
& + n_{\beta}
\sum_{i \in T \setminus (G^{-} \cup G^{+})} f_{i}^{\text{eq}}(\rho^{\text{G}},\boldsymbol{u})
(c_{i,\alpha}c_{i,\beta} - c_{i,\alpha}u_{\beta} - c_{i,\beta}u_{\alpha} + u_{\alpha}u_{\beta} )
- 2 n_{\beta}
\sum_{i \in G^{+}}
f_{i}^{\text{eq}}(\rho^{\text{G}},\boldsymbol{u})(c_{i,\alpha}u_{\beta} + c_{i,\beta}u_{\alpha})
\\
& - n_{\alpha} p^{\text{G}},
\end{split}
\end{equation}
where the balance of the forces is disturbed by
\begin{equation}
\begin{split}
&- n_{\beta}
\sum_{i \in N^{-}}
f_{i}^{\star}(\boldsymbol{x}, t)(c_{i,\alpha}c_{i,\beta} - c_{i,\alpha}u_{\beta} - c_{i,\beta}u_{\alpha} + u_{\alpha}u_{\beta})
+ n_{\beta}
\sum_{i \in G^{+}}
f_{i}^{\star}(\boldsymbol{x}, t)(c_{i,\alpha}c_{i,\beta} + c_{i,\alpha}u_{\beta} + c_{i,\beta}u_{\alpha} + u_{\alpha}u_{\beta})
\\
& + n_{\beta}
\sum_{i \in T \setminus (G^{-} \cup G^{+})} f_{i}^{\text{eq}}(\rho^{\text{G}},\boldsymbol{u})
(c_{i,\alpha}c_{i,\beta} - c_{i,\alpha}u_{\beta} - c_{i,\beta}u_{\alpha} + u_{\alpha}u_{\beta} )
- 2 n_{\beta}
\sum_{i \in G^{+}}
f_{i}^{\text{eq}}(\rho^{\text{G}},\boldsymbol{u})(c_{i,\alpha}u_{\beta} + c_{i,\beta}u_{\alpha}).
\end{split}
\end{equation}

\section{Numerical experiments}\label{app:ne}

This section extends \Cref{sec:ne} with additional results and figures.

\subsection{Gravity wave}\label{app:ne-gw}

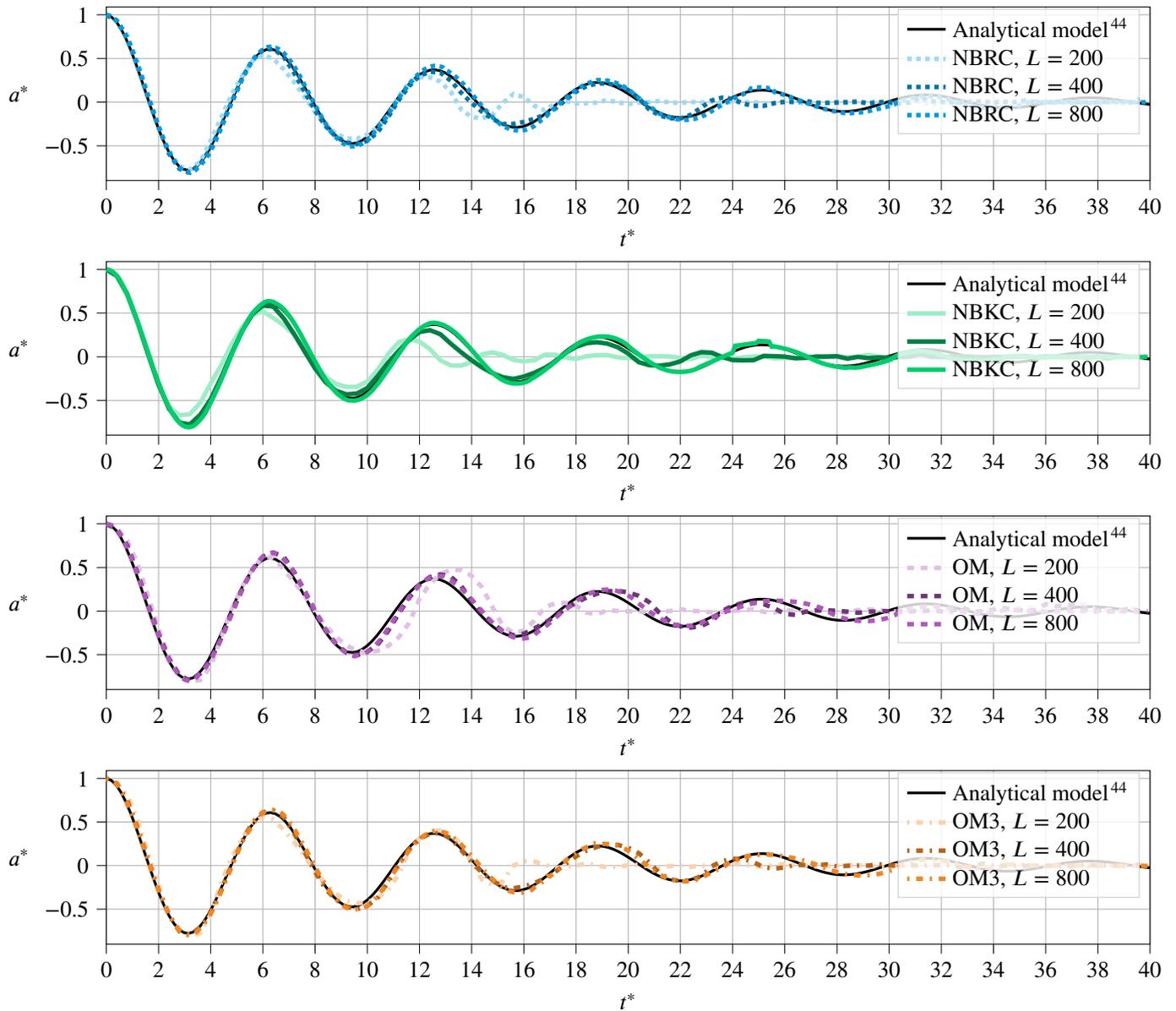
\begin{figure}[h!]
\centering
\setlength{\figureheight}{0.24\textwidth}
\setlength{\figurewidth}{1\textwidth}
	\begin{subfigure}{\textwidth}
		\input{figures/gravity-wave/convergence-nbrc.tex}%
	\end{subfigure}
	\hfill
	\begin{subfigure}{\textwidth}
		\input{figures/gravity-wave/convergence-nbkc.tex}%
	\end{subfigure}
	\begin{subfigure}{\textwidth}
		\input{figures/gravity-wave/convergence-om.tex}%
	\end{subfigure}
	\hfill
	\begin{subfigure}{\textwidth}
		\input{figures/gravity-wave/convergence-om3.tex}%
	\end{subfigure}
	\caption{\label{fig:app-ne-gravity-wave-convergence}
		Simulated surface elevation of the gravity wave in terms of non-dimensional amplitude $a^{*}(0,t^{*})$ and time $t^{*}$.
		The simulations were performed with computational domain resolutions, that is, wavelengths of $L \in \{200, 400, 800\}$ lattice cells.
		A higher computational domain resolution captures more of the standing wave's oscillations.
		}
\end{figure}

\FloatBarrier
\newpage

\subsection{Rectangular dam break}\label{app:ne-rdb}
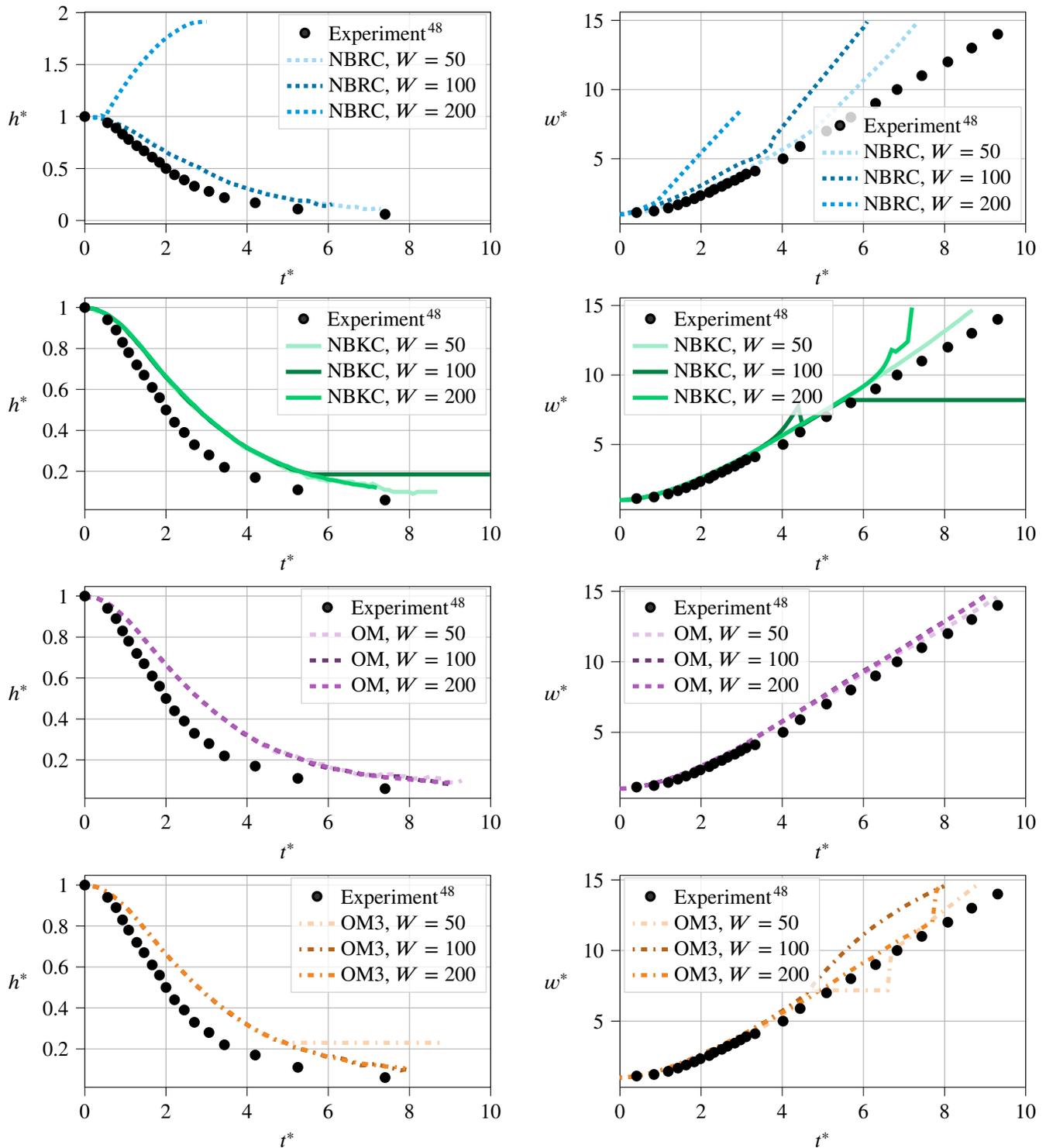
\begin{figure}[h!]
	\centering
	\setlength{\figureheight}{0.29\textwidth}
	\setlength{\figurewidth}{0.475\textwidth}
	\begin{subfigure}[t]{0.49\textwidth}
		\centering
		\input{figures/dam-break-rectangular/convergence-y-nbrc.tex}%
	\end{subfigure}
	\hfill
	\begin{subfigure}[t]{0.49\textwidth}
		\centering
		\input{figures/dam-break-rectangular/convergence-x-nbrc.tex}%
	\end{subfigure}
	\begin{subfigure}[t]{0.49\textwidth}
		\centering
		\input{figures/dam-break-rectangular/convergence-y-nbkc.tex}%
	\end{subfigure}
	\hfill
	\begin{subfigure}[t]{0.49\textwidth}
		\centering
		\input{figures/dam-break-rectangular/convergence-x-nbkc.tex}%
	\end{subfigure}
	\begin{subfigure}[t]{0.49\textwidth}
		\centering
		\input{figures/dam-break-rectangular/convergence-y-om.tex}%
	\end{subfigure}
	\hfill
	\begin{subfigure}[t]{0.49\textwidth}
		\centering
		\input{figures/dam-break-rectangular/convergence-x-om.tex}%
	\end{subfigure}
	\begin{subfigure}[t]{0.49\textwidth}
		\centering
		\input{figures/dam-break-rectangular/convergence-y-om3.tex}%
	\end{subfigure}
	\hfill
	\begin{subfigure}[t]{0.49\textwidth}
		\centering
		\input{figures/dam-break-rectangular/convergence-x-om3.tex}%
	\end{subfigure}
	\caption{\label{fig:dam-break-rectangular-convergence}
		Simulated rectangular dam break with non-dimensionalized residual dam height $h^{*}(t^{*})$, width $w^{*}(t^{*})$, and time $t^{*}$.
		The simulations were performed with computational domain resolutions, that is, initial dam widths of $W \in \{50, 100, 200\}$ lattice cells.
		Only the OM variant converged well. The other variants led to splashing as described in \Cref{sec:ne-rdb-rad}.
		These splash droplets disturbed the evaluation algorithm and even led to numerical instabilities as observed for the NBKC variant with $W=100$.
	}
\end{figure}

\FloatBarrier
\newpage

\subsection{Cylindrical dam break}\label{app:ne-cdb}
\begin{figure}[h!]
	\centering
	\setlength{\figureheight}{0.35\textwidth}
	\setlength{\figurewidth}{0.475\textwidth}
	\begin{subfigure}[t]{0.49\textwidth}
		\centering
		\input{figures/dam-break-cylindrical/convergence-nbrc.tex}%
	\end{subfigure}
	\hfill
	\begin{subfigure}[t]{0.49\textwidth}
		\centering
		\input{figures/dam-break-cylindrical/convergence-nbkc.tex}%
	\end{subfigure}
	\hfill
	\begin{subfigure}[t]{0.49\textwidth}
		\centering
		\input{figures/dam-break-cylindrical/convergence-om.tex}%
	\end{subfigure}
	\hfill
	\begin{subfigure}[t]{0.49\textwidth}
		\centering
		\input{figures/dam-break-cylindrical/convergence-om3.tex}%
	\end{subfigure}
	\caption{\label{fig:dam-break-cylindrical-convergence}
		Simulated cylindrical dam break with non-dimensionalized liquid column radius $r^{*}(t^{*})$ and time $t^{*}$.
		The simulations were performed with computational domain resolutions, that is, initial column diameters of $D \in \{50, 100, 200\}$ lattice cells.
		The markers represent the mean values of $r^{*}(t^{*})$.
		All variants but the NBRC converged well.
	}
\end{figure}
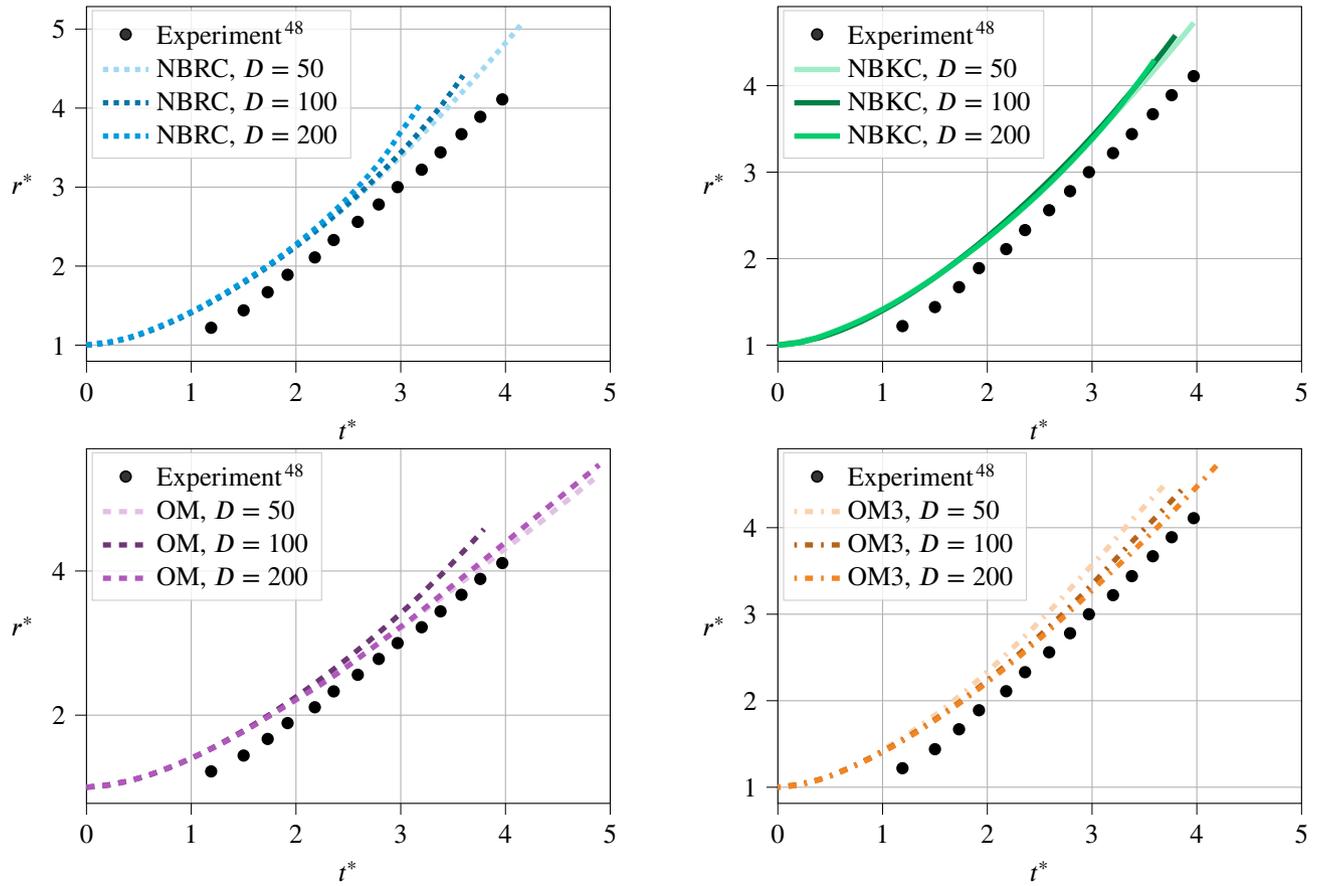

\FloatBarrier
\newpage

\subsection{Taylor bubble}\label{app:ne-tb}
\enlargethispage{2\baselineskip}
\begin{figure}[h!]
	\centering
	\setlength{\figureheight}{0.3\textwidth}
	\setlength{\figurewidth}{0.48\textwidth}
	\begin{subfigure}[t]{0.49\textwidth}
		\centering
		\input{figures/taylor-bubble/convergence-shape-front-nbrc.tex}%
	\end{subfigure}
	\hfill
	\begin{subfigure}[t]{0.49\textwidth}
		\centering
		\input{figures/taylor-bubble/convergence-shape-tail-nbrc.tex}%
	\end{subfigure}
	\begin{subfigure}[t]{0.49\textwidth}
		\centering
		\input{figures/taylor-bubble/convergence-shape-front-nbkc.tex}%
	\end{subfigure}
	\hfill
	\begin{subfigure}[t]{0.49\textwidth}
		\centering
		\input{figures/taylor-bubble/convergence-shape-tail-nbkc.tex}%
	\end{subfigure}
	\begin{subfigure}[t]{0.49\textwidth}
		\centering
		\input{figures/taylor-bubble/convergence-shape-front-om.tex}%
	\end{subfigure}
	\hfill
	\begin{subfigure}[t]{0.49\textwidth}
		\centering
		\input{figures/taylor-bubble/convergence-shape-tail-om.tex}%
	\end{subfigure}
	\begin{subfigure}[t]{0.49\textwidth}
		\centering
		\input{figures/taylor-bubble/convergence-shape-front-om3.tex}%
		\caption*{Bubble front}
	\end{subfigure}
	\hfill
	\begin{subfigure}[t]{0.49\textwidth}
		\centering
		\input{figures/taylor-bubble/convergence-shape-tail-om3.tex}%
		\caption*{Bubble tail}
	\end{subfigure}
	\caption{\label{fig:taylor-bubble-shape-convergence}
		Simulated shape of the Taylor bubble's front and tail.
		The simulations were performed with computational domain resolutions, that is, tube diameters of $D \in \{32, 64, 128\}$ lattice cells.
		The comparison with experimental data~\cite{bugg2002VelocityFieldTaylor} was drawn in terms of the non-dimensionalized axial location $z^{*}$ and radial location $r^{*}$ at time $t^{*}=15$.
		All boundary condition variants converged well.
	}
\end{figure}
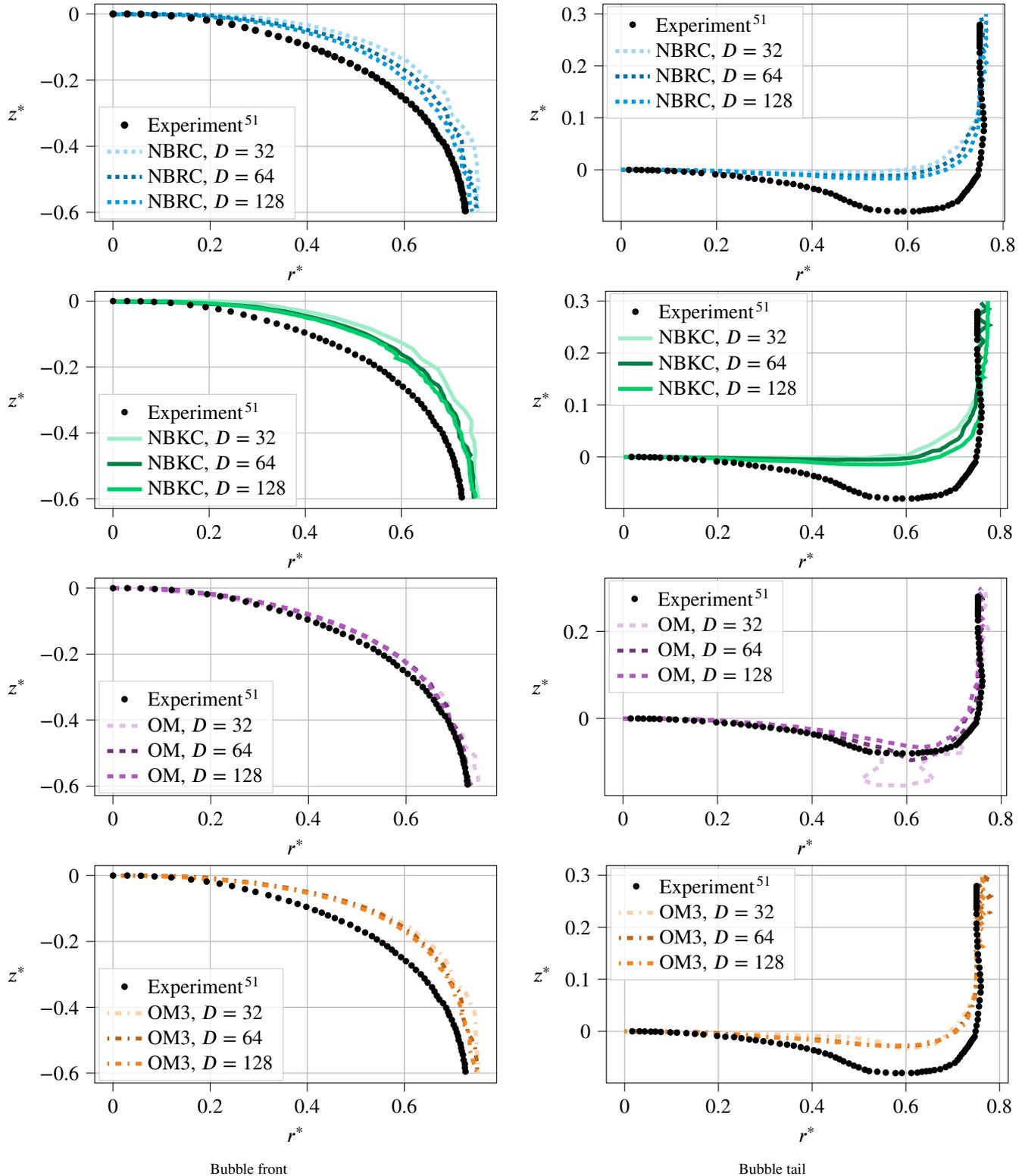

\FloatBarrier

%% file: figures/gravity-wave/convergence-nbrc.tex
% This file was created with tikzplotlib v0.10.1.
\begin{tikzpicture}

\definecolor{darkgray176}{RGB}{176,176,176}
\definecolor{dodgerblue0154222}{RGB}{0,154,222}
\definecolor{lightgray204}{RGB}{204,204,204}
\definecolor{mediumturquoise64179230}{RGB}{0,116,166}
\definecolor{skyblue128205238}{RGB}{159,217,243}

\begin{axis}[
height=\figureheight,
legend cell align={left},
legend style={
	fill opacity=0.8,
	draw opacity=1,
	text opacity=1,
	at={(0.99,0.99)},
	anchor=north east,
	draw=lightgray204
},
tick align=outside,
tick pos=left,
width=\figurewidth,
x grid style={darkgray176},
xlabel={\(\displaystyle t^{*}\)},
xmajorgrids,
xmin=0, xmax=40,
xtick style={color=black},
y grid style={darkgray176},
ylabel style={rotate=-90.0},
ylabel={\(\displaystyle a^{*}\)},
ymajorgrids,
ymin=-0.8969269, ymax=1.0890329,
ytick style={color=black}
]
\addplot [very thick, black]
table {%
0 0.998026728630066
0.185185194015503 0.9781174659729
0.370370388031006 0.925613880157471
0.555555582046509 0.843235969543457
0.740740776062012 0.734638094902039
0.925925970077515 0.604258418083191
1.11111116409302 0.457145810127258
1.29629623889923 0.298771142959595
1.66666662693024 -0.0289593935012817
1.85185182094574 -0.187046051025391
2.03703713417053 -0.33423376083374
2.22222232818604 -0.465845823287964
2.40740752220154 -0.577867984771729
2.59259247779846 -0.667066097259521
2.77777767181396 -0.731073379516602
2.96296286582947 -0.768445372581482
3.14814805984497 -0.778682231903076
3.33333325386047 -0.762219071388245
3.51851844787598 -0.720383405685425
3.70370364189148 -0.655325293540955
3.88888883590698 -0.569920063018799
4.07407426834106 -0.467650175094604
4.25925922393799 -0.352469682693481
4.44444465637207 -0.228656530380249
4.81481504440308 0.0270652770996094
5 0.150189638137817
5.18518495559692 0.264672756195068
5.37037038803101 0.366881251335144
5.55555534362793 0.453703284263611
5.74074077606201 0.522638082504272
5.92592573165894 0.571863412857056
6.11111116409302 0.600278258323669
6.29629611968994 0.607518672943115
6.48148155212402 0.593949556350708
6.66666650772095 0.560630798339844
6.85185194015503 0.509261131286621
7.03703689575195 0.442102909088135
7.22222232818604 0.361888408660889
7.40740728378296 0.271714210510254
7.59259271621704 0.174924850463867
7.96296310424805 -0.0246013402938843
8.14814853668213 -0.1204913854599
8.33333301544189 -0.209532022476196
8.51851844787598 -0.288901329040527
8.70370388031006 -0.356186509132385
8.88888931274414 -0.409454345703125
9.07407379150391 -0.447302341461182
9.25925922393799 -0.468891620635986
9.44444465637207 -0.473958611488342
9.62963008880615 -0.462807536125183
9.81481456756592 -0.436284065246582
10 -0.39573061466217
10.1851854324341 -0.342926740646362
10.3703699111938 -0.28001594543457
10.5555553436279 -0.209422945976257
10.740740776062 -0.133762836456299
11.1111106872559 0.0219100713729858
11.2962961196899 0.0965865850448608
11.481481552124 0.165835618972778
11.6666669845581 0.227465391159058
11.8518514633179 0.279605627059937
12.037036895752 0.320761799812317
12.222222328186 0.349854826927185
12.4074077606201 0.366245865821838
12.5925922393799 0.369745016098022
12.777777671814 0.360605359077454
12.962963104248 0.339500904083252
13.1481485366821 0.307492256164551
13.3333330154419 0.265979051589966
13.518518447876 0.2166428565979
13.7037038803101 0.16138219833374
13.8888893127441 0.10224175453186
14.259259223938 -0.0192111730575562
14.4444446563721 -0.077364444732666
14.6296300888062 -0.131218194961548
14.8148145675659 -0.179070353507996
15 -0.219471216201782
15.1851854324341 -0.251265406608582
15.3703699111938 -0.273622870445251
15.5555553436279 -0.286057472229004
15.740740776062 -0.28843355178833
15.9259262084961 -0.280960321426392
16.1111106872559 -0.264175057411194
16.2962970733643 -0.23891544342041
16.481481552124 -0.206282138824463
16.6666660308838 -0.167594313621521
16.8518524169922 -0.124338269233704
17.037036895752 -0.0781127214431763
17.4074077606201 0.0166382789611816
17.5925922393799 0.0619223117828369
17.7777786254883 0.103801369667053
17.962963104248 0.140953660011292
18.1481475830078 0.172255396842957
18.3333339691162 0.196813941001892
18.518518447876 0.213990688323975
18.7037029266357 0.223415970802307
18.8888893127441 0.224993705749512
19.0740737915039 0.218896269798279
19.2592601776123 0.205551862716675
19.4444446563721 0.185621857643127
19.6296291351318 0.159971714019775
19.8148155212402 0.129636168479919
20 0.0957789421081543
20.1851844787598 0.0596495866775513
20.5555553436279 -0.0142668485641479
20.7407398223877 -0.0495278835296631
20.9259262084961 -0.0820932388305664
21.1111106872559 -0.110936284065247
21.2962970733643 -0.135186195373535
21.481481552124 -0.154153108596802
21.6666660308838 -0.167346119880676
21.8518524169922 -0.174484133720398
22.037036895752 -0.175499677658081
22.2222213745117 -0.170534729957581
22.4074077606201 -0.159930109977722
22.5925922393799 -0.144207835197449
22.7777786254883 -0.124048590660095
22.962963104248 -0.100263833999634
23.1481475830078 -0.0737646818161011
23.518518447876 -0.0165609121322632
23.7037029266357 0.0121328830718994
23.8888893127441 0.0395880937576294
24.0740737915039 0.064909815788269
24.2592601776123 0.0873005390167236
24.4444446563721 0.106085658073425
24.6296291351318 0.120732069015503
24.8148155212402 0.130862236022949
25 0.136263012886047
25.1851844787598 0.136887431144714
25.3703708648682 0.132851958274841
25.5555553436279 0.124427795410156
25.7407398223877 0.112027168273926
25.9259262084961 0.0961849689483643
26.1111106872559 0.0775377750396729
26.2962970733643 0.0567986965179443
26.6666660308838 0.0121220350265503
26.8518524169922 -0.010246753692627
27.037036895752 -0.0316232442855835
27.2222213745117 -0.0513113737106323
27.4074077606201 -0.0686922073364258
27.5925922393799 -0.083242654800415
27.7777786254883 -0.0945510864257812
27.962963104248 -0.102327346801758
28.1481475830078 -0.106409549713135
28.3333339691162 -0.106765747070312
28.518518447876 -0.103491306304932
28.7037029266357 -0.0968018770217896
28.8888893127441 -0.0870226621627808
29.0740737915039 -0.0745742321014404
29.2592601776123 -0.0599559545516968
29.4444446563721 -0.0437257289886475
30.1851844787598 0.0252456665039062
30.3703708648682 0.0405528545379639
30.5555553436279 0.0540436506271362
30.7407398223877 0.0653131008148193
30.9259262084961 0.0740430355072021
31.1111106872559 0.0800105333328247
31.2962970733643 0.0830929279327393
31.481481552124 0.0832687616348267
31.6666660308838 0.0806158781051636
31.8518524169922 0.0753058195114136
32.037036895752 0.0675951242446899
32.2222213745117 0.0578144788742065
32.4074058532715 0.0463552474975586
32.7777786254883 0.0201783180236816
33.1481475830078 -0.00718581676483154
33.3333320617676 -0.020142674446106
33.5185203552246 -0.0320430994033813
33.7037048339844 -0.0425139665603638
33.8888893127441 -0.051241397857666
34.0740737915039 -0.0579798221588135
34.2592582702637 -0.0625578165054321
34.4444427490234 -0.0648825168609619
34.6296310424805 -0.0649402141571045
34.8148155212402 -0.0627939701080322
35 -0.0585802793502808
35.1851844787598 -0.0525015592575073
35.3703689575195 -0.0448178052902222
35.5555572509766 -0.0358356237411499
35.9259262084961 -0.0153669118881226
36.4814796447754 0.0160623788833618
36.6666679382324 0.0253137350082397
36.8518524169922 0.033440113067627
37.037036895752 0.0401983261108398
37.2222213745117 0.0453987121582031
37.4074058532715 0.0489096641540527
37.5925941467285 0.0506608486175537
37.7777786254883 0.0506438016891479
37.962963104248 0.0489097833633423
38.1481475830078 0.0455672740936279
38.3333320617676 0.0407758951187134
38.5185203552246 0.0347399711608887
38.8888893127441 0.0199233293533325
39.6296310424805 -0.0128018856048584
39.8148155212402 -0.0199935436248779
40 -0.0262999534606934
};
\addlegendentry{Analytical model~\cite{dingemans1997WaterWavePropagation}}
\addplot [line width=2pt, skyblue128205238, dotted]
table {%
0 0.995025992393494
0.399999976158142 0.935208082199097
0.799999952316284 0.767857074737549
1.20000004768372 0.467030048370361
1.60000002384186 0.0924720764160156
2 -0.275756001472473
2.40000009536743 -0.574639081954956
2.79999995231628 -0.73415207862854
3.20000004768372 -0.771013021469116
3.59999990463257 -0.680015087127686
4 -0.438637971878052
4.40000009536743 -0.177571058273315
4.80000019073486 0.0850093364715576
5.19999980926514 0.298750996589661
5.59999990463257 0.470911979675293
6 0.539927005767822
6.40000009536743 0.49440598487854
6.80000019073486 0.375720977783203
7.19999980926514 0.226593017578125
7.59999990463257 0.0592267513275146
8 -0.116307020187378
8.39999961853027 -0.256070971488953
8.80000019073486 -0.359977006912231
9.19999980926514 -0.41559100151062
9.60000038146973 -0.419651985168457
10 -0.371721982955933
10.3999996185303 -0.28407096862793
10.8000001907349 -0.154155969619751
11.1999998092651 0.0643396377563477
11.6000003814697 0.219594955444336
12 0.282770991325378
12.3999996185303 0.285156965255737
12.8000001907349 0.235059022903442
13.1999998092651 0.118633031845093
13.6000003814697 -0.0850067138671875
14 -0.169780015945435
14.3999996185303 -0.184782981872559
14.8000001907349 -0.143133997917175
15.1999998092651 -0.00863587856292725
15.6000003814697 0.0932892560958862
16 0.0573194026947021
16.3999996185303 -0.041633129119873
16.7999992370605 -0.0943056344985962
17.2000007629395 -0.0168744325637817
17.6000003814697 0.0109604597091675
18 -0.0139733552932739
18.3999996185303 -0.0137654542922974
18.7999992370605 -0.00365889072418213
19.2000007629395 0.0232893228530884
19.6000003814697 -0.00311803817749023
20 -0.0192924737930298
20.3999996185303 -0.00655722618103027
20.7999992370605 -0.00105416774749756
21.2000007629395 0.0130815505981445
21.6000003814697 -0.00417256355285645
22 -0.00815236568450928
22.3999996185303 0.0030820369720459
22.7999992370605 -0.00475835800170898
23.2000007629395 0.00922369956970215
23.6000003814697 -0.0128821134567261
24 0.00335574150085449
24.3999996185303 -0.00221538543701172
24.7999992370605 0.00200915336608887
25.2000007629395 -0.0150072574615479
25.6000003814697 -0.0137543678283691
26 0.0129028558731079
26.3999996185303 -0.00934433937072754
26.7999992370605 -0.00937187671661377
27.2000007629395 -0.00357651710510254
27.6000003814697 -0.0091785192489624
28 -0.00447404384613037
28.3999996185303 -0.0101529359817505
28.7999992370605 0.00318920612335205
29.2000007629395 -0.00280582904815674
29.6000003814697 0.00963568687438965
30 -0.0260696411132812
30.3999996185303 -0.00630331039428711
30.7999992370605 0.00920367240905762
31.2000007629395 0.00547266006469727
31.6000003814697 -0.0310415029525757
32 0.00228762626647949
32.4000015258789 -0.0126080513000488
32.7999992370605 0.00121772289276123
33.2000007629395 0.0105049610137939
33.5999984741211 -0.00377845764160156
34.4000015258789 0.0207148790359497
34.7999992370605 0.00149309635162354
35.2000007629395 -0.00257456302642822
35.5999984741211 0.00547480583190918
36 -0.00127482414245605
36.4000015258789 0.0134059190750122
36.7999992370605 0.0125515460968018
37.2000007629395 -0.0132999420166016
37.5999984741211 -0.00485694408416748
38 -0.00466394424438477
38.4000015258789 0.010765552520752
38.7999992370605 -0.0166751146316528
39.2000007629395 -0.0114587545394897
39.5999984741211 -0.00120866298675537
};
\addlegendentry{NBRC, $L=200$}
\addplot [line width=2pt, mediumturquoise64179230, dotted]
table {%
0 0.997520923614502
0.399999976158142 0.933876037597656
0.799999952316284 0.735665082931519
1.20000004768372 0.420966029167175
1.60000002384186 0.0528105497360229
2 -0.286443948745728
2.40000009536743 -0.58842396736145
2.79999995231628 -0.763504981994629
3.20000004768372 -0.806656002998352
3.59999990463257 -0.706192970275879
4 -0.528813004493713
4.40000009536743 -0.277652978897095
4.80000019073486 0.00165247917175293
5.19999980926514 0.272518992424011
5.59999990463257 0.482282996177673
6 0.595530986785889
6.40000009536743 0.610607981681824
6.80000019073486 0.517693042755127
7.19999980926514 0.331699013710022
7.59999990463257 0.117066025733948
8 -0.0954737663269043
8.39999961853027 -0.273880958557129
8.80000019073486 -0.395832061767578
9.19999980926514 -0.463178038597107
9.60000038146973 -0.479303002357483
10 -0.442353010177612
10.3999996185303 -0.343644022941589
10.8000001907349 -0.16943895816803
11.1999998092651 0.0206210613250732
11.6000003814697 0.18202805519104
12 0.300122976303101
12.3999996185303 0.348397016525269
12.8000001907349 0.337417960166931
13.1999998092651 0.245305061340332
13.6000003814697 0.10748302936554
14 -0.00969088077545166
14.3999996185303 -0.11087703704834
14.8000001907349 -0.19336199760437
15.1999998092651 -0.249812006950378
15.6000003814697 -0.248748064041138
16 -0.226166963577271
16.3999996185303 -0.19058895111084
16.7999992370605 -0.143456935882568
17.2000007629395 -0.0751285552978516
17.6000003814697 0.0236650705337524
18 0.11955201625824
18.3999996185303 0.182132959365845
18.7999992370605 0.213389039039612
19.2000007629395 0.213618993759155
19.6000003814697 0.188823938369751
20 0.137946963310242
20.3999996185303 0.0486909151077271
20.7999992370605 -0.0826553106307983
21.2000007629395 -0.151389002799988
21.6000003814697 -0.185407996177673
22 -0.187262058258057
22.3999996185303 -0.143787980079651
23.2000007629395 0.0196247100830078
23.6000003814697 0.0515344142913818
24 0.0372631549835205
24.3999996185303 -0.0166867971420288
24.7999992370605 -0.0424220561981201
25.2000007629395 -0.0430850982666016
25.6000003814697 -0.0110718011856079
26 0.00231385231018066
26.3999996185303 0.00623273849487305
26.7999992370605 0.0149085521697998
27.2000007629395 -0.00879442691802979
27.6000003814697 -0.00817549228668213
28.3999996185303 0.000119566917419434
28.7999992370605 0.00636482238769531
29.2000007629395 -0.0163619518280029
29.6000003814697 0.000645041465759277
30 -0.00134515762329102
30.3999996185303 -0.0162274837493896
30.7999992370605 -0.00550532341003418
31.2000007629395 0.0517901182174683
31.6000003814697 0.00880038738250732
32 0.00499129295349121
32.4000015258789 -0.00354087352752686
32.7999992370605 -0.00429201126098633
33.2000007629395 0.00378203392028809
33.5999984741211 0.0102930068969727
34 -0.00857651233673096
34.4000015258789 0.000809669494628906
34.7999992370605 -0.00243532657623291
35.2000007629395 0.0196868181228638
35.5999984741211 -0.00272202491760254
36 -0.00708043575286865
36.4000015258789 0.00273072719573975
36.7999992370605 0.0101170539855957
37.2000007629395 0.00213992595672607
37.5999984741211 -0.00159609317779541
38 -0.00420093536376953
38.4000015258789 0.00919854640960693
38.7999992370605 0.0156954526901245
39.2000007629395 -0.00696003437042236
39.5999984741211 0.0403460264205933
};
\addlegendentry{NBRC, $L=400$}
\addplot [line width=2pt, dodgerblue0154222, dotted]
table {%
0 0.998762011528015
0.100000023841858 0.994367003440857
0.200000047683716 0.981643915176392
0.299999952316284 0.960475921630859
0.399999976158142 0.930549025535583
0.5 0.891003012657166
0.600000023841858 0.83961296081543
0.700000047683716 0.783674955368042
0.799999952316284 0.720283031463623
0.899999976158142 0.651365041732788
1 0.575496912002563
1.10000002384186 0.496871948242188
1.20000004768372 0.413831949234009
1.29999995231628 0.328215956687927
1.5 0.153197050094604
1.60000002384186 0.0648585557937622
1.79999995231628 -0.108137011528015
1.89999997615814 -0.191972970962524
2 -0.277402997016907
2.09999990463257 -0.355728030204773
2.20000004768372 -0.427456974983215
2.29999995231628 -0.497040987014771
2.40000009536743 -0.558259963989258
2.5 -0.616128921508789
2.59999990463257 -0.666401028633118
2.70000004768372 -0.709881067276001
2.79999995231628 -0.746645927429199
2.90000009536743 -0.773880004882812
3 -0.792703986167908
3.09999990463257 -0.802834033966064
3.20000004768372 -0.80439305305481
3.29999995231628 -0.797539949417114
3.40000009536743 -0.782029986381531
3.5 -0.75064492225647
3.59999990463257 -0.724220991134644
3.70000004768372 -0.688231945037842
3.79999995231628 -0.645879030227661
3.90000009536743 -0.607663989067078
4 -0.554832935333252
4.09999990463257 -0.49658203125
4.19999980926514 -0.434682965278625
4.40000009536743 -0.297172069549561
4.80000019073486 -0.00578880310058594
4.90000009536743 0.0548908710479736
5 0.123865008354187
5.09999990463257 0.189892053604126
5.19999980926514 0.257880926132202
5.30000019073486 0.318364024162292
5.40000009536743 0.375046014785767
5.5 0.426530003547668
5.59999990463257 0.473263025283813
5.69999980926514 0.515673995018005
5.80000019073486 0.550190925598145
5.90000009536743 0.578727960586548
6 0.601814031600952
6.09999990463257 0.618965983390808
6.19999980926514 0.632695913314819
6.30000019073486 0.634734988212585
6.40000009536743 0.624302983283997
6.5 0.620525002479553
6.59999990463257 0.60738205909729
6.69999980926514 0.586600065231323
6.80000019073486 0.560958027839661
6.90000009536743 0.529919981956482
7 0.493939995765686
7.09999990463257 0.452415943145752
7.19999980926514 0.407372951507568
7.30000019073486 0.358417987823486
7.40000009536743 0.306532025337219
7.5 0.252303957939148
7.59999990463257 0.192759990692139
7.69999980926514 0.135807991027832
7.80000019073486 0.0813119411468506
8 -0.0337109565734863
8.19999980926514 -0.144119024276733
8.30000019073486 -0.195466995239258
8.5 -0.293516039848328
8.60000038146973 -0.335559964179993
8.69999980926514 -0.374536991119385
8.80000019073486 -0.401265978813171
8.89999961853027 -0.430490016937256
9 -0.45536994934082
9.10000038146973 -0.476189017295837
9.19999980926514 -0.4930499792099
9.30000019073486 -0.505691051483154
9.39999961853027 -0.507343053817749
9.60000038146973 -0.503028988838196
9.80000019073486 -0.474954009056091
9.89999961853027 -0.457939028739929
10 -0.437425971031189
10.1000003814697 -0.413219928741455
10.1999998092651 -0.384850978851318
10.3000001907349 -0.329653978347778
10.3999996185303 -0.294266939163208
10.5 -0.256901025772095
10.6000003814697 -0.222504019737244
10.8000001907349 -0.136623024940491
11 -0.0463142395019531
11.1000003814697 -0.00152432918548584
11.1999998092651 0.0400917530059814
11.3000001907349 0.0836590528488159
11.3999996185303 0.125350952148438
11.6000003814697 0.203909039497375
11.6999998092651 0.240823030471802
11.8000001907349 0.272696971893311
11.8999996185303 0.30342698097229
12 0.33098304271698
12.1000003814697 0.355550050735474
12.1999998092651 0.37616503238678
12.3000001907349 0.392691969871521
12.3999996185303 0.404950976371765
12.5 0.412485003471375
12.6000003814697 0.415351986885071
12.6999998092651 0.413924932479858
12.8000001907349 0.408146977424622
12.8999996185303 0.397454977035522
13 0.363980054855347
13.1000003814697 0.352823972702026
13.1999998092651 0.333135962486267
13.3000001907349 0.311156034469604
13.3999996185303 0.286586046218872
13.6000003814697 0.23196005821228
13.6999998092651 0.200814008712769
13.8000001907349 0.168243050575256
13.8999996185303 0.13397204875946
14 0.0964373350143433
14.3000001907349 -0.0107197761535645
14.5 -0.0800118446350098
14.6000003814697 -0.113441944122314
14.8000001907349 -0.172340989112854
14.8999996185303 -0.199813008308411
15 -0.225394010543823
15.1000003814697 -0.252457022666931
15.1999998092651 -0.270807027816772
15.3000001907349 -0.287270069122314
15.3999996185303 -0.30054497718811
15.5 -0.310860991477966
15.6000003814697 -0.318374991416931
15.6999998092651 -0.323128938674927
15.8000001907349 -0.325099945068359
15.8999996185303 -0.324283003807068
16 -0.320649981498718
16.1000003814697 -0.313835024833679
16.2000007629395 -0.303648948669434
16.2999992370605 -0.289885997772217
16.3999996185303 -0.27233898639679
16.5 -0.246535062789917
16.7999992370605 -0.175488948822021
16.8999996185303 -0.150248050689697
17 -0.119328022003174
17.3999996185303 -0.00913119316101074
17.5 0.0149139165878296
17.7000007629395 0.0671563148498535
17.8999996185303 0.116572022438049
18 0.138736009597778
18.1000003814697 0.158818006515503
18.2000007629395 0.177024960517883
18.2999992370605 0.193441987037659
18.3999996185303 0.208073973655701
18.5 0.220953941345215
18.6000003814697 0.232208967208862
18.7000007629395 0.241935968399048
18.7999992370605 0.250710964202881
18.8999996185303 0.252591013908386
19 0.251224994659424
19.1000003814697 0.243520975112915
19.2000007629395 0.247230052947998
19.3999996185303 0.232108950614929
19.5 0.22271203994751
19.6000003814697 0.211395978927612
19.7000007629395 0.198037028312683
19.7999992370605 0.182512044906616
19.8999996185303 0.164708971977234
20 0.144516944885254
20.1000003814697 0.12227201461792
20.2000007629395 0.0968508720397949
20.6000003814697 5.60283660888672e-06
20.7999992370605 -0.045606255531311
21 -0.0884525775909424
21.1000003814697 -0.108675003051758
21.2000007629395 -0.13256299495697
21.2999992370605 -0.140550971031189
21.3999996185303 -0.154459953308105
21.5 -0.166314005851746
21.6000003814697 -0.176346063613892
21.7000007629395 -0.184690952301025
21.7999992370605 -0.191457033157349
21.8999996185303 -0.196630954742432
22 -0.200199961662292
22.1000003814697 -0.202216029167175
22.2000007629395 -0.202604055404663
22.2999992370605 -0.20131504535675
22.3999996185303 -0.198333978652954
22.5 -0.193554043769836
22.6000003814697 -0.186821937561035
22.7000007629395 -0.177971005439758
22.7999992370605 -0.166926980018616
22.8999996185303 -0.15351402759552
23 -0.13714599609375
23.1000003814697 -0.125179052352905
23.2000007629395 -0.101829051971436
23.3999996185303 -0.0654304027557373
23.7000007629395 -0.0129814147949219
23.7999992370605 0.00387394428253174
24 0.035030722618103
24.2000007629395 0.0639920234680176
24.3999996185303 0.0908190011978149
24.5 0.103355050086975
24.6000003814697 0.111999034881592
24.7000007629395 0.155984044075012
24.8999996185303 0.165863990783691
25 0.168362021446228
25.1000003814697 0.168807983398438
25.2000007629395 0.167340040206909
25.2999992370605 0.163787961006165
25.3999996185303 0.157879948616028
25.5 0.171869993209839
25.6000003814697 0.11680793762207
25.7999992370605 0.0995162725448608
26 0.0807079076766968
26.2000007629395 0.0603018999099731
26.3999996185303 0.0379823446273804
26.6000003814697 0.013505220413208
26.7000007629395 0.000583171844482422
26.7999992370605 -0.00620841979980469
27 -0.0275906324386597
27.2000007629395 -0.0479245185852051
27.3999996185303 -0.0666649341583252
27.6000003814697 -0.0834012031555176
27.7999992370605 -0.0978734493255615
28 -0.10997200012207
28.2999992370605 -0.125705003738403
28.3999996185303 -0.126402020454407
28.5 -0.126068949699402
28.7999992370605 -0.120252966880798
28.8999996185303 -0.117483019828796
29.1000003814697 -0.108196020126343
29.2999992370605 -0.0966571569442749
29.5 -0.0835280418395996
29.7000007629395 -0.06903076171875
29.8999996185303 -0.053193211555481
30.1000003814697 -0.035069465637207
30.2000007629395 -0.02496337890625
30.2999992370605 0.000608563423156738
30.3999996185303 0.00353908538818359
30.5 0.0144640207290649
30.6000003814697 0.0243786573410034
30.7000007629395 0.0332489013671875
30.7999992370605 0.041107177734375
30.8999996185303 0.0479716062545776
31 0.0537054538726807
31.1000003814697 0.0583609342575073
31.2000007629395 0.0619730949401855
31.2999992370605 0.064501166343689
31.3999996185303 0.0659350156784058
31.5 0.0662859678268433
31.6000003814697 0.0655844211578369
31.7000007629395 0.0638351440429688
31.7999992370605 0.0609784126281738
31.8999996185303 0.0570963621139526
32 0.0522241592407227
32.0999984741211 0.046440601348877
32.2000007629395 0.0396133661270142
32.2999992370605 0.0315885543823242
32.4000015258789 0.0221890211105347
32.5 0.011144757270813
32.5999984741211 -0.00104498863220215
32.9000015258789 -0.0299485921859741
33 -0.0371654033660889
33.0999984741211 -0.0432857275009155
33.2000007629395 -0.0484000444412231
33.4000015258789 -0.0561373233795166
33.5999984741211 -0.061437726020813
33.7999992370605 -0.0644944906234741
34 -0.0654195547103882
34.2000007629395 -0.0642522573471069
34.4000015258789 -0.0612411499023438
34.5999984741211 -0.055902361869812
34.7999992370605 -0.0480849742889404
35 -0.0376094579696655
35.2000007629395 -0.024646520614624
35.2999992370605 -0.00266170501708984
35.4000015258789 0.00139200687408447
35.5 0.000756263732910156
35.7000007629395 0.00558209419250488
35.7999992370605 0.00643563270568848
35.9000015258789 0.00608968734741211
36 0.0045548677444458
36.0999984741211 0.000996828079223633
36.2999992370605 -0.00491464138031006
36.4000015258789 -0.00833463668823242
36.5 -0.00442290306091309
36.5999984741211 -0.00252401828765869
36.9000015258789 -0.00464224815368652
37.2000007629395 -0.00473380088806152
37.4000015258789 -0.00384986400604248
37.5 -0.00486290454864502
37.5999984741211 -0.00502371788024902
37.7000007629395 -0.0110584497451782
37.7999992370605 -0.0011516809463501
37.9000015258789 -0.0112971067428589
38 -0.00144290924072266
38.0999984741211 -0.0046316385269165
38.2000007629395 -0.00254476070404053
38.2999992370605 0.00196075439453125
38.4000015258789 -0.00363743305206299
38.5 0.000872015953063965
38.5999984741211 -0.00332236289978027
38.7000007629395 0.0012589693069458
38.7999992370605 0.00156867504119873
38.9000015258789 -0.0047527551651001
39 0.00158774852752686
39.0999984741211 0.00414717197418213
39.2000007629395 -0.00859248638153076
39.2999992370605 -0.00771605968475342
39.4000015258789 -0.00221812725067139
39.5 0.00138306617736816
39.5999984741211 -0.00956976413726807
39.7000007629395 -0.00160586833953857
39.7999992370605 -0.00128853321075439
39.9000015258789 0.00148415565490723
};
\addlegendentry{NBRC, $L=800$}
\end{axis}

\end{tikzpicture}

%% file: figures/gravity-wave/convergence-nbkc.tex
% This file was created with tikzplotlib v0.10.1.
\begin{tikzpicture}

\definecolor{darkgray176}{RGB}{176,176,176}
\definecolor{lightgray204}{RGB}{204,204,204}
\definecolor{mediumaquamarine128230181}{RGB}{159,236,200}
\definecolor{mediumaquamarine64218145}{RGB}{0,128,68}
\definecolor{springgreen0205108}{RGB}{0,205,108}

\begin{axis}[
height=\figureheight,
legend cell align={left},
legend style={
	fill opacity=0.8,
	draw opacity=1,
	text opacity=1,
	at={(0.99,0.99)},
	anchor=north east,
	draw=lightgray204
},
tick align=outside,
tick pos=left,
width=\figurewidth,
x grid style={darkgray176},
xlabel={\(\displaystyle t^{*}\)},
xmajorgrids,
xmin=0, xmax=40,
xtick style={color=black},
y grid style={darkgray176},
ylabel style={rotate=-90.0},
ylabel={\(\displaystyle a^{*}\)},
ymajorgrids,
ymin=-0.89411605, ymax=1.08889905,
ytick style={color=black}
]
\addplot [very thick, black]
table {%
0 0.998026728630066
0.185185194015503 0.9781174659729
0.370370388031006 0.925613880157471
0.555555582046509 0.843235969543457
0.740740776062012 0.734638094902039
0.925925970077515 0.604258418083191
1.11111116409302 0.457145810127258
1.29629623889923 0.298771142959595
1.66666662693024 -0.0289593935012817
1.85185182094574 -0.187046051025391
2.03703713417053 -0.33423376083374
2.22222232818604 -0.465845823287964
2.40740752220154 -0.577867984771729
2.59259247779846 -0.667066097259521
2.77777767181396 -0.731073379516602
2.96296286582947 -0.768445372581482
3.14814805984497 -0.778682231903076
3.33333325386047 -0.762219071388245
3.51851844787598 -0.720383405685425
3.70370364189148 -0.655325293540955
3.88888883590698 -0.569920063018799
4.07407426834106 -0.467650175094604
4.25925922393799 -0.352469682693481
4.44444465637207 -0.228656530380249
4.81481504440308 0.0270652770996094
5 0.150189638137817
5.18518495559692 0.264672756195068
5.37037038803101 0.366881251335144
5.55555534362793 0.453703284263611
5.74074077606201 0.522638082504272
5.92592573165894 0.571863412857056
6.11111116409302 0.600278258323669
6.29629611968994 0.607518672943115
6.48148155212402 0.593949556350708
6.66666650772095 0.560630798339844
6.85185194015503 0.509261131286621
7.03703689575195 0.442102909088135
7.22222232818604 0.361888408660889
7.40740728378296 0.271714210510254
7.59259271621704 0.174924850463867
7.96296310424805 -0.0246013402938843
8.14814853668213 -0.1204913854599
8.33333301544189 -0.209532022476196
8.51851844787598 -0.288901329040527
8.70370388031006 -0.356186509132385
8.88888931274414 -0.409454345703125
9.07407379150391 -0.447302341461182
9.25925922393799 -0.468891620635986
9.44444465637207 -0.473958611488342
9.62963008880615 -0.462807536125183
9.81481456756592 -0.436284065246582
10 -0.39573061466217
10.1851854324341 -0.342926740646362
10.3703699111938 -0.28001594543457
10.5555553436279 -0.209422945976257
10.740740776062 -0.133762836456299
11.1111106872559 0.0219100713729858
11.2962961196899 0.0965865850448608
11.481481552124 0.165835618972778
11.6666669845581 0.227465391159058
11.8518514633179 0.279605627059937
12.037036895752 0.320761799812317
12.222222328186 0.349854826927185
12.4074077606201 0.366245865821838
12.5925922393799 0.369745016098022
12.777777671814 0.360605359077454
12.962963104248 0.339500904083252
13.1481485366821 0.307492256164551
13.3333330154419 0.265979051589966
13.518518447876 0.2166428565979
13.7037038803101 0.16138219833374
13.8888893127441 0.10224175453186
14.259259223938 -0.0192111730575562
14.4444446563721 -0.077364444732666
14.6296300888062 -0.131218194961548
14.8148145675659 -0.179070353507996
15 -0.219471216201782
15.1851854324341 -0.251265406608582
15.3703699111938 -0.273622870445251
15.5555553436279 -0.286057472229004
15.740740776062 -0.28843355178833
15.9259262084961 -0.280960321426392
16.1111106872559 -0.264175057411194
16.2962970733643 -0.23891544342041
16.481481552124 -0.206282138824463
16.6666660308838 -0.167594313621521
16.8518524169922 -0.124338269233704
17.037036895752 -0.0781127214431763
17.4074077606201 0.0166382789611816
17.5925922393799 0.0619223117828369
17.7777786254883 0.103801369667053
17.962963104248 0.140953660011292
18.1481475830078 0.172255396842957
18.3333339691162 0.196813941001892
18.518518447876 0.213990688323975
18.7037029266357 0.223415970802307
18.8888893127441 0.224993705749512
19.0740737915039 0.218896269798279
19.2592601776123 0.205551862716675
19.4444446563721 0.185621857643127
19.6296291351318 0.159971714019775
19.8148155212402 0.129636168479919
20 0.0957789421081543
20.1851844787598 0.0596495866775513
20.5555553436279 -0.0142668485641479
20.7407398223877 -0.0495278835296631
20.9259262084961 -0.0820932388305664
21.1111106872559 -0.110936284065247
21.2962970733643 -0.135186195373535
21.481481552124 -0.154153108596802
21.6666660308838 -0.167346119880676
21.8518524169922 -0.174484133720398
22.037036895752 -0.175499677658081
22.2222213745117 -0.170534729957581
22.4074077606201 -0.159930109977722
22.5925922393799 -0.144207835197449
22.7777786254883 -0.124048590660095
22.962963104248 -0.100263833999634
23.1481475830078 -0.0737646818161011
23.518518447876 -0.0165609121322632
23.7037029266357 0.0121328830718994
23.8888893127441 0.0395880937576294
24.0740737915039 0.064909815788269
24.2592601776123 0.0873005390167236
24.4444446563721 0.106085658073425
24.6296291351318 0.120732069015503
24.8148155212402 0.130862236022949
25 0.136263012886047
25.1851844787598 0.136887431144714
25.3703708648682 0.132851958274841
25.5555553436279 0.124427795410156
25.7407398223877 0.112027168273926
25.9259262084961 0.0961849689483643
26.1111106872559 0.0775377750396729
26.2962970733643 0.0567986965179443
26.6666660308838 0.0121220350265503
26.8518524169922 -0.010246753692627
27.037036895752 -0.0316232442855835
27.2222213745117 -0.0513113737106323
27.4074077606201 -0.0686922073364258
27.5925922393799 -0.083242654800415
27.7777786254883 -0.0945510864257812
27.962963104248 -0.102327346801758
28.1481475830078 -0.106409549713135
28.3333339691162 -0.106765747070312
28.518518447876 -0.103491306304932
28.7037029266357 -0.0968018770217896
28.8888893127441 -0.0870226621627808
29.0740737915039 -0.0745742321014404
29.2592601776123 -0.0599559545516968
29.4444446563721 -0.0437257289886475
30.1851844787598 0.0252456665039062
30.3703708648682 0.0405528545379639
30.5555553436279 0.0540436506271362
30.7407398223877 0.0653131008148193
30.9259262084961 0.0740430355072021
31.1111106872559 0.0800105333328247
31.2962970733643 0.0830929279327393
31.481481552124 0.0832687616348267
31.6666660308838 0.0806158781051636
31.8518524169922 0.0753058195114136
32.037036895752 0.0675951242446899
32.2222213745117 0.0578144788742065
32.4074058532715 0.0463552474975586
32.7777786254883 0.0201783180236816
33.1481475830078 -0.00718581676483154
33.3333320617676 -0.020142674446106
33.5185203552246 -0.0320430994033813
33.7037048339844 -0.0425139665603638
33.8888893127441 -0.051241397857666
34.0740737915039 -0.0579798221588135
34.2592582702637 -0.0625578165054321
34.4444427490234 -0.0648825168609619
34.6296310424805 -0.0649402141571045
34.8148155212402 -0.0627939701080322
35 -0.0585802793502808
35.1851844787598 -0.0525015592575073
35.3703689575195 -0.0448178052902222
35.5555572509766 -0.0358356237411499
35.9259262084961 -0.0153669118881226
36.4814796447754 0.0160623788833618
36.6666679382324 0.0253137350082397
36.8518524169922 0.033440113067627
37.037036895752 0.0401983261108398
37.2222213745117 0.0453987121582031
37.4074058532715 0.0489096641540527
37.5925941467285 0.0506608486175537
37.7777786254883 0.0506438016891479
37.962963104248 0.0489097833633423
38.1481475830078 0.0455672740936279
38.3333320617676 0.0407758951187134
38.5185203552246 0.0347399711608887
38.8888893127441 0.0199233293533325
39.6296310424805 -0.0128018856048584
39.8148155212402 -0.0199935436248779
40 -0.0262999534606934
};
\addlegendentry{Analytical model~\cite{dingemans1997WaterWavePropagation}}
\addplot [line width=2pt, mediumaquamarine128230181]
table {%
0 0.995038986206055
0.399999976158142 0.920225024223328
0.799999952316284 0.727980017662048
1.20000004768372 0.397395014762878
1.60000002384186 0.0219324827194214
2 -0.325511932373047
2.40000009536743 -0.569786071777344
2.79999995231628 -0.668873071670532
3.20000004768372 -0.655750036239624
3.59999990463257 -0.52581799030304
4 -0.304158926010132
4.40000009536743 -0.0961446762084961
4.80000019073486 0.142972946166992
5.19999980926514 0.339563965797424
5.59999990463257 0.478644013404846
6 0.519602060317993
6.40000009536743 0.443634986877441
6.80000019073486 0.36130702495575
7.19999980926514 0.25019896030426
7.59999990463257 0.0931713581085205
8 -0.0969374179840088
8.39999961853027 -0.230674028396606
8.80000019073486 -0.309062957763672
9.19999980926514 -0.345342040061951
9.60000038146973 -0.341933965682983
10 -0.281877994537354
10.3999996185303 -0.148370981216431
10.8000001907349 0.031089186668396
11.1999998092651 0.165945053100586
11.6000003814697 0.200917959213257
12 0.173493027687073
12.3999996185303 0.0882219076156616
12.8000001907349 -0.0401997566223145
13.1999998092651 -0.098825216293335
13.6000003814697 -0.0998681783676147
14 -0.0618011951446533
14.3999996185303 0.0292229652404785
14.8000001907349 0.0508885383605957
15.1999998092651 0.0202517509460449
15.6000003814697 -0.0357588529586792
16 -0.0529768466949463
16.3999996185303 -0.0396126508712769
16.7999992370605 0.0231403112411499
17.2000007629395 0.024045467376709
17.6000003814697 -0.00864124298095703
18 -0.0249123573303223
18.3999996185303 0.00592410564422607
18.7999992370605 0.0193737745285034
19.2000007629395 0.0192480087280273
19.6000003814697 -0.00615465641021729
20 -0.0155134201049805
20.3999996185303 -0.00991034507751465
20.7999992370605 0.00680446624755859
21.6000003814697 0.00330698490142822
22 -0.0127958059310913
22.3999996185303 -0.0262435674667358
22.7999992370605 0.0128341913223267
23.2000007629395 -0.000998497009277344
23.6000003814697 -0.00409221649169922
24 -0.0213526487350464
24.3999996185303 -0.0076897144317627
24.7999992370605 0.00382649898529053
25.2000007629395 0.00156497955322266
25.6000003814697 -0.0185303688049316
26.3999996185303 -0.0113900899887085
26.7999992370605 0.00977563858032227
27.2000007629395 -0.013696551322937
27.6000003814697 0.00260508060455322
28 -0.00511395931243896
28.3999996185303 -0.00618612766265869
28.7999992370605 -0.000108957290649414
29.2000007629395 0.0212589502334595
29.6000003814697 0.000240325927734375
30 -0.016457200050354
30.3999996185303 -0.00424408912658691
30.7999992370605 0.0108158588409424
31.2000007629395 8.91685485839844e-05
31.6000003814697 -0.00302231311798096
32 -0.0194871425628662
32.4000015258789 -0.00351524353027344
32.7999992370605 -0.00343871116638184
33.2000007629395 -0.000146269798278809
33.5999984741211 -0.00599098205566406
34 -0.0177583694458008
34.4000015258789 -0.00580799579620361
34.7999992370605 0.0161985158920288
35.2000007629395 0.0150725841522217
35.5999984741211 -0.00583088397979736
36 -0.0129892826080322
36.4000015258789 -0.00132918357849121
36.7999992370605 -0.0117100477218628
37.2000007629395 -0.00367355346679688
37.5999984741211 -0.000607490539550781
38 -0.0243839025497437
38.4000015258789 -0.00554239749908447
38.7999992370605 -0.0126694440841675
39.2000007629395 -0.00345683097839355
39.5999984741211 0.0142440795898438
};
\addlegendentry{NBKC, $L=200$}
\addplot [line width=2pt, mediumaquamarine64218145]
table {%
0 0.997523069381714
0.399999976158142 0.926706075668335
0.799999952316284 0.715119004249573
1.20000004768372 0.393481016159058
1.60000002384186 0.030180811882019
2 -0.317767977714539
2.40000009536743 -0.593737006187439
2.79999995231628 -0.752192974090576
3.20000004768372 -0.772753000259399
3.59999990463257 -0.661412954330444
4 -0.482748031616211
4.40000009536743 -0.234815955162048
4.80000019073486 0.0362381935119629
5.19999980926514 0.294286012649536
5.59999990463257 0.487928032875061
6 0.586570024490356
6.40000009536743 0.574218034744263
6.80000019073486 0.438706040382385
7.19999980926514 0.258899927139282
7.59999990463257 0.0592162609100342
8 -0.134451985359192
8.39999961853027 -0.290019035339355
8.80000019073486 -0.387294054031372
9.19999980926514 -0.429059982299805
9.60000038146973 -0.419757962226868
10 -0.356564998626709
10.3999996185303 -0.219372034072876
10.8000001907349 -0.0663889646530151
11.1999998092651 0.0673072338104248
11.6000003814697 0.189988017082214
12 0.283091068267822
12.3999996185303 0.304698944091797
12.8000001907349 0.258239030838013
13.1999998092651 0.150200009346008
13.6000003814697 0.0511796474456787
14 -0.0404312610626221
14.3999996185303 -0.122213006019592
14.8000001907349 -0.188190937042236
15.1999998092651 -0.233289003372192
15.6000003814697 -0.253193020820618
16 -0.225904941558838
16.3999996185303 -0.183153033256531
16.7999992370605 -0.126402974128723
17.2000007629395 -0.0534727573394775
17.6000003814697 0.0439345836639404
18 0.112820982933044
18.3999996185303 0.155197978019714
18.7999992370605 0.1666020154953
19.2000007629395 0.150797009468079
19.6000003814697 0.109042048454285
20 0.0370738506317139
20.3999996185303 -0.061229944229126
20.7999992370605 -0.0952216386795044
21.2000007629395 -0.101256966590881
21.6000003814697 -0.0865254402160645
22 -0.0524215698242188
22.3999996185303 0.0170563459396362
22.7999992370605 0.0528163909912109
23.2000007629395 0.0469357967376709
23.6000003814697 -0.00246345996856689
24 -0.0373326539993286
24.3999996185303 -0.0406860113143921
24.7999992370605 -0.0385055541992188
25.2000007629395 0.0106483697891235
25.6000003814697 0.00714325904846191
26 0.00174200534820557
26.3999996185303 -0.0169298648834229
26.7999992370605 -0.0123463869094849
27.2000007629395 0.00279009342193604
27.6000003814697 0.00209188461303711
28 0.00665628910064697
28.3999996185303 -0.0216799974441528
28.7999992370605 -0.0012742280960083
29.2000007629395 -0.015299916267395
29.6000003814697 -0.00713920593261719
30 -0.0112127065658569
30.3999996185303 -0.00230669975280762
30.7999992370605 0.0008544921875
31.2000007629395 0.0293866395950317
31.6000003814697 0.00163042545318604
32 0.000270843505859375
32.4000015258789 0.000759243965148926
32.7999992370605 0.00515341758728027
33.2000007629395 0.00136196613311768
33.5999984741211 -0.0039525032043457
34 -0.00115311145782471
34.7999992370605 -0.000960350036621094
35.2000007629395 0.00755965709686279
35.5999984741211 -0.00488615036010742
36 0.00503551959991455
36.4000015258789 0.00144648551940918
36.7999992370605 0.00565886497497559
37.2000007629395 0.00331854820251465
37.5999984741211 -0.0121958255767822
38 0.0022505521774292
38.4000015258789 -0.0209492444992065
38.7999992370605 0.00305700302124023
39.5999984741211 -0.00504863262176514
};
\addlegendentry{NBKC, $L=400$}
\addplot [line width=2pt, springgreen0205108]
table {%
0 0.998762011528015
0.100000023841858 0.993924975395203
0.200000047683716 0.980396032333374
0.299999952316284 0.958212971687317
0.399999976158142 0.927106976509094
0.5 0.886186003684998
0.600000023841858 0.833953022956848
0.700000047683716 0.776895999908447
0.799999952316284 0.710017919540405
0.899999976158142 0.640094041824341
1 0.565475940704346
1.10000002384186 0.484501004219055
1.20000004768372 0.40138304233551
1.29999995231628 0.315608978271484
1.39999997615814 0.228484034538269
1.60000002384186 0.0504564046859741
1.70000004768372 -0.040787935256958
1.79999995231628 -0.130362987518311
1.89999997615814 -0.209709048271179
2 -0.2915940284729
2.09999990463257 -0.369996070861816
2.20000004768372 -0.443076014518738
2.29999995231628 -0.511745929718018
2.40000009536743 -0.574030995368958
2.5 -0.63051700592041
2.59999990463257 -0.679280996322632
2.70000004768372 -0.720803022384644
2.79999995231628 -0.752135992050171
2.90000009536743 -0.779577970504761
3 -0.795516014099121
3.09999990463257 -0.803802013397217
3.20000004768372 -0.80397891998291
3.29999995231628 -0.795619010925293
3.40000009536743 -0.778027057647705
3.5 -0.748877048492432
3.59999990463257 -0.718366026878357
3.70000004768372 -0.680436015129089
3.79999995231628 -0.638143062591553
3.90000009536743 -0.57837700843811
4 -0.522871017456055
4.09999990463257 -0.461437940597534
4.30000019073486 -0.334123969078064
4.40000009536743 -0.265931010246277
4.5 -0.195724010467529
4.59999990463257 -0.124001979827881
4.69999980926514 -0.0505863428115845
4.80000019073486 0.0240367650985718
4.90000009536743 0.0941057205200195
5 0.157526016235352
5.09999990463257 0.222831010818481
5.19999980926514 0.286229014396667
5.30000019073486 0.344751000404358
5.40000009536743 0.404348015785217
5.5 0.453873991966248
5.59999990463257 0.498762965202332
5.80000019073486 0.563148975372314
5.90000009536743 0.589807987213135
6 0.610852003097534
6.09999990463257 0.630346059799194
6.19999980926514 0.636101007461548
6.30000019073486 0.634618997573853
6.40000009536743 0.627028942108154
6.5 0.611170053482056
6.59999990463257 0.594181060791016
6.69999980926514 0.572251081466675
6.80000019073486 0.544957995414734
6.90000009536743 0.511865019798279
7 0.472394943237305
7.09999990463257 0.43026602268219
7.19999980926514 0.384369015693665
7.30000019073486 0.335041046142578
7.40000009536743 0.283257007598877
7.59999990463257 0.171623945236206
7.69999980926514 0.116651058197021
7.90000009536743 0.00307643413543701
8.10000038146973 -0.106420040130615
8.19999980926514 -0.159232020378113
8.30000019073486 -0.209192037582397
8.39999961853027 -0.255519032478333
8.5 -0.299150943756104
8.60000038146973 -0.33938193321228
8.69999980926514 -0.371883988380432
8.80000019073486 -0.407840967178345
8.89999961853027 -0.434664964675903
9 -0.457165956497192
9.10000038146973 -0.475643038749695
9.19999980926514 -0.490007996559143
9.30000019073486 -0.498708009719849
9.39999961853027 -0.499263048171997
9.5 -0.502537965774536
9.60000038146973 -0.496127009391785
9.69999980926514 -0.487748026847839
9.80000019073486 -0.474359035491943
9.89999961853027 -0.457546949386597
10 -0.436883926391602
10.1000003814697 -0.411455988883972
10.1999998092651 -0.382459044456482
10.3000001907349 -0.344409942626953
10.3999996185303 -0.309258937835693
10.5 -0.271751999855042
10.6999998092651 -0.187494993209839
10.8000001907349 -0.144402980804443
10.8999996185303 -0.102529048919678
11.1000003814697 -0.0121328830718994
11.3999996185303 0.115885019302368
11.5 0.16100800037384
11.6000003814697 0.19842803478241
11.6999998092651 0.233674049377441
11.8999996185303 0.291278958320618
12 0.316380977630615
12.1000003814697 0.338325977325439
12.1999998092651 0.357182025909424
12.3000001907349 0.37176501750946
12.3999996185303 0.383810043334961
12.5 0.388674020767212
12.6000003814697 0.388669967651367
12.6999998092651 0.38373601436615
12.8999996185303 0.365951061248779
13 0.351354956626892
13.1000003814697 0.334733009338379
13.1999998092651 0.315567016601562
13.3000001907349 0.293581008911133
13.3999996185303 0.268612027168274
13.5 0.2411789894104
13.6000003814697 0.211642026901245
13.6999998092651 0.180506944656372
13.8999996185303 0.114606976509094
14.1999998092651 0.00935351848602295
14.3000001907349 -0.0272868871688843
14.3999996185303 -0.060938835144043
14.5 -0.0936496257781982
14.6000003814697 -0.125383973121643
14.6999998092651 -0.15587306022644
14.8000001907349 -0.182808041572571
14.8999996185303 -0.207772016525269
15 -0.230914950370789
15.1000003814697 -0.251906037330627
15.1999998092651 -0.268902063369751
15.3000001907349 -0.282729029655457
15.3999996185303 -0.293110013008118
15.5 -0.300298929214478
15.6000003814697 -0.304661989212036
15.6999998092651 -0.306321978569031
15.8000001907349 -0.305428981781006
15.8999996185303 -0.301991939544678
16 -0.295763969421387
16.1000003814697 -0.286234974861145
16.2000007629395 -0.273162961006165
16.2999992370605 -0.250740051269531
16.3999996185303 -0.23511803150177
16.6000003814697 -0.195652961730957
16.7000007629395 -0.174587965011597
16.7999992370605 -0.152248024940491
16.8999996185303 -0.128497004508972
17 -0.108057975769043
17.2999992370605 -0.0324866771697998
17.3999996185303 -0.00196564197540283
17.5 0.0190070867538452
17.7000007629395 0.0667251348495483
18 0.135601043701172
18.1000003814697 0.153985977172852
18.2000007629395 0.170287013053894
18.2999992370605 0.184605002403259
18.3999996185303 0.196974992752075
18.5 0.207458972930908
18.6000003814697 0.216174006462097
18.7000007629395 0.223091006278992
18.7999992370605 0.228188991546631
18.8999996185303 0.231485962867737
19 0.233013033866882
19.1000003814697 0.232738971710205
19.2000007629395 0.230435013771057
19.2999992370605 0.225973963737488
19.3999996185303 0.219388961791992
19.5 0.210572004318237
19.6000003814697 0.199342966079712
19.7000007629395 0.185547947883606
19.7999992370605 0.169028997421265
19.8999996185303 0.149317979812622
20 0.124758958816528
20.1000003814697 0.0979059934616089
20.2000007629395 0.0741568803787231
20.3999996185303 0.0292340517044067
20.5 0.00722372531890869
20.6000003814697 -0.0123136043548584
20.7000007629395 -0.0339659452438354
20.7999992370605 -0.0543218851089478
20.8999996185303 -0.0734884738922119
21 -0.0915184020996094
21.1000003814697 -0.108443021774292
21.2000007629395 -0.124209046363831
21.2999992370605 -0.136304974555969
21.3999996185303 -0.146916031837463
21.5 -0.155526041984558
21.6000003814697 -0.162233948707581
21.7000007629395 -0.167261004447937
21.7999992370605 -0.170704960823059
21.8999996185303 -0.17270302772522
22 -0.173213958740234
22.1000003814697 -0.172207951545715
22.2000007629395 -0.169548034667969
22.2999992370605 -0.165171980857849
22.3999996185303 -0.159028053283691
22.5 -0.150894999504089
22.6000003814697 -0.14036500453949
22.7000007629395 -0.123026013374329
22.7999992370605 -0.108147978782654
23.3999996185303 -0.0299581289291382
23.7000007629395 0.0103321075439453
23.8999996185303 0.0345462560653687
24 0.0461388826370239
24.1000003814697 0.125445008277893
24.5 0.141463994979858
24.7999992370605 0.153195977210999
24.8999996185303 0.156666040420532
25 0.183375954627991
25.1000003814697 0.17936098575592
25.2000007629395 0.179237008094788
25.2999992370605 0.177569031715393
25.3999996185303 0.173097014427185
25.5 0.115828037261963
25.6000003814697 0.118738055229187
25.8999996185303 0.0997495651245117
26.1000003814697 0.085237979888916
26.2999992370605 0.0677081346511841
26.3999996185303 0.0576968193054199
26.5 0.0467685461044312
26.6000003814697 0.034893274307251
26.7000007629395 0.0220909118652344
26.8999996185303 -0.00589561462402344
27.1000003814697 -0.0341165065765381
27.2000007629395 -0.0470696687698364
27.2999992370605 -0.0590270757675171
27.3999996185303 -0.0699112415313721
27.5 -0.0796679258346558
27.6000003814697 -0.0883356332778931
27.7000007629395 -0.0960044860839844
27.7999992370605 -0.102750062942505
28 -0.113721013069153
28.2999992370605 -0.126628041267395
28.5 -0.121371030807495
28.6000003814697 -0.122684001922607
28.7000007629395 -0.125283002853394
28.8999996185303 -0.118293046951294
29.5 -0.0808078050613403
29.7000007629395 -0.0658186674118042
29.7999992370605 -0.0573736429214478
29.8999996185303 -0.0480606555938721
30 -0.0376938581466675
30.1000003814697 -0.0260821580886841
30.2000007629395 -0.000257134437561035
30.2999992370605 0.0157550573348999
30.3999996185303 0.0263158082962036
30.5 0.0355441570281982
30.6000003814697 0.0433038473129272
30.7000007629395 0.0496978759765625
30.7999992370605 0.0549230575561523
30.8999996185303 0.0590360164642334
31 0.0619974136352539
31.1000003814697 0.0637211799621582
31.2000007629395 0.0641303062438965
31.2999992370605 0.0630905628204346
31.3999996185303 0.0606414079666138
31.5 0.0569267272949219
31.6000003814697 0.0520449876785278
31.7000007629395 0.0459506511688232
31.7999992370605 0.0385825634002686
31.8999996185303 0.0300352573394775
32 0.0201364755630493
32.0999984741211 0.00782239437103271
32.2000007629395 -0.00325071811676025
32.2999992370605 -0.0124863386154175
32.4000015258789 -0.0193560123443604
32.5 -0.024630069732666
32.7000007629395 -0.0323570966720581
32.9000015258789 -0.0371735095977783
33.0999984741211 -0.0397306680679321
33.2999992370605 -0.0405914783477783
33.5 -0.0401650667190552
33.7000007629395 -0.0383708477020264
33.9000015258789 -0.0345425605773926
34.0999984741211 -0.0279934406280518
34.2000007629395 -0.0234514474868774
34.2999992370605 -0.0448869466781616
34.4000015258789 -0.0041128396987915
34.5 -0.00176322460174561
34.5999984741211 0.00233137607574463
34.7000007629395 -0.000951528549194336
34.7999992370605 0.00338649749755859
34.9000015258789 0.00374352931976318
35.0999984741211 -0.00196206569671631
35.4000015258789 0.00171482563018799
35.5 -0.00184512138366699
35.5999984741211 -0.00650441646575928
35.7000007629395 -0.000193953514099121
35.7999992370605 0.00149405002593994
35.9000015258789 -0.00750303268432617
36 -0.0109421014785767
36.0999984741211 -0.000189542770385742
36.2000007629395 -0.00521659851074219
36.4000015258789 -0.00149047374725342
36.5 -0.00452220439910889
36.5999984741211 -0.00329732894897461
36.7999992370605 -0.00539565086364746
36.9000015258789 0.000769495964050293
37 -0.0031360387802124
37.0999984741211 -0.00315344333648682
37.2000007629395 -0.0013052225112915
37.2999992370605 -0.00499856472015381
37.4000015258789 0.000505685806274414
37.5 0.00273072719573975
37.5999984741211 0.00104987621307373
37.7000007629395 -0.00295805931091309
37.7999992370605 -0.000927090644836426
37.9000015258789 -0.00186026096343994
38 0.000921964645385742
38.0999984741211 -0.00199079513549805
38.2999992370605 0.00403988361358643
38.4000015258789 -0.00186371803283691
38.5 0.000823616981506348
38.5999984741211 -0.00903034210205078
38.7000007629395 -0.00118076801300049
38.7999992370605 0.0014193058013916
38.9000015258789 -0.00106644630432129
39 0.00108814239501953
39.0999984741211 0.00216567516326904
39.2000007629395 -0.00937020778656006
39.2999992370605 0.000502943992614746
39.4000015258789 -0.000337600708007812
39.5 0.000411391258239746
39.7000007629395 -0.0026390552520752
39.7999992370605 0.000881791114807129
39.9000015258789 -0.00191390514373779
};
\addlegendentry{NBKC, $L=800$}
\end{axis}

\end{tikzpicture}

%% file: figures/gravity-wave/convergence-om.tex
% This file was created with tikzplotlib v0.10.1.
\begin{tikzpicture}

\definecolor{darkgray176}{RGB}{176,176,176}
\definecolor{lightgray204}{RGB}{204,204,204}
\definecolor{mediumorchid17588186}{RGB}{175,88,186}
\definecolor{orchid195130203}{RGB}{109,55,116}
\definecolor{plum215172220}{RGB}{225,192,229}

\begin{axis}[
height=\figureheight,
legend cell align={left},
legend style={
	fill opacity=0.8,
	draw opacity=1,
	text opacity=1,
	at={(0.99,0.99)},
	anchor=north east,
	draw=lightgray204
},
tick align=outside,
tick pos=left,
width=\figurewidth,
x grid style={darkgray176},
xlabel={\(\displaystyle t^{*}\)},
xmajorgrids,
xmin=0, xmax=40,
xtick style={color=black},
y grid style={darkgray176},
ylabel style={rotate=-90.0},
ylabel={\(\displaystyle a^{*}\)},
ymajorgrids,
ymin=-0.89866675, ymax=1.08911575,
ytick style={color=black}
]
\addplot [very thick, black]
table {%
0 0.998026728630066
0.185185194015503 0.9781174659729
0.370370388031006 0.925613880157471
0.555555582046509 0.843235969543457
0.740740776062012 0.734638094902039
0.925925970077515 0.604258418083191
1.11111116409302 0.457145810127258
1.29629623889923 0.298771142959595
1.66666662693024 -0.0289593935012817
1.85185182094574 -0.187046051025391
2.03703713417053 -0.33423376083374
2.22222232818604 -0.465845823287964
2.40740752220154 -0.577867984771729
2.59259247779846 -0.667066097259521
2.77777767181396 -0.731073379516602
2.96296286582947 -0.768445372581482
3.14814805984497 -0.778682231903076
3.33333325386047 -0.762219071388245
3.51851844787598 -0.720383405685425
3.70370364189148 -0.655325293540955
3.88888883590698 -0.569920063018799
4.07407426834106 -0.467650175094604
4.25925922393799 -0.352469682693481
4.44444465637207 -0.228656530380249
4.81481504440308 0.0270652770996094
5 0.150189638137817
5.18518495559692 0.264672756195068
5.37037038803101 0.366881251335144
5.55555534362793 0.453703284263611
5.74074077606201 0.522638082504272
5.92592573165894 0.571863412857056
6.11111116409302 0.600278258323669
6.29629611968994 0.607518672943115
6.48148155212402 0.593949556350708
6.66666650772095 0.560630798339844
6.85185194015503 0.509261131286621
7.03703689575195 0.442102909088135
7.22222232818604 0.361888408660889
7.40740728378296 0.271714210510254
7.59259271621704 0.174924850463867
7.96296310424805 -0.0246013402938843
8.14814853668213 -0.1204913854599
8.33333301544189 -0.209532022476196
8.51851844787598 -0.288901329040527
8.70370388031006 -0.356186509132385
8.88888931274414 -0.409454345703125
9.07407379150391 -0.447302341461182
9.25925922393799 -0.468891620635986
9.44444465637207 -0.473958611488342
9.62963008880615 -0.462807536125183
9.81481456756592 -0.436284065246582
10 -0.39573061466217
10.1851854324341 -0.342926740646362
10.3703699111938 -0.28001594543457
10.5555553436279 -0.209422945976257
10.740740776062 -0.133762836456299
11.1111106872559 0.0219100713729858
11.2962961196899 0.0965865850448608
11.481481552124 0.165835618972778
11.6666669845581 0.227465391159058
11.8518514633179 0.279605627059937
12.037036895752 0.320761799812317
12.222222328186 0.349854826927185
12.4074077606201 0.366245865821838
12.5925922393799 0.369745016098022
12.777777671814 0.360605359077454
12.962963104248 0.339500904083252
13.1481485366821 0.307492256164551
13.3333330154419 0.265979051589966
13.518518447876 0.2166428565979
13.7037038803101 0.16138219833374
13.8888893127441 0.10224175453186
14.259259223938 -0.0192111730575562
14.4444446563721 -0.077364444732666
14.6296300888062 -0.131218194961548
14.8148145675659 -0.179070353507996
15 -0.219471216201782
15.1851854324341 -0.251265406608582
15.3703699111938 -0.273622870445251
15.5555553436279 -0.286057472229004
15.740740776062 -0.28843355178833
15.9259262084961 -0.280960321426392
16.1111106872559 -0.264175057411194
16.2962970733643 -0.23891544342041
16.481481552124 -0.206282138824463
16.6666660308838 -0.167594313621521
16.8518524169922 -0.124338269233704
17.037036895752 -0.0781127214431763
17.4074077606201 0.0166382789611816
17.5925922393799 0.0619223117828369
17.7777786254883 0.103801369667053
17.962963104248 0.140953660011292
18.1481475830078 0.172255396842957
18.3333339691162 0.196813941001892
18.518518447876 0.213990688323975
18.7037029266357 0.223415970802307
18.8888893127441 0.224993705749512
19.0740737915039 0.218896269798279
19.2592601776123 0.205551862716675
19.4444446563721 0.185621857643127
19.6296291351318 0.159971714019775
19.8148155212402 0.129636168479919
20 0.0957789421081543
20.1851844787598 0.0596495866775513
20.5555553436279 -0.0142668485641479
20.7407398223877 -0.0495278835296631
20.9259262084961 -0.0820932388305664
21.1111106872559 -0.110936284065247
21.2962970733643 -0.135186195373535
21.481481552124 -0.154153108596802
21.6666660308838 -0.167346119880676
21.8518524169922 -0.174484133720398
22.037036895752 -0.175499677658081
22.2222213745117 -0.170534729957581
22.4074077606201 -0.159930109977722
22.5925922393799 -0.144207835197449
22.7777786254883 -0.124048590660095
22.962963104248 -0.100263833999634
23.1481475830078 -0.0737646818161011
23.518518447876 -0.0165609121322632
23.7037029266357 0.0121328830718994
23.8888893127441 0.0395880937576294
24.0740737915039 0.064909815788269
24.2592601776123 0.0873005390167236
24.4444446563721 0.106085658073425
24.6296291351318 0.120732069015503
24.8148155212402 0.130862236022949
25 0.136263012886047
25.1851844787598 0.136887431144714
25.3703708648682 0.132851958274841
25.5555553436279 0.124427795410156
25.7407398223877 0.112027168273926
25.9259262084961 0.0961849689483643
26.1111106872559 0.0775377750396729
26.2962970733643 0.0567986965179443
26.6666660308838 0.0121220350265503
26.8518524169922 -0.010246753692627
27.037036895752 -0.0316232442855835
27.2222213745117 -0.0513113737106323
27.4074077606201 -0.0686922073364258
27.5925922393799 -0.083242654800415
27.7777786254883 -0.0945510864257812
27.962963104248 -0.102327346801758
28.1481475830078 -0.106409549713135
28.3333339691162 -0.106765747070312
28.518518447876 -0.103491306304932
28.7037029266357 -0.0968018770217896
28.8888893127441 -0.0870226621627808
29.0740737915039 -0.0745742321014404
29.2592601776123 -0.0599559545516968
29.4444446563721 -0.0437257289886475
30.1851844787598 0.0252456665039062
30.3703708648682 0.0405528545379639
30.5555553436279 0.0540436506271362
30.7407398223877 0.0653131008148193
30.9259262084961 0.0740430355072021
31.1111106872559 0.0800105333328247
31.2962970733643 0.0830929279327393
31.481481552124 0.0832687616348267
31.6666660308838 0.0806158781051636
31.8518524169922 0.0753058195114136
32.037036895752 0.0675951242446899
32.2222213745117 0.0578144788742065
32.4074058532715 0.0463552474975586
32.7777786254883 0.0201783180236816
33.1481475830078 -0.00718581676483154
33.3333320617676 -0.020142674446106
33.5185203552246 -0.0320430994033813
33.7037048339844 -0.0425139665603638
33.8888893127441 -0.051241397857666
34.0740737915039 -0.0579798221588135
34.2592582702637 -0.0625578165054321
34.4444427490234 -0.0648825168609619
34.6296310424805 -0.0649402141571045
34.8148155212402 -0.0627939701080322
35 -0.0585802793502808
35.1851844787598 -0.0525015592575073
35.3703689575195 -0.0448178052902222
35.5555572509766 -0.0358356237411499
35.9259262084961 -0.0153669118881226
36.4814796447754 0.0160623788833618
36.6666679382324 0.0253137350082397
36.8518524169922 0.033440113067627
37.037036895752 0.0401983261108398
37.2222213745117 0.0453987121582031
37.4074058532715 0.0489096641540527
37.5925941467285 0.0506608486175537
37.7777786254883 0.0506438016891479
37.962963104248 0.0489097833633423
38.1481475830078 0.0455672740936279
38.3333320617676 0.0407758951187134
38.5185203552246 0.0347399711608887
38.8888893127441 0.0199233293533325
39.6296310424805 -0.0128018856048584
39.8148155212402 -0.0199935436248779
40 -0.0262999534606934
};
\addlegendentry{Analytical model~\cite{dingemans1997WaterWavePropagation}}
\addplot [line width=2pt, plum215172220, dashed]
table {%
0 0.995046019554138
0.399999976158142 0.950763940811157
0.799999952316284 0.808948993682861
1.20000004768372 0.543231010437012
1.60000002384186 0.16669499874115
2 -0.201701998710632
2.40000009536743 -0.525989055633545
2.79999995231628 -0.725576996803284
3.20000004768372 -0.808313012123108
3.59999990463257 -0.789221048355103
4 -0.633036971092224
4.40000009536743 -0.301779985427856
4.80000019073486 0.0194422006607056
5.19999980926514 0.275645971298218
5.59999990463257 0.485623002052307
6 0.62247097492218
6.40000009536743 0.611737012863159
6.80000019073486 0.47677206993103
7.19999980926514 0.33498203754425
7.59999990463257 0.175333976745605
8 -0.00271856784820557
8.39999961853027 -0.167657017707825
8.80000019073486 -0.29525101184845
9.19999980926514 -0.392742037773132
9.60000038146973 -0.454290986061096
10 -0.474174022674561
10.3999996185303 -0.455407977104187
10.8000001907349 -0.398062944412231
11.1999998092651 -0.301569938659668
11.6000003814697 -0.137910962104797
12 0.139304995536804
12.3999996185303 0.311249971389771
12.8000001907349 0.42243504524231
13.1999998092651 0.47664201259613
13.6000003814697 0.470803022384644
14 0.404440999031067
14.3999996185303 0.270715951919556
14.8000001907349 0.0453760623931885
15.1999998092651 -0.172307968139648
15.6000003814697 -0.247395992279053
16 -0.224733948707581
16.3999996185303 -0.109186053276062
16.7999992370605 0.0426548719406128
17.2000007629395 0.0895310640335083
17.6000003814697 0.0946066379547119
18 0.0622830390930176
18.3999996185303 -0.0173989534378052
18.7999992370605 -0.03529953956604
19.2000007629395 0.00898635387420654
20 -0.00375866889953613
20.3999996185303 -0.00128829479217529
20.7999992370605 -0.0167582035064697
21.2000007629395 -0.002585768699646
21.6000003814697 0.00105714797973633
22 0.0215905904769897
22.3999996185303 0.00147700309753418
22.7999992370605 -0.000331521034240723
23.2000007629395 -0.0143656730651855
23.6000003814697 -0.0106267929077148
24 0.00220322608947754
24.3999996185303 0.0108509063720703
24.7999992370605 0.0180836915969849
25.2000007629395 -0.00853359699249268
25.6000003814697 -0.0141220092773438
26 0.0110913515090942
26.3999996185303 0.00150704383850098
26.7999992370605 -0.000947833061218262
27.2000007629395 0.00486266613006592
27.6000003814697 -0.00696957111358643
28 0.00424027442932129
28.3999996185303 -0.0244995355606079
28.7999992370605 0.0125217437744141
29.2000007629395 0.0165884494781494
29.6000003814697 0.0161346197128296
30 -0.00257468223571777
30.3999996185303 -0.00175082683563232
30.7999992370605 -0.0109543800354004
31.2000007629395 -0.00730323791503906
31.6000003814697 0.010907769203186
32 0.0126515626907349
32.4000015258789 -0.0254710912704468
32.7999992370605 -2.68220901489258e-05
33.2000007629395 -0.013791561126709
33.5999984741211 0.00317752361297607
34 0.0125253200531006
34.4000015258789 0.0097200870513916
34.7999992370605 0.0252904891967773
35.2000007629395 0.0477265119552612
35.5999984741211 0.0319423675537109
36 -0.014556884765625
36.4000015258789 -0.00151681900024414
36.7999992370605 0.0482884645462036
37.2000007629395 0.0865025520324707
37.5999984741211 0.0783455371856689
38 -0.0034254789352417
38.4000015258789 0.016453742980957
38.7999992370605 -0.00344789028167725
39.2000007629395 0.0378127098083496
39.5999984741211 0.00469601154327393
};
\addlegendentry{OM, $L=200$}
\addplot [line width=2pt, orchid195130203, dashed]
table {%
0 0.99752402305603
0.399999976158142 0.940743923187256
0.799999952316284 0.752541065216064
1.20000004768372 0.434407949447632
1.60000002384186 0.080593466758728
2 -0.261647939682007
2.40000009536743 -0.547101020812988
2.79999995231628 -0.735967993736267
3.20000004768372 -0.797615051269531
3.59999990463257 -0.713258028030396
4 -0.553308010101318
4.40000009536743 -0.311136960983276
4.80000019073486 -0.0256695747375488
5.19999980926514 0.249719977378845
5.59999990463257 0.475378036499023
6 0.624511003494263
6.40000009536743 0.671405076980591
6.80000019073486 0.622491002082825
7.19999980926514 0.447697997093201
7.59999990463257 0.227390050888062
8 -0.00846827030181885
8.39999961853027 -0.214282035827637
8.80000019073486 -0.37451696395874
9.19999980926514 -0.476173043251038
9.60000038146973 -0.503876924514771
10 -0.466395974159241
10.3999996185303 -0.382113933563232
10.8000001907349 -0.227064967155457
11.1999998092651 -0.0323957204818726
11.6000003814697 0.149013996124268
12 0.303880929946899
12.3999996185303 0.391466021537781
12.8000001907349 0.422256946563721
13.1999998092651 0.394827008247375
13.6000003814697 0.288275003433228
14 0.116431951522827
14.3999996185303 -0.0272527933120728
14.8000001907349 -0.139441013336182
15.1999998092651 -0.230175018310547
15.6000003814697 -0.276528000831604
16 -0.259408950805664
16.3999996185303 -0.198441982269287
16.7999992370605 -0.135676026344299
17.2000007629395 -0.057805061340332
17.6000003814697 0.0268372297286987
18 0.0976587533950806
18.3999996185303 0.155583024024963
18.7999992370605 0.199249982833862
19.2000007629395 0.228217959403992
19.6000003814697 0.23213005065918
20 0.215347051620483
20.3999996185303 0.175052046775818
20.7999992370605 0.103469014167786
21.2000007629395 -0.020656943321228
21.6000003814697 -0.121075034141541
22 -0.174350023269653
22.3999996185303 -0.189087986946106
22.7999992370605 -0.16939902305603
23.2000007629395 -0.115959048271179
23.6000003814697 0.000845074653625488
24 0.0626298189163208
24.3999996185303 0.0929969549179077
24.7999992370605 0.0987915992736816
25.2000007629395 0.0791739225387573
25.6000003814697 0.0323765277862549
26 -0.0306706428527832
26.3999996185303 -0.0437588691711426
26.7999992370605 -0.0062713623046875
27.2000007629395 0.00838160514831543
27.6000003814697 0.00841176509857178
28 0.00423872470855713
28.3999996185303 -0.00609433650970459
28.7999992370605 -0.0189429521560669
29.2000007629395 -0.00660169124603271
29.6000003814697 -0.00297844409942627
30 -0.000678896903991699
30.3999996185303 -0.00400376319885254
31.2000007629395 0.00398111343383789
31.6000003814697 0.00501871109008789
32 -0.00317442417144775
32.4000015258789 0.00409352779388428
32.7999992370605 0.0399256944656372
33.2000007629395 -0.00494122505187988
33.5999984741211 -0.00235223770141602
34 0.0179318189620972
34.4000015258789 0.0364905595779419
34.7999992370605 0.0166112184524536
35.2000007629395 0.00553047657012939
35.5999984741211 0.00766134262084961
36 0.0806676149368286
36.4000015258789 0.057212233543396
36.7999992370605 0.0675287246704102
37.2000007629395 0.0636905431747437
37.5999984741211 0.0201187133789062
38 0.0124517679214478
38.4000015258789 -0.00137007236480713
38.7999992370605 -0.00243079662322998
39.2000007629395 0.0298724174499512
39.5999984741211 -3.39746475219727e-05
};
\addlegendentry{OM, $L=400$}
\addplot [line width=2pt, mediumorchid17588186, dashed]
table {%
0 0.998762011528015
0.100000023841858 0.994828939437866
0.200000047683716 0.982807993888855
0.299999952316284 0.962276935577393
0.399999976158142 0.932793974876404
0.5 0.893494009971619
0.600000023841858 0.842328071594238
0.700000047683716 0.786297082901001
0.799999952316284 0.722010016441345
0.899999976158142 0.652584075927734
1 0.576744079589844
1.10000002384186 0.497925043106079
1.20000004768372 0.413704991340637
1.39999997615814 0.239748001098633
1.70000004768372 -0.0260088443756104
1.79999995231628 -0.112491965293884
1.89999997615814 -0.195806980133057
2 -0.27603006362915
2.09999990463257 -0.353016972541809
2.20000004768372 -0.428619980812073
2.29999995231628 -0.496137022972107
2.40000009536743 -0.556262969970703
2.5 -0.61202597618103
2.59999990463257 -0.65842604637146
2.70000004768372 -0.700309038162231
2.79999995231628 -0.735682010650635
2.90000009536743 -0.764733076095581
3 -0.784049034118652
3.09999990463257 -0.795300960540771
3.20000004768372 -0.798488020896912
3.29999995231628 -0.793385028839111
3.40000009536743 -0.779205083847046
3.5 -0.74266505241394
3.59999990463257 -0.715782999992371
3.70000004768372 -0.6839200258255
3.79999995231628 -0.646121978759766
3.90000009536743 -0.606420993804932
4 -0.557600021362305
4.09999990463257 -0.499666929244995
4.19999980926514 -0.445410013198853
4.30000019073486 -0.383062958717346
4.5 -0.249894976615906
4.59999990463257 -0.179123997688293
4.69999980926514 -0.110198020935059
4.80000019073486 -0.0379635095596313
4.90000009536743 0.0320632457733154
5 0.103232979774475
5.09999990463257 0.171885967254639
5.19999980926514 0.239356994628906
5.30000019073486 0.305757999420166
5.40000009536743 0.366426944732666
5.59999990463257 0.470193028450012
5.69999980926514 0.519294023513794
5.80000019073486 0.559345960617065
5.90000009536743 0.593632936477661
6 0.622719049453735
6.09999990463257 0.647341966629028
6.19999980926514 0.660584926605225
6.30000019073486 0.667017936706543
6.40000009536743 0.666648983955383
6.5 0.65939199924469
6.59999990463257 0.644887924194336
6.69999980926514 0.619534969329834
6.80000019073486 0.592074990272522
6.90000009536743 0.561034917831421
7 0.525666952133179
7.09999990463257 0.486235022544861
7.19999980926514 0.442214965820312
7.30000019073486 0.395133018493652
7.40000009536743 0.343513011932373
7.5 0.290817975997925
7.59999990463257 0.236253976821899
7.69999980926514 0.180225968360901
8 0.00807738304138184
8.19999980926514 -0.100646018981934
8.30000019073486 -0.156430006027222
8.39999961853027 -0.206692934036255
8.5 -0.25316596031189
8.80000019073486 -0.375056982040405
8.89999961853027 -0.405925989151001
9 -0.434309959411621
9.10000038146973 -0.458480000495911
9.19999980926514 -0.478670001029968
9.30000019073486 -0.494931936264038
9.39999961853027 -0.505424976348877
9.5 -0.517091035842896
9.69999980926514 -0.502695083618164
9.80000019073486 -0.496636986732483
9.89999961853027 -0.483924031257629
10 -0.467611074447632
10.1000003814697 -0.447211027145386
10.1999998092651 -0.422731041908264
10.3000001907349 -0.393355011940002
10.3999996185303 -0.362823009490967
10.5 -0.326390981674194
10.6000003814697 -0.287392973899841
10.8999996185303 -0.154026985168457
11.1000003814697 -0.0595624446868896
11.1999998092651 -0.0125722885131836
11.3000001907349 0.0288463830947876
11.5 0.117655038833618
11.6000003814697 0.159544944763184
11.6999998092651 0.198439002037048
11.8000001907349 0.235764980316162
11.8999996185303 0.266941070556641
12 0.296841979026794
12.1000003814697 0.323804020881653
12.1999998092651 0.348186016082764
12.3000001907349 0.369080066680908
12.3999996185303 0.385233998298645
12.5 0.395622968673706
12.6000003814697 0.402047991752625
12.6999998092651 0.4045569896698
12.8000001907349 0.402822017669678
12.8999996185303 0.39647102355957
13 0.38490903377533
13.1000003814697 0.374253034591675
13.3999996185303 0.308836936950684
13.5 0.283990979194641
13.6999998092651 0.22978401184082
13.8000001907349 0.199445009231567
13.8999996185303 0.167914986610413
14 0.135055065155029
14.3999996185303 -0.0014568567276001
14.5 -0.036340594291687
14.6999998092651 -0.101043939590454
14.8000001907349 -0.132197976112366
14.8999996185303 -0.161097049713135
15 -0.187917947769165
15.1000003814697 -0.212857007980347
15.1999998092651 -0.236021995544434
15.3000001907349 -0.256849050521851
15.3999996185303 -0.274073958396912
15.5 -0.287925004959106
15.6000003814697 -0.298709034919739
15.6999998092651 -0.306486964225769
15.8000001907349 -0.311406016349792
15.8999996185303 -0.313542008399963
16 -0.31292200088501
16.1000003814697 -0.309325933456421
16.2000007629395 -0.302356004714966
16.2999992370605 -0.291509032249451
16.3999996185303 -0.275830030441284
16.5 -0.24447500705719
16.7000007629395 -0.203472971916199
16.7999992370605 -0.181800007820129
16.8999996185303 -0.15890896320343
17 -0.134266972541809
17.1000003814697 -0.107012987136841
17.3999996185303 -0.0276778936386108
17.5 -0.000401973724365234
17.6000003814697 0.0158282518386841
17.7999992370605 0.0651739835739136
17.8999996185303 0.0888755321502686
18 0.111502051353455
18.1000003814697 0.133098959922791
18.2000007629395 0.152121067047119
18.2999992370605 0.169412016868591
18.3999996185303 0.185004949569702
18.5 0.19897997379303
18.6000003814697 0.211370944976807
18.7000007629395 0.22191596031189
18.7999992370605 0.2308030128479
18.8999996185303 0.238139986991882
19 0.243108987808228
19.1000003814697 0.246075987815857
19.2000007629395 0.247045040130615
19.2999992370605 0.245898008346558
19.3999996185303 0.242202043533325
19.5 0.236523985862732
19.6000003814697 0.228986024856567
19.7000007629395 0.219367980957031
19.7999992370605 0.207612037658691
19.8999996185303 0.193876981735229
20 0.178099036216736
20.1000003814697 0.160084009170532
20.2000007629395 0.139196038246155
20.2999992370605 0.112213969230652
20.5 0.0654237270355225
20.7999992370605 -0.00205838680267334
20.8999996185303 -0.0231677293777466
21 -0.0428003072738647
21.2000007629395 -0.0792694091796875
21.5 -0.130988955497742
21.6000003814697 -0.137817025184631
21.7000007629395 -0.147956013679504
21.7999992370605 -0.155954957008362
21.8999996185303 -0.162062048912048
22 -0.166389942169189
22.1000003814697 -0.169082999229431
22.2000007629395 -0.170220017433167
22.2999992370605 -0.169775009155273
22.3999996185303 -0.167713046073914
22.5 -0.16398298740387
22.6000003814697 -0.158447027206421
22.7000007629395 -0.150534987449646
22.7999992370605 -0.138818979263306
22.8999996185303 -0.117249965667725
23.5 -0.0389975309371948
23.7999992370605 0.00166559219360352
23.8999996185303 0.00553524494171143
24.1000003814697 0.028159499168396
24.2999992370605 0.0487804412841797
24.5 0.0667276382446289
24.7000007629395 0.0823664665222168
24.8999996185303 0.0971131324768066
25 0.103832960128784
25.1000003814697 0.109161019325256
25.2000007629395 0.113576054573059
25.2999992370605 0.117146015167236
25.3999996185303 0.119632959365845
25.5 0.120445966720581
25.6000003814697 0.122576951980591
25.7000007629395 0.119349002838135
26.1000003814697 0.114053010940552
26.2999992370605 0.109364032745361
26.5 0.101999998092651
26.7000007629395 0.0919234752655029
26.8999996185303 0.0794086456298828
27.1000003814697 0.0642776489257812
27.2000007629395 0.0556421279907227
27.2999992370605 0.0459563732147217
27.3999996185303 0.0348124504089355
27.5 0.0218281745910645
27.6000003814697 0.00599157810211182
27.7000007629395 -0.0128645896911621
27.7999992370605 -0.0288549661636353
27.8999996185303 -0.0429201126098633
28 -0.0554053783416748
28.1000003814697 -0.0665074586868286
28.2000007629395 -0.0763607025146484
28.2999992370605 -0.0850144624710083
28.3999996185303 -0.092464804649353
28.5 -0.0987446308135986
28.6000003814697 -0.103950023651123
28.7000007629395 -0.108119010925293
28.7999992370605 -0.11127495765686
28.8999996185303 -0.113450050354004
29 -0.114673972129822
29.1000003814697 -0.114956974983215
29.2000007629395 -0.114254951477051
29.2999992370605 -0.112573027610779
29.3999996185303 -0.109930038452148
29.5 -0.106307983398438
29.6000003814697 -0.101668953895569
29.7000007629395 -0.0960001945495605
29.7999992370605 -0.0892230272293091
29.8999996185303 -0.0812511444091797
30 -0.0719293355941772
30.1000003814697 -0.0610668659210205
30.2000007629395 -0.0485037565231323
30.2999992370605 -0.0340604782104492
30.3999996185303 -0.0165894031524658
30.5 -0.000951766967773438
30.7999992370605 0.0261560678482056
30.8999996185303 0.0331501960754395
31 0.0389782190322876
31.1000003814697 0.0437874794006348
31.2000007629395 0.0476791858673096
31.2999992370605 0.0506008863449097
31.5 0.0538513660430908
31.7000007629395 0.0539159774780273
31.7999992370605 0.0526415109634399
32 0.0474172830581665
32.2000007629395 0.0394314527511597
32.4000015258789 0.0284824371337891
32.5 0.0216799974441528
32.5999984741211 0.0137943029403687
32.7999992370605 -0.00456297397613525
32.9000015258789 -0.0119657516479492
33 -0.0184558629989624
33.0999984741211 -0.023560643196106
33.2000007629395 -0.0274616479873657
33.2999992370605 -0.0302454233169556
33.4000015258789 -0.0318955183029175
33.5 -0.0324150323867798
33.5999984741211 -0.0316379070281982
33.7000007629395 -0.0294075012207031
33.7999992370605 -0.0255899429321289
33.9000015258789 -0.0200518369674683
34 -0.0123590230941772
34.0999984741211 0.000766992568969727
34.2000007629395 0.000928044319152832
34.2999992370605 0.00508284568786621
34.4000015258789 0.00789904594421387
34.5999984741211 0.0109307765960693
34.7999992370605 0.0104807615280151
35 0.00753843784332275
35.0999984741211 0.00506675243377686
35.2000007629395 0.00175273418426514
35.4000015258789 0.00214505195617676
35.5 -0.000946879386901855
35.5999984741211 -0.000678896903991699
35.7999992370605 -0.00448191165924072
35.9000015258789 -0.00782084465026855
36.0999984741211 -0.0033113956451416
36.2000007629395 0.000921845436096191
36.2999992370605 -0.00333166122436523
36.5 0.000268459320068359
36.5999984741211 -0.000584959983825684
36.7000007629395 -0.0108448266983032
36.7999992370605 -8.51154327392578e-05
36.9000015258789 0.00359046459197998
37 -0.00536882877349854
37.0999984741211 0.00745189189910889
37.2000007629395 -2.74181365966797e-06
37.2999992370605 -0.000896096229553223
37.4000015258789 0.000311374664306641
37.5 0.00434517860412598
37.5999984741211 0.00233304500579834
37.7000007629395 -0.00250864028930664
37.7999992370605 -0.00160181522369385
37.9000015258789 -0.00769424438476562
38 0.000309109687805176
38.0999984741211 0.00297319889068604
38.2000007629395 -0.00939822196960449
38.4000015258789 -0.00294351577758789
38.5 -0.00571906566619873
38.5999984741211 0.00448822975158691
38.7000007629395 0.000719189643859863
38.7999992370605 0.000595569610595703
38.9000015258789 -0.00103390216827393
39 -0.00420486927032471
39.0999984741211 -0.00103592872619629
39.2000007629395 0.00027620792388916
39.2999992370605 -0.000887751579284668
39.4000015258789 0.00146150588989258
39.5 -0.00924575328826904
39.7000007629395 0.00267326831817627
39.7999992370605 -0.000923991203308105
39.9000015258789 0.000490427017211914
};
\addlegendentry{OM, $L=800$}
\end{axis}

\end{tikzpicture}

%% file: figures/gravity-wave/convergence-om3.tex
% This file was created with tikzplotlib v0.10.1.
\begin{tikzpicture}

\definecolor{burlywood248194145}{RGB}{250,209,172}
\definecolor{darkgray176}{RGB}{176,176,176}
\definecolor{darkorange24213334}{RGB}{242,133,34}
\definecolor{lightgray204}{RGB}{204,204,204}
\definecolor{sandybrown24516489}{RGB}{181,100,26}

\begin{axis}[
height=\figureheight,
legend cell align={left},
legend style={
	fill opacity=0.8,
	draw opacity=1,
	text opacity=1,
	at={(0.99,0.99)},
	anchor=north east,
	draw=lightgray204
},
tick align=outside,
tick pos=left,
width=\figurewidth,
x grid style={darkgray176},
xlabel={\(\displaystyle t^{*}\)},
xmajorgrids,
xmin=0, xmax=40,
xtick style={color=black},
y grid style={darkgray176},
ylabel style={rotate=-90.0},
ylabel={\(\displaystyle a^{*}\)},
ymajorgrids,
ymin=-0.90082345, ymax=1.08921845,
ytick style={color=black}
]
\addplot [very thick, black]
table {%
0 0.998026728630066
0.185185194015503 0.9781174659729
0.370370388031006 0.925613880157471
0.555555582046509 0.843235969543457
0.740740776062012 0.734638094902039
0.925925970077515 0.604258418083191
1.11111116409302 0.457145810127258
1.29629623889923 0.298771142959595
1.66666662693024 -0.0289593935012817
1.85185182094574 -0.187046051025391
2.03703713417053 -0.33423376083374
2.22222232818604 -0.465845823287964
2.40740752220154 -0.577867984771729
2.59259247779846 -0.667066097259521
2.77777767181396 -0.731073379516602
2.96296286582947 -0.768445372581482
3.14814805984497 -0.778682231903076
3.33333325386047 -0.762219071388245
3.51851844787598 -0.720383405685425
3.70370364189148 -0.655325293540955
3.88888883590698 -0.569920063018799
4.07407426834106 -0.467650175094604
4.25925922393799 -0.352469682693481
4.44444465637207 -0.228656530380249
4.81481504440308 0.0270652770996094
5 0.150189638137817
5.18518495559692 0.264672756195068
5.37037038803101 0.366881251335144
5.55555534362793 0.453703284263611
5.74074077606201 0.522638082504272
5.92592573165894 0.571863412857056
6.11111116409302 0.600278258323669
6.29629611968994 0.607518672943115
6.48148155212402 0.593949556350708
6.66666650772095 0.560630798339844
6.85185194015503 0.509261131286621
7.03703689575195 0.442102909088135
7.22222232818604 0.361888408660889
7.40740728378296 0.271714210510254
7.59259271621704 0.174924850463867
7.96296310424805 -0.0246013402938843
8.14814853668213 -0.1204913854599
8.33333301544189 -0.209532022476196
8.51851844787598 -0.288901329040527
8.70370388031006 -0.356186509132385
8.88888931274414 -0.409454345703125
9.07407379150391 -0.447302341461182
9.25925922393799 -0.468891620635986
9.44444465637207 -0.473958611488342
9.62963008880615 -0.462807536125183
9.81481456756592 -0.436284065246582
10 -0.39573061466217
10.1851854324341 -0.342926740646362
10.3703699111938 -0.28001594543457
10.5555553436279 -0.209422945976257
10.740740776062 -0.133762836456299
11.1111106872559 0.0219100713729858
11.2962961196899 0.0965865850448608
11.481481552124 0.165835618972778
11.6666669845581 0.227465391159058
11.8518514633179 0.279605627059937
12.037036895752 0.320761799812317
12.222222328186 0.349854826927185
12.4074077606201 0.366245865821838
12.5925922393799 0.369745016098022
12.777777671814 0.360605359077454
12.962963104248 0.339500904083252
13.1481485366821 0.307492256164551
13.3333330154419 0.265979051589966
13.518518447876 0.2166428565979
13.7037038803101 0.16138219833374
13.8888893127441 0.10224175453186
14.259259223938 -0.0192111730575562
14.4444446563721 -0.077364444732666
14.6296300888062 -0.131218194961548
14.8148145675659 -0.179070353507996
15 -0.219471216201782
15.1851854324341 -0.251265406608582
15.3703699111938 -0.273622870445251
15.5555553436279 -0.286057472229004
15.740740776062 -0.28843355178833
15.9259262084961 -0.280960321426392
16.1111106872559 -0.264175057411194
16.2962970733643 -0.23891544342041
16.481481552124 -0.206282138824463
16.6666660308838 -0.167594313621521
16.8518524169922 -0.124338269233704
17.037036895752 -0.0781127214431763
17.4074077606201 0.0166382789611816
17.5925922393799 0.0619223117828369
17.7777786254883 0.103801369667053
17.962963104248 0.140953660011292
18.1481475830078 0.172255396842957
18.3333339691162 0.196813941001892
18.518518447876 0.213990688323975
18.7037029266357 0.223415970802307
18.8888893127441 0.224993705749512
19.0740737915039 0.218896269798279
19.2592601776123 0.205551862716675
19.4444446563721 0.185621857643127
19.6296291351318 0.159971714019775
19.8148155212402 0.129636168479919
20 0.0957789421081543
20.1851844787598 0.0596495866775513
20.5555553436279 -0.0142668485641479
20.7407398223877 -0.0495278835296631
20.9259262084961 -0.0820932388305664
21.1111106872559 -0.110936284065247
21.2962970733643 -0.135186195373535
21.481481552124 -0.154153108596802
21.6666660308838 -0.167346119880676
21.8518524169922 -0.174484133720398
22.037036895752 -0.175499677658081
22.2222213745117 -0.170534729957581
22.4074077606201 -0.159930109977722
22.5925922393799 -0.144207835197449
22.7777786254883 -0.124048590660095
22.962963104248 -0.100263833999634
23.1481475830078 -0.0737646818161011
23.518518447876 -0.0165609121322632
23.7037029266357 0.0121328830718994
23.8888893127441 0.0395880937576294
24.0740737915039 0.064909815788269
24.2592601776123 0.0873005390167236
24.4444446563721 0.106085658073425
24.6296291351318 0.120732069015503
24.8148155212402 0.130862236022949
25 0.136263012886047
25.1851844787598 0.136887431144714
25.3703708648682 0.132851958274841
25.5555553436279 0.124427795410156
25.7407398223877 0.112027168273926
25.9259262084961 0.0961849689483643
26.1111106872559 0.0775377750396729
26.2962970733643 0.0567986965179443
26.6666660308838 0.0121220350265503
26.8518524169922 -0.010246753692627
27.037036895752 -0.0316232442855835
27.2222213745117 -0.0513113737106323
27.4074077606201 -0.0686922073364258
27.5925922393799 -0.083242654800415
27.7777786254883 -0.0945510864257812
27.962963104248 -0.102327346801758
28.1481475830078 -0.106409549713135
28.3333339691162 -0.106765747070312
28.518518447876 -0.103491306304932
28.7037029266357 -0.0968018770217896
28.8888893127441 -0.0870226621627808
29.0740737915039 -0.0745742321014404
29.2592601776123 -0.0599559545516968
29.4444446563721 -0.0437257289886475
30.1851844787598 0.0252456665039062
30.3703708648682 0.0405528545379639
30.5555553436279 0.0540436506271362
30.7407398223877 0.0653131008148193
30.9259262084961 0.0740430355072021
31.1111106872559 0.0800105333328247
31.2962970733643 0.0830929279327393
31.481481552124 0.0832687616348267
31.6666660308838 0.0806158781051636
31.8518524169922 0.0753058195114136
32.037036895752 0.0675951242446899
32.2222213745117 0.0578144788742065
32.4074058532715 0.0463552474975586
32.7777786254883 0.0201783180236816
33.1481475830078 -0.00718581676483154
33.3333320617676 -0.020142674446106
33.5185203552246 -0.0320430994033813
33.7037048339844 -0.0425139665603638
33.8888893127441 -0.051241397857666
34.0740737915039 -0.0579798221588135
34.2592582702637 -0.0625578165054321
34.4444427490234 -0.0648825168609619
34.6296310424805 -0.0649402141571045
34.8148155212402 -0.0627939701080322
35 -0.0585802793502808
35.1851844787598 -0.0525015592575073
35.3703689575195 -0.0448178052902222
35.5555572509766 -0.0358356237411499
35.9259262084961 -0.0153669118881226
36.4814796447754 0.0160623788833618
36.6666679382324 0.0253137350082397
36.8518524169922 0.033440113067627
37.037036895752 0.0401983261108398
37.2222213745117 0.0453987121582031
37.4074058532715 0.0489096641540527
37.5925941467285 0.0506608486175537
37.7777786254883 0.0506438016891479
37.962963104248 0.0489097833633423
38.1481475830078 0.0455672740936279
38.3333320617676 0.0407758951187134
38.5185203552246 0.0347399711608887
38.8888893127441 0.0199233293533325
39.6296310424805 -0.0128018856048584
39.8148155212402 -0.0199935436248779
40 -0.0262999534606934
};
\addlegendentry{Analytical model~\cite{dingemans1997WaterWavePropagation}}
\addplot [line width=2pt, burlywood248194145, dash pattern=on 1pt off 3pt on 3pt off 3pt]
table {%
0 0.995046019554138
0.399999976158142 0.949944019317627
0.799999952316284 0.803389072418213
1.20000004768372 0.523195028305054
1.60000002384186 0.137160062789917
2 -0.235365033149719
2.40000009536743 -0.550785064697266
2.79999995231628 -0.737192988395691
3.20000004768372 -0.810366988182068
3.59999990463257 -0.772027015686035
4 -0.559551954269409
4.40000009536743 -0.195927977561951
4.80000019073486 0.0879695415496826
5.19999980926514 0.310343980789185
5.59999990463257 0.498208045959473
6 0.557893037796021
6.40000009536743 0.526841998100281
6.80000019073486 0.409819006919861
7.19999980926514 0.268725991249084
7.59999990463257 0.107570052146912
8 -0.0632288455963135
8.39999961853027 -0.209022998809814
8.80000019073486 -0.325008988380432
9.19999980926514 -0.401533007621765
9.60000038146973 -0.433825016021729
10 -0.422065019607544
10.3999996185303 -0.369261026382446
10.8000001907349 -0.275148987770081
11.1999998092651 -0.103603005409241
11.6000003814697 0.161666989326477
12 0.310222029685974
12.3999996185303 0.390182018280029
12.8000001907349 0.405912041664124
13.1999998092651 0.359235048294067
13.6000003814697 0.245625972747803
14 0.0414289236068726
14.3999996185303 -0.161679029464722
14.8000001907349 -0.22038996219635
15.1999998092651 -0.179851055145264
15.6000003814697 0.000514984130859375
16 0.0482476949691772
16.3999996185303 0.0452908277511597
16.7999992370605 0.0124971866607666
17.2000007629395 -0.0270090103149414
17.6000003814697 -0.0152190923690796
18 -0.00115692615509033
18.3999996185303 0.0232628583908081
19.2000007629395 -0.0205943584442139
19.6000003814697 -0.0139771699905396
20 0.00273358821868896
20.3999996185303 0.0104334354400635
20.7999992370605 -0.0100741386413574
21.2000007629395 -0.0262686014175415
21.6000003814697 -0.00538098812103271
22 0.000182628631591797
22.3999996185303 -0.0137373208999634
22.7999992370605 0.00150430202484131
23.2000007629395 -0.0152237415313721
23.6000003814697 -0.0103449821472168
24 0.0207935571670532
24.3999996185303 0.0134716033935547
24.7999992370605 -0.00290381908416748
25.2000007629395 -0.0137250423431396
25.6000003814697 0.00602900981903076
26 0.012487530708313
26.3999996185303 0.00156450271606445
26.7999992370605 0.00129413604736328
27.2000007629395 -0.0109661817550659
27.6000003814697 -0.0066065788269043
28 -0.00361216068267822
28.7999992370605 0.00853800773620605
29.2000007629395 -0.0127769708633423
29.6000003814697 -0.00650501251220703
30 -0.018506646156311
30.3999996185303 0.0133918523788452
30.7999992370605 -0.011059045791626
31.2000007629395 -0.0135440826416016
31.6000003814697 -0.00114345550537109
32 0.0617556571960449
32.4000015258789 0.0210291147232056
32.7999992370605 -0.0122568607330322
33.2000007629395 -0.0097963809967041
33.5999984741211 -0.00154554843902588
34 -0.00199258327484131
34.4000015258789 0.0204547643661499
34.7999992370605 -0.0100036859512329
35.2000007629395 -0.0119529962539673
35.5999984741211 0.00370204448699951
36 -0.00209927558898926
36.4000015258789 -0.00284934043884277
36.7999992370605 0.0038609504699707
37.2000007629395 -0.0327873229980469
37.5999984741211 2.38418579101562e-07
38 0.0110154151916504
38.4000015258789 -0.0165649652481079
38.7999992370605 -0.000970244407653809
39.2000007629395 -0.0100425481796265
39.5999984741211 0.00461471080780029
};
\addlegendentry{OM3, $L=200$}
\addplot [line width=2pt, sandybrown24516489, dash pattern=on 1pt off 3pt on 3pt off 3pt]
table {%
0 0.99752402305603
0.399999976158142 0.940001010894775
0.799999952316284 0.747719049453735
1.20000004768372 0.429832935333252
1.60000002384186 0.0698345899581909
2 -0.278888940811157
2.40000009536743 -0.566012978553772
2.79999995231628 -0.755007982254028
3.20000004768372 -0.795858979225159
3.59999990463257 -0.710413932800293
4 -0.541055917739868
4.40000009536743 -0.284405946731567
4.80000019073486 0.00470662117004395
5.19999980926514 0.275218963623047
5.59999990463257 0.494853973388672
6 0.610775947570801
6.40000009536743 0.633132934570312
6.80000019073486 0.555504083633423
7.19999980926514 0.366982936859131
7.59999990463257 0.15138304233551
8 -0.0732003450393677
8.39999961853027 -0.270403027534485
8.80000019073486 -0.390362024307251
9.19999980926514 -0.471864938735962
9.60000038146973 -0.500911951065063
10 -0.464017987251282
10.3999996185303 -0.367220997810364
10.8000001907349 -0.183297038078308
11.1999998092651 0.00957179069519043
11.6000003814697 0.185994029045105
12 0.30299699306488
12.3999996185303 0.366788983345032
12.8000001907349 0.371757030487061
13.1999998092651 0.312415957450867
13.6000003814697 0.177546977996826
14 0.0383977890014648
14.3999996185303 -0.0826201438903809
14.8000001907349 -0.172382950782776
15.1999998092651 -0.247422933578491
15.6000003814697 -0.26000702381134
16 -0.241253018379211
16.3999996185303 -0.201854944229126
16.7999992370605 -0.147657036781311
17.2000007629395 -0.0757585763931274
17.6000003814697 0.0188338756561279
18 0.104874968528748
18.3999996185303 0.170048952102661
18.7999992370605 0.213933944702148
19.2000007629395 0.24439001083374
19.6000003814697 0.23760199546814
20 0.205724954605103
20.3999996185303 0.14305305480957
20.7999992370605 0.03905189037323
21.2000007629395 -0.0892738103866577
21.6000003814697 -0.154973030090332
22 -0.18028199672699
22.3999996185303 -0.168956995010376
22.7999992370605 -0.120903968811035
23.2000007629395 -0.00810575485229492
23.6000003814697 0.0518449544906616
24 0.077884316444397
24.3999996185303 0.0753177404403687
24.7999992370605 0.0399560928344727
25.2000007629395 -0.0208990573883057
25.6000003814697 -0.036096453666687
26 -0.021417498588562
26.3999996185303 0.00854921340942383
26.7999992370605 0.00573849678039551
27.2000007629395 0.00150036811828613
27.6000003814697 0.0129575729370117
28 -0.00925886631011963
28.3999996185303 -0.0029151439666748
28.7999992370605 -6.90221786499023e-05
29.2000007629395 -0.000584006309509277
29.6000003814697 0.0157401561737061
30 0.000497341156005859
30.3999996185303 -0.000502824783325195
30.7999992370605 0.00633299350738525
31.2000007629395 0.000846266746520996
31.6000003814697 0.00424098968505859
32 0.00335454940795898
32.4000015258789 0.00039374828338623
32.7999992370605 -0.0225052833557129
33.2000007629395 -0.00476014614105225
33.5999984741211 0.0237722396850586
34 -0.00324082374572754
34.4000015258789 -0.00192868709564209
34.7999992370605 0.000577211380004883
35.2000007629395 -0.000600337982177734
35.5999984741211 -0.00277268886566162
36 -0.00302207469940186
36.4000015258789 -0.00230693817138672
36.7999992370605 -0.0118366479873657
37.2000007629395 0.00729334354400635
37.5999984741211 0.0107196569442749
38 -0.00188493728637695
38.4000015258789 0.00317811965942383
38.7999992370605 -0.00197279453277588
39.2000007629395 0.00570082664489746
39.5999984741211 0.00984275341033936
};
\addlegendentry{OM3, $L=400$}
\addplot [line width=2pt, darkorange24213334, dash pattern=on 1pt off 3pt on 3pt off 3pt]
table {%
0 0.998762011528015
0.100000023841858 0.994807958602905
0.200000047683716 0.982699990272522
0.299999952316284 0.96199893951416
0.399999976158142 0.932250022888184
0.5 0.892561912536621
0.600000023841858 0.840934991836548
0.700000047683716 0.784502029418945
0.799999952316284 0.7187819480896
0.899999976158142 0.648963928222656
1 0.572751045227051
1.20000004768372 0.406301975250244
1.29999995231628 0.322068929672241
1.39999997615814 0.234745025634766
1.5 0.145895957946777
1.60000002384186 0.0540558099746704
1.79999995231628 -0.119176983833313
1.89999997615814 -0.202139019966125
2 -0.282387971878052
2.09999990463257 -0.359488964080811
2.20000004768372 -0.432026028633118
2.29999995231628 -0.499680995941162
2.40000009536743 -0.560293912887573
2.5 -0.61597204208374
2.59999990463257 -0.668796062469482
2.70000004768372 -0.710446000099182
2.79999995231628 -0.745651960372925
2.90000009536743 -0.767174959182739
3 -0.785771012306213
3.09999990463257 -0.796287059783936
3.20000004768372 -0.799028992652893
3.29999995231628 -0.793969988822937
3.40000009536743 -0.780483961105347
3.5 -0.749683022499084
3.59999990463257 -0.71984601020813
3.70000004768372 -0.685963988304138
3.79999995231628 -0.645732998847961
3.90000009536743 -0.601537942886353
4 -0.549947023391724
4.09999990463257 -0.497164964675903
4.19999980926514 -0.436149001121521
4.30000019073486 -0.367514967918396
4.40000009536743 -0.299983978271484
4.59999990463257 -0.156116008758545
4.69999980926514 -0.087916374206543
4.90000009536743 0.0573333501815796
5 0.128257036209106
5.09999990463257 0.197039008140564
5.19999980926514 0.262158989906311
5.30000019073486 0.324880957603455
5.40000009536743 0.368353009223938
5.5 0.414183974266052
5.59999990463257 0.462506055831909
5.69999980926514 0.508413076400757
5.80000019073486 0.546187996864319
5.90000009536743 0.577316045761108
6 0.602517008781433
6.09999990463257 0.623250007629395
6.19999980926514 0.637187957763672
6.30000019073486 0.641128063201904
6.40000009536743 0.637995004653931
6.5 0.627946972846985
6.59999990463257 0.624405980110168
6.69999980926514 0.604234933853149
6.80000019073486 0.577566981315613
6.90000009536743 0.54633903503418
7 0.51004695892334
7.19999980926514 0.420789957046509
7.30000019073486 0.366837978363037
7.40000009536743 0.319414019584656
7.5 0.266134023666382
7.59999990463257 0.211045026779175
7.80000019073486 0.0978691577911377
7.90000009536743 0.0409723520278931
8.10000038146973 -0.0693823099136353
8.19999980926514 -0.123329997062683
8.39999961853027 -0.222475051879883
8.5 -0.269762992858887
8.60000038146973 -0.312286019325256
8.69999980926514 -0.351680994033813
8.80000019073486 -0.385553956031799
8.89999961853027 -0.416177034378052
9 -0.442332983016968
9.10000038146973 -0.464514017105103
9.19999980926514 -0.482792973518372
9.30000019073486 -0.496798992156982
9.39999961853027 -0.505537986755371
9.5 -0.506421089172363
9.60000038146973 -0.500606060028076
9.69999980926514 -0.498080015182495
9.80000019073486 -0.492519974708557
9.89999961853027 -0.478451013565063
10 -0.460667014122009
10.1000003814697 -0.438353061676025
10.1999998092651 -0.41144597530365
10.3999996185303 -0.347156047821045
10.5 -0.308933019638062
10.6000003814697 -0.26833701133728
10.6999998092651 -0.216434001922607
10.8999996185303 -0.126757025718689
11 -0.0843393802642822
11.1000003814697 -0.0381700992584229
11.1999998092651 0.00355100631713867
11.3000001907349 0.0492968559265137
11.3999996185303 0.0932012796401978
11.5 0.129492044448853
11.6000003814697 0.169429063796997
11.6999998092651 0.207247972488403
11.8000001907349 0.247244954109192
11.8999996185303 0.272371053695679
12 0.30076003074646
12.1000003814697 0.326151967048645
12.1999998092651 0.348237037658691
12.3000001907349 0.366914987564087
12.3999996185303 0.384320020675659
12.5 0.392737984657288
12.6000003814697 0.396777033805847
12.6999998092651 0.396542072296143
12.8000001907349 0.39183497428894
12.8999996185303 0.382475018501282
13 0.375769972801208
13.1000003814697 0.358579039573669
13.1999998092651 0.339143037796021
13.3000001907349 0.317775011062622
13.3999996185303 0.294096946716309
13.5 0.267812967300415
13.6000003814697 0.239766955375671
13.6999998092651 0.209200024604797
13.8000001907349 0.177495002746582
13.8999996185303 0.144484043121338
14.3000001907349 0.00489270687103271
14.3999996185303 -0.0337077379226685
14.5 -0.0669139623641968
14.6000003814697 -0.0992451906204224
14.6999998092651 -0.12679398059845
14.8000001907349 -0.155573010444641
14.8999996185303 -0.182530999183655
15 -0.207520961761475
15.1000003814697 -0.230716943740845
15.1999998092651 -0.260733008384705
15.3000001907349 -0.272341966629028
15.3999996185303 -0.286494970321655
15.5 -0.297507047653198
15.6000003814697 -0.305546998977661
15.6999998092651 -0.310766935348511
15.8000001907349 -0.313210010528564
15.8999996185303 -0.312900066375732
16 -0.309836030006409
16.1000003814697 -0.303753972053528
16.2000007629395 -0.294206976890564
16.2999992370605 -0.280622959136963
16.3999996185303 -0.251546025276184
16.5 -0.233206987380981
16.7000007629395 -0.190804004669189
16.7999992370605 -0.168308019638062
16.8999996185303 -0.120553970336914
17 -0.124380946159363
17.2000007629395 -0.0729910135269165
17.3999996185303 -0.0207669734954834
17.5 0.0302325487136841
17.7000007629395 0.0801446437835693
17.7999992370605 0.104225993156433
17.8999996185303 0.122733950614929
18.1000003814697 0.162706971168518
18.2000007629395 0.180333018302917
18.2999992370605 0.19625997543335
18.3999996185303 0.210531949996948
18.5 0.223250031471252
18.6000003814697 0.251080989837646
18.7000007629395 0.255291938781738
18.7999992370605 0.258553028106689
18.8999996185303 0.260959029197693
19 0.248502969741821
19.1000003814697 0.249202013015747
19.3999996185303 0.239897966384888
19.5 0.233958959579468
19.6000003814697 0.225888967514038
19.7000007629395 0.215564012527466
19.7999992370605 0.20302402973175
19.8999996185303 0.188091993331909
20 0.170253992080688
20.1000003814697 0.148995041847229
20.2000007629395 0.126376032829285
20.2999992370605 0.0995985269546509
20.5 0.0518759489059448
20.7000007629395 0.00542974472045898
20.7999992370605 -0.0196075439453125
20.8999996185303 -0.0413068532943726
21 -0.0619356632232666
21.1000003814697 -0.0816020965576172
21.2999992370605 -0.118404984474182
21.5 -0.144389033317566
21.6000003814697 -0.154888987541199
21.7000007629395 -0.163454055786133
21.7999992370605 -0.17023503780365
21.8999996185303 -0.175274968147278
22 -0.178673028945923
22.1000003814697 -0.180516958236694
22.2000007629395 -0.180783033370972
22.2999992370605 -0.17938494682312
22.3999996185303 -0.176193952560425
22.5 -0.171069025993347
22.6000003814697 -0.163710951805115
22.7000007629395 -0.153517007827759
22.7999992370605 -0.139461040496826
22.8999996185303 -0.123553991317749
23.1000003814697 -0.0938853025436401
23.6000003814697 -0.0208319425582886
23.7000007629395 -0.00535786151885986
23.7999992370605 0.0232516527175903
24 0.047074556350708
24.2000007629395 0.0691407918930054
24.3999996185303 0.0896408557891846
24.6000003814697 0.10822606086731
24.7999992370605 0.124341011047363
24.8999996185303 0.129142045974731
25 0.123839020729065
25.1000003814697 0.121168971061707
25.2000007629395 0.123456954956055
25.2999992370605 0.127887010574341
25.3999996185303 0.124212026596069
25.5 0.124783992767334
25.6000003814697 0.123860955238342
25.7000007629395 0.117612957954407
26 0.104400992393494
26.2000007629395 0.0941083431243896
26.3999996185303 0.0821942090988159
26.6000003814697 0.06834876537323
26.7999992370605 0.052198052406311
27 0.0332722663879395
27.2000007629395 0.0113120079040527
27.2999992370605 -0.00348877906799316
27.3999996185303 -0.0120190382003784
27.6000003814697 -0.0357820987701416
27.7999992370605 -0.0565446615219116
28 -0.0742150545120239
28.2000007629395 -0.0886218547821045
28.2999992370605 -0.0945967435836792
28.3999996185303 -0.099734902381897
28.6000003814697 -0.107545018196106
28.7000007629395 -0.110232949256897
28.7999992370605 -0.112069010734558
29 -0.113193035125732
29.1000003814697 -0.112516045570374
29.2000007629395 -0.110970020294189
29.2999992370605 -0.108540058135986
29.3999996185303 -0.105247974395752
29.5 -0.10106098651886
29.6000003814697 -0.095879077911377
29.7000007629395 -0.0896408557891846
29.7999992370605 -0.082334041595459
29.8999996185303 -0.0739307403564453
30 -0.064340353012085
30.1000003814697 -0.0534757375717163
30.2000007629395 -0.0411235094070435
30.2999992370605 -0.0269098281860352
30.3999996185303 -0.00652801990509033
30.5 0.00824284553527832
30.6000003814697 0.0207183361053467
30.7000007629395 0.0313423871994019
30.7999992370605 0.040341854095459
30.8999996185303 0.0478663444519043
31 0.0540610551834106
31.1000003814697 0.0590590238571167
31.2000007629395 0.0629664659500122
31.2999992370605 0.0658656358718872
31.3999996185303 0.0677726268768311
31.5 0.0686789751052856
31.6000003814697 0.0686744451522827
31.7000007629395 0.0678322315216064
31.7999992370605 0.0661091804504395
31.8999996185303 0.0634458065032959
32 0.0599100589752197
32.0999984741211 0.0555307865142822
32.2000007629395 0.0503083467483521
32.2999992370605 0.0442050695419312
32.4000015258789 0.0371872186660767
32.5 0.0291934013366699
32.5999984741211 0.0199762582778931
32.7000007629395 0.00918245315551758
32.7999992370605 -0.010563850402832
32.9000015258789 -0.0118937492370605
33 -0.0182690620422363
33.0999984741211 -0.0232416391372681
33.2999992370605 -0.0304569005966187
33.5 -0.03510582447052
33.5999984741211 -0.0364562273025513
33.7000007629395 -0.0369712114334106
33.7999992370605 -0.0365946292877197
33.9000015258789 -0.0352193117141724
34 -0.0326247215270996
34.0999984741211 -0.0285029411315918
34.2000007629395 -0.0226297378540039
34.2999992370605 -0.014912486076355
34.4000015258789 0.000582575798034668
34.5 0.00668728351593018
34.5999984741211 0.0103667974472046
34.7000007629395 0.0125894546508789
34.9000015258789 0.0143022537231445
35.0999984741211 0.012836217880249
35.2999992370605 0.00809586048126221
35.5 0.000971078872680664
35.5999984741211 0.00213229656219482
35.7000007629395 -0.00271332263946533
35.7999992370605 0.00083458423614502
35.9000015258789 -0.00843405723571777
36.2999992370605 -0.00490903854370117
36.4000015258789 -4.63724136352539e-05
36.5 -0.00679874420166016
36.5999984741211 0.00114071369171143
36.7000007629395 -0.0100762844085693
36.7999992370605 0.00093543529510498
36.9000015258789 -0.00126659870147705
37 0.001090407371521
37.0999984741211 -0.00412201881408691
37.2000007629395 0.000274181365966797
37.4000015258789 -0.0103039741516113
37.5 -0.00130581855773926
37.5999984741211 -0.00907039642333984
37.7000007629395 -0.00168871879577637
37.7999992370605 0.00207102298736572
37.9000015258789 -0.00374734401702881
38 -0.00154685974121094
38.0999984741211 -0.00202381610870361
38.2000007629395 0.00174915790557861
38.2999992370605 -0.00423479080200195
38.4000015258789 -0.00335049629211426
38.5 0.000832915306091309
38.5999984741211 -0.00953400135040283
38.7000007629395 0.00228655338287354
38.7999992370605 -0.00124013423919678
38.9000015258789 -0.00192630290985107
39 -0.00675654411315918
39.2000007629395 -0.000333666801452637
39.2999992370605 0.00115847587585449
39.4000015258789 -0.000847101211547852
39.7000007629395 -0.00171816349029541
39.7999992370605 -0.00465643405914307
39.9000015258789 -0.00162529945373535
};
\addlegendentry{OM3, $L=800$}
\end{axis}

\end{tikzpicture}

%% file: figures/dam-break-rectangular/convergence-y-nbrc.tex
% This file was created with tikzplotlib v0.10.1.
\begin{tikzpicture}

\definecolor{darkgray176}{RGB}{176,176,176}
\definecolor{dodgerblue0154222}{RGB}{0,154,222}
\definecolor{lightgray204}{RGB}{204,204,204}
\definecolor{mediumturquoise64179230}{RGB}{0,116,166}
\definecolor{skyblue128205238}{RGB}{159,217,243}

\begin{axis}[
height=\figureheight,
legend cell align={left},
legend style={
	fill opacity=0.8,
	draw opacity=1,
	text opacity=1,
	at={(0.99,0.99)},
	anchor=north east,
	draw=lightgray204
},
tick align=outside,
tick pos=left,
width=\figurewidth,
x grid style={darkgray176},
xlabel={\(\displaystyle t^{*}\)},
xmajorgrids,
xmin=0, xmax=10,
xtick style={color=black},
y grid style={darkgray176},
ylabel style={rotate=-90.0},
ylabel={\(\displaystyle h^{*}\)},
ymajorgrids,
ymin=-0.032625, ymax=2.005125,
ytick style={color=black}
]
\addplot [very thick, black, mark=*, mark size=2, mark options={solid}, only marks]
table {%
0 1
0.559999942779541 0.940000057220459
0.769999980926514 0.889999985694885
0.930000066757202 0.829999923706055
1.08000004291534 0.779999971389771
1.27999997138977 0.720000028610229
1.46000003814697 0.670000076293945
1.6599999666214 0.610000014305115
1.8400000333786 0.559999942779541
2 0.5
2.21000003814697 0.440000057220459
2.45000004768372 0.389999985694885
2.70000004768372 0.330000042915344
3.05999994277954 0.279999971389771
3.44000005722046 0.220000028610229
4.19999980926514 0.169999957084656
5.25 0.110000014305115
7.40000009536743 0.059999942779541
};
\addlegendentry{Experiment~\cite{martin1952PartIVExperimental}}
\addplot [line width=2pt, skyblue128205238, dotted]
table {%
0 1
0.0998772382736206 1
0.199753999710083 0.990000009536743
0.29963207244873 0.990000009536743
0.599262952804565 0.960000038146973
0.699140071868896 0.940000057220459
0.799018025398254 0.930000066757202
1.49816000461578 0.789999961853027
1.59803998470306 0.759999990463257
1.69790995121002 0.740000009536743
1.79779005050659 0.710000038146973
1.99753999710083 0.670000076293945
2.09741997718811 0.639999985694885
2.39704990386963 0.579999923706055
2.49692988395691 0.569999933242798
2.59681010246277 0.539999961853027
2.99632000923157 0.460000038146973
3.09618997573853 0.450000047683716
3.39582991600037 0.389999985694885
3.49569988250732 0.379999995231628
3.59558010101318 0.360000014305115
3.79533004760742 0.340000033378601
3.8952100276947 0.319999933242798
4.99385976791382 0.210000038146973
5.09373998641968 0.210000038146973
5.19361019134521 0.190000057220459
5.39337015151978 0.190000057220459
5.49324989318848 0.180000066757202
5.59312009811401 0.180000066757202
5.79288005828857 0.159999966621399
5.89275979995728 0.159999966621399
5.99263000488281 0.149999976158142
6.19238996505737 0.149999976158142
6.29226016998291 0.139999985694885
6.39213991165161 0.139999985694885
6.49202013015747 0.129999995231628
6.69177007675171 0.129999995231628
6.89153003692627 0.110000014305115
7.09127998352051 0.110000014305115
7.19116020202637 0.120000004768372
7.29103994369507 0.110000014305115
};
\addlegendentry{NBRC, $W=50$}
\addplot [line width=2pt, mediumturquoise64179230, dotted]
table {%
0 1
0.0998772382736206 0.995000004768372
0.199753999710083 0.995000004768372
0.29963207244873 0.990000009536743
0.599262952804565 0.960000038146973
0.898895025253296 0.914999961853027
1.19852995872498 0.855000019073486
1.3982800245285 0.805000066757202
1.49816000461578 0.774999976158142
1.69790995121002 0.725000023841858
1.89767003059387 0.684999942779541
2.09741997718811 0.634999990463257
2.19729995727539 0.620000004768372
2.29717993736267 0.600000023841858
2.39704990386963 0.585000038146973
2.59681010246277 0.545000076293945
2.79656004905701 0.514999985694885
2.89644002914429 0.495000004768372
2.99632000923157 0.470000028610229
3.29594993591309 0.409999966621399
3.39582991600037 0.394999980926514
3.59558010101318 0.355000019073486
3.79533004760742 0.335000038146973
3.8952100276947 0.319999933242798
3.99509000778198 0.309999942779541
4.09497022628784 0.294999957084656
4.39459991455078 0.264999985694885
4.49447011947632 0.25
4.59434986114502 0.245000004768372
4.79410982131958 0.225000023841858
4.99385976791382 0.215000033378601
5.09373998641968 0.200000047683716
5.29348993301392 0.190000057220459
5.49324989318848 0.190000057220459
5.59312009811401 0.180000066757202
5.69299983978271 0.159999966621399
5.79288005828857 0.144999980926514
5.89275979995728 0.139999985694885
5.99263000488281 0.139999985694885
6.09251022338867 0.159999966621399
};
\addlegendentry{NBRC, $W=100$}
\addplot [line width=2pt, dodgerblue0154222, dotted]
table {%
0 1
0.099940299987793 0.997499942779541
0.199880957603455 0.990000009536743
0.299821019172668 0.995000004768372
0.399760961532593 1.00999999046326
0.499701023101807 0.995000004768372
0.599642038345337 1.07000005245209
0.799521923065186 1.19500005245209
0.899461984634399 1.25499999523163
0.99940299987793 1.30999994277954
1.09933996200562 1.36749994754791
1.19928002357483 1.41999995708466
1.29921996593475 1.47000002861023
1.39916002750397 1.51750004291534
1.49909996986389 1.5625
1.69897997379303 1.64750003814697
1.89885997772217 1.7224999666214
2.09875011444092 1.78250002861023
2.19868993759155 1.80999994277954
2.29862999916077 1.83249998092651
2.39857006072998 1.85749995708466
2.49850988388062 1.87750005722046
2.59844994544983 1.89250004291534
2.69839000701904 1.90499997138977
2.79833006858826 1.90750002861023
2.89826989173889 1.91250002384186
2.99820995330811 1.91250002384186
};
\addlegendentry{NBRC, $W=200$}
\end{axis}

\end{tikzpicture}

%% file: figures/dam-break-rectangular/convergence-x-nbrc.tex
% This file was created with tikzplotlib v0.10.1.
\begin{tikzpicture}

\definecolor{darkgray176}{RGB}{176,176,176}
\definecolor{dodgerblue0154222}{RGB}{0,154,222}
\definecolor{lightgray204}{RGB}{204,204,204}
\definecolor{mediumturquoise64179230}{RGB}{0,116,166}
\definecolor{skyblue128205238}{RGB}{159,217,243}

\begin{axis}[
height=\figureheight,
legend cell align={left},
legend style={
	fill opacity=0.8,
	draw opacity=1,
	text opacity=1,
	at={(0.99,0.01)},
	anchor=south east,
	draw=lightgray204
},
tick align=outside,
tick pos=left,
width=\figurewidth,
x grid style={darkgray176},
xlabel={\(\displaystyle t^{*}\)},
xmajorgrids,
xmin=0, xmax=10,
xtick style={color=black},
y grid style={darkgray176},
ylabel style={rotate=-90.0},
ylabel={\(\displaystyle w^{*}\)},
ymajorgrids,
ymin=0.2945, ymax=15.5955,
ytick style={color=black}
]
\addplot [very thick, black, mark=*, mark size=2, mark options={solid}, only marks]
table {%
0.409999966621399 1.11000001430511
0.839999914169312 1.22000002861023
1.19000005722046 1.44000005722046
1.42999994754791 1.66999995708466
1.62999999523163 1.88999998569489
1.83000004291534 2.10999989509583
1.98000001907349 2.32999992370605
2.20000004768372 2.55999994277954
2.3199999332428 2.77999997138977
2.50999999046326 3
2.65000009536743 3.22000002861023
2.82999992370605 3.44000005722046
2.97000002861023 3.67000007629395
3.10999989509583 3.89000010490417
3.32999992370605 4.1100001335144
4.01999998092651 5
4.44000005722046 5.8899998664856
5.09000015258789 7
5.69000005722046 8
6.30000019073486 9
6.82999992370605 10
7.44000005722046 11
8.07999992370605 12
8.67000007629395 13
9.3100004196167 14
};
\addlegendentry{Experiment~\cite{martin1952PartIVExperimental}}
\addplot [line width=2pt, skyblue128205238, dotted]
table {%
0 1
0.0998772382736206 1
0.199753999710083 1.01999998092651
0.29963207244873 1.05999994277954
0.399508953094482 1.08000004291534
0.499386072158813 1.13999998569489
0.599262952804565 1.17999994754791
0.699140071868896 1.25999999046326
0.799018025398254 1.32000005245209
0.898895025253296 1.41999995708466
0.998772025108337 1.5
1.3982800245285 1.89999997615814
1.49816000461578 2.01999998092651
1.59803998470306 2.11999988555908
1.89767003059387 2.48000001907349
1.99753999710083 2.61999988555908
2.19729995727539 2.85999989509583
2.69668006896973 3.55999994277954
2.79656004905701 3.72000002861023
2.89644002914429 3.85999989509583
3.19606995582581 4.34000015258789
3.29594993591309 4.48000001907349
3.39582991600037 4.65999984741211
3.69546008110046 5.1399998664856
3.8952100276947 5.5
3.99509000778198 5.65999984741211
4.09497022628784 5.84000015258789
4.19483995437622 6
4.29472017288208 6.17999982833862
4.39459991455078 6.34000015258789
4.69423007965088 6.94000005722046
4.79410982131958 7.15999984741211
4.89398002624512 7.42000007629395
5.29348993301392 8.53999996185303
5.59312009811401 9.4399995803833
5.69299983978271 9.72000026702881
6.09251022338867 10.9200000762939
6.29226016998291 11.4799995422363
6.59189987182617 12.3800001144409
6.89153003692627 13.3400001525879
6.99139976501465 13.6800003051758
7.09127998352051 14.039999961853
7.29103994369507 14.7200002670288
};
\addlegendentry{NBRC, $W=50$}
\addplot [line width=2pt, mediumturquoise64179230, dotted]
table {%
0 1
0.0998772382736206 1.00999999046326
0.199753999710083 1.07000005245209
0.29963207244873 1.14999997615814
0.399508953094482 1.22000002861023
0.499386072158813 1.27999997138977
0.599262952804565 1.36000001430511
0.799018025398254 1.53999996185303
1.0986499786377 1.87000000476837
1.29840004444122 2.10999989509583
1.3982800245285 2.24000000953674
1.49816000461578 2.35999989509583
1.59803998470306 2.5
1.69790995121002 2.63000011444092
1.79779005050659 2.76999998092651
2.09741997718811 3.22000002861023
2.19729995727539 3.39000010490417
2.29717993736267 3.57999992370605
2.39704990386963 3.75999999046326
2.49692988395691 3.9300000667572
2.69668006896973 4.25
2.79656004905701 4.40000009536743
2.89644002914429 4.55999994277954
2.99632000923157 4.69000005722046
3.29594993591309 5.01999998092651
3.39582991600037 5.15999984741211
3.49569988250732 5.34000015258789
3.59558010101318 5.55999994277954
3.69546008110046 5.82999992370605
3.79533004760742 6.53999996185303
4.09497022628784 7.63000011444092
4.49447011947632 9.10000038146973
4.79410982131958 10.1700000762939
5.29348993301392 11.9300003051758
5.69299983978271 13.3800001144409
5.89275979995728 14.1400003433228
5.99263000488281 14.4300003051758
6.09251022338867 14.8999996185303
};
\addlegendentry{NBRC, $W=100$}
\addplot [line width=2pt, dodgerblue0154222, dotted]
table {%
0 1
0.099940299987793 0.990000009536743
0.199880957603455 1.08000004291534
0.299821019172668 1.15499997138977
0.399760961532593 1.25
0.499701023101807 1.375
0.599642038345337 1.49000000953674
0.699581980705261 1.61000001430511
0.799521923065186 1.72500002384186
0.899461984634399 1.89499998092651
0.99940299987793 2.21000003814697
1.59904003143311 4.15500020980835
1.9988100528717 5.42999982833862
2.69839000701904 7.63000011444092
2.99820995330811 8.54500007629395
};
\addlegendentry{NBRC, $W=200$}
\end{axis}

\end{tikzpicture}

%% file: figures/dam-break-rectangular/convergence-y-nbkc.tex
% This file was created with tikzplotlib v0.10.1.
\begin{tikzpicture}

\definecolor{darkgray176}{RGB}{176,176,176}
\definecolor{lightgray204}{RGB}{204,204,204}
\definecolor{mediumaquamarine128230181}{RGB}{159,236,200}
\definecolor{mediumaquamarine64218145}{RGB}{0,128,68}
\definecolor{springgreen0205108}{RGB}{0,205,108}

\begin{axis}[
height=\figureheight,
legend cell align={left},
legend style={
	fill opacity=0.8,
	draw opacity=1,
	text opacity=1,
	at={(0.99,0.99)},
	anchor=north east,
	draw=lightgray204
},
tick align=outside,
tick pos=left,
width=\figurewidth,
x grid style={darkgray176},
xlabel={\(\displaystyle t^{*}\)},
xmajorgrids,
xmin=0, xmax=10,
xtick style={color=black},
y grid style={darkgray176},
ylabel style={rotate=-90.0},
ylabel={\(\displaystyle h^{*}\)},
ymajorgrids,
ymin=0.013, ymax=1.047,
ytick style={color=black}
]
\addplot [very thick, black, mark=*, mark size=2, mark options={solid}, only marks]
table {%
0 1
0.559999942779541 0.940000057220459
0.769999980926514 0.889999985694885
0.930000066757202 0.829999923706055
1.08000004291534 0.779999971389771
1.27999997138977 0.720000028610229
1.46000003814697 0.670000076293945
1.6599999666214 0.610000014305115
1.8400000333786 0.559999942779541
2 0.5
2.21000003814697 0.440000057220459
2.45000004768372 0.389999985694885
2.70000004768372 0.330000042915344
3.05999994277954 0.279999971389771
3.44000005722046 0.220000028610229
4.19999980926514 0.169999957084656
5.25 0.110000014305115
7.40000009536743 0.059999942779541
};
\addlegendentry{Experiment~\cite{martin1952PartIVExperimental}}
\addplot [line width=2pt, mediumaquamarine128230181]
table {%
0 1
0.0998772382736206 1
0.29963207244873 0.980000019073486
0.399508953094482 0.980000019073486
0.599262952804565 0.960000038146973
0.699140071868896 0.940000057220459
0.799018025398254 0.930000066757202
1.29840004444122 0.829999923706055
1.3982800245285 0.799999952316284
1.49816000461578 0.779999971389771
1.59803998470306 0.75
1.79779005050659 0.710000038146973
1.89767003059387 0.680000066757202
1.99753999710083 0.660000085830688
2.09741997718811 0.629999995231628
2.69668006896973 0.509999990463257
2.79656004905701 0.5
2.99632000923157 0.460000038146973
3.09618997573853 0.450000047683716
3.29594993591309 0.409999966621399
3.39582991600037 0.399999976158142
3.49569988250732 0.379999995231628
3.59558010101318 0.370000004768372
3.69546008110046 0.350000023841858
3.79533004760742 0.340000033378601
3.8952100276947 0.319999933242798
5.09373998641968 0.200000047683716
5.29348993301392 0.200000047683716
5.39337015151978 0.190000057220459
5.49324989318848 0.169999957084656
5.69299983978271 0.169999957084656
5.79288005828857 0.159999966621399
5.89275979995728 0.159999966621399
5.99263000488281 0.149999976158142
6.09251022338867 0.159999966621399
6.19238996505737 0.149999976158142
6.59189987182617 0.149999976158142
6.69177007675171 0.139999985694885
6.89153003692627 0.139999985694885
6.99139976501465 0.129999995231628
7.09127998352051 0.139999985694885
7.39091014862061 0.110000014305115
7.49078989028931 0.110000014305115
7.59067010879517 0.100000023841858
7.99018001556396 0.100000023841858
8.09004974365234 0.0900000333786011
8.1899299621582 0.100000023841858
8.6893196105957 0.100000023841858
};
\addlegendentry{NBKC, $W=50$}
\addplot [line width=2pt, mediumaquamarine64218145]
table {%
0 1
0.0998772382736206 0.995000004768372
0.199753999710083 0.995000004768372
0.29963207244873 0.990000009536743
0.599262952804565 0.960000038146973
0.898895025253296 0.914999961853027
0.998772025108337 0.894999980926514
1.0986499786377 0.870000004768372
1.19852995872498 0.850000023841858
1.29840004444122 0.825000047683716
1.3982800245285 0.805000066757202
1.79779005050659 0.704999923706055
1.89767003059387 0.684999942779541
1.99753999710083 0.660000085830688
2.29717993736267 0.600000023841858
2.39704990386963 0.575000047683716
2.59681010246277 0.535000085830688
2.69668006896973 0.519999980926514
2.89644002914429 0.480000019073486
2.99632000923157 0.465000033378601
3.09618997573853 0.444999933242798
3.79533004760742 0.340000033378601
3.8952100276947 0.330000042915344
3.99509000778198 0.315000057220459
4.79410982131958 0.235000014305115
4.89398002624512 0.230000019073486
5.09373998641968 0.210000038146973
5.59312009811401 0.184999942779541
9.98771953582764 0.184999942779541
};
\addlegendentry{NBKC, $W=100$}
\addplot [line width=2pt, springgreen0205108]
table {%
0 1
0.199880957603455 0.995000004768372
0.499701023101807 0.972500085830688
0.599642038345337 0.960000038146973
0.799521923065186 0.930000066757202
0.99940299987793 0.894999980926514
1.09933996200562 0.875
1.39916002750397 0.807500004768372
1.49909996986389 0.782500028610229
1.59904003143311 0.759999990463257
1.89885997772217 0.684999942779541
1.9988100528717 0.664999961853027
2.19868993759155 0.620000004768372
2.59844994544983 0.539999961853027
2.69839000701904 0.522500038146973
2.89826989173889 0.482499957084656
2.99820995330811 0.465000033378601
3.09815001487732 0.450000047683716
3.19809007644653 0.432500004768372
3.29802989959717 0.417500019073486
3.39796996116638 0.399999976158142
3.4979100227356 0.387500047683716
3.59785008430481 0.370000004768372
3.79772996902466 0.340000033378601
3.89767003059387 0.327499985694885
3.99761009216309 0.317499995231628
4.09754991531372 0.305000066757202
4.49731016159058 0.264999985694885
4.59724998474121 0.257499933242798
4.69718980789185 0.247499942779541
4.79713010787964 0.240000009536743
4.89706993103027 0.230000019073486
4.99701023101807 0.225000023841858
5.0969500541687 0.215000033378601
5.39677000045776 0.192499995231628
5.4967098236084 0.1875
5.59665012359619 0.180000066757202
5.69658994674683 0.177500009536743
5.89648008346558 0.167500019073486
5.99641990661621 0.159999966621399
6.19630002975464 0.154999971389771
6.39618015289307 0.144999980926514
6.4961199760437 0.142500042915344
6.59605979919434 0.137500047683716
6.89588022232056 0.129999995231628
6.99582004547119 0.125
7.09575986862183 0.125
7.19570016860962 0.120000004768372
};
\addlegendentry{NBKC, $W=200$}
\end{axis}

\end{tikzpicture}

%% file: figures/dam-break-rectangular/convergence-x-nbkc.tex
% This file was created with tikzplotlib v0.10.1.
\begin{tikzpicture}

\definecolor{darkgray176}{RGB}{176,176,176}
\definecolor{lightgray204}{RGB}{204,204,204}
\definecolor{mediumaquamarine128230181}{RGB}{159,236,200}
\definecolor{mediumaquamarine64218145}{RGB}{0,128,68}
\definecolor{springgreen0205108}{RGB}{0,205,108}

\begin{axis}[
height=\figureheight,
legend cell align={left},
legend style={
	fill opacity=0.8,
	draw opacity=1,
	text opacity=1,
	at={(0.01,0.99)},
	anchor=north west,
	draw=lightgray204
},
tick align=outside,
tick pos=left,
width=\figurewidth,
x grid style={darkgray176},
xlabel={\(\displaystyle t^{*}\)},
xmajorgrids,
xmin=0, xmax=10,
xtick style={color=black},
y grid style={darkgray176},
ylabel style={rotate=-90.0},
ylabel={\(\displaystyle w^{*}\)},
ymajorgrids,
ymin=0.30725, ymax=15.54775,
ytick style={color=black}
]
\addplot [very thick, black, mark=*, mark size=2, mark options={solid}, only marks]
table {%
0.409999966621399 1.11000001430511
0.839999914169312 1.22000002861023
1.19000005722046 1.44000005722046
1.42999994754791 1.66999995708466
1.62999999523163 1.88999998569489
1.83000004291534 2.10999989509583
1.98000001907349 2.32999992370605
2.20000004768372 2.55999994277954
2.3199999332428 2.77999997138977
2.50999999046326 3
2.65000009536743 3.22000002861023
2.82999992370605 3.44000005722046
2.97000002861023 3.67000007629395
3.10999989509583 3.89000010490417
3.32999992370605 4.1100001335144
4.01999998092651 5
4.44000005722046 5.8899998664856
5.09000015258789 7
5.69000005722046 8
6.30000019073486 9
6.82999992370605 10
7.44000005722046 11
8.07999992370605 12
8.67000007629395 13
9.3100004196167 14
};
\addlegendentry{Experiment~\cite{martin1952PartIVExperimental}}
\addplot [line width=2pt, mediumaquamarine128230181]
table {%
0 1
0.0998772382736206 1
0.199753999710083 1.01999998092651
0.29963207244873 1.05999994277954
0.399508953094482 1.08000004291534
0.499386072158813 1.13999998569489
0.599262952804565 1.17999994754791
0.699140071868896 1.25999999046326
0.799018025398254 1.32000005245209
1.0986499786377 1.55999994277954
1.49816000461578 1.96000003814697
1.99753999710083 2.55999994277954
2.09741997718811 2.70000004768372
2.19729995727539 2.8199999332428
2.59681010246277 3.38000011444092
2.69668006896973 3.53999996185303
2.79656004905701 3.6800000667572
3.59558010101318 4.96000003814697
3.69546008110046 5.1399998664856
3.99509000778198 5.61999988555908
4.09497022628784 5.80000019073486
4.29472017288208 6.11999988555908
4.39459991455078 6.30000019073486
4.59434986114502 6.61999988555908
4.79410982131958 6.98000001907349
4.89398002624512 7.1399998664856
4.99385976791382 7.32000017166138
5.09373998641968 7.48000001907349
5.39337015151978 8.02000045776367
5.49324989318848 8.22000026702881
5.59312009811401 8.39999961853027
5.69299983978271 8.60000038146973
5.79288005828857 8.77999973297119
6.09251022338867 9.38000011444092
6.59189987182617 10.2799997329712
6.69177007675171 10.4799995422363
6.79164981842041 10.6599998474121
7.59067010879517 12.2600002288818
7.89029979705811 12.9200000762939
7.99018001556396 13.1199998855591
8.38969039916992 14
8.4895601272583 14.2399997711182
8.6893196105957 14.6800003051758
};
\addlegendentry{NBKC, $W=50$}
\addplot [line width=2pt, mediumaquamarine64218145]
table {%
0 1
0.0998772382736206 1
0.199753999710083 1.01999998092651
0.29963207244873 1.04999995231628
0.399508953094482 1.0900000333786
0.499386072158813 1.13999998569489
0.699140071868896 1.25999999046326
0.799018025398254 1.33000004291534
0.898895025253296 1.4099999666214
0.998772025108337 1.5
1.0986499786377 1.58000004291534
1.3982800245285 1.87999999523163
1.59803998470306 2.09999990463257
1.89767003059387 2.46000003814697
1.99753999710083 2.58999991416931
2.09741997718811 2.71000003814697
2.19729995727539 2.84999990463257
2.29717993736267 2.98000001907349
2.79656004905701 3.6800000667572
3.09618997573853 4.13000011444092
3.19606995582581 4.28999996185303
3.29594993591309 4.46000003814697
3.49569988250732 4.82000017166138
3.69546008110046 5.21999979019165
3.79533004760742 5.44999980926514
3.8952100276947 5.71999979019165
3.99509000778198 6.03000020980835
4.09497022628784 6.40000009536743
4.19483995437622 6.82000017166138
4.39459991455078 7.73000001907349
4.49447011947632 6.40999984741211
4.59434986114502 6.57999992370605
4.69423007965088 6.76000022888184
4.79410982131958 6.92999982833862
5.49324989318848 8.1899995803833
5.59312009811401 8.19999980926514
9.98771953582764 8.19999980926514
};
\addlegendentry{NBKC, $W=100$}
\addplot [line width=2pt, springgreen0205108]
table {%
0 1
0.099940299987793 1.00499999523163
0.199880957603455 1.02499997615814
0.299821019172668 1.05499994754791
0.499701023101807 1.14499998092651
0.599642038345337 1.21000003814697
0.699581980705261 1.26999998092651
0.799521923065186 1.3400000333786
0.899461984634399 1.41999995708466
0.99940299987793 1.50499999523163
1.09933996200562 1.59500002861023
1.19928002357483 1.69000005722046
1.39916002750397 1.88999998569489
1.49909996986389 1.99500000476837
1.79892003536224 2.33999991416931
1.9988100528717 2.58999991416931
2.19868993759155 2.84999990463257
2.39857006072998 3.11999988555908
2.49850988388062 3.26500010490417
2.69839000701904 3.54500007629395
3.09815001487732 4.15000009536743
3.59785008430481 4.95499992370605
3.79772996902466 5.28999996185303
3.89767003059387 5.46000003814697
4.09754991531372 5.81500005722046
4.29743003845215 6.15999984741211
4.79713010787964 7.00500011444092
4.99701023101807 7.36999988555908
5.39677000045776 8.07499980926514
5.89648008346558 8.97500038146973
5.99641990661621 9.17500019073486
6.096360206604 9.39000034332275
6.19630002975464 9.63500022888184
6.29623985290527 9.89000034332275
6.39618015289307 10.1850004196167
6.4961199760437 10.5450000762939
6.59605979919434 11.0600004196167
6.69600009918213 11.8050003051758
6.79593992233276 11.6750001907349
6.89588022232056 11.9049997329712
6.99582004547119 12.1499996185303
7.09575986862183 12.414999961853
7.19570016860962 14.8549995422363
};
\addlegendentry{NBKC, $W=200$}
\end{axis}

\end{tikzpicture}

%% file: figures/dam-break-rectangular/convergence-y-om.tex
% This file was created with tikzplotlib v0.10.1.
\begin{tikzpicture}

\definecolor{darkgray176}{RGB}{176,176,176}
\definecolor{lightgray204}{RGB}{204,204,204}
\definecolor{mediumorchid17588186}{RGB}{175,88,186}
\definecolor{orchid195130203}{RGB}{109,55,116}
\definecolor{plum215172220}{RGB}{225,192,229}

\begin{axis}[
height=\figureheight,
legend cell align={left},
legend style={
	fill opacity=0.8,
	draw opacity=1,
	text opacity=1,
	at={(0.99,0.99)},
	anchor=north east,
	draw=lightgray204
},
tick align=outside,
tick pos=left,
width=\figurewidth,
x grid style={darkgray176},
xlabel={\(\displaystyle t^{*}\)},
xmajorgrids,
xmin=0, xmax=10,
xtick style={color=black},
y grid style={darkgray176},
ylabel style={rotate=-90.0},
ylabel={\(\displaystyle h^{*}\)},
ymajorgrids,
ymin=0.013, ymax=1.047,
ytick style={color=black}
]
\addplot [very thick, black, mark=*, mark size=2, mark options={solid}, only marks]
table {%
0 1
0.559999942779541 0.940000057220459
0.769999980926514 0.889999985694885
0.930000066757202 0.829999923706055
1.08000004291534 0.779999971389771
1.27999997138977 0.720000028610229
1.46000003814697 0.670000076293945
1.6599999666214 0.610000014305115
1.8400000333786 0.559999942779541
2 0.5
2.21000003814697 0.440000057220459
2.45000004768372 0.389999985694885
2.70000004768372 0.330000042915344
3.05999994277954 0.279999971389771
3.44000005722046 0.220000028610229
4.19999980926514 0.169999957084656
5.25 0.110000014305115
7.40000009536743 0.059999942779541
};
\addlegendentry{Experiment~\cite{martin1952PartIVExperimental}}
\addplot [line width=2pt, plum215172220, dashed]
table {%
0 1
0.0998772382736206 0.990000009536743
0.29963207244873 0.990000009536743
0.599262952804565 0.960000038146973
0.699140071868896 0.940000057220459
0.799018025398254 0.930000066757202
1.29840004444122 0.829999923706055
1.3982800245285 0.799999952316284
1.49816000461578 0.779999971389771
1.59803998470306 0.75
1.79779005050659 0.710000038146973
1.89767003059387 0.680000066757202
2.89644002914429 0.480000019073486
2.99632000923157 0.470000028610229
3.19606995582581 0.430000066757202
3.29594993591309 0.419999957084656
3.39582991600037 0.399999976158142
3.49569988250732 0.389999985694885
3.59558010101318 0.370000004768372
3.79533004760742 0.350000023841858
3.8952100276947 0.330000042915344
4.19483995437622 0.299999952316284
4.29472017288208 0.279999971389771
4.79410982131958 0.230000019073486
4.89398002624512 0.240000009536743
4.99385976791382 0.230000019073486
5.09373998641968 0.230000019073486
5.39337015151978 0.200000047683716
5.59312009811401 0.200000047683716
5.79288005828857 0.180000066757202
5.89275979995728 0.180000066757202
6.09251022338867 0.159999966621399
6.19238996505737 0.159999966621399
6.29226016998291 0.149999976158142
6.39213991165161 0.149999976158142
6.49202013015747 0.139999985694885
6.59189987182617 0.139999985694885
6.69177007675171 0.129999995231628
6.79164981842041 0.129999995231628
6.89153003692627 0.120000004768372
6.99139976501465 0.129999995231628
7.09127998352051 0.129999995231628
7.19116020202637 0.120000004768372
7.29103994369507 0.129999995231628
7.39091014862061 0.120000004768372
7.49078989028931 0.129999995231628
7.69053983688354 0.129999995231628
7.7904200553894 0.120000004768372
7.89029979705811 0.120000004768372
7.99018001556396 0.110000014305115
8.38969039916992 0.110000014305115
8.4895601272583 0.120000004768372
8.58944034576416 0.120000004768372
8.6893196105957 0.110000014305115
8.7891902923584 0.110000014305115
8.9889497756958 0.0900000333786011
9.18869972229004 0.0900000333786011
9.2885799407959 0.100000023841858
};
\addlegendentry{OM, $W=50$}
\addplot [line width=2pt, orchid195130203, dashed]
table {%
0 1
0.0998772382736206 0.995000004768372
0.199753999710083 0.995000004768372
0.29963207244873 0.990000009536743
0.599262952804565 0.960000038146973
0.799018025398254 0.930000066757202
0.898895025253296 0.910000085830688
0.998772025108337 0.894999980926514
1.0986499786377 0.875
1.19852995872498 0.850000023841858
1.29840004444122 0.829999923706055
1.49816000461578 0.779999971389771
1.59803998470306 0.759999990463257
1.79779005050659 0.710000038146973
1.89767003059387 0.690000057220459
1.99753999710083 0.664999961853027
2.09741997718811 0.644999980926514
2.19729995727539 0.620000004768372
2.69668006896973 0.519999980926514
2.79656004905701 0.504999995231628
2.89644002914429 0.485000014305115
2.99632000923157 0.470000028610229
3.09618997573853 0.450000047683716
3.19606995582581 0.434999942779541
3.29594993591309 0.414999961853027
3.39582991600037 0.404999971389771
3.8952100276947 0.330000042915344
4.09497022628784 0.309999942779541
4.19483995437622 0.294999957084656
4.39459991455078 0.274999976158142
4.49447011947632 0.269999980926514
4.79410982131958 0.240000009536743
4.89398002624512 0.235000014305115
4.99385976791382 0.225000023841858
5.19361019134521 0.215000033378601
5.29348993301392 0.205000042915344
5.39337015151978 0.200000047683716
5.59312009811401 0.180000066757202
6.09251022338867 0.154999971389771
6.29226016998291 0.154999971389771
6.59189987182617 0.139999985694885
6.69177007675171 0.139999985694885
6.79164981842041 0.129999995231628
6.89153003692627 0.129999995231628
6.99139976501465 0.125
7.19116020202637 0.125
7.39091014862061 0.115000009536743
7.49078989028931 0.115000009536743
7.59067010879517 0.120000004768372
7.7904200553894 0.120000004768372
7.99018001556396 0.110000014305115
8.09004974365234 0.110000014305115
8.28981018066406 0.100000023841858
8.4895601272583 0.100000023841858
8.7891902923584 0.0850000381469727
8.88906955718994 0.0850000381469727
8.9889497756958 0.0800000429153442
};
\addlegendentry{OM, $W=100$}
\addplot [line width=2pt, mediumorchid17588186, dashed]
table {%
0 1
0.199880957603455 0.995000004768372
0.299821019172668 0.990000009536743
0.399760961532593 0.982500076293945
0.499701023101807 0.972500085830688
0.599642038345337 0.960000038146973
0.799521923065186 0.930000066757202
0.99940299987793 0.894999980926514
1.09933996200562 0.872499942779541
1.19928002357483 0.852499961853027
1.29921996593475 0.829999923706055
1.39916002750397 0.805000066757202
1.49909996986389 0.782500028610229
1.59904003143311 0.757499933242798
1.69897997379303 0.735000014305115
1.79892003536224 0.710000038146973
2.09875011444092 0.642499923706055
2.19868993759155 0.622499942779541
2.29862999916077 0.600000023841858
2.49850988388062 0.559999942779541
2.59844994544983 0.542500019073486
2.79833006858826 0.502500057220459
2.99820995330811 0.467499971389771
3.09815001487732 0.452499985694885
3.19809007644653 0.434999942779541
3.69778990745544 0.360000014305115
4.19749021530151 0.297500014305115
4.49731016159058 0.267500042915344
4.59724998474121 0.259999990463257
4.69718980789185 0.25
5.19688987731934 0.212499976158142
5.29683017730713 0.207499980926514
5.59665012359619 0.184999942779541
5.69658994674683 0.182500004768372
5.79653978347778 0.174999952316284
6.39618015289307 0.144999980926514
6.4961199760437 0.144999980926514
6.59605979919434 0.139999985694885
6.69600009918213 0.137500047683716
6.79593992233276 0.132499933242798
6.99582004547119 0.127500057220459
7.09575986862183 0.122499942779541
7.19570016860962 0.122499942779541
7.29563999176025 0.120000004768372
7.39557981491089 0.115000009536743
7.49552011489868 0.115000009536743
7.89527988433838 0.105000019073486
7.99522018432617 0.105000019073486
8.19509983062744 0.100000023841858
8.29504013061523 0.100000023841858
8.4949197769165 0.0950000286102295
8.5948600769043 0.0950000286102295
8.69480037689209 0.0924999713897705
8.79473972320557 0.0924999713897705
8.89468002319336 0.0900000333786011
8.99462032318115 0.0900000333786011
};
\addlegendentry{OM, $W=200$}
\end{axis}

\end{tikzpicture}

%% file: figures/dam-break-rectangular/convergence-x-om.tex
% This file was created with tikzplotlib v0.10.1.
\begin{tikzpicture}

\definecolor{darkgray176}{RGB}{176,176,176}
\definecolor{lightgray204}{RGB}{204,204,204}
\definecolor{mediumorchid17588186}{RGB}{175,88,186}
\definecolor{orchid195130203}{RGB}{109,55,116}
\definecolor{plum215172220}{RGB}{225,192,229}

\begin{axis}[
height=\figureheight,
legend cell align={left},
legend style={
  fill opacity=0.8,
  draw opacity=1,
  text opacity=1,
  at={(0.01,0.99)},
  anchor=north west,
  draw=lightgray204
},
tick align=outside,
tick pos=left,
width=\figurewidth,
x grid style={darkgray176},
xlabel={\(\displaystyle t^{*}\)},
xmajorgrids,
xmin=0, xmax=10,
xtick style={color=black},
y grid style={darkgray176},
ylabel style={rotate=-90.0},
ylabel={\(\displaystyle w^{*}\)},
ymajorgrids,
ymin=0.317, ymax=15.343,
ytick style={color=black}
]
\addplot [very thick, black, mark=*, mark size=2, mark options={solid}, only marks]
table {%
0.409999966621399 1.11000001430511
0.839999914169312 1.22000002861023
1.19000005722046 1.44000005722046
1.42999994754791 1.66999995708466
1.62999999523163 1.88999998569489
1.83000004291534 2.10999989509583
1.98000001907349 2.32999992370605
2.20000004768372 2.55999994277954
2.3199999332428 2.77999997138977
2.50999999046326 3
2.65000009536743 3.22000002861023
2.82999992370605 3.44000005722046
2.97000002861023 3.67000007629395
3.10999989509583 3.89000010490417
3.32999992370605 4.1100001335144
4.01999998092651 5
4.44000005722046 5.8899998664856
5.09000015258789 7
5.69000005722046 8
6.30000019073486 9
6.82999992370605 10
7.44000005722046 11
8.07999992370605 12
8.67000007629395 13
9.3100004196167 14
};
\addlegendentry{Experiment~\cite{martin1952PartIVExperimental}}
\addplot [line width=2pt, plum215172220, dashed]
table {%
0 1
0.0998772382736206 1
0.199753999710083 1.01999998092651
0.499386072158813 1.13999998569489
0.699140071868896 1.25999999046326
1.0986499786377 1.58000004291534
1.49816000461578 1.98000001907349
1.59803998470306 2.09999990463257
1.69790995121002 2.20000004768372
1.99753999710083 2.55999994277954
2.09741997718811 2.70000004768372
2.19729995727539 2.8199999332428
2.59681010246277 3.38000011444092
2.69668006896973 3.53999996185303
2.79656004905701 3.6800000667572
2.89644002914429 3.83999991416931
2.99632000923157 3.98000001907349
3.19606995582581 4.30000019073486
3.29594993591309 4.48000001907349
3.39582991600037 4.6399998664856
4.19483995437622 6.07999992370605
4.29472017288208 6.23999977111816
4.39459991455078 6.42000007629395
4.59434986114502 6.73999977111816
4.69423007965088 6.92000007629395
4.79410982131958 7.05999994277954
4.89398002624512 7.21999979019165
4.99385976791382 7.40000009536743
5.29348993301392 7.88000011444092
5.99263000488281 9.14000034332275
6.09251022338867 9.30000019073486
6.29226016998291 9.65999984741211
6.39213991165161 9.81999969482422
6.49202013015747 10
6.69177007675171 10.3199996948242
6.79164981842041 10.5
6.89153003692627 10.6400003433228
7.09127998352051 10.960000038147
7.19116020202637 11.1000003814697
7.49078989028931 11.5799999237061
7.59067010879517 11.7200002670288
8.28981018066406 12.8400001525879
8.38969039916992 13.0200004577637
8.58944034576416 13.3400001525879
8.6893196105957 13.5200004577637
8.7891902923584 13.6800003051758
8.88906955718994 13.8599996566772
8.9889497756958 14.0200004577637
9.08882999420166 14.2200002670288
9.2885799407959 14.5799999237061
};
\addlegendentry{OM, $W=50$}
\addplot [line width=2pt, orchid195130203, dashed]
table {%
0 1
0.0998772382736206 1
0.199753999710083 1.01999998092651
0.29963207244873 1.04999995231628
0.499386072158813 1.14999997615814
0.699140071868896 1.26999998092651
0.799018025398254 1.3400000333786
0.898895025253296 1.41999995708466
1.19852995872498 1.69000005722046
1.3982800245285 1.88999998569489
1.59803998470306 2.10999989509583
1.99753999710083 2.58999991416931
2.19729995727539 2.84999990463257
2.29717993736267 2.99000000953674
2.39704990386963 3.11999988555908
2.49692988395691 3.26999998092651
2.59681010246277 3.41000008583069
2.89644002914429 3.85999989509583
3.09618997573853 4.17999982833862
3.19606995582581 4.34999990463257
3.39582991600037 4.71000003814697
3.49569988250732 4.88000011444092
3.79533004760742 5.42000007629395
3.8952100276947 5.59000015258789
3.99509000778198 5.76999998092651
4.19483995437622 6.1100001335144
4.39459991455078 6.46999979019165
4.49447011947632 6.6399998664856
4.79410982131958 7.17999982833862
4.89398002624512 7.34999990463257
4.99385976791382 7.53000020980835
5.09373998641968 7.71999979019165
5.19361019134521 7.90000009536743
5.29348993301392 8.10000038146973
5.39337015151978 8.27999973297119
5.49324989318848 8.47000026702881
5.79288005828857 9.01000022888184
5.99263000488281 9.35000038146973
6.09251022338867 9.5
6.39213991165161 9.97999954223633
6.49202013015747 10.1599998474121
6.59189987182617 10.3299999237061
6.69177007675171 10.5100002288818
6.79164981842041 10.6800003051758
7.09127998352051 11.2200002670288
7.19116020202637 11.4099998474121
7.29103994369507 11.5900001525879
7.49078989028931 11.9700002670288
7.59067010879517 12.1499996185303
7.69053983688354 12.3400001525879
7.89029979705811 12.6800003051758
7.99018001556396 12.8400001525879
8.1899299621582 13.1800003051758
8.58944034576416 13.8999996185303
8.6893196105957 14.0900001525879
8.7891902923584 14.2700004577637
8.88906955718994 14.4700002670288
8.9889497756958 14.6599998474121
};
\addlegendentry{OM, $W=100$}
\addplot [line width=2pt, mediumorchid17588186, dashed]
table {%
0 1
0.099940299987793 1.00499999523163
0.199880957603455 1.02499997615814
0.299821019172668 1.05999994277954
0.399760961532593 1.10000002384186
0.499701023101807 1.14999997615814
0.599642038345337 1.21000003814697
0.699581980705261 1.27499997615814
0.799521923065186 1.35000002384186
0.99940299987793 1.50999999046326
1.09933996200562 1.60000002384186
1.19928002357483 1.69500005245209
1.39916002750397 1.89499998092651
1.59904003143311 2.11500000953674
1.69897997379303 2.23000001907349
1.89885997772217 2.48000001907349
1.9988100528717 2.60999989509583
2.09875011444092 2.74499988555908
2.19868993759155 2.875
2.29862999916077 3.00999999046326
2.39857006072998 3.15000009536743
2.59844994544983 3.42000007629395
2.69839000701904 3.55999994277954
2.79833006858826 3.71000003814697
2.89826989173889 3.86999988555908
2.99820995330811 4.03999996185303
3.09815001487732 4.19500017166138
3.19809007644653 4.3600001335144
3.39796996116638 4.7350001335144
3.79772996902466 5.40999984741211
3.89767003059387 5.57000017166138
4.09754991531372 5.91499996185303
4.19749021530151 6.07999992370605
4.39736986160278 6.43499994277954
4.49731016159058 6.61499977111816
4.79713010787964 7.13000011444092
4.89706993103027 7.30499982833862
5.0969500541687 7.67500019073486
5.19688987731934 7.86999988555908
5.29683017730713 8.05500030517578
5.4967098236084 8.40999984741211
5.79653978347778 8.94499969482422
5.99641990661621 9.27999973297119
6.096360206604 9.4350004196167
6.19630002975464 9.60499954223633
6.29623985290527 9.78499984741211
6.39618015289307 9.97999954223633
6.4961199760437 10.164999961853
6.69600009918213 10.5100002288818
6.89588022232056 10.8299999237061
7.29563999176025 11.5349998474121
7.49552011489868 11.8950004577637
7.69540023803711 12.2799997329712
7.99522018432617 12.8249998092651
8.09515953063965 13.0150003433228
8.4949197769165 13.7299995422363
8.5948600769043 13.8950004577637
8.69480037689209 14.0699996948242
8.79473972320557 14.2550001144409
8.89468002319336 14.4300003051758
8.99462032318115 14.5950002670288
};
\addlegendentry{OM, $W=200$}
\end{axis}

\end{tikzpicture}

%% file: figures/dam-break-rectangular/convergence-y-om3.tex
% This file was created with tikzplotlib v0.10.1.
\begin{tikzpicture}

\definecolor{burlywood248194145}{RGB}{250,209,172}
\definecolor{darkgray176}{RGB}{176,176,176}
\definecolor{darkorange24213334}{RGB}{242,133,34}
\definecolor{lightgray204}{RGB}{204,204,204}
\definecolor{sandybrown24516489}{RGB}{181,100,26}

\begin{axis}[
height=\figureheight,
legend cell align={left},
legend style={
	fill opacity=0.8,
	draw opacity=1,
	text opacity=1,
	at={(0.99,0.99)},
	anchor=north east,
	draw=lightgray204
},
tick align=outside,
tick pos=left,
width=\figurewidth,
x grid style={darkgray176},
xlabel={\(\displaystyle t^{*}\)},
xmajorgrids,
xmin=0, xmax=10,
xtick style={color=black},
y grid style={darkgray176},
ylabel style={rotate=-90.0},
ylabel={\(\displaystyle h^{*}\)},
ymajorgrids,
ymin=0.013, ymax=1.047,
ytick style={color=black}
]
\addplot [very thick, black, mark=*, mark size=2, mark options={solid}, only marks]
table {%
0 1
0.559999942779541 0.940000057220459
0.769999980926514 0.889999985694885
0.930000066757202 0.829999923706055
1.08000004291534 0.779999971389771
1.27999997138977 0.720000028610229
1.46000003814697 0.670000076293945
1.6599999666214 0.610000014305115
1.8400000333786 0.559999942779541
2 0.5
2.21000003814697 0.440000057220459
2.45000004768372 0.389999985694885
2.70000004768372 0.330000042915344
3.05999994277954 0.279999971389771
3.44000005722046 0.220000028610229
4.19999980926514 0.169999957084656
5.25 0.110000014305115
7.40000009536743 0.059999942779541
};
\addlegendentry{Experiment~\cite{martin1952PartIVExperimental}}
\addplot [line width=2pt, burlywood248194145, dash pattern=on 1pt off 3pt on 3pt off 3pt]
table {%
0 1
0.0998772382736206 0.990000009536743
0.29963207244873 0.990000009536743
0.599262952804565 0.960000038146973
0.699140071868896 0.940000057220459
0.799018025398254 0.930000066757202
1.29840004444122 0.829999923706055
1.3982800245285 0.799999952316284
1.49816000461578 0.779999971389771
1.59803998470306 0.75
1.79779005050659 0.710000038146973
1.89767003059387 0.680000066757202
2.99632000923157 0.460000038146973
3.09618997573853 0.450000047683716
3.29594993591309 0.409999966621399
3.39582991600037 0.399999976158142
3.49569988250732 0.379999995231628
3.59558010101318 0.370000004768372
3.69546008110046 0.350000023841858
3.99509000778198 0.319999933242798
4.09497022628784 0.299999952316284
4.79410982131958 0.230000019073486
8.7891902923584 0.230000019073486
};
\addlegendentry{OM3, $W=50$}
\addplot [line width=2pt, sandybrown24516489, dash pattern=on 1pt off 3pt on 3pt off 3pt]
table {%
0 1
0.0998772382736206 0.995000004768372
0.199753999710083 0.995000004768372
0.29963207244873 0.990000009536743
0.599262952804565 0.960000038146973
0.799018025398254 0.930000066757202
0.898895025253296 0.910000085830688
0.998772025108337 0.894999980926514
1.0986499786377 0.875
1.19852995872498 0.850000023841858
1.29840004444122 0.829999923706055
1.49816000461578 0.779999971389771
1.59803998470306 0.759999990463257
1.89767003059387 0.684999942779541
1.99753999710083 0.664999961853027
2.09741997718811 0.639999985694885
2.79656004905701 0.5
2.89644002914429 0.485000014305115
2.99632000923157 0.465000033378601
3.09618997573853 0.450000047683716
3.19606995582581 0.430000066757202
3.69546008110046 0.355000019073486
3.79533004760742 0.345000028610229
3.8952100276947 0.330000042915344
3.99509000778198 0.319999933242798
4.09497022628784 0.305000066757202
4.69423007965088 0.245000004768372
4.89398002624512 0.235000014305115
4.99385976791382 0.225000023841858
5.09373998641968 0.210000038146973
5.79288005828857 0.174999952316284
5.89275979995728 0.164999961853027
5.99263000488281 0.159999966621399
6.09251022338867 0.159999966621399
6.19238996505737 0.164999961853027
6.29226016998291 0.154999971389771
6.49202013015747 0.144999980926514
6.69177007675171 0.125
6.79164981842041 0.120000004768372
6.89153003692627 0.120000004768372
6.99139976501465 0.125
7.19116020202637 0.125
7.7904200553894 0.0950000286102295
7.89029979705811 0.100000023841858
7.99018001556396 0.0950000286102295
};
\addlegendentry{OM3, $W=100$}
\addplot [line width=2pt, darkorange24213334, dash pattern=on 1pt off 3pt on 3pt off 3pt]
table {%
0 1
0.199880957603455 0.995000004768372
0.299821019172668 0.990000009536743
0.399760961532593 0.980000019073486
0.499701023101807 0.972500085830688
0.599642038345337 0.960000038146973
0.799521923065186 0.930000066757202
0.99940299987793 0.894999980926514
1.09933996200562 0.872499942779541
1.19928002357483 0.852499961853027
1.29921996593475 0.829999923706055
1.39916002750397 0.805000066757202
1.49909996986389 0.782500028610229
1.69897997379303 0.732500076293945
2.19868993759155 0.620000004768372
2.69839000701904 0.519999980926514
2.89826989173889 0.485000014305115
2.99820995330811 0.465000033378601
3.09815001487732 0.450000047683716
3.19809007644653 0.432500004768372
3.4979100227356 0.387500047683716
3.59785008430481 0.375
3.89767003059387 0.330000042915344
4.09754991531372 0.305000066757202
4.49731016159058 0.264999985694885
4.59724998474121 0.257499933242798
4.69718980789185 0.247499942779541
5.0969500541687 0.217499971389771
5.19688987731934 0.207499980926514
5.39677000045776 0.197499990463257
5.4967098236084 0.190000057220459
5.59665012359619 0.184999942779541
5.69658994674683 0.177500009536743
6.096360206604 0.157500028610229
6.19630002975464 0.154999971389771
6.4961199760437 0.139999985694885
6.59605979919434 0.137500047683716
6.79593992233276 0.127500057220459
6.89588022232056 0.127500057220459
6.99582004547119 0.122499942779541
7.09575986862183 0.122499942779541
7.49552011489868 0.112499952316284
7.59545993804932 0.112499952316284
7.89527988433838 0.105000019073486
};
\addlegendentry{OM3, $W=200$}
\end{axis}

\end{tikzpicture}

%% file: figures/dam-break-rectangular/convergence-x-om3.tex
% This file was created with tikzplotlib v0.10.1.
\begin{tikzpicture}

\definecolor{burlywood248194145}{RGB}{250,209,172}
\definecolor{darkgray176}{RGB}{176,176,176}
\definecolor{darkorange24213334}{RGB}{242,133,34}
\definecolor{lightgray204}{RGB}{204,204,204}
\definecolor{sandybrown24516489}{RGB}{181,100,26}

\begin{axis}[
height=\figureheight,
legend cell align={left},
legend style={
	fill opacity=0.8,
	draw opacity=1,
	text opacity=1,
	at={(0.01,0.99)},
	anchor=north west,
	draw=lightgray204
},
tick align=outside,
tick pos=left,
width=\figurewidth,
x grid style={darkgray176},
xlabel={\(\displaystyle t^{*}\)},
xmajorgrids,
xmin=0, xmax=10,
xtick style={color=black},
y grid style={darkgray176},
ylabel style={rotate=-90.0},
ylabel={\(\displaystyle w^{*}\)},
ymajorgrids,
ymin=0.31825, ymax=15.31675,
ytick style={color=black}
]
\addplot [very thick, black, mark=*, mark size=2, mark options={solid}, only marks]
table {%
0.409999966621399 1.11000001430511
0.839999914169312 1.22000002861023
1.19000005722046 1.44000005722046
1.42999994754791 1.66999995708466
1.62999999523163 1.88999998569489
1.83000004291534 2.10999989509583
1.98000001907349 2.32999992370605
2.20000004768372 2.55999994277954
2.3199999332428 2.77999997138977
2.50999999046326 3
2.65000009536743 3.22000002861023
2.82999992370605 3.44000005722046
2.97000002861023 3.67000007629395
3.10999989509583 3.89000010490417
3.32999992370605 4.1100001335144
4.01999998092651 5
4.44000005722046 5.8899998664856
5.09000015258789 7
5.69000005722046 8
6.30000019073486 9
6.82999992370605 10
7.44000005722046 11
8.07999992370605 12
8.67000007629395 13
9.3100004196167 14
};
\addlegendentry{Experiment~\cite{martin1952PartIVExperimental}}
\addplot [line width=2pt, burlywood248194145, dash pattern=on 1pt off 3pt on 3pt off 3pt]
table {%
0 1
0.0998772382736206 1
0.199753999710083 1.01999998092651
0.499386072158813 1.13999998569489
0.599262952804565 1.20000004768372
0.699140071868896 1.27999997138977
0.799018025398254 1.3400000333786
0.998772025108337 1.5
1.0986499786377 1.60000002384186
1.19852995872498 1.67999994754791
1.29840004444122 1.77999997138977
1.3982800245285 1.89999997615814
1.69790995121002 2.20000004768372
1.79779005050659 2.3199999332428
1.89767003059387 2.42000007629395
2.19729995727539 2.77999997138977
2.29717993736267 2.92000007629395
2.59681010246277 3.27999997138977
3.09618997573853 3.98000001907349
3.19606995582581 4.09999990463257
3.49569988250732 4.51999998092651
3.8952100276947 5.23999977111816
4.09497022628784 5.6399998664856
4.19483995437622 5.82000017166138
4.39459991455078 6.09999990463257
4.49447011947632 6.57999992370605
4.69423007965088 7.17999982833862
6.59189987182617 7.17999982833862
6.69177007675171 9.9399995803833
6.89153003692627 10.3800001144409
6.99139976501465 10.5799999237061
7.09127998352051 10.8000001907349
7.19116020202637 11.039999961853
7.29103994369507 11.2600002288818
7.39091014862061 11.5
7.59067010879517 11.9399995803833
7.69053983688354 12.1800003051758
8.1899299621582 13.2799997329712
8.28981018066406 13.5200004577637
8.7891902923584 14.6199998855591
};
\addlegendentry{OM3, $W=50$}
\addplot [line width=2pt, sandybrown24516489, dash pattern=on 1pt off 3pt on 3pt off 3pt]
table {%
0 1
0.0998772382736206 1
0.199753999710083 1.01999998092651
0.399508953094482 1.10000002384186
0.499386072158813 1.14999997615814
0.699140071868896 1.26999998092651
0.998772025108337 1.50999999046326
1.0986499786377 1.60000002384186
1.19852995872498 1.70000004768372
1.29840004444122 1.78999996185303
1.3982800245285 1.89999997615814
1.59803998470306 2.14000010490417
1.69790995121002 2.25
2.49692988395691 3.28999996185303
2.69668006896973 3.52999997138977
2.79656004905701 3.64000010490417
2.89644002914429 3.76999998092651
2.99632000923157 3.9300000667572
3.09618997573853 4.13000011444092
3.19606995582581 4.30999994277954
3.29594993591309 4.48000001907349
3.39582991600037 4.65999984741211
3.49569988250732 4.82000017166138
3.59558010101318 4.96000003814697
3.99509000778198 5.71999979019165
4.09497022628784 5.90000009536743
4.19483995437622 6.09000015258789
4.39459991455078 6.48999977111816
4.49447011947632 6.69999980926514
4.59434986114502 6.96999979019165
4.69423007965088 7.30000019073486
4.79410982131958 7.69000005722046
4.89398002624512 8.02999973297119
4.99385976791382 8.35000038146973
5.19361019134521 8.94999980926514
5.39337015151978 9.52999973297119
5.49324989318848 9.8100004196167
5.69299983978271 10.3500003814697
5.79288005828857 10.6099996566772
5.99263000488281 11.1099996566772
6.09251022338867 11.3400001525879
6.19238996505737 11.5600004196167
6.29226016998291 11.7700004577637
6.49202013015747 12.1700000762939
6.59189987182617 12.3599996566772
6.69177007675171 12.539999961853
6.79164981842041 12.7299995422363
6.89153003692627 12.9099998474121
7.09127998352051 13.25
7.19116020202637 13.4099998474121
7.29103994369507 13.5600004196167
7.39091014862061 13.7200002670288
7.59067010879517 14.0200004577637
7.69053983688354 14.1599998474121
7.7904200553894 14.289999961853
7.99018001556396 14.5699996948242
};
\addlegendentry{OM3, $W=100$}
\addplot [line width=2pt, darkorange24213334, dash pattern=on 1pt off 3pt on 3pt off 3pt]
table {%
0 1
0.099940299987793 1.00499999523163
0.199880957603455 1.02499997615814
0.299821019172668 1.05999994277954
0.399760961532593 1.10000002384186
0.499701023101807 1.14999997615814
0.599642038345337 1.21000003814697
0.699581980705261 1.27999997138977
0.899461984634399 1.42999994754791
0.99940299987793 1.51499998569489
1.09933996200562 1.60500001907349
1.29921996593475 1.80499994754791
1.39916002750397 1.9099999666214
1.49909996986389 2.01999998092651
1.59904003143311 2.13499999046326
1.69897997379303 2.23499989509583
1.79892003536224 2.35500001907349
1.89885997772217 2.49000000953674
1.9988100528717 2.61500000953674
2.09875011444092 2.73499989509583
2.19868993759155 2.875
2.29862999916077 2.99499988555908
2.39857006072998 3.11999988555908
2.49850988388062 3.26500010490417
2.59844994544983 3.43499994277954
2.69839000701904 3.58500003814697
2.79833006858826 3.74499988555908
2.89826989173889 3.875
2.99820995330811 4.01000022888184
3.19809007644653 4.32000017166138
3.29802989959717 4.45499992370605
3.39796996116638 4.60500001907349
3.69778990745544 5.1100001335144
3.79772996902466 5.25500011444092
3.89767003059387 5.44500017166138
3.99761009216309 5.59999990463257
4.09754991531372 5.7649998664856
4.29743003845215 6.07999992370605
4.49731016159058 6.39499998092651
4.59724998474121 6.54500007629395
4.69718980789185 6.7350001335144
4.79713010787964 6.94999980926514
4.89706993103027 7.13500022888184
4.99701023101807 7.30499982833862
5.0969500541687 7.46500015258789
5.19688987731934 7.6399998664856
5.39677000045776 8.04500007629395
5.59665012359619 8.40999984741211
5.69658994674683 8.60000038146973
5.79653978347778 8.77000045776367
5.89648008346558 8.96500015258789
6.29623985290527 9.60000038146973
6.39618015289307 9.83500003814697
6.4961199760437 10.0450000762939
6.79593992233276 10.5799999237061
6.89588022232056 10.7449998855591
7.49552011489868 11.6800003051758
7.59545993804932 11.8500003814697
7.69540023803711 12.0100002288818
7.79534006118774 14.2049999237061
7.89527988433838 14.6350002288818
};
\addlegendentry{OM3, $W=200$}
\end{axis}

\end{tikzpicture}

%% file: figures/dam-break-cylindrical/convergence-nbrc.tex
% This file was created with tikzplotlib v0.10.1.
\begin{tikzpicture}

\definecolor{darkgray176}{RGB}{176,176,176}
\definecolor{dodgerblue0154222}{RGB}{0,154,222}
\definecolor{lightgray204}{RGB}{204,204,204}
\definecolor{mediumturquoise64179230}{RGB}{0,116,166}
\definecolor{skyblue128205238}{RGB}{159,217,243}

\begin{axis}[
height=\figureheight,
legend cell align={left},
legend style={
	fill opacity=0.8,
	draw opacity=1,
	text opacity=1,
	at={(0.01,0.99)},
	anchor=north west,
	draw=lightgray204
},
tick align=outside,
tick pos=left,
width=\figurewidth,
x grid style={darkgray176},
xlabel={\(\displaystyle t^{*}\)},
xmajorgrids,
xmin=0, xmax=5,
xtick style={color=black},
y grid style={darkgray176},
ylabel style={rotate=-90.0},
ylabel={\(\displaystyle r^{*}\)},
ymajorgrids,
ymin=0.7959545, ymax=5.2849555,
ytick style={color=black}
]
\addplot [semithick, black, mark=*, mark size=2, mark options={solid}, only marks]
table {%
1.19000005722046 1.22000002861023
1.5 1.44000005722046
1.73000001907349 1.66999995708466
1.91999995708466 1.88999998569489
2.1800000667572 2.10999989509583
2.35999989509583 2.32999992370605
2.58999991416931 2.55999994277954
2.78999996185303 2.77999997138977
2.97000002861023 3
3.20000004768372 3.22000002861023
3.38000011444092 3.44000005722046
3.57999992370605 3.67000007629395
3.75999999046326 3.89000010490417
3.97000002861023 4.1100001335144
};
\addlegendentry{Experiment~\cite{martin1952PartIVExperimental}}
\addplot [line width=2pt, skyblue128205238, dotted]
table {%
0 1
0.0968506336212158 1.03032994270325
0.193701028823853 1.03997004032135
0.290552020072937 1.07033002376556
0.387402057647705 1.08608996868134
0.4842529296875 1.12413001060486
0.581104040145874 1.17045998573303
0.677953958511353 1.21352994441986
0.774805068969727 1.27215003967285
0.871655941009521 1.32599997520447
0.96850597858429 1.38503003120422
1.06535995006561 1.45064997673035
1.25906002521515 1.58815002441406
1.35590994358063 1.66266000270844
1.54961001873016 1.82261002063751
1.64646005630493 1.90604996681213
1.74330997467041 1.99195003509521
2.03385996818542 2.27272009849548
2.1307098865509 2.37107992172241
2.32440996170044 2.57652997970581
2.42126989364624 2.68415999412537
2.61496996879578 2.9087700843811
2.71181988716125 3.02557992935181
2.80867004394531 3.14562010765076
2.90551996231079 3.26806998252869
3.09922003746033 3.52220988273621
3.29292011260986 3.78707003593445
3.38977003097534 3.92127990722656
3.58347010612488 4.19726991653442
3.68032002449036 4.33826017379761
3.87402009963989 4.62908983230591
3.97088003158569 4.77672004699707
4.16457986831665 5.08091020584106
};
\addlegendentry{NBRC, $D=50$}
\addplot [line width=2pt, mediumturquoise64179230, dotted]
table {%
0 1
0.0998772382736206 1.01600003242493
0.199753999710083 1.02908003330231
0.29963207244873 1.05735003948212
0.399508953094482 1.08768999576569
0.499386072158813 1.12902998924255
0.599262952804565 1.17719995975494
0.699140071868896 1.22811996936798
0.799018025398254 1.28515005111694
0.898895025253296 1.34586000442505
0.998772025108337 1.41066002845764
1.0986499786377 1.47801005840302
1.19852995872498 1.54989004135132
1.29840004444122 1.62555003166199
1.3982800245285 1.70527005195618
1.49816000461578 1.788370013237
1.59803998470306 1.87441003322601
1.69790995121002 1.96364998817444
1.79779005050659 2.05627989768982
1.89767003059387 2.15107011795044
1.99753999710083 2.24920988082886
2.09741997718811 2.35074996948242
2.19729995727539 2.45555996894836
2.29717993736267 2.56303000450134
2.39704990386963 2.67439007759094
2.49692988395691 2.78957009315491
2.59681010246277 2.90869998931885
2.69668006896973 3.03180003166199
2.79656004905701 3.16032004356384
2.89644002914429 3.29321002960205
2.99632000923157 3.43172001838684
3.09618997573853 3.57386994361877
3.19606995582581 3.72328996658325
3.29594993591309 3.88220000267029
3.39582991600037 4.04562997817993
3.49569988250732 4.22493982315063
3.59558010101318 4.41247987747192
};
\addlegendentry{NBRC, $D=100$}
\addplot [line width=2pt, dodgerblue0154222, dotted]
table {%
0 1
0.0998772382736206 1.00959002971649
0.199753999710083 1.02849996089935
0.29963207244873 1.05751001834869
0.399508953094482 1.09397995471954
0.499386072158813 1.13763999938965
0.599262952804565 1.18703997135162
0.699140071868896 1.2406200170517
0.799018025398254 1.29674005508423
0.898895025253296 1.35748994350433
0.998772025108337 1.42174994945526
1.0986499786377 1.49075996875763
1.19852995872498 1.56394994258881
1.29840004444122 1.64120995998383
1.3982800245285 1.72121000289917
1.49816000461578 1.80343997478485
1.59803998470306 1.88923001289368
1.69790995121002 1.97821998596191
1.79779005050659 2.07202005386353
1.89767003059387 2.16972994804382
1.99753999710083 2.27239990234375
2.09741997718811 2.38058996200562
2.19729995727539 2.49265003204346
2.29717993736267 2.61149001121521
2.39704990386963 2.73761010169983
2.49692988395691 2.86681008338928
2.59681010246277 3.00845003128052
2.69668006896973 3.1548900604248
2.79656004905701 3.31029009819031
2.89644002914429 3.48676991462708
2.99632000923157 3.69159007072449
3.09618997573853 3.87050008773804
3.19606995582581 4.0701699256897
};
\addlegendentry{NBRC, $D=200$}
\end{axis}

\end{tikzpicture}

%% file: figures/dam-break-cylindrical/convergence-nbkc.tex
% This file was created with tikzplotlib v0.10.1.
\begin{tikzpicture}

\definecolor{darkgray176}{RGB}{176,176,176}
\definecolor{lightgray204}{RGB}{204,204,204}
\definecolor{mediumaquamarine128230181}{RGB}{159,236,200}
\definecolor{mediumaquamarine64218145}{RGB}{0,128,68}
\definecolor{springgreen0205108}{RGB}{0,205,108}

\begin{axis}[
height=\figureheight,
legend cell align={left},
legend style={
	fill opacity=0.8,
	draw opacity=1,
	text opacity=1,
	at={(0.01,0.99)},
	anchor=north west,
	draw=lightgray204
},
tick align=outside,
tick pos=left,
width=\figurewidth,
x grid style={darkgray176},
xlabel={\(\displaystyle t^{*}\)},
xmajorgrids,
xmin=0, xmax=5,
xtick style={color=black},
y grid style={darkgray176},
ylabel style={rotate=-90.0},
ylabel={\(\displaystyle r^{*}\)},
ymajorgrids,
ymin=0.8135965, ymax=4.9144735,
ytick style={color=black}
]
\addplot [semithick, black, mark=*, mark size=2, mark options={solid}, only marks]
table {%
1.19000005722046 1.22000002861023
1.5 1.44000005722046
1.73000001907349 1.66999995708466
1.91999995708466 1.88999998569489
2.1800000667572 2.10999989509583
2.35999989509583 2.32999992370605
2.58999991416931 2.55999994277954
2.78999996185303 2.77999997138977
2.97000002861023 3
3.20000004768372 3.22000002861023
3.38000011444092 3.44000005722046
3.57999992370605 3.67000007629395
3.75999999046326 3.89000010490417
3.97000002861023 4.1100001335144
};
\addlegendentry{Experiment~\cite{martin1952PartIVExperimental}}
\addplot [line width=2pt, mediumaquamarine128230181]
table {%
0 1
0.0968506336212158 1.02464997768402
0.193701028823853 1.03804004192352
0.290552020072937 1.07054996490479
0.387402057647705 1.08272004127502
0.4842529296875 1.12231004238129
0.581104040145874 1.1708300113678
0.774805068969727 1.25978994369507
0.871655941009521 1.31780004501343
0.96850597858429 1.38564002513885
1.06535995006561 1.4483300447464
1.16220998764038 1.51500999927521
1.25906002521515 1.58431005477905
1.35590994358063 1.66504001617432
1.4527599811554 1.73878002166748
1.54961001873016 1.82041001319885
1.64646005630493 1.90985000133514
1.74330997467041 2.00282001495361
1.84016001224518 2.09274005889893
1.93701004981995 2.18537998199463
2.03385996818542 2.28819990158081
2.1307098865509 2.38440990447998
2.32440996170044 2.59227991104126
2.5181200504303 2.81055998802185
2.71181988716125 3.03746008872986
2.90551996231079 3.27536010742188
3.09922003746033 3.52100992202759
3.19606995582581 3.64556002616882
3.48661994934082 4.03141021728516
3.68032002449036 4.30201005935669
3.87402009963989 4.58336019515991
3.97088003158569 4.72806978225708
};
\addlegendentry{NBKC, $D=50$}
\addplot [line width=2pt, mediumaquamarine64218145]
table {%
0 1
0.199753999710083 1.02860999107361
0.399508953094482 1.08675003051758
0.499386072158813 1.12607002258301
0.599262952804565 1.17005002498627
0.699140071868896 1.22022998332977
0.799018025398254 1.27520000934601
0.898895025253296 1.33589005470276
0.998772025108337 1.40039002895355
1.0986499786377 1.4675999879837
1.19852995872498 1.54121994972229
1.29840004444122 1.61765003204346
1.3982800245285 1.6981600522995
1.49816000461578 1.7820600271225
1.59803998470306 1.87033998966217
1.69790995121002 1.96103000640869
1.79779005050659 2.05397009849548
1.89767003059387 2.1508800983429
1.99753999710083 2.25089001655579
2.09741997718811 2.35339999198914
2.19729995727539 2.45902991294861
2.29717993736267 2.56807994842529
2.39704990386963 2.67926001548767
2.49692988395691 2.79317998886108
2.69668006896973 3.02970004081726
2.89644002914429 3.27740001678467
3.09618997573853 3.53986001014709
3.19606995582581 3.67710995674133
3.29594993591309 3.82133007049561
3.39582991600037 3.96157002449036
3.59558010101318 4.26762008666992
3.69546008110046 4.42120981216431
3.79533004760742 4.58034992218018
};
\addlegendentry{NBKC, $D=100$}
\addplot [line width=2pt, springgreen0205108]
table {%
0 1
0.0998772382736206 1.00853002071381
0.199753999710083 1.02604997158051
0.29963207244873 1.05704998970032
0.399508953094482 1.09686994552612
0.499386072158813 1.14031994342804
0.599262952804565 1.18720996379852
0.699140071868896 1.23769998550415
0.799018025398254 1.29200005531311
0.898895025253296 1.350909948349
0.998772025108337 1.41391003131866
1.0986499786377 1.48038995265961
1.19852995872498 1.55093002319336
1.29840004444122 1.62583994865417
1.3982800245285 1.70369005203247
1.49816000461578 1.78443002700806
1.59803998470306 1.86799001693726
1.69790995121002 1.95476996898651
1.79779005050659 2.04407000541687
1.89767003059387 2.13595008850098
1.99753999710083 2.23130989074707
2.09741997718811 2.33071994781494
2.19729995727539 2.43276000022888
2.29717993736267 2.53732991218567
2.39704990386963 2.64541006088257
2.49692988395691 2.75644993782043
2.59681010246277 2.87204003334045
2.69668006896973 2.99218988418579
2.79656004905701 3.11581993103027
2.99632000923157 3.37477993965149
3.09618997573853 3.51165008544922
3.19606995582581 3.65511989593506
3.29594993591309 3.80441999435425
3.49569988250732 4.1263599395752
3.59558010101318 4.29679012298584
};
\addlegendentry{NBKC, $D=200$}
\end{axis}

\end{tikzpicture}

%% file: figures/dam-break-cylindrical/convergence-om.tex
% This file was created with tikzplotlib v0.10.1.
\begin{tikzpicture}

\definecolor{darkgray176}{RGB}{176,176,176}
\definecolor{lightgray204}{RGB}{204,204,204}
\definecolor{mediumorchid17588186}{RGB}{175,88,186}
\definecolor{orchid195130203}{RGB}{109,55,116}
\definecolor{plum215172220}{RGB}{225,192,229}

\begin{axis}[
height=\figureheight,
legend cell align={left},
legend style={
	fill opacity=0.8,
	draw opacity=1,
	text opacity=1,
	at={(0.01,0.99)},
	anchor=north west,
	draw=lightgray204
},
tick align=outside,
tick pos=left,
width=\figurewidth,
x grid style={darkgray176},
xlabel={\(\displaystyle t^{*}\)},
xmajorgrids,
xmin=0, xmax=5,
xtick style={color=black},
y grid style={darkgray176},
ylabel style={rotate=-90.0},
ylabel={\(\displaystyle r^{*}\)},
ymajorgrids,
ymin=0.776347, ymax=5.696713,
ytick style={color=black}
]
\addplot [semithick, black, mark=*, mark size=2, mark options={solid}, only marks]
table {%
1.19000005722046 1.22000002861023
1.5 1.44000005722046
1.73000001907349 1.66999995708466
1.91999995708466 1.88999998569489
2.1800000667572 2.10999989509583
2.35999989509583 2.32999992370605
2.58999991416931 2.55999994277954
2.78999996185303 2.77999997138977
2.97000002861023 3
3.20000004768372 3.22000002861023
3.38000011444092 3.44000005722046
3.57999992370605 3.67000007629395
3.75999999046326 3.89000010490417
3.97000002861023 4.1100001335144
};
\addlegendentry{Experiment~\cite{martin1952PartIVExperimental}}
\addplot [line width=2pt, plum215172220, dashed]
table {%
0 1
0.0968506336212158 1.02286005020142
0.193701028823853 1.03618001937866
0.290552020072937 1.06209003925323
0.387402057647705 1.07738995552063
0.4842529296875 1.11576998233795
0.581104040145874 1.15838003158569
0.677953958511353 1.21512997150421
0.871655941009521 1.32088994979858
0.96850597858429 1.38689994812012
1.06535995006561 1.45061004161835
1.25906002521515 1.59710001945496
1.4527599811554 1.7537100315094
1.54961001873016 1.82871997356415
1.64646005630493 1.9172500371933
1.84016001224518 2.08051991462708
1.93701004981995 2.17027997970581
2.22756004333496 2.43247008323669
2.32440996170044 2.52584004402161
2.42126989364624 2.62300992012024
2.5181200504303 2.72603988647461
2.61496996879578 2.81713008880615
2.71181988716125 2.91303992271423
2.80867004394531 3.01963996887207
2.90551996231079 3.11072993278503
3.00237011909485 3.21027994155884
3.09922003746033 3.31354999542236
3.19606995582581 3.4242000579834
3.29292011260986 3.52501010894775
3.38977003097534 3.62908005714417
3.58347010612488 3.84839010238647
3.77716994285583 4.05600023269653
3.87402009963989 4.1699800491333
3.97088003158569 4.28163003921509
4.06772994995117 4.38526010513306
4.35828018188477 4.70484018325806
4.45513010025024 4.82678985595703
4.55198001861572 4.93756008148193
4.6488299369812 5.04385995864868
4.74567985534668 5.16454982757568
4.84252977371216 5.27234983444214
};
\addlegendentry{OM, $D=50$}
\addplot [line width=2pt, orchid195130203, dashed]
table {%
0 1
0.199753999710083 1.02860999107361
0.399508953094482 1.08675003051758
0.499386072158813 1.12607002258301
0.599262952804565 1.17005002498627
0.699140071868896 1.22022998332977
0.799018025398254 1.27520000934601
0.898895025253296 1.33589005470276
0.998772025108337 1.40039002895355
1.0986499786377 1.4675999879837
1.19852995872498 1.54121994972229
1.29840004444122 1.61765003204346
1.3982800245285 1.6981600522995
1.49816000461578 1.7820600271225
1.59803998470306 1.87033998966217
1.69790995121002 1.96103000640869
1.79779005050659 2.05397009849548
1.89767003059387 2.1508800983429
1.99753999710083 2.25089001655579
2.09741997718811 2.35339999198914
2.19729995727539 2.45902991294861
2.29717993736267 2.56807994842529
2.39704990386963 2.67926001548767
2.49692988395691 2.79317998886108
2.69668006896973 3.02970004081726
2.89644002914429 3.27740001678467
3.09618997573853 3.53986001014709
3.19606995582581 3.67710995674133
3.29594993591309 3.82133007049561
3.39582991600037 3.96157002449036
3.59558010101318 4.26762008666992
3.69546008110046 4.42120981216431
3.79533004760742 4.58034992218018
};
\addlegendentry{OM, $D=100$}
\addplot [line width=2pt, mediumorchid17588186, dashed]
table {%
0 1
0.0998772382736206 1.00802004337311
0.199753999710083 1.02504003047943
0.29963207244873 1.05154001712799
0.399508953094482 1.08580994606018
0.499386072158813 1.1258499622345
0.599262952804565 1.17148995399475
0.699140071868896 1.22225999832153
0.799018025398254 1.27874004840851
0.898895025253296 1.33844995498657
0.998772025108337 1.40128004550934
1.0986499786377 1.46823000907898
1.19852995872498 1.5382000207901
1.29840004444122 1.61097002029419
1.3982800245285 1.68839001655579
1.49816000461578 1.77014994621277
1.59803998470306 1.85602998733521
2.19729995727539 2.38798999786377
2.29717993736267 2.48127007484436
2.39704990386963 2.5792601108551
2.49692988395691 2.68318009376526
3.09618997573853 3.33053994178772
3.19606995582581 3.44525003433228
3.8952100276947 4.27276992797852
4.59434986114502 5.11206007003784
4.89398002624512 5.473060131073
};
\addlegendentry{OM, $D=200$}
\end{axis}

\end{tikzpicture}

%% file: figures/dam-break-cylindrical/convergence-om3.tex
% This file was created with tikzplotlib v0.10.1.
\begin{tikzpicture}

\definecolor{burlywood248194145}{RGB}{250,209,172}
\definecolor{darkgray176}{RGB}{176,176,176}
\definecolor{darkorange24213334}{RGB}{242,133,34}
\definecolor{lightgray204}{RGB}{204,204,204}
\definecolor{sandybrown24516489}{RGB}{181,100,26}

\begin{axis}[
height=\figureheight,
legend cell align={left},
legend style={
	fill opacity=0.8,
	draw opacity=1,
	text opacity=1,
	at={(0.01,0.99)},
	anchor=north west,
	draw=lightgray204
},
tick align=outside,
tick pos=left,
width=\figurewidth,
x grid style={darkgray176},
xlabel={\(\displaystyle t^{*}\)},
xmajorgrids,
xmin=0, xmax=5,
xtick style={color=black},
y grid style={darkgray176},
ylabel style={rotate=-90.0},
ylabel={\(\displaystyle r^{*}\)},
ymajorgrids,
ymin=0.8136815, ymax=4.9126885,
ytick style={color=black}
]
\addplot [semithick, black, mark=*, mark size=2, mark options={solid}, only marks]
table {%
1.19000005722046 1.22000002861023
1.5 1.44000005722046
1.73000001907349 1.66999995708466
1.91999995708466 1.88999998569489
2.1800000667572 2.10999989509583
2.35999989509583 2.32999992370605
2.58999991416931 2.55999994277954
2.78999996185303 2.77999997138977
2.97000002861023 3
3.20000004768372 3.22000002861023
3.38000011444092 3.44000005722046
3.57999992370605 3.67000007629395
3.75999999046326 3.89000010490417
3.97000002861023 4.1100001335144
};
\addlegendentry{Experiment~\cite{martin1952PartIVExperimental}}
\addplot [line width=2pt, burlywood248194145, dash pattern=on 1pt off 3pt on 3pt off 3pt]
table {%
0 1
0.0968506336212158 1.02464997768402
0.193701028823853 1.03804004192352
0.290552020072937 1.06836998462677
0.387402057647705 1.08525002002716
0.4842529296875 1.12050998210907
0.581104040145874 1.16839003562927
0.677953958511353 1.22851002216339
0.774805068969727 1.28083002567291
0.871655941009521 1.33496999740601
0.96850597858429 1.40901005268097
1.06535995006561 1.47959995269775
1.16220998764038 1.54714000225067
1.25906002521515 1.62685000896454
1.35590994358063 1.70864999294281
1.54961001873016 1.87600004673004
1.64646005630493 1.96784996986389
1.84016001224518 2.16126990318298
1.93701004981995 2.26719999313354
2.03385996818542 2.37089991569519
2.1307098865509 2.4784300327301
2.22756004333496 2.59483003616333
2.32440996170044 2.70531010627747
2.42126989364624 2.82032990455627
2.5181200504303 2.94455003738403
2.61496996879578 3.06130003929138
2.71181988716125 3.18747997283936
3.00237011909485 3.5717499256134
3.09922003746033 3.69692993164062
3.19606995582581 3.82902002334595
3.29292011260986 3.96595001220703
3.38977003097534 4.09887981414795
3.58347010612488 4.35440015792847
3.68032002449036 4.47852993011475
};
\addlegendentry{OM3, $D=50$}
\addplot [line width=2pt, sandybrown24516489, dash pattern=on 1pt off 3pt on 3pt off 3pt]
table {%
0 1
0.0998772382736206 1.01350998878479
0.199753999710083 1.02990996837616
0.29963207244873 1.05736994743347
0.399508953094482 1.09081995487213
0.499386072158813 1.13163995742798
0.699140071868896 1.23091995716095
0.799018025398254 1.28925001621246
0.898895025253296 1.34921002388
0.998772025108337 1.41305005550385
1.19852995872498 1.55153000354767
1.3982800245285 1.70214998722076
1.49816000461578 1.78137004375458
1.59803998470306 1.86357998847961
1.69790995121002 1.94968998432159
1.79779005050659 2.03989005088806
1.89767003059387 2.13316988945007
1.99753999710083 2.23000001907349
2.19729995727539 2.43050003051758
2.29717993736267 2.53472995758057
2.49692988395691 2.74925994873047
2.59681010246277 2.860680103302
2.69668006896973 2.9745500087738
2.79656004905701 3.09238004684448
2.89644002914429 3.21357989311218
2.99632000923157 3.33867001533508
3.39582991600037 3.85648989677429
3.69546008110046 4.22905015945435
3.79533004760742 4.35977983474731
3.8952100276947 4.4879298210144
};
\addlegendentry{OM3, $D=100$}
\addplot [line width=2pt, darkorange24213334, dash pattern=on 1pt off 3pt on 3pt off 3pt]
table {%
0 1
0.0998772382736206 1.00839996337891
0.199753999710083 1.02583003044128
0.29963207244873 1.05349004268646
0.399508953094482 1.08822000026703
0.499386072158813 1.12920999526978
0.599262952804565 1.17536997795105
0.699140071868896 1.22662997245789
0.799018025398254 1.28251004219055
0.898895025253296 1.34174001216888
0.998772025108337 1.40468001365662
1.0986499786377 1.47126996517181
1.19852995872498 1.54129004478455
1.29840004444122 1.61483001708984
1.3982800245285 1.69135999679565
1.49816000461578 1.7711900472641
1.59803998470306 1.853639960289
1.69790995121002 1.93918001651764
1.79779005050659 2.02765989303589
1.89767003059387 2.11833000183105
1.99753999710083 2.21150994300842
2.09741997718811 2.306960105896
2.19729995727539 2.40439009666443
2.29717993736267 2.50430011749268
2.39704990386963 2.60747003555298
2.59681010246277 2.82088994979858
2.69668006896973 2.93095993995667
2.89644002914429 3.15610003471375
3.29594993591309 3.6172399520874
3.49569988250732 3.85111999511719
3.69546008110046 4.09104013442993
3.8952100276947 4.33989000320435
4.09497022628784 4.59554004669189
4.19483995437622 4.72636985778809
};
\addlegendentry{OM3, $D=200$}
\end{axis}

\end{tikzpicture}

%% file: figures/taylor-bubble/convergence-shape-front-nbrc.tex
% This file was created with tikzplotlib v0.10.1.
\begin{tikzpicture}

\definecolor{darkgray176}{RGB}{176,176,176}
\definecolor{dodgerblue0154222}{RGB}{0,154,222}
\definecolor{lightgray204}{RGB}{204,204,204}
\definecolor{mediumturquoise64179230}{RGB}{0,116,166}
\definecolor{skyblue128205238}{RGB}{159,217,243}

\begin{axis}[
height=\figureheight,
legend cell align={left},
legend style={
	fill opacity=0.8,
	draw opacity=1,
	text opacity=1,
	at={(0.01,0.01)},
	anchor=south west,
	draw=lightgray204
},
tick align=outside,
tick pos=left,
width=\figurewidth,
x grid style={darkgray176},
xlabel={\(\displaystyle r^{*}\)},
xmajorgrids,
xmin=-0.0376370728015625, xmax=0.790378528832812,
xtick style={color=black},
y grid style={darkgray176},
ylabel style={rotate=-90.0},
ylabel={\(\displaystyle z^{*}\)},
ymajorgrids,
ymin=-0.627388358149219, ymax=0.0298756361023437,
ytick style={color=black}
]
\addplot [ultra thick, black, mark=*, mark size=1, mark options={solid}, only marks]
table {%
0 0
0.0293359756469727 0
0.0573979616165161 -0.000633955001831055
0.0854599475860596 -0.00253605842590332
0.119897961616516 -0.00570595264434814
0.160714030265808 -0.0120450258255005
0.192602038383484 -0.0183850526809692
0.219388008117676 -0.0247249603271484
0.244897961616516 -0.0316979885101318
0.271683931350708 -0.0412089824676514
0.293367981910706 -0.0494489669799805
0.320153951644897 -0.0595920085906982
0.340561985969543 -0.0684679746627808
0.362244009971619 -0.0773429870605469
0.380102038383484 -0.0862189531326294
0.39795994758606 -0.0950939655303955
0.415816068649292 -0.10523796081543
0.429846048355103 -0.112844944000244
0.442602038383484 -0.121086955070496
0.456632018089294 -0.129328012466431
0.470664024353027 -0.138838052749634
0.48852002620697 -0.149615049362183
0.503826022148132 -0.162294030189514
0.519132018089294 -0.173071980476379
0.533164024353027 -0.183848977088928
0.543367981910706 -0.193992018699646
0.554846048355103 -0.204769968986511
0.565052032470703 -0.214913964271545
0.57397997379303 -0.225057005882263
0.581632018089294 -0.235200047492981
0.593111991882324 -0.247879028320312
0.603316068649292 -0.259925007820129
0.614795923233032 -0.273237943649292
0.623723983764648 -0.286550998687744
0.633928060531616 -0.299864053726196
0.641582012176514 -0.312543034553528
0.64795994758606 -0.325222969055176
0.656888008117676 -0.339169979095459
0.66198992729187 -0.351848959922791
0.667092084884644 -0.365162014961243
0.672194004058838 -0.376574039459229
0.679846048355103 -0.387984991073608
0.686223983764648 -0.401932001113892
0.690052032470703 -0.414610981941223
0.692601919174194 -0.424754977226257
0.697704076766968 -0.438068032264709
0.701529979705811 -0.453283071517944
0.705358028411865 -0.464694023132324
0.707906007766724 -0.476740002632141
0.71045994758606 -0.48751699924469
0.711734056472778 -0.49956202507019
0.715561985969543 -0.511608004570007
0.716835975646973 -0.523019075393677
0.718111991882324 -0.533161997795105
0.720664024353027 -0.544573068618774
0.721935987472534 -0.560423016548157
0.72448992729187 -0.572468042373657
0.72448992729187 -0.583878993988037
0.725765943527222 -0.595291018486023
};
\addlegendentry{Experiment~\cite{bugg2002VelocityFieldTaylor}}
\addplot [line width=2pt, skyblue128205238, dotted]
table {%
0 -6.38961791992188e-05
0.0625 -9.20295715332031e-05
0.0944106578826904 -0.000155925750732422
0.125 -0.000767707824707031
0.1875 -0.00139141082763672
0.21941065788269 -0.00209331512451172
0.28191065788269 -0.00703191757202148
0.308988332748413 -0.0127062797546387
0.337912201881409 -0.0185141563415527
0.34441065788269 -0.0204935073852539
0.40691065788269 -0.0355119705200195
0.446361064910889 -0.0497641563415527
0.46941065788269 -0.059135913848877
0.515622973442078 -0.0810141563415527
0.53191065788269 -0.090059757232666
0.566073536872864 -0.112264156341553
0.59441065788269 -0.134482860565186
0.607128262519836 -0.143514156341553
0.631168365478516 -0.174764156341553
0.65691065788269 -0.205703735351562
0.657277703285217 -0.206030368804932
0.670657873153687 -0.221969604492188
0.682979226112366 -0.237264156341553
0.687657356262207 -0.253219604492188
0.692557811737061 -0.268514156341553
0.695029616355896 -0.299764156341553
0.705898404121399 -0.315719604492188
0.724968791007996 -0.349340438842773
0.731456160545349 -0.362264156341553
0.743310332298279 -0.393514156341553
0.744593024253845 -0.409469604492188
0.746587514877319 -0.424764156341553
0.749995827674866 -0.456014156341553
0.749494791030884 -0.471969604492188
0.749565124511719 -0.487264156341553
0.751913785934448 -0.503219604492188
0.752152323722839 -0.518514156341553
0.749595522880554 -0.549764156341553
0.74940037727356 -0.581014156341553
0.752741456031799 -0.596969604492188
};
\addlegendentry{NBRC, $D=32$}
\addplot [line width=2pt, mediumturquoise64179230, dotted]
table {%
0 -0.000384330749511719
0.03125 -0.000501155853271484
0.0469405651092529 -0.000109195709228516
0.0625 0
0.09375 -0.000790119171142578
0.125 -0.00117969512939453
0.171940565109253 -0.00272798538208008
0.203190565109253 -0.00470447540283203
0.234440565109253 -0.0100235939025879
0.28125 -0.0169916152954102
0.296940565109253 -0.0193138122558594
0.319182634353638 -0.024712085723877
0.328190565109253 -0.0274543762207031
0.359440565109253 -0.0346417427062988
0.390690565109253 -0.0451107025146484
0.422263383865356 -0.055962085723877
0.453190565109253 -0.0681028366088867
0.46069073677063 -0.071587085723877
0.492510557174683 -0.087212085723877
0.51995849609375 -0.102837085723877
0.539501428604126 -0.118462085723877
0.546940565109253 -0.123250961303711
0.562103271484375 -0.134087085723877
0.579556703567505 -0.148778438568115
0.580970287322998 -0.149712085723877
0.595662593841553 -0.165337085723877
0.609440565109253 -0.178031921386719
0.613648891448975 -0.180962085723877
0.619073152542114 -0.188807487487793
0.624770879745483 -0.196587085723877
0.629084348678589 -0.204432487487793
0.632704973220825 -0.212212085723877
0.649622917175293 -0.227837085723877
0.656936645507812 -0.243462085723877
0.661819458007812 -0.251307487487793
0.664746284484863 -0.259087085723877
0.667482614517212 -0.265049457550049
0.671940565109253 -0.271242141723633
0.673393726348877 -0.272829532623291
0.674372911453247 -0.274712085723877
0.685342788696289 -0.290337085723877
0.687236547470093 -0.298182487487793
0.68863844871521 -0.305962085723877
0.690483093261719 -0.321587085723877
0.693017244338989 -0.329432487487793
0.698240518569946 -0.337212085723877
0.699429273605347 -0.344991683959961
0.703190565109253 -0.347652912139893
0.709035158157349 -0.352837085723877
0.714418172836304 -0.360616683959961
0.711424350738525 -0.368462085723877
0.716553449630737 -0.376307487487793
0.718507766723633 -0.384087085723877
0.71876049041748 -0.391932487487793
0.719398260116577 -0.399712085723877
0.719079732894897 -0.407557487487793
0.719101667404175 -0.415337085723877
0.719482898712158 -0.423182487487793
0.720726251602173 -0.430962085723877
0.722779989242554 -0.438807487487793
0.723041534423828 -0.446587085723877
0.72291111946106 -0.454432487487793
0.723903894424438 -0.462212085723877
0.727594614028931 -0.469991683959961
0.732967138290405 -0.475836277008057
0.735433101654053 -0.477837085723877
0.73066520690918 -0.485682487487793
0.732503890991211 -0.493462085723877
0.734857559204102 -0.495180130004883
0.740823030471802 -0.501307487487793
0.740429401397705 -0.509087085723877
0.736710786819458 -0.516932487487793
0.7447509765625 -0.532557487487793
0.745757579803467 -0.540337085723877
0.74517560005188 -0.548182487487793
0.74820351600647 -0.563807487487793
0.746492147445679 -0.571587085723877
0.747218608856201 -0.579432487487793
0.750126600265503 -0.587212085723877
0.749560117721558 -0.595057487487793
};
\addlegendentry{NBRC, $D=64$}
\addplot [line width=2pt, dodgerblue0154222, dotted]
table {%
0.00718438625335693 -0.000845909118652344
0.0143687725067139 -0.00139141082763672
0.046875 -0.00121593475341797
0.0540593862533569 -0.00100040435791016
0.0696843862533569 -0.00146961212158203
0.078125 -0.0013427734375
0.09375 -0.00200939178466797
0.100934386253357 -0.00190162658691406
0.116559386253357 -0.00247287750244141
0.179059386253357 -0.00548362731933594
0.195866942405701 -0.00776767730712891
0.196861147880554 -0.00798320770263672
0.210309386253357 -0.0103273391723633
0.234375 -0.0128173828125
0.241559386253357 -0.0137519836425781
0.250243306159973 -0.0157957077026367
0.257184386253357 -0.0175189971923828
0.265625 -0.0189542770385742
0.272809386253357 -0.0198879241943359
0.3125 -0.0289936065673828
0.320938587188721 -0.0308094024658203
0.32243800163269 -0.0314207077026367
0.335309386253357 -0.035212516784668
0.34375 -0.0367317199707031
0.34846019744873 -0.0392332077026367
0.352316975593567 -0.0408134460449219
0.359375 -0.0435142517089844
0.366559386253357 -0.0438385009765625
0.370067358016968 -0.0453577041625977
0.377251744270325 -0.0481595993041992
0.382184386253357 -0.0500583648681641
0.390049338340759 -0.0523796081542969
0.397233724594116 -0.0556488037109375
0.397809386253357 -0.0562810897827148
0.40625 -0.0592355728149414
0.414773106575012 -0.0631380081176758
0.421875 -0.0658807754516602
0.429059386253357 -0.0682163238525391
0.432668566703796 -0.0696926116943359
0.433956027030945 -0.0704832077026367
0.439852952957153 -0.0730695724487305
0.450053930282593 -0.0782957077026367
0.463809728622437 -0.0849227905273438
0.486507058143616 -0.0976772308349609
0.491994023323059 -0.101733207702637
0.498781204223633 -0.105325698852539
0.504284262657166 -0.109725475311279
0.507184386253357 -0.111666679382324
0.511088967323303 -0.113945484161377
0.522809386253357 -0.119936943054199
0.525839567184448 -0.1217942237854
0.529499292373657 -0.125170707702637
0.530149221420288 -0.129391193389893
0.538434386253357 -0.134541511535645
0.539688229560852 -0.134677410125732
0.544450759887695 -0.137203693389893
0.545729517936707 -0.140795707702637
0.550062894821167 -0.145016193389893
0.554059386253357 -0.147465705871582
0.555920481681824 -0.148608207702637
0.561533212661743 -0.152828693389893
0.564598083496094 -0.156420707702637
0.566142678260803 -0.157570362091064
0.569684386253357 -0.159511566162109
0.572773575782776 -0.161790370941162
0.575428247451782 -0.164233207702637
0.576350092887878 -0.167825222015381
0.579346537590027 -0.172045707702637
0.582954406738281 -0.175098896026611
0.586212515830994 -0.179319381713867
0.587304592132568 -0.179858207702637
0.592197179794312 -0.183450222015381
0.59589409828186 -0.187670707702637
0.596835255622864 -0.191262722015381
0.597355842590332 -0.191550731658936
0.600934386253357 -0.194067478179932
0.603610038757324 -0.195626735687256
0.607202291488647 -0.199075222015381
0.608983278274536 -0.203295707702637
0.611418724060059 -0.206887722015381
0.612608790397644 -0.211108207702637
0.613765478134155 -0.212473392486572
0.616559386253357 -0.214203357696533
0.620062232017517 -0.21669340133667
0.62200927734375 -0.218920707702637
0.627128601074219 -0.226733207702637
0.628400206565857 -0.230325222015381
0.630802750587463 -0.234545707702637
0.632547378540039 -0.235372066497803
0.636462807655334 -0.238137722015381
0.639839053153992 -0.242358207702637
0.640837669372559 -0.245950222015381
0.640576004981995 -0.250170707702637
0.642084717750549 -0.253762722015381
0.644861698150635 -0.257983207702637
0.645623207092285 -0.258378505706787
0.647809386253357 -0.260150909423828
0.650324702262878 -0.262598514556885
0.654074907302856 -0.265795707702637
0.654024600982666 -0.269387722015381
0.655476450920105 -0.273608207702637
0.657028317451477 -0.277200222015381
0.659357905387878 -0.281420707702637
0.661445021629333 -0.285317420959473
0.665713310241699 -0.289376735687256
0.668895959854126 -0.292825222015381
0.670094609260559 -0.297045707702637
0.671589016914368 -0.300637722015381
0.672677755355835 -0.304858207702637
0.673116087913513 -0.308450222015381
0.675124883651733 -0.312670707702637
0.676433205604553 -0.316891193389893
0.683257699012756 -0.324703693389893
0.685060977935791 -0.328295707702637
0.686607599258423 -0.336108207702637
0.686629056930542 -0.339700222015381
0.687751770019531 -0.343920707702637
0.688580513000488 -0.351733207702637
0.691662669181824 -0.355016708374023
0.695064187049866 -0.359236717224121
0.695538401603699 -0.359545707702637
0.69703996181488 -0.363137722015381
0.699538350105286 -0.367358207702637
0.699215054512024 -0.370950222015381
0.699924111366272 -0.375170707702637
0.701619625091553 -0.378762722015381
0.703484654426575 -0.394387722015381
0.705299377441406 -0.398608207702637
0.707375764846802 -0.402200222015381
0.706995964050293 -0.406420707702637
0.708030462265015 -0.408256053924561
0.712900638580322 -0.412476539611816
0.713899254798889 -0.414233207702637
0.714739799499512 -0.417825222015381
0.713026404380798 -0.422045707702637
0.713960409164429 -0.425637722015381
0.716577053070068 -0.429858207702637
0.717926979064941 -0.437670707702637
0.718185663223267 -0.441262722015381
0.717712879180908 -0.445483207702637
0.718509078025818 -0.456887722015381
0.718981862068176 -0.461108207702637
0.719657182693481 -0.464700222015381
0.721117377281189 -0.468920707702637
0.72104549407959 -0.472512722015381
0.722337007522583 -0.476733207702637
0.724837183952332 -0.47892427444458
0.726069450378418 -0.483144760131836
0.727894186973572 -0.484545707702637
0.726191520690918 -0.488137722015381
0.728225827217102 -0.492358207702637
0.730474472045898 -0.495950222015381
0.731310129165649 -0.500170707702637
0.731539964675903 -0.503762722015381
0.731421828269958 -0.507983207702637
0.732370257377625 -0.511575222015381
0.732640266418457 -0.515795707702637
0.732000827789307 -0.519387722015381
0.733501791954041 -0.527200222015381
0.733608245849609 -0.535012722015381
0.733912110328674 -0.539233207702637
0.733792185783386 -0.550637722015381
0.734832406044006 -0.558450222015381
0.734418749809265 -0.562670707702637
0.735434770584106 -0.570483207702637
0.7350252866745 -0.574075222015381
0.735776543617249 -0.578295707702637
0.737069725990295 -0.581887722015381
0.738217711448669 -0.586108207702637
0.739514589309692 -0.597512722015381
};
\addlegendentry{NBRC, $D=128$}
\end{axis}

\end{tikzpicture}

%% file: figures/taylor-bubble/convergence-shape-tail-nbrc.tex
% This file was created with tikzplotlib v0.10.1.
\begin{tikzpicture}

\definecolor{darkgray176}{RGB}{176,176,176}
\definecolor{dodgerblue0154222}{RGB}{0,154,222}
\definecolor{lightgray204}{RGB}{204,204,204}
\definecolor{mediumturquoise64179230}{RGB}{0,116,166}
\definecolor{skyblue128205238}{RGB}{159,217,243}

\begin{axis}[
height=\figureheight,
legend cell align={left},
legend style={
  fill opacity=0.8,
  draw opacity=1,
  text opacity=1,
  at={(0.01,0.99)},
  anchor=north west,
  draw=lightgray204
},
tick align=outside,
tick pos=left,
width=\figurewidth,
x grid style={darkgray176},
xlabel={\(\displaystyle r^{*}\)},
xmajorgrids,
xmin=-0.038249188661875, xmax=0.803232961899375,
xtick style={color=black},
y grid style={darkgray176},
ylabel style={rotate=-90.0},
ylabel={\(\displaystyle z^{*}\)},
ymajorgrids,
ymin=-0.0993263449601563, ymax=0.319011244163282,
ytick style={color=black}
]
\addplot [very thick, black, mark=*, mark size=1, mark options={solid}, only marks]
table {%
0.751070022583008 0.279245018959045
0.751060009002686 0.276661038398743
0.75104808807373 0.272783041000366
0.751029968261719 0.267291069030762
0.751013994216919 0.262120962142944
0.750993967056274 0.256628036499023
0.750977993011475 0.250813007354736
0.750959992408752 0.245319962501526
0.750944018363953 0.240473031997681
0.750927925109863 0.235626935958862
0.750891923904419 0.225288033485413
0.75083601474762 0.207193970680237
0.750802040100098 0.196854948997498
0.752063989639282 0.185546040534973
0.752671957015991 0.172943949699402
0.753286004066467 0.16228199005127
0.753888010978699 0.148064970970154
0.755147933959961 0.136109948158264
0.757050037384033 0.122215986251831
0.75830602645874 0.109614968299866
0.759566068649292 0.0979830026626587
0.760175943374634 0.085705041885376
0.759490013122559 0.0747189521789551
0.75749397277832 0.059533953666687
0.756160020828247 0.0479030609130859
0.754166007041931 0.0336869955062866
0.753479957580566 0.0223790407180786
0.750848054885864 0.0110709667205811
0.748866081237793 8.59498977661133e-05
0.746886014938354 -0.0102519989013672
0.743628025054932 -0.0141290426254272
0.740370035171509 -0.0180050134658813
0.737761974334717 -0.0218809843063354
0.735153913497925 -0.0257569551467896
0.731894016265869 -0.0302799940109253
0.727335929870605 -0.0351250171661377
0.724077939987183 -0.0396469831466675
0.719521999359131 -0.0438460111618042
0.714959979057312 -0.0496599674224854
0.710399985313416 -0.0551519393920898
0.704541921615601 -0.0606429576873779
0.693494081497192 -0.0651619434356689
0.68504798412323 -0.0687140226364136
0.673354029655457 -0.072909951210022
0.661018013954163 -0.0745220184326172
0.651278018951416 -0.0761339664459229
0.643490076065063 -0.0767780542373657
0.63440203666687 -0.0777440071105957
0.624661922454834 -0.0793570280075073
0.611681938171387 -0.0799989700317383
0.591563940048218 -0.0799920558929443
0.577935934066772 -0.0803110599517822
0.55717396736145 -0.0793349742889404
0.539010047912598 -0.0777130126953125
0.521492004394531 -0.0764150619506836
0.512416005134583 -0.0738279819488525
0.499451994895935 -0.0693000555038452
0.48907995223999 -0.0660660266876221
0.478709936141968 -0.0618619918823242
0.467043995857239 -0.0576579570770264
0.458620071411133 -0.0537780523300171
0.450191974639893 -0.0502209663391113
0.439177989959717 -0.0456939935684204
0.428156018257141 -0.0427830219268799
0.413892030715942 -0.0389009714126587
0.40351402759552 -0.0366359949111938
0.388602018356323 -0.0330770015716553
0.375632047653198 -0.0304880142211914
0.360715985298157 -0.0275750160217285
0.34644603729248 -0.0256320238113403
0.330878019332886 -0.0240110158920288
0.312067985534668 -0.02109694480896
0.295850038528442 -0.0191529989242554
0.27963399887085 -0.0172100067138672
0.266011953353882 -0.0155899524688721
0.249145984649658 -0.0133229494094849
0.233579993247986 -0.0110559463500977
0.214761972427368 -0.0104039907455444
0.197247982025146 -0.0087820291519165
0.168053984642029 -0.0061880350112915
0.149888038635254 -0.00488996505737305
0.131721973419189 -0.00359201431274414
0.111608028411865 -0.00261604785919189
0.0973340272903442 -0.00196504592895508
0.0772199630737305 -0.0013120174407959
0.0655399560928345 -0.000661969184875488
0.055806040763855 -0.000658988952636719
0.0428299903869629 -0.000331997871398926
0.0317980051040649 -5.00679016113281e-06
0.0162240266799927 0
};
\addlegendentry{Experiment~\cite{bugg2002VelocityFieldTaylor}}
\addplot [line width=2pt, skyblue128205238, dotted]
table {%
0 0
0.0625 -4.38690185546875e-05
0.0944106578826904 -0.000139713287353516
0.21941065788269 -0.00146722793579102
0.375 -0.00352048873901367
0.40691065788269 -0.00390338897705078
0.4375 -0.00324583053588867
0.46941065788269 -0.00315046310424805
0.5 -0.00493955612182617
0.53191065788269 -0.00552988052368164
0.5625 -0.00143098831176758
0.59441065788269 -9.10758972167969e-05
0.605541110038757 0.00356292724609375
0.627134084701538 0.00495481491088867
0.636130452156067 0.00639247894287109
0.65691065788269 0.0124235153198242
0.675118923187256 0.0223479270935059
0.690251469612122 0.0362048149108887
0.705708265304565 0.0449380874633789
0.719981908798218 0.0608935356140137
0.727167248725891 0.0674548149108887
0.733557462692261 0.0827493667602539
0.745405793190002 0.0987048149108887
0.746886372566223 0.113999366760254
0.750447630882263 0.129954814910889
0.750511288642883 0.161204814910889
0.750700950622559 0.176499366760254
0.751935958862305 0.192454814910889
0.752635836601257 0.223704814910889
0.754042744636536 0.286204814910889
0.754327654838562 0.254954814910889
0.757450819015503 0.270249366760254
};
\addlegendentry{NBRC, $D=32$}
\addplot [line width=2pt, mediumturquoise64179230, dotted]
table {%
0 0
0.0469405651092529 -0.000101566314697266
0.109440565109253 -0.000632762908935547
0.15625 -0.00157690048217773
0.1875 -0.00238895416259766
0.217553377151489 -0.00344276428222656
0.25 -0.00479412078857422
0.28125 -0.00618505477905273
0.359440565109253 -0.00887489318847656
0.40625 -0.0100221633911133
0.4375 -0.0107254981994629
0.46875 -0.0108966827392578
0.484440565109253 -0.0113363265991211
0.5 -0.0113906860351562
0.515690565109253 -0.0112495422363281
0.546940565109253 -0.0114059448242188
0.5625 -0.0110950469970703
0.578190565109253 -0.0111103057861328
0.59375 -0.00929784774780273
0.609440565109253 -0.00857639312744141
0.621106386184692 -0.00561857223510742
0.636665821075439 -0.00269317626953125
0.640690565109253 -0.00134372711181641
0.65625 0.00333976745605469
0.671940565109253 0.00828218460083008
0.690539360046387 0.0179805755615234
0.703190565109253 0.0240139961242676
0.7081139087677 0.0272030830383301
0.714775085449219 0.0349826812744141
0.718330383300781 0.0428280830383301
0.719344854354858 0.0506076812744141
0.727384805679321 0.0584530830383301
0.726344347000122 0.0662984848022461
0.73272705078125 0.0740780830383301
0.734440565109253 0.0758819580078125
0.739448308944702 0.0819234848022461
0.744061470031738 0.0897030830383301
0.745180368423462 0.0974826812744141
0.747730016708374 0.10532808303833
0.749612092971802 0.12095308303833
0.751115083694458 0.128732681274414
0.751359939575195 0.13657808303833
0.750320672988892 0.144357681274414
0.750174522399902 0.15220308303833
0.750211954116821 0.159982681274414
0.751114130020142 0.16782808303833
0.751103401184082 0.175607681274414
0.75020432472229 0.18345308303833
0.750945806503296 0.19907808303833
0.750693798065186 0.206857681274414
0.752473115921021 0.21470308303833
0.752488613128662 0.222482681274414
0.758630037307739 0.23032808303833
0.758235931396484 0.23085355758667
0.752024173736572 0.238173484802246
0.751824855804443 0.24595308303833
0.756662368774414 0.253732681274414
0.757647752761841 0.26157808303833
0.753522872924805 0.269357681274414
0.755159378051758 0.27720308303833
0.75615668296814 0.284982681274414
0.753721714019775 0.29282808303833
};
\addlegendentry{NBRC, $D=64$}
\addplot [line width=2pt, dodgerblue0154222, dotted]
table {%
0 0
0.0625 -0.000128746032714844
0.116559386253357 -0.000597953796386719
0.163434386253357 -0.00116825103759766
0.203125 -0.00206565856933594
0.225934386253357 -0.00287437438964844
0.24935507774353 -0.00423908233642578
0.28125 -0.00640106201171875
0.304059386253357 -0.00749540328979492
0.34375 -0.00874233245849609
0.375 -0.00987672805786133
0.390625 -0.0107383728027344
0.397809386253357 -0.0111908912658691
0.429059386253357 -0.0139174461364746
0.453125 -0.0151605606079102
0.46875 -0.0157217979431152
0.491559386253357 -0.0162811279296875
0.522809386253357 -0.0167636871337891
0.554059386253357 -0.0169081687927246
0.569684386253357 -0.016873836517334
0.578125 -0.0168862342834473
0.585309386253357 -0.0166134834289551
0.59375 -0.0159635543823242
0.609375 -0.0157699584960938
0.625 -0.0132536888122559
0.63247013092041 -0.0125041007995605
0.647809386253357 -0.00908851623535156
0.654993772506714 -0.0075836181640625
0.664114356040955 -0.00469160079956055
0.664705991744995 -0.00440883636474609
0.670618772506714 -0.00196599960327148
0.679059386253357 -0.000290870666503906
0.681954741477966 0.000858783721923828
0.686740517616272 0.00312089920043945
0.690395355224609 0.00604963302612305
0.694684386253357 0.00768041610717773
0.698176026344299 0.00964164733886719
0.701417207717896 0.0109333992004395
0.706616640090942 0.0137395858764648
0.710309386253357 0.0166959762573242
0.711408615112305 0.0173320770263672
0.71329927444458 0.0187458992004395
0.716928720474243 0.0229663848876953
0.720621585845947 0.0265583992004395
0.723541975021362 0.0296435356140137
0.727730512619019 0.0332355499267578
0.729106307029724 0.0343708992004395
0.732659816741943 0.0385913848876953
0.734024882316589 0.0421833992004395
0.733611345291138 0.0464038848876953
0.73504102230072 0.0499958992004395
0.737066745758057 0.0542163848876953
0.737928867340088 0.0578083992004395
0.73950719833374 0.0620288848876953
0.741573691368103 0.0659499168395996
0.744266271591187 0.0698413848876953
0.746328234672546 0.0734333992004395
0.747037291526794 0.0776538848876953
0.748064637184143 0.0812458992004395
0.749098658561707 0.0890583992004395
0.749712944030762 0.101091384887695
0.74963390827179 0.104683399200439
0.749330043792725 0.108903884887695
0.749554753303528 0.120308399200439
0.749538779258728 0.128120899200439
0.749567866325378 0.135933399200439
0.750200867652893 0.140153884887695
0.753103375434875 0.143745899200439
0.752672910690308 0.147966384887695
0.749914050102234 0.151558399200439
0.753822088241577 0.15489673614502
0.761951327323914 0.159370899200439
0.757184386253357 0.161758899688721
0.752972483634949 0.165189743041992
0.749380350112915 0.167183399200439
0.757184386253357 0.171984672546387
0.763198375701904 0.174995899200439
0.757184386253357 0.178383350372314
0.750148296356201 0.182808399200439
0.760582566261292 0.188244819641113
0.764743804931641 0.190620899200439
0.761345624923706 0.193142890930176
0.757184386253357 0.195889949798584
0.752306342124939 0.198433399200439
0.757184386253357 0.201093673706055
0.760625720024109 0.203633308410645
0.764508366584778 0.206245899200439
0.761067032814026 0.209356784820557
0.757184386253357 0.21312141418457
0.755508542060852 0.214058399200439
0.757184386253357 0.214925765991211
0.760769367218018 0.218072414398193
0.764829397201538 0.221870899200439
0.761368632316589 0.226091384887695
0.757520794868469 0.229683399200439
0.761224985122681 0.233275413513184
0.764820694923401 0.237495899200439
0.76028573513031 0.245308399200439
0.764573693275452 0.253120899200439
0.762598633766174 0.257341384887695
0.761650323867798 0.260933399200439
0.763870239257812 0.265153884887695
0.764983773231506 0.268745899200439
0.762471079826355 0.276558399200439
0.764811396598816 0.284370899200439
0.763788461685181 0.292183399200439
0.764109253883362 0.296403884887695
0.764590620994568 0.299995899200439
};
\addlegendentry{NBRC, $D=128$}
\end{axis}

\end{tikzpicture}

%% file: figures/taylor-bubble/convergence-shape-front-nbkc.tex
% This file was created with tikzplotlib v0.10.1.
\begin{tikzpicture}

\definecolor{darkgray176}{RGB}{176,176,176}
\definecolor{lightgray204}{RGB}{204,204,204}
\definecolor{mediumaquamarine128230181}{RGB}{159,236,200}
\definecolor{mediumaquamarine64218145}{RGB}{0,128,68}
\definecolor{springgreen0205108}{RGB}{0,205,108}

\begin{axis}[
height=\figureheight,
legend cell align={left},
legend style={
	fill opacity=0.8,
	draw opacity=1,
	text opacity=1,
	at={(0.01,0.01)},
	anchor=south west,
	draw=lightgray204
},
tick align=outside,
tick pos=left,
width=\figurewidth,
x grid style={darkgray176},
xlabel={\(\displaystyle r^{*}\)},
xmajorgrids,
xmin=-0.0380160838365625, xmax=0.798337760567813,
xtick style={color=black},
y grid style={darkgray176},
ylabel style={rotate=-90.0},
ylabel={\(\displaystyle z^{*}\)},
ymajorgrids,
ymin=-0.629401087760156, ymax=0.0299714803695312,
ytick style={color=black}
]
\addplot [very thick, black, mark=*, mark size=1, mark options={solid}, only marks]
table {%
0 0
0.0293359756469727 0
0.0573979616165161 -0.000633955001831055
0.0854599475860596 -0.00253605842590332
0.119897961616516 -0.00570595264434814
0.160714030265808 -0.0120450258255005
0.192602038383484 -0.0183850526809692
0.219388008117676 -0.0247249603271484
0.244897961616516 -0.0316979885101318
0.271683931350708 -0.0412089824676514
0.293367981910706 -0.0494489669799805
0.320153951644897 -0.0595920085906982
0.340561985969543 -0.0684679746627808
0.362244009971619 -0.0773429870605469
0.380102038383484 -0.0862189531326294
0.39795994758606 -0.0950939655303955
0.415816068649292 -0.10523796081543
0.429846048355103 -0.112844944000244
0.442602038383484 -0.121086955070496
0.456632018089294 -0.129328012466431
0.470664024353027 -0.138838052749634
0.48852002620697 -0.149615049362183
0.503826022148132 -0.162294030189514
0.519132018089294 -0.173071980476379
0.533164024353027 -0.183848977088928
0.543367981910706 -0.193992018699646
0.554846048355103 -0.204769968986511
0.565052032470703 -0.214913964271545
0.57397997379303 -0.225057005882263
0.581632018089294 -0.235200047492981
0.593111991882324 -0.247879028320312
0.603316068649292 -0.259925007820129
0.614795923233032 -0.273237943649292
0.623723983764648 -0.286550998687744
0.633928060531616 -0.299864053726196
0.641582012176514 -0.312543034553528
0.64795994758606 -0.325222969055176
0.656888008117676 -0.339169979095459
0.66198992729187 -0.351848959922791
0.667092084884644 -0.365162014961243
0.672194004058838 -0.376574039459229
0.679846048355103 -0.387984991073608
0.686223983764648 -0.401932001113892
0.690052032470703 -0.414610981941223
0.692601919174194 -0.424754977226257
0.697704076766968 -0.438068032264709
0.701529979705811 -0.453283071517944
0.705358028411865 -0.464694023132324
0.707906007766724 -0.476740002632141
0.71045994758606 -0.48751699924469
0.711734056472778 -0.49956202507019
0.715561985969543 -0.511608004570007
0.716835975646973 -0.523019075393677
0.718111991882324 -0.533161997795105
0.720664024353027 -0.544573068618774
0.721935987472534 -0.560423016548157
0.72448992729187 -0.572468042373657
0.72448992729187 -0.583878993988037
0.725765943527222 -0.595291018486023
};
\addlegendentry{Experiment~\cite{bugg2002VelocityFieldTaylor}}
\addplot [line width=2pt, mediumaquamarine128230181]
table {%
0 -0.0010075569152832
0.0625 -0.000523567199707031
0.125 -0.000101566314697266
0.158813953399658 0
0.1875 -0.000573635101318359
0.221313953399658 -0.000996589660644531
0.283813953399658 -0.00450420379638672
0.348020792007446 -0.0181851387023926
0.350465536117554 -0.0190134048461914
0.408813953399658 -0.0333685874938965
0.460730791091919 -0.0502634048461914
0.471313953399658 -0.0545735359191895
0.533813953399658 -0.0834612846374512
0.596313953399658 -0.123354434967041
0.622787237167358 -0.144013404846191
0.641761064529419 -0.175263404846191
0.676064491271973 -0.206513404846191
0.683195829391479 -0.2234206199646
0.688838481903076 -0.237763404846191
0.694400787353516 -0.2546706199646
0.71074652671814 -0.300263404846191
0.716474771499634 -0.308395862579346
0.726001262664795 -0.322738647460938
0.732108116149902 -0.331513404846191
0.739351034164429 -0.3484206199646
0.746011972427368 -0.362763404846191
0.746015310287476 -0.3796706199646
0.744535207748413 -0.394013404846191
0.747331619262695 -0.4109206199646
0.751539707183838 -0.425263404846191
0.752442598342896 -0.4421706199646
0.754029035568237 -0.456513404846191
0.753170013427734 -0.4734206199646
0.751569509506226 -0.487763404846191
0.752083301544189 -0.5046706199646
0.751790761947632 -0.519013404846191
0.752639532089233 -0.5359206199646
0.754570007324219 -0.550263404846191
0.757383346557617 -0.5671706199646
0.760321617126465 -0.5984206199646
};
\addlegendentry{NBKC, $D=32$}
\addplot [line width=2pt, mediumaquamarine64218145]
table {%
0 -0.00110149383544922
0.0156266689300537 -0.00160932540893555
0.03125 -0.00176572799682617
0.0468766689300537 -0.00146865844726562
0.0781266689300537 -0.00281238555908203
0.09375 -0.00320291519165039
0.125 -0.00356245040893555
0.140626668930054 -0.00420331954956055
0.148799180984497 -0.00511693954467773
0.164425849914551 -0.00630474090576172
0.171876668930054 -0.00714015960693359
0.203126668930054 -0.009124755859375
0.21875 -0.00996828079223633
0.25 -0.0124058723449707
0.265626668930054 -0.0136795043945312
0.296876668930054 -0.0176486968994141
0.31952428817749 -0.023460865020752
0.328126668930054 -0.0256171226501465
0.359376668930054 -0.0317811965942383
0.40625 -0.0467495918273926
0.423749923706055 -0.0525546073913574
0.453126668930054 -0.0635466575622559
0.484376668930054 -0.0776872634887695
0.4968421459198 -0.0838046073913574
0.520569801330566 -0.0966482162475586
0.525407791137695 -0.0994296073913574
0.547134399414062 -0.115054607391357
0.566598415374756 -0.130679607391357
0.588260889053345 -0.146304607391357
0.600414037704468 -0.161929607391357
0.61461353302002 -0.173202514648438
0.619853258132935 -0.177554607391357
0.628745317459106 -0.193179607391357
0.640626668930054 -0.205163955688477
0.645323514938354 -0.208804607391357
0.655009508132935 -0.224429607391357
0.657815933227539 -0.232243061065674
0.660248517990112 -0.240054607391357
0.665889739990234 -0.246625423431396
0.671876668930054 -0.252289295196533
0.676359176635742 -0.255679607391357
0.680506467819214 -0.263493061065674
0.682384490966797 -0.271304607391357
0.684698820114136 -0.279118061065674
0.688209295272827 -0.286929607391357
0.691346883773804 -0.302554607391357
0.696456670761108 -0.310368061065674
0.701070308685303 -0.318179607391357
0.702029705047607 -0.323469161987305
0.703126668930054 -0.325664520263672
0.705656051635742 -0.331280708312988
0.706129550933838 -0.333804607391357
0.711423873901367 -0.341618061065674
0.714400053024292 -0.349429607391357
0.715843915939331 -0.357243061065674
0.718226432800293 -0.365054607391357
0.718920469284058 -0.372868061065674
0.71889066696167 -0.380679607391357
0.719357967376709 -0.396304607391357
0.720165729522705 -0.404118061065674
0.720637559890747 -0.411929607391357
0.724206686019897 -0.419743061065674
0.730652570724487 -0.436104774475098
0.736870050430298 -0.444179534912109
0.738571882247925 -0.450993061065674
0.740127801895142 -0.458804607391357
0.7439284324646 -0.466618061065674
0.740185022354126 -0.474429607391357
0.740339756011963 -0.482243061065674
0.745628118515015 -0.490054607391357
0.74795937538147 -0.521304607391357
0.749028205871582 -0.529118061065674
0.74955940246582 -0.536929607391357
0.748771905899048 -0.544743061065674
0.749029636383057 -0.552554607391357
0.749901533126831 -0.560368061065674
0.749490737915039 -0.568179607391357
0.750103235244751 -0.591618061065674
0.750628232955933 -0.599429607391357
};
\addlegendentry{NBKC, $D=64$}
\addplot [line width=2pt, springgreen0205108]
table {%
0 -0.000968456268310547
0.0124937295913696 -0.00109338760375977
0.015625 -0.000874519348144531
0.03125 -0.000796318054199219
0.046875 -0.00156164169311523
0.0625 -0.00143718719482422
0.0749937295913696 0
0.078125 0
0.09375 -0.00128078460693359
0.11250627040863 -0.00109386444091797
0.12813127040863 -0.00256252288818359
0.140625 -0.00287485122680664
0.15625 -0.00399923324584961
0.203125 -0.00679636001586914
0.21875 -0.00815582275390625
0.234375 -0.0113430023193359
0.25 -0.0130777359008789
0.265625 -0.0152177810668945
0.278953671455383 -0.0179996490478516
0.28125 -0.0186243057250977
0.296875 -0.0213742256164551
0.3125 -0.024014949798584
0.319997787475586 -0.0261569023132324
0.328125 -0.0282807350158691
0.34375 -0.0316557884216309
0.351272821426392 -0.0340476036071777
0.359375 -0.0364212989807129
0.376500010490417 -0.0412650108337402
0.390625 -0.0456085205078125
0.40625 -0.0516395568847656
0.41874372959137 -0.054450511932373
0.4375 -0.0631241798400879
0.453125 -0.0694522857666016
0.461733460426331 -0.073422908782959
0.476136207580566 -0.0803275108337402
0.477930665016174 -0.0810947418212891
0.490574598312378 -0.0881400108337402
0.492131471633911 -0.0892047882080078
0.5 -0.0923905372619629
0.507194638252258 -0.0959525108337402
0.508575558662415 -0.0975179672241211
0.515625 -0.100738048553467
0.521156549453735 -0.10376501083374
0.522778987884521 -0.105330467224121
0.52647864818573 -0.106781959533691
0.534939527511597 -0.11157751083374
0.536903977394104 -0.113142967224121
0.548475861549377 -0.120955467224121
0.55476975440979 -0.128767967224121
0.557235836982727 -0.12973690032959
0.5625 -0.132851123809814
0.566823601722717 -0.13501501083374
0.568242788314819 -0.136580467224121
0.569164752960205 -0.14282751083374
0.571670651435852 -0.144252300262451
0.578125 -0.149999141693115
0.579318523406982 -0.15064001083374
0.581097364425659 -0.152205467224121
0.587587833404541 -0.15845251083374
0.587669253349304 -0.160017967224121
0.591080069541931 -0.16626501083374
0.588781237602234 -0.172512054443359
0.59375 -0.174985885620117
0.59688127040863 -0.177257061004639
0.598902583122253 -0.178313732147217
0.609375 -0.183172702789307
0.616554021835327 -0.188137054443359
0.617646813392639 -0.18970251083374
0.618294954299927 -0.191267967224121
0.623100996017456 -0.19851541519165
0.625 -0.199889659881592
0.630389332771301 -0.20532751083374
0.631457090377808 -0.206892967224121
0.634543061256409 -0.21314001083374
0.635826826095581 -0.214705467224121
0.639812350273132 -0.22095251083374
0.640048503875732 -0.222630977630615
0.640625 -0.22295618057251
0.64850652217865 -0.230330467224121
0.649718403816223 -0.23657751083374
0.650580406188965 -0.242824554443359
0.652050018310547 -0.24439001083374
0.659538984298706 -0.250637054443359
0.661258101463318 -0.253767967224121
0.664070725440979 -0.261580467224121
0.667943835258484 -0.26782751083374
0.670255064964294 -0.274074554443359
0.67245352268219 -0.27564001083374
0.678243398666382 -0.281887054443359
0.678619146347046 -0.285017967224121
0.680630683898926 -0.29126501083374
0.681563854217529 -0.292830467224121
0.683050513267517 -0.29907751083374
0.683726906776428 -0.300570964813232
0.691438674926758 -0.30689001083374
0.69228732585907 -0.308455467224121
0.696905732154846 -0.33032751083374
0.699779391288757 -0.336574554443359
0.701897144317627 -0.33814001083374
0.706337571144104 -0.344387054443359
0.706243634223938 -0.34595251083374
0.706857323646545 -0.347517967224121
0.709780931472778 -0.35376501083374
0.710719347000122 -0.363142967224121
0.711838006973267 -0.370955467224121
0.714481711387634 -0.378767967224121
0.716893076896667 -0.38501501083374
0.717246890068054 -0.386580467224121
0.720561504364014 -0.39315938949585
0.721428871154785 -0.394392967224121
0.724152565002441 -0.40064001083374
0.724484443664551 -0.402205467224121
0.726063013076782 -0.417830467224121
0.726978421211243 -0.425642967224121
0.727759718894958 -0.433455467224121
0.727947115898132 -0.43970251083374
0.72828209400177 -0.441267967224121
0.731244444847107 -0.449080467224121
0.732763290405273 -0.456047534942627
0.736466646194458 -0.464705467224121
0.737753510475159 -0.47095251083374
0.737584471702576 -0.472517967224121
0.740543723106384 -0.480330467224121
0.740247368812561 -0.488142967224121
0.741063117980957 -0.495955467224121
0.741912603378296 -0.50220251083374
0.741815567016602 -0.503767967224121
0.742087364196777 -0.511580467224121
0.741971969604492 -0.519392967224121
0.742497205734253 -0.527205467224121
0.743478298187256 -0.535017967224121
0.744359135627747 -0.54907751083374
0.745048046112061 -0.550642967224121
0.746809601783752 -0.55689001083374
0.746408820152283 -0.558455467224121
0.746896147727966 -0.56470251083374
0.746608734130859 -0.570949554443359
0.75 -0.575780391693115
0.752124786376953 -0.578762054443359
0.752453446388245 -0.581892967224121
0.754406332969666 -0.589705467224121
0.754631161689758 -0.59595251083374
0.754953622817993 -0.597517967224121
};
\addlegendentry{NBKC, $D=128$}
\end{axis}

\end{tikzpicture}

%% file: figures/taylor-bubble/convergence-shape-tail-nbkc.tex
% This file was created with tikzplotlib v0.10.1.
\begin{tikzpicture}

\definecolor{darkgray176}{RGB}{176,176,176}
\definecolor{lightgray204}{RGB}{204,204,204}
\definecolor{mediumaquamarine128230181}{RGB}{159,236,200}
\definecolor{mediumaquamarine64218145}{RGB}{0,128,68}
\definecolor{springgreen0205108}{RGB}{0,205,108}

\begin{axis}[
height=\figureheight,
legend cell align={left},
legend style={
	fill opacity=0.8,
	draw opacity=1,
	text opacity=1,
	at={(0.01,0.99)},
	anchor=north west,
	draw=lightgray204
},
tick align=outside,
tick pos=left,
width=\figurewidth,
x grid style={darkgray176},
xlabel={\(\displaystyle r^{*}\)},
xmajorgrids,
xmin=-0.038814213871875, xmax=0.815098491309375,
xtick style={color=black},
y grid style={darkgray176},
ylabel style={rotate=-90.0},
ylabel={\(\displaystyle z^{*}\)},
ymajorgrids,
ymin=-0.0993188347746094, ymax=0.318853530266797,
ytick style={color=black}
]
\addplot [very thick, black, mark=*, mark size=1, mark options={solid}, only marks]
table {%
0.751070022583008 0.279245018959045
0.751060009002686 0.276661038398743
0.75104808807373 0.272783041000366
0.751029968261719 0.267291069030762
0.751013994216919 0.262120962142944
0.750993967056274 0.256628036499023
0.750977993011475 0.250813007354736
0.750959992408752 0.245319962501526
0.750944018363953 0.240473031997681
0.750927925109863 0.235626935958862
0.750891923904419 0.225288033485413
0.75083601474762 0.207193970680237
0.750802040100098 0.196854948997498
0.752063989639282 0.185546040534973
0.752671957015991 0.172943949699402
0.753286004066467 0.16228199005127
0.753888010978699 0.148064970970154
0.755147933959961 0.136109948158264
0.757050037384033 0.122215986251831
0.75830602645874 0.109614968299866
0.759566068649292 0.0979830026626587
0.760175943374634 0.085705041885376
0.759490013122559 0.0747189521789551
0.75749397277832 0.059533953666687
0.756160020828247 0.0479030609130859
0.754166007041931 0.0336869955062866
0.753479957580566 0.0223790407180786
0.750848054885864 0.0110709667205811
0.748866081237793 8.59498977661133e-05
0.746886014938354 -0.0102519989013672
0.743628025054932 -0.0141290426254272
0.740370035171509 -0.0180050134658813
0.737761974334717 -0.0218809843063354
0.735153913497925 -0.0257569551467896
0.731894016265869 -0.0302799940109253
0.727335929870605 -0.0351250171661377
0.724077939987183 -0.0396469831466675
0.719521999359131 -0.0438460111618042
0.714959979057312 -0.0496599674224854
0.710399985313416 -0.0551519393920898
0.704541921615601 -0.0606429576873779
0.693494081497192 -0.0651619434356689
0.68504798412323 -0.0687140226364136
0.673354029655457 -0.072909951210022
0.661018013954163 -0.0745220184326172
0.651278018951416 -0.0761339664459229
0.643490076065063 -0.0767780542373657
0.63440203666687 -0.0777440071105957
0.624661922454834 -0.0793570280075073
0.611681938171387 -0.0799989700317383
0.591563940048218 -0.0799920558929443
0.577935934066772 -0.0803110599517822
0.55717396736145 -0.0793349742889404
0.539010047912598 -0.0777130126953125
0.521492004394531 -0.0764150619506836
0.512416005134583 -0.0738279819488525
0.499451994895935 -0.0693000555038452
0.48907995223999 -0.0660660266876221
0.478709936141968 -0.0618619918823242
0.467043995857239 -0.0576579570770264
0.458620071411133 -0.0537780523300171
0.450191974639893 -0.0502209663391113
0.439177989959717 -0.0456939935684204
0.428156018257141 -0.0427830219268799
0.413892030715942 -0.0389009714126587
0.40351402759552 -0.0366359949111938
0.388602018356323 -0.0330770015716553
0.375632047653198 -0.0304880142211914
0.360715985298157 -0.0275750160217285
0.34644603729248 -0.0256320238113403
0.330878019332886 -0.0240110158920288
0.312067985534668 -0.02109694480896
0.295850038528442 -0.0191529989242554
0.27963399887085 -0.0172100067138672
0.266011953353882 -0.0155899524688721
0.249145984649658 -0.0133229494094849
0.233579993247986 -0.0110559463500977
0.214761972427368 -0.0104039907455444
0.197247982025146 -0.0087820291519165
0.168053984642029 -0.0061880350112915
0.149888038635254 -0.00488996505737305
0.131721973419189 -0.00359201431274414
0.111608028411865 -0.00261604785919189
0.0973340272903442 -0.00196504592895508
0.0772199630737305 -0.0013120174407959
0.0655399560928345 -0.000661969184875488
0.055806040763855 -0.000658988952636719
0.0428299903869629 -0.000331997871398926
0.0317980051040649 -5.00679016113281e-06
0.0162240266799927 0
};
\addlegendentry{Experiment~\cite{bugg2002VelocityFieldTaylor}}
\addplot [line width=2pt, mediumaquamarine128230181]
table {%
0 0
0.0625 -0.000988483428955078
0.125 -0.000471591949462891
0.158813953399658 -0.000623703002929688
0.1875 -0.000996112823486328
0.25 -0.000866413116455078
0.283813953399658 -0.00123834609985352
0.3125 -0.00123834609985352
0.375 -0.00164461135864258
0.408813953399658 -0.00159358978271484
0.4375 -0.00123500823974609
0.471313953399658 -0.0017085075378418
0.5 -0.00325775146484375
0.533813953399658 -0.000874042510986328
0.5625 0.00249671936035156
0.596313953399658 0.000992298126220703
0.623831987380981 0.0133681297302246
0.64859938621521 0.0220885276794434
0.658813953399658 0.027198314666748
0.68321418762207 0.0409603118896484
0.706535816192627 0.0533385276794434
0.711900234222412 0.0588316917419434
0.721313953399658 0.0707511901855469
0.726010799407959 0.0757389068603516
0.733767509460449 0.0845885276794434
0.738853454589844 0.0989313125610352
0.749711275100708 0.115838527679443
0.751162767410278 0.130181312561035
0.752407073974609 0.147088527679443
0.752650737762451 0.161431312561035
0.753114223480225 0.178338527679443
0.753114223480225 0.192681312561035
0.754256963729858 0.209588527679443
0.754707813262939 0.240838527679443
0.755593776702881 0.223931312561035
0.75638222694397 0.272088527679443
0.75779914855957 0.286431312561035
};
\addlegendentry{NBKC, $D=32$}
\addplot [line width=2pt, mediumaquamarine64218145]
table {%
0 0
0.0625 -0.000187873840332031
0.109376668930054 -0.000586509704589844
0.234376668930054 -0.00246143341064453
0.3125 -0.00366449356079102
0.390626668930054 -0.00467205047607422
0.421876668930054 -0.00497722625732422
0.4375 -0.00535202026367188
0.453126668930054 -0.00547695159912109
0.46875 -0.00518798828125
0.484376668930054 -0.00510215759277344
0.5 -0.00532054901123047
0.515626668930054 -0.00535202026367188
0.53125 -0.00514125823974609
0.546876668930054 -0.00513315200805664
0.578126668930054 -0.00435209274291992
0.609376668930054 -0.0019688606262207
0.625 0.000210285186767578
0.640626668930054 0.00561761856079102
0.65625 0.0113978385925293
0.671876668930054 0.0129837989807129
0.676537990570068 0.0166091918945312
0.682535886764526 0.0191011428833008
0.692161560058594 0.0254364013671875
0.710956573486328 0.0347261428833008
0.717751264572144 0.0425376892089844
0.72038745880127 0.0503511428833008
0.723857402801514 0.0581626892089844
0.733976602554321 0.0692334175109863
0.734376668930054 0.0701165199279785
0.737051963806152 0.0770468711853027
0.740806102752686 0.0816011428833008
0.743245124816895 0.0894126892089844
0.747476816177368 0.0972261428833008
0.748353242874146 0.105037689208984
0.748726606369019 0.112851142883301
0.749437570571899 0.120662689208984
0.750693798065186 0.128476142883301
0.752665519714355 0.136287689208984
0.752048254013062 0.144101142883301
0.75006103515625 0.151912689208984
0.751000165939331 0.167537689208984
0.751071929931641 0.175351142883301
0.750779628753662 0.183162689208984
0.758352994918823 0.190976142883301
0.750292301177979 0.198789596557617
0.75151252746582 0.206601142883301
0.757242679595947 0.214414596557617
0.763671636581421 0.220944881439209
0.765552520751953 0.222226142883301
0.756364583969116 0.230037689208984
0.750918865203857 0.237851142883301
0.758225917816162 0.24243688583374
0.765626668930054 0.247952938079834
0.776284217834473 0.253476142883301
0.765626668930054 0.258890151977539
0.757950782775879 0.263514518737793
0.750187635421753 0.269101142883301
0.757863283157349 0.276553630828857
0.774193048477173 0.284726142883301
0.765626668930054 0.289468765258789
0.758357048034668 0.296882152557373
};
\addlegendentry{NBKC, $D=64$}
\addplot [line width=2pt, springgreen0205108]
table {%
0 0
0.03125 -0.000116825103759766
0.125 -0.000846385955810547
0.206410884857178 -0.00332498550415039
0.234375 -0.00458765029907227
0.25 -0.00553560256958008
0.29374372959137 -0.00668573379516602
0.3125 -0.00732612609863281
0.359375 -0.00809526443481445
0.375 -0.00866842269897461
0.40311872959137 -0.0106415748596191
0.421875 -0.011782169342041
0.4375 -0.0127663612365723
0.46875 -0.0138339996337891
0.515625 -0.014493465423584
0.53125 -0.0143780708312988
0.546875 -0.0145244598388672
0.5625 -0.0143089294433594
0.578125 -0.0137500762939453
0.59061872959137 -0.0134000778198242
0.59375 -0.0130524635314941
0.60787832736969 -0.0117135047912598
0.62186872959137 -0.00911855697631836
0.625 -0.00896215438842773
0.640625 -0.00688266754150391
0.643258333206177 -0.00604391098022461
0.654477834701538 -0.00327634811401367
0.65625 -0.00255584716796875
0.671875 0.000343799591064453
0.679831385612488 0.00453615188598633
0.698524832725525 0.0123486518859863
0.705291867256165 0.0157256126403809
0.715111970901489 0.0201611518859863
0.720384478569031 0.023406982421875
0.725706934928894 0.0279736518859863
0.726924896240234 0.0295391082763672
0.734375 0.0354657173156738
0.735016584396362 0.0357861518859863
0.736508131027222 0.0373516082763672
0.741668105125427 0.0435986518859863
0.742717623710632 0.0498456954956055
0.74673318862915 0.0576581954956055
0.746974349021912 0.0592236518859863
0.749110698699951 0.0630092620849609
0.750720143318176 0.0645751953125
0.752207040786743 0.0670361518859863
0.755005598068237 0.0732831954956055
0.754989862442017 0.0748486518859863
0.756451606750488 0.0810956954956055
0.757159352302551 0.0826611518859863
0.758240342140198 0.0982861518859863
0.757765650749207 0.104533195495605
0.758490681648254 0.112345695495605
0.758318543434143 0.113911151885986
0.75830614566803 0.121723651885986
0.757793784141541 0.129536151885986
0.758350014686584 0.137348651885986
0.758015632629395 0.145161151885986
0.760464429855347 0.151408195495605
0.764200806617737 0.152973651885986
0.761431932449341 0.154539108276367
0.761506915092468 0.160786151885986
0.764268040657043 0.167033195495605
0.764108300209045 0.168598651885986
0.762933969497681 0.174845695495605
0.763839483261108 0.176411151885986
0.765834808349609 0.178809642791748
0.768733382225037 0.182658195495605
0.769171714782715 0.184223651885986
0.767125010490417 0.192036151885986
0.770443558692932 0.199848651885986
0.769181609153748 0.206095695495605
0.769206643104553 0.207661151885986
0.771193265914917 0.213908195495605
0.771374821662903 0.215473651885986
0.770812630653381 0.221720695495605
0.770975470542908 0.223286151885986
0.772412180900574 0.229533195495605
0.772556185722351 0.231098651885986
0.772553324699402 0.238911151885986
0.772937536239624 0.246723651885986
0.773756146430969 0.254536151885986
0.773415565490723 0.262348651885986
0.77267849445343 0.268595695495605
0.772716045379639 0.270161151885986
0.774165272712708 0.276408195495605
0.774149656295776 0.277973651885986
0.773599982261658 0.284220695495605
0.773759603500366 0.285786151885986
0.773553133010864 0.293598651885986
0.77336573600769 0.299845695495605
};
\addlegendentry{NBKC, $D=128$}
\end{axis}

\end{tikzpicture}

%% file: figures/taylor-bubble/convergence-shape-front-om.tex
% This file was created with tikzplotlib v0.10.1.
\begin{tikzpicture}

\definecolor{darkgray176}{RGB}{176,176,176}
\definecolor{lightgray204}{RGB}{204,204,204}
\definecolor{mediumorchid17588186}{RGB}{175,88,186}
\definecolor{orchid195130203}{RGB}{109,55,116}
\definecolor{plum215172220}{RGB}{225,192,229}

\begin{axis}[
height=\figureheight,
legend cell align={left},
legend style={
	fill opacity=0.8,
	draw opacity=1,
	text opacity=1,
	at={(0.01,0.01)},
	anchor=south west,
	draw=lightgray204
},
tick align=outside,
tick pos=left,
width=\figurewidth,
x grid style={darkgray176},
xlabel={\(\displaystyle r^{*}\)},
xmajorgrids,
xmin=-0.0373953402040625, xmax=0.785302144285313,
xtick style={color=black},
y grid style={darkgray176},
ylabel style={rotate=-90.0},
ylabel={\(\displaystyle z^{*}\)},
ymajorgrids,
ymin=-0.628900408746093, ymax=0.0299476385117187,
ytick style={color=black}
]
\addplot [very thick, black, mark=*, mark size=1, mark options={solid}, only marks]
table {%
0 0
0.0293359756469727 0
0.0573979616165161 -0.000633955001831055
0.0854599475860596 -0.00253605842590332
0.119897961616516 -0.00570595264434814
0.160714030265808 -0.0120450258255005
0.192602038383484 -0.0183850526809692
0.219388008117676 -0.0247249603271484
0.244897961616516 -0.0316979885101318
0.271683931350708 -0.0412089824676514
0.293367981910706 -0.0494489669799805
0.320153951644897 -0.0595920085906982
0.340561985969543 -0.0684679746627808
0.362244009971619 -0.0773429870605469
0.380102038383484 -0.0862189531326294
0.39795994758606 -0.0950939655303955
0.415816068649292 -0.10523796081543
0.429846048355103 -0.112844944000244
0.442602038383484 -0.121086955070496
0.456632018089294 -0.129328012466431
0.470664024353027 -0.138838052749634
0.48852002620697 -0.149615049362183
0.503826022148132 -0.162294030189514
0.519132018089294 -0.173071980476379
0.533164024353027 -0.183848977088928
0.543367981910706 -0.193992018699646
0.554846048355103 -0.204769968986511
0.565052032470703 -0.214913964271545
0.57397997379303 -0.225057005882263
0.581632018089294 -0.235200047492981
0.593111991882324 -0.247879028320312
0.603316068649292 -0.259925007820129
0.614795923233032 -0.273237943649292
0.623723983764648 -0.286550998687744
0.633928060531616 -0.299864053726196
0.641582012176514 -0.312543034553528
0.64795994758606 -0.325222969055176
0.656888008117676 -0.339169979095459
0.66198992729187 -0.351848959922791
0.667092084884644 -0.365162014961243
0.672194004058838 -0.376574039459229
0.679846048355103 -0.387984991073608
0.686223983764648 -0.401932001113892
0.690052032470703 -0.414610981941223
0.692601919174194 -0.424754977226257
0.697704076766968 -0.438068032264709
0.701529979705811 -0.453283071517944
0.705358028411865 -0.464694023132324
0.707906007766724 -0.476740002632141
0.71045994758606 -0.48751699924469
0.711734056472778 -0.49956202507019
0.715561985969543 -0.511608004570007
0.716835975646973 -0.523019075393677
0.718111991882324 -0.533161997795105
0.720664024353027 -0.544573068618774
0.721935987472534 -0.560423016548157
0.72448992729187 -0.572468042373657
0.72448992729187 -0.583878993988037
0.725765943527222 -0.595291018486023
};
\addlegendentry{Experiment~\cite{bugg2002VelocityFieldTaylor}}
\addplot [line width=2pt, plum215172220, dashed]
table {%
0 0
0.0312505960464478 0
0.0937505960464478 -0.00412511825561523
0.138541579246521 -0.0124845504760742
0.156250596046448 -0.0156874656677246
0.218750596046448 -0.0253124237060547
0.273225545883179 -0.0405311584472656
0.281250596046448 -0.0436563491821289
0.343750596046448 -0.0611872673034668
0.387679100036621 -0.0797343254089355
0.406250596046448 -0.0876874923706055
0.437456846237183 -0.103031158447266
0.468750596046448 -0.120093822479248
0.493356823921204 -0.134281158447266
0.531250596046448 -0.161593437194824
0.537619352340698 -0.165531158447266
0.568906903266907 -0.196781158447266
0.593750596046448 -0.219124794006348
0.606175541877747 -0.228031158447266
0.6273193359375 -0.259281158447266
0.639644384384155 -0.290531158447266
0.673694372177124 -0.321781158447266
0.685744404792786 -0.353031158447266
0.684706807136536 -0.384281158447266
0.697588086128235 -0.415531158447266
0.715663075447083 -0.446781158447266
0.718944311141968 -0.478031158447266
0.741144299507141 -0.509281158447266
0.739306807518005 -0.540531158447266
0.746813058853149 -0.571781158447266
0.747906804084778 -0.587406635284424
};
\addlegendentry{OM, $D=32$}
\addplot [line width=2pt, orchid195130203, dashed]
table {%
0 0
0.03125 -0.000226974487304688
0.0625 -0.000905990600585938
0.0781247615814209 -0.00135898590087891
0.109374761581421 -0.00318717956542969
0.140624761581421 -0.00659370422363281
0.171874761581421 -0.0136241912841797
0.203124761581421 -0.0178747177124023
0.234374761581421 -0.0243434906005859
0.238043546676636 -0.025609016418457
0.265624761581421 -0.0325307846069336
0.296874761581421 -0.0413436889648438
0.328124761581421 -0.0507335662841797
0.359374761581421 -0.0627651214599609
0.386962175369263 -0.0741090774536133
0.404212474822998 -0.0819215774536133
0.421874761581421 -0.0901088714599609
0.453124761581421 -0.106546401977539
0.484374761581421 -0.125015258789062
0.500346660614014 -0.134984016418457
0.523965358734131 -0.150609016418457
0.539359092712402 -0.166234016418457
0.546874761581421 -0.171359062194824
0.560437202453613 -0.181859016418457
0.574087381362915 -0.197484016418457
0.578124761581421 -0.200140476226807
0.591987371444702 -0.213109016418457
0.600852966308594 -0.228734016418457
0.609374761581421 -0.236405849456787
0.618909120559692 -0.244359016418457
0.634812355041504 -0.275609016418457
0.640624761581421 -0.28068733215332
0.65069055557251 -0.291234016418457
0.655452966690063 -0.306859016418457
0.660893678665161 -0.322484016418457
0.667709112167358 -0.338109016418457
0.671874761581421 -0.341796398162842
0.682249784469604 -0.353734016418457
0.6869215965271 -0.369359016418457
0.686674833297729 -0.384984016418457
0.690446615219116 -0.400609016418457
0.70590615272522 -0.431859016418457
0.710468530654907 -0.447484016418457
0.715768575668335 -0.463109016418457
0.714524745941162 -0.478734016418457
0.718634128570557 -0.494359016418457
0.719837427139282 -0.509984016418457
0.722021579742432 -0.525609016418457
0.720168590545654 -0.556859016418457
0.728184223175049 -0.572484016418457
0.728515386581421 -0.588109016418457
0.734374761581421 -0.598952770233154
};
\addlegendentry{OM, $D=64$}
\addplot [line width=2pt, mediumorchid17588186, dashed]
table {%
0 0
0.03125 -0.000117301940917969
0.0546867847442627 -0.000624656677246094
0.0703117847442627 -0.00140666961669922
0.0859367847442627 -0.00265598297119141
0.109375 -0.00574207305908203
0.117186784744263 -0.00656223297119141
0.140625 -0.00847721099853516
0.148436784744263 -0.00921916961669922
0.171875 -0.0136327743530273
0.179686784744263 -0.0148439407348633
0.195311784744263 -0.0167970657348633
0.21875 -0.0219535827636719
0.226561784744263 -0.0233592987060547
0.243249416351318 -0.0269927978515625
0.244311809539795 -0.02734375
0.28237509727478 -0.0370311737060547
0.302778959274292 -0.04296875
0.304686784744263 -0.0437498092651367
0.320311784744263 -0.0481252670288086
0.370997667312622 -0.06640625
0.386256694793701 -0.0726957321166992
0.410544157028198 -0.0836715698242188
0.423317909240723 -0.08984375
0.445311784744263 -0.101093292236328
0.467406988143921 -0.11328125
0.480708599090576 -0.12109375
0.497566223144531 -0.132851600646973
0.505377769470215 -0.138280868530273
0.514861822128296 -0.14453125
0.525735139846802 -0.15234375
0.533960103988647 -0.16015625
0.544907093048096 -0.16796875
0.552094459533691 -0.17578125
0.554686784744263 -0.177499771118164
0.561839818954468 -0.18359375
0.5696702003479 -0.19179630279541
0.570311784744263 -0.192187309265137
0.577710151672363 -0.19921875
0.583280563354492 -0.20703125
0.585936784744263 -0.208906173706055
0.592608690261841 -0.21484375
0.596686840057373 -0.22265625
0.605717897415161 -0.23046875
0.610264778137207 -0.23828125
0.616608619689941 -0.246562480926514
0.617186784744263 -0.247031211853027
0.623249292373657 -0.25390625
0.62621808052063 -0.26171875
0.630749225616455 -0.26953125
0.632811784744263 -0.271015644073486
0.638952255249023 -0.27734375
0.640139818191528 -0.28515625
0.648139953613281 -0.29296875
0.649077415466309 -0.30078125
0.656139850616455 -0.30859375
0.657389879226685 -0.31640625
0.659405469894409 -0.32421875
0.665991544723511 -0.330351829528809
0.66792106628418 -0.33203125
0.671366214752197 -0.343749523162842
0.67238974571228 -0.34765625
0.672999143600464 -0.35546875
0.677014827728271 -0.36328125
0.681780576705933 -0.368594169616699
0.683874130249023 -0.37109375
0.685921192169189 -0.37890625
0.687374114990234 -0.38671875
0.687843084335327 -0.39453125
0.689014911651611 -0.40234375
0.692155361175537 -0.41015625
0.699264764785767 -0.41796875
0.699671030044556 -0.42578125
0.701749324798584 -0.43359375
0.703264951705933 -0.44140625
0.703999280929565 -0.44921875
0.703686714172363 -0.45703125
0.705296039581299 -0.46484375
0.708788156509399 -0.476562023162842
0.709811687469482 -0.48046875
0.710936784744263 -0.481468677520752
0.716311693191528 -0.48828125
0.715546131134033 -0.49609375
0.717741250991821 -0.507812023162842
0.718389749526978 -0.51171875
0.71870231628418 -0.523437023162842
0.718733549118042 -0.53515625
0.720389842987061 -0.55078125
0.723702430725098 -0.55859375
0.723889827728271 -0.56640625
0.728842973709106 -0.58203125
0.729233503341675 -0.58984375
0.732483625411987 -0.59765625
};
\addlegendentry{OM, $D=128$}
\end{axis}

\end{tikzpicture}

%% file: figures/taylor-bubble/convergence-shape-tail-om.tex
% This file was created with tikzplotlib v0.10.1.
\begin{tikzpicture}

\definecolor{darkgray176}{RGB}{176,176,176}
\definecolor{lightgray204}{RGB}{204,204,204}
\definecolor{mediumorchid17588186}{RGB}{175,88,186}
\definecolor{orchid195130203}{RGB}{109,55,116}
\definecolor{plum215172220}{RGB}{225,192,229}

\begin{axis}[
height=\figureheight,
legend cell align={left},
legend style={
	fill opacity=0.8,
	draw opacity=1,
	text opacity=1,
	at={(0.01,0.99)},
	anchor=north west,
	draw=lightgray204
},
tick align=outside,
tick pos=left,
width=\figurewidth,
x grid style={darkgray176},
xlabel={\(\displaystyle r^{*}\)},
xmajorgrids,
xmin=-0.03864940404875, xmax=0.81163748502375,
xtick style={color=black},
y grid style={darkgray176},
ylabel style={rotate=-90.0},
ylabel={\(\displaystyle z^{*}\)},
ymajorgrids,
ymin=-0.176890158633594, ymax=0.321882510180469,
ytick style={color=black}
]
\addplot [very thick, black, mark=*, mark size=1, mark options={solid}, only marks]
table {%
0.751070022583008 0.279245018959045
0.751060009002686 0.276661038398743
0.75104808807373 0.272783041000366
0.751029968261719 0.267291069030762
0.751013994216919 0.262120962142944
0.750993967056274 0.256628036499023
0.750977993011475 0.250813007354736
0.750959992408752 0.245319962501526
0.750944018363953 0.240473031997681
0.750927925109863 0.235626935958862
0.750891923904419 0.225288033485413
0.75083601474762 0.207193970680237
0.750802040100098 0.196854948997498
0.752063989639282 0.185546040534973
0.752671957015991 0.172943949699402
0.753286004066467 0.16228199005127
0.753888010978699 0.148064970970154
0.755147933959961 0.136109948158264
0.757050037384033 0.122215986251831
0.75830602645874 0.109614968299866
0.759566068649292 0.0979830026626587
0.760175943374634 0.085705041885376
0.759490013122559 0.0747189521789551
0.75749397277832 0.059533953666687
0.756160020828247 0.0479030609130859
0.754166007041931 0.0336869955062866
0.753479957580566 0.0223790407180786
0.750848054885864 0.0110709667205811
0.748866081237793 8.59498977661133e-05
0.746886014938354 -0.0102519989013672
0.743628025054932 -0.0141290426254272
0.740370035171509 -0.0180050134658813
0.737761974334717 -0.0218809843063354
0.735153913497925 -0.0257569551467896
0.731894016265869 -0.0302799940109253
0.727335929870605 -0.0351250171661377
0.724077939987183 -0.0396469831466675
0.719521999359131 -0.0438460111618042
0.714959979057312 -0.0496599674224854
0.710399985313416 -0.0551519393920898
0.704541921615601 -0.0606429576873779
0.693494081497192 -0.0651619434356689
0.68504798412323 -0.0687140226364136
0.673354029655457 -0.072909951210022
0.661018013954163 -0.0745220184326172
0.651278018951416 -0.0761339664459229
0.643490076065063 -0.0767780542373657
0.63440203666687 -0.0777440071105957
0.624661922454834 -0.0793570280075073
0.611681938171387 -0.0799989700317383
0.591563940048218 -0.0799920558929443
0.577935934066772 -0.0803110599517822
0.55717396736145 -0.0793349742889404
0.539010047912598 -0.0777130126953125
0.521492004394531 -0.0764150619506836
0.512416005134583 -0.0738279819488525
0.499451994895935 -0.0693000555038452
0.48907995223999 -0.0660660266876221
0.478709936141968 -0.0618619918823242
0.467043995857239 -0.0576579570770264
0.458620071411133 -0.0537780523300171
0.450191974639893 -0.0502209663391113
0.439177989959717 -0.0456939935684204
0.428156018257141 -0.0427830219268799
0.413892030715942 -0.0389009714126587
0.40351402759552 -0.0366359949111938
0.388602018356323 -0.0330770015716553
0.375632047653198 -0.0304880142211914
0.360715985298157 -0.0275750160217285
0.34644603729248 -0.0256320238113403
0.330878019332886 -0.0240110158920288
0.312067985534668 -0.02109694480896
0.295850038528442 -0.0191529989242554
0.27963399887085 -0.0172100067138672
0.266011953353882 -0.0155899524688721
0.249145984649658 -0.0133229494094849
0.233579993247986 -0.0110559463500977
0.214761972427368 -0.0104039907455444
0.197247982025146 -0.0087820291519165
0.168053984642029 -0.0061880350112915
0.149888038635254 -0.00488996505737305
0.131721973419189 -0.00359201431274414
0.111608028411865 -0.00261604785919189
0.0973340272903442 -0.00196504592895508
0.0772199630737305 -0.0013120174407959
0.0655399560928345 -0.000661969184875488
0.055806040763855 -0.000658988952636719
0.0428299903869629 -0.000331997871398926
0.0317980051040649 -5.00679016113281e-06
0.0162240266799927 0
};
\addlegendentry{Experiment~\cite{bugg2002VelocityFieldTaylor}}
\addplot [line width=2pt, plum215172220, dashed]
table {%
0 0
0.0312505960464478 0
0.0937505960464478 -0.000718593597412109
0.156250596046448 -0.00271844863891602
0.218750596046448 -0.00771856307983398
0.267588376998901 -0.0161561965942383
0.281250596046448 -0.0185937881469727
0.343750596046448 -0.0265622138977051
0.406250596046448 -0.0363125801086426
0.43446934223175 -0.0449686050415039
0.468750596046448 -0.054905891418457
0.531250596046448 -0.0689687728881836
0.545313119888306 -0.0762186050415039
0.548213124275208 -0.107468605041504
0.531250596046448 -0.117656230926514
0.504169344902039 -0.138718605041504
0.531250596046448 -0.152249813079834
0.593750596046448 -0.154218673706055
0.656250596046448 -0.140218734741211
0.659225583076477 -0.138718605041504
0.656250596046448 -0.137218475341797
0.624900579452515 -0.107468605041504
0.656250596046448 -0.0841250419616699
0.718750596046448 -0.0784997940063477
0.72333812713623 -0.0762186050415039
0.718750596046448 -0.0698437690734863
0.703981876373291 -0.0449686050415039
0.718750596046448 -0.0333747863769531
0.740263104438782 -0.0137186050415039
0.749119400978088 0.0175313949584961
0.752138137817383 0.0487813949584961
0.750463128089905 0.0800313949584961
0.751506805419922 0.111281394958496
0.764494299888611 0.142531394958496
0.756144404411316 0.173781394958496
0.772988080978394 0.205031394958496
0.756619334220886 0.236281394958496
0.770156860351562 0.267531394958496
0.764775633811951 0.298781394958496
};
\addlegendentry{OM, $D=32$}
\addplot [line width=2pt, orchid195130203, dashed]
table {%
0 0
0.0625 -0.000171661376953125
0.0781247615814209 -0.000265598297119141
0.140624761581421 -0.00170278549194336
0.171874761581421 -0.00307798385620117
0.203124761581421 -0.00528097152709961
0.25 -0.0113749504089355
0.265624761581421 -0.0131402015686035
0.296874761581421 -0.0155467987060547
0.328124761581421 -0.0183906555175781
0.390624761581421 -0.0299530029296875
0.421874761581421 -0.0342812538146973
0.441205978393555 -0.0394372940063477
0.453124761581421 -0.0430779457092285
0.484374761581421 -0.0489840507507324
0.509618520736694 -0.0571246147155762
0.515624761581421 -0.0591874122619629
0.546874761581421 -0.0665154457092285
0.559212207794189 -0.0706872940063477
0.578124761581421 -0.0780935287475586
0.590062379837036 -0.0863122940063477
0.609374761581421 -0.0958123207092285
0.640624761581421 -0.0905623435974121
0.647130966186523 -0.0863122940063477
0.671874761581421 -0.0738277435302734
0.679990530014038 -0.0706872940063477
0.689028024673462 -0.0550622940063477
0.703124761581421 -0.0447497367858887
0.712424755096436 -0.0394372940063477
0.72404670715332 -0.0238122940063477
0.721293687820435 -0.00818729400634766
0.729781150817871 0.00743770599365234
0.734374761581421 0.013359546661377
0.741246700286865 0.0230627059936523
0.746456146240234 0.0386877059936523
0.749381065368652 0.0543127059936523
0.751193523406982 0.0699377059936523
0.750093460083008 0.0855627059936523
0.750215530395508 0.101187705993652
0.754074811935425 0.116812705993652
0.750030994415283 0.132437705993652
0.757646799087524 0.148062705993652
0.752456188201904 0.163687705993652
0.758977890014648 0.179312705993652
0.753796577453613 0.194937705993652
0.761605978012085 0.210562705993652
0.75330924987793 0.226187705993652
0.761387348175049 0.241812705993652
0.756456136703491 0.257437705993652
0.761384248733521 0.273062705993652
0.753546714782715 0.288687705993652
0.759535789489746 0.296500205993652
};
\addlegendentry{OM, $D=64$}
\addplot [line width=2pt, mediumorchid17588186, dashed]
table {%
0 0
0.046875 -0.000183582305908203
0.0859367847442627 -0.000726699829101562
0.15625 -0.00202322006225586
0.179686784744263 -0.00276565551757812
0.203125 -0.00415658950805664
0.214087009429932 -0.00511741638183594
0.226561784744263 -0.00649213790893555
0.25 -0.00826549530029297
0.296875 -0.0111522674560547
0.304686784744263 -0.0119137763977051
0.320311784744263 -0.0145468711853027
0.335936784744263 -0.0162811279296875
0.359375 -0.0182380676269531
0.367186784744263 -0.0189919471740723
0.382811784744263 -0.0222654342651367
0.398436784744263 -0.0243358612060547
0.414061784744263 -0.0259842872619629
0.445311784744263 -0.0319609642028809
0.460936784744263 -0.0339765548706055
0.474828720092773 -0.0372495651245117
0.476561784744263 -0.0377731323242188
0.507811784744263 -0.0431637763977051
0.512266397476196 -0.0445389747619629
0.523436784744263 -0.0472731590270996
0.539061784744263 -0.0496950149536133
0.551988124847412 -0.053187370300293
0.554686784744263 -0.054023265838623
0.570311784744263 -0.056617259979248
0.601561784744263 -0.0631952285766602
0.632811784744263 -0.0666637420654297
0.648436784744263 -0.0632500648498535
0.664061784744263 -0.0610857009887695
0.66624927520752 -0.0601639747619629
0.679686784744263 -0.0487656593322754
0.692257404327393 -0.0430507659912109
0.695311784744263 -0.041562557220459
0.70015549659729 -0.0367264747619629
0.702608585357666 -0.0289139747619629
0.70810866355896 -0.0211014747619629
0.710936784744263 -0.0183749198913574
0.715405464172363 -0.0132889747619629
0.717296123504639 -0.00547647476196289
0.724202394485474 0.00233602523803711
0.726717948913574 0.0101485252380371
0.732717990875244 0.0179610252380371
0.733999252319336 0.0335860252380371
0.735202312469482 0.0413985252380371
0.738030433654785 0.0492110252380371
0.740499258041382 0.0648360252380371
0.742186784744263 0.0669922828674316
0.745967864990234 0.0726485252380371
0.74898362159729 0.0804610252380371
0.749733686447144 0.0882735252380371
0.749835252761841 0.0999927520751953
0.750335216522217 0.115617752075195
0.750702381134033 0.139055252075195
0.750921010971069 0.142961025238037
0.752561807632446 0.158586025238037
0.753389835357666 0.166398525238037
0.751514911651611 0.174211025238037
0.754358530044556 0.182023525238037
0.756639957427979 0.189836025238037
0.7538743019104 0.197648525238037
0.752811670303345 0.205461025238037
0.758655309677124 0.212472438812256
0.759499311447144 0.213273525238037
0.757811784744263 0.214515686035156
0.751718044281006 0.221086025238037
0.757811784744263 0.227351665496826
0.759936809539795 0.228898525238037
0.757811784744263 0.23118782043457
0.753499269485474 0.236711025238037
0.757811784744263 0.24469518661499
0.75373363494873 0.252336025238037
0.761311769485474 0.260148525238037
0.757811784744263 0.262726783752441
0.75204610824585 0.267961025238037
0.763202428817749 0.275773525238037
0.754725456237793 0.281425952911377
0.751639842987061 0.283586025238037
0.760358333587646 0.289734363555908
0.762905359268188 0.291398525238037
0.752858638763428 0.299211025238037
};
\addlegendentry{OM, $D=128$}
\end{axis}

\end{tikzpicture}

%% file: figures/taylor-bubble/convergence-shape-front-om3.tex
% This file was created with tikzplotlib v0.10.1.
\begin{tikzpicture}

\definecolor{burlywood248194145}{RGB}{250,209,172}
\definecolor{darkgray176}{RGB}{176,176,176}
\definecolor{darkorange24213334}{RGB}{242,133,34}
\definecolor{lightgray204}{RGB}{204,204,204}
\definecolor{sandybrown24516489}{RGB}{181,100,26}

\begin{axis}[
height=\figureheight,
legend cell align={left},
legend style={
	fill opacity=0.8,
	draw opacity=1,
	text opacity=1,
	at={(0.01,0.01)},
	anchor=south west,
	draw=lightgray204
},
tick align=outside,
tick pos=left,
width=\figurewidth,
x grid style={darkgray176},
xlabel={\(\displaystyle r^{*}\)},
xmajorgrids,
xmin=-0.037621843815, xmax=0.790058720115,
xtick style={color=black},
y grid style={darkgray176},
ylabel style={rotate=-90.0},
ylabel={\(\displaystyle z^{*}\)},
ymajorgrids,
ymin=-0.629934310917187, ymax=0.0299968719484375,
ytick style={color=black}
]
\addplot [very thick, black, mark=*, mark size=1, mark options={solid}, only marks]
table {%
0 0
0.0293359756469727 0
0.0573979616165161 -0.000633955001831055
0.0854599475860596 -0.00253605842590332
0.119897961616516 -0.00570595264434814
0.160714030265808 -0.0120450258255005
0.192602038383484 -0.0183850526809692
0.219388008117676 -0.0247249603271484
0.244897961616516 -0.0316979885101318
0.271683931350708 -0.0412089824676514
0.293367981910706 -0.0494489669799805
0.320153951644897 -0.0595920085906982
0.340561985969543 -0.0684679746627808
0.362244009971619 -0.0773429870605469
0.380102038383484 -0.0862189531326294
0.39795994758606 -0.0950939655303955
0.415816068649292 -0.10523796081543
0.429846048355103 -0.112844944000244
0.442602038383484 -0.121086955070496
0.456632018089294 -0.129328012466431
0.470664024353027 -0.138838052749634
0.48852002620697 -0.149615049362183
0.503826022148132 -0.162294030189514
0.519132018089294 -0.173071980476379
0.533164024353027 -0.183848977088928
0.543367981910706 -0.193992018699646
0.554846048355103 -0.204769968986511
0.565052032470703 -0.214913964271545
0.57397997379303 -0.225057005882263
0.581632018089294 -0.235200047492981
0.593111991882324 -0.247879028320312
0.603316068649292 -0.259925007820129
0.614795923233032 -0.273237943649292
0.623723983764648 -0.286550998687744
0.633928060531616 -0.299864053726196
0.641582012176514 -0.312543034553528
0.64795994758606 -0.325222969055176
0.656888008117676 -0.339169979095459
0.66198992729187 -0.351848959922791
0.667092084884644 -0.365162014961243
0.672194004058838 -0.376574039459229
0.679846048355103 -0.387984991073608
0.686223983764648 -0.401932001113892
0.690052032470703 -0.414610981941223
0.692601919174194 -0.424754977226257
0.697704076766968 -0.438068032264709
0.701529979705811 -0.453283071517944
0.705358028411865 -0.464694023132324
0.707906007766724 -0.476740002632141
0.71045994758606 -0.48751699924469
0.711734056472778 -0.49956202507019
0.715561985969543 -0.511608004570007
0.716835975646973 -0.523019075393677
0.718111991882324 -0.533161997795105
0.720664024353027 -0.544573068618774
0.721935987472534 -0.560423016548157
0.72448992729187 -0.572468042373657
0.72448992729187 -0.583878993988037
0.725765943527222 -0.595291018486023
};
\addlegendentry{Experiment~\cite{bugg2002VelocityFieldTaylor}}
\addplot [line width=2pt, burlywood248194145, dash pattern=on 1pt off 3pt on 3pt off 3pt]
table {%
0 -0.000796794891357422
0.0312494039535522 0
0.0937494039535522 -0.00137472152709961
0.156249403953552 -0.0078125
0.218749403953552 -0.0157184600830078
0.281249403953552 -0.0223126411437988
0.343749403953552 -0.0344061851501465
0.352799415588379 -0.0374374389648438
0.406249403953552 -0.0512499809265137
0.458661913871765 -0.0686874389648438
0.468749403953552 -0.0728750228881836
0.531249403953552 -0.10003137588501
0.577955603599548 -0.131187438964844
0.593749403953552 -0.142499923706055
0.618805646896362 -0.162437438964844
0.637974381446838 -0.193687438964844
0.656249403953552 -0.213375091552734
0.668455600738525 -0.224937438964844
0.684786915779114 -0.256187438964844
0.69508695602417 -0.287437438964844
0.697811961174011 -0.318687438964844
0.718749403953552 -0.344749927520752
0.72443687915802 -0.349937438964844
0.733636856079102 -0.381187438964844
0.743949413299561 -0.412437438964844
0.747136950492859 -0.443687438964844
0.746436953544617 -0.474937438964844
0.748449444770813 -0.506187438964844
0.749380707740784 -0.537437438964844
0.752436876296997 -0.568687438964844
0.75206196308136 -0.599937438964844
};
\addlegendentry{OM3, $D=32$}
\addplot [line width=2pt, sandybrown24516489, dash pattern=on 1pt off 3pt on 3pt off 3pt]
table {%
0 -0.000140190124511719
0.03125 -2.33650207519531e-05
0.0625 -0.000281333923339844
0.09375 -0.0009765625
0.140624642372131 -0.00257778167724609
0.171874642372131 -0.00445270538330078
0.203124642372131 -0.00764036178588867
0.234374642372131 -0.0139374732971191
0.265624642372131 -0.0178279876708984
0.296874642372131 -0.0225625038146973
0.319116592407227 -0.0285310745239258
0.328124642372131 -0.0308279991149902
0.359374642372131 -0.0370464324951172
0.390624642372131 -0.0467653274536133
0.421874642372131 -0.0557026863098145
0.426318407058716 -0.0574841499328613
0.453124642372131 -0.0669999122619629
0.484374642372131 -0.0804529190063477
0.501608967781067 -0.0887341499328613
0.515624642372131 -0.0961089134216309
0.530046463012695 -0.104359149932861
0.55481219291687 -0.119984149932861
0.571180939674377 -0.135609149932861
0.578124642372131 -0.140281200408936
0.592005848884583 -0.151234149932861
0.604877710342407 -0.166859149932861
0.609374642372131 -0.170062065124512
0.622471570968628 -0.182484149932861
0.6308434009552 -0.198109149932861
0.640624642372131 -0.208374977111816
0.64662778377533 -0.213734149932861
0.656537175178528 -0.229359149932861
0.660040259361267 -0.244984149932861
0.671874642372131 -0.257905960083008
0.675405859947205 -0.260609149932861
0.68187153339386 -0.276234149932861
0.688994884490967 -0.299671649932861
0.69147777557373 -0.307484149932861
0.693884015083313 -0.323109149932861
0.703873157501221 -0.337898254394531
0.704621553421021 -0.338734149932861
0.714080929756165 -0.354359149932861
0.71708083152771 -0.377796649932861
0.718165278434753 -0.385609149932861
0.719271540641785 -0.416859149932861
0.721149682998657 -0.432484149932861
0.724565267562866 -0.448109149932861
0.729505896568298 -0.463734149932861
0.730268359184265 -0.479359149932861
0.738249659538269 -0.494984149932861
0.742724657058716 -0.510609149932861
0.745799660682678 -0.526234149932861
0.746752738952637 -0.541859149932861
0.748502731323242 -0.557484149932861
0.749774694442749 -0.573109149932861
0.749918341636658 -0.588734149932861
0.74958872795105 -0.596546649932861
};
\addlegendentry{OM3, $D=64$}
\addplot [line width=2pt, darkorange24213334, dash pattern=on 1pt off 3pt on 3pt off 3pt]
table {%
0 -0.000233650207519531
0.015625 -3.814697265625e-05
0.0390616655349731 -7.72476196289062e-05
0.058336615562439 -0.000546455383300781
0.140625 -0.00449180603027344
0.164061665534973 -0.00593662261962891
0.181364178657532 -0.00804710388183594
0.189175844192505 -0.00933551788330078
0.195311665534973 -0.0103120803833008
0.226561665534973 -0.0133590698242188
0.250316381454468 -0.0175771713256836
0.257811665534973 -0.0189838409423828
0.273436665534973 -0.0210151672363281
0.304686665534973 -0.0275774002075195
0.322160363197327 -0.0312108993530273
0.324008584022522 -0.0317964553833008
0.335936665534973 -0.0348434448242188
0.353542327880859 -0.038945198059082
0.355522632598877 -0.0396089553833008
0.390833497047424 -0.049687385559082
0.403174638748169 -0.053593635559082
0.437865614891052 -0.0657806396484375
0.448040843009949 -0.0696873664855957
0.47269880771637 -0.0804290771484375
0.485555410385132 -0.0864839553833008
0.496659755706787 -0.0919919013977051
0.507811665534973 -0.0978903770446777
0.523436665534973 -0.106952667236328
0.528367877006531 -0.109921455383301
0.544518113136292 -0.121639728546143
0.552131652832031 -0.127264976501465
0.56100070476532 -0.133358955383301
0.570311665534973 -0.141561985015869
0.587077379226685 -0.156015396118164
0.588217973709106 -0.156796455383301
0.594952344894409 -0.164608955383301
0.602647662162781 -0.171640396118164
0.603733539581299 -0.172421455383301
0.609905362129211 -0.180233955383301
0.617186665534973 -0.188358783721924
0.624030470848083 -0.195858955383301
0.627905368804932 -0.203671455383301
0.636795997619629 -0.211483955383301
0.64071786403656 -0.219296455383301
0.643889784812927 -0.227108955383301
0.652671098709106 -0.234921455383301
0.655874133110046 -0.242733955383301
0.658030390739441 -0.250546455383301
0.664764881134033 -0.257804393768311
0.665467858314514 -0.258358955383301
0.670592784881592 -0.270077228546143
0.672436714172363 -0.273983955383301
0.673405408859253 -0.281796455383301
0.685811638832092 -0.297421455383301
0.68759286403656 -0.313046455383301
0.694905400276184 -0.328671455383301
0.695311665534973 -0.329311847686768
0.698905467987061 -0.336483955383301
0.702577233314514 -0.352108955383301
0.703928709030151 -0.363827228546143
0.704530358314514 -0.367733955383301
0.709374070167542 -0.378674983978271
0.711842894554138 -0.383358955383301
0.714171051979065 -0.391171455383301
0.717545986175537 -0.406796455383301
0.718436717987061 -0.414608955383301
0.717421054840088 -0.422421455383301
0.721936702728271 -0.438046455383301
0.724905371665955 -0.445858955383301
0.724077343940735 -0.453671455383301
0.729561686515808 -0.461483955383301
0.732545971870422 -0.481014728546143
0.733264803886414 -0.484921455383301
0.733546018600464 -0.492733955383301
0.734436631202698 -0.500546455383301
0.734256982803345 -0.512264728546143
0.734092950820923 -0.516171455383301
0.735233545303345 -0.523983955383301
0.734733581542969 -0.531796455383301
0.736780405044556 -0.539608955383301
0.736967921257019 -0.547421455383301
0.739577293395996 -0.555233955383301
0.742624163627625 -0.568714618682861
0.743061661720276 -0.570858955383301
0.745796084403992 -0.578671455383301
0.744046092033386 -0.586483955383301
0.748952269554138 -0.594296455383301
0.747069716453552 -0.598202228546143
};
\addlegendentry{OM3, $D=128$}
\end{axis}

\end{tikzpicture}

%% file: figures/taylor-bubble/convergence-shape-tail-om3.tex
% This file was created with tikzplotlib v0.10.1.
\begin{tikzpicture}

\definecolor{burlywood248194145}{RGB}{250,209,172}
\definecolor{darkgray176}{RGB}{176,176,176}
\definecolor{darkorange24213334}{RGB}{242,133,34}
\definecolor{lightgray204}{RGB}{204,204,204}
\definecolor{sandybrown24516489}{RGB}{181,100,26}

\begin{axis}[
height=\figureheight,
legend cell align={left},
legend style={
	fill opacity=0.8,
	draw opacity=1,
	text opacity=1,
	at={(0.01,0.99)},
	anchor=north west,
	draw=lightgray204
},
tick align=outside,
tick pos=left,
width=\figurewidth,
x grid style={darkgray176},
xlabel={\(\displaystyle r^{*}\)},
xmajorgrids,
xmin=-0.0389827966690625, xmax=0.818638730050312,
xtick style={color=black},
y grid style={darkgray176},
ylabel style={rotate=-90.0},
ylabel={\(\displaystyle z^{*}\)},
ymajorgrids,
ymin=-0.0992933716703125, ymax=0.318318805076563,
ytick style={color=black}
]
\addplot [very thick, black, mark=*, mark size=1, mark options={solid}, only marks]
table {%
0.751070022583008 0.279245018959045
0.751060009002686 0.276661038398743
0.75104808807373 0.272783041000366
0.751029968261719 0.267291069030762
0.751013994216919 0.262120962142944
0.750993967056274 0.256628036499023
0.750977993011475 0.250813007354736
0.750959992408752 0.245319962501526
0.750944018363953 0.240473031997681
0.750927925109863 0.235626935958862
0.750891923904419 0.225288033485413
0.75083601474762 0.207193970680237
0.750802040100098 0.196854948997498
0.752063989639282 0.185546040534973
0.752671957015991 0.172943949699402
0.753286004066467 0.16228199005127
0.753888010978699 0.148064970970154
0.755147933959961 0.136109948158264
0.757050037384033 0.122215986251831
0.75830602645874 0.109614968299866
0.759566068649292 0.0979830026626587
0.760175943374634 0.085705041885376
0.759490013122559 0.0747189521789551
0.75749397277832 0.059533953666687
0.756160020828247 0.0479030609130859
0.754166007041931 0.0336869955062866
0.753479957580566 0.0223790407180786
0.750848054885864 0.0110709667205811
0.748866081237793 8.59498977661133e-05
0.746886014938354 -0.0102519989013672
0.743628025054932 -0.0141290426254272
0.740370035171509 -0.0180050134658813
0.737761974334717 -0.0218809843063354
0.735153913497925 -0.0257569551467896
0.731894016265869 -0.0302799940109253
0.727335929870605 -0.0351250171661377
0.724077939987183 -0.0396469831466675
0.719521999359131 -0.0438460111618042
0.714959979057312 -0.0496599674224854
0.710399985313416 -0.0551519393920898
0.704541921615601 -0.0606429576873779
0.693494081497192 -0.0651619434356689
0.68504798412323 -0.0687140226364136
0.673354029655457 -0.072909951210022
0.661018013954163 -0.0745220184326172
0.651278018951416 -0.0761339664459229
0.643490076065063 -0.0767780542373657
0.63440203666687 -0.0777440071105957
0.624661922454834 -0.0793570280075073
0.611681938171387 -0.0799989700317383
0.591563940048218 -0.0799920558929443
0.577935934066772 -0.0803110599517822
0.55717396736145 -0.0793349742889404
0.539010047912598 -0.0777130126953125
0.521492004394531 -0.0764150619506836
0.512416005134583 -0.0738279819488525
0.499451994895935 -0.0693000555038452
0.48907995223999 -0.0660660266876221
0.478709936141968 -0.0618619918823242
0.467043995857239 -0.0576579570770264
0.458620071411133 -0.0537780523300171
0.450191974639893 -0.0502209663391113
0.439177989959717 -0.0456939935684204
0.428156018257141 -0.0427830219268799
0.413892030715942 -0.0389009714126587
0.40351402759552 -0.0366359949111938
0.388602018356323 -0.0330770015716553
0.375632047653198 -0.0304880142211914
0.360715985298157 -0.0275750160217285
0.34644603729248 -0.0256320238113403
0.330878019332886 -0.0240110158920288
0.312067985534668 -0.02109694480896
0.295850038528442 -0.0191529989242554
0.27963399887085 -0.0172100067138672
0.266011953353882 -0.0155899524688721
0.249145984649658 -0.0133229494094849
0.233579993247986 -0.0110559463500977
0.214761972427368 -0.0104039907455444
0.197247982025146 -0.0087820291519165
0.168053984642029 -0.0061880350112915
0.149888038635254 -0.00488996505737305
0.131721973419189 -0.00359201431274414
0.111608028411865 -0.00261604785919189
0.0973340272903442 -0.00196504592895508
0.0772199630737305 -0.0013120174407959
0.0655399560928345 -0.000661969184875488
0.055806040763855 -0.000658988952636719
0.0428299903869629 -0.000331997871398926
0.0317980051040649 -5.00679016113281e-06
0.0162240266799927 0
};
\addlegendentry{Experiment~\cite{bugg2002VelocityFieldTaylor}}
\addplot [line width=2pt, burlywood248194145, dash pattern=on 1pt off 3pt on 3pt off 3pt]
table {%
0 0
0.0625 4.67300415039062e-05
0.0937494039535522 -1.57356262207031e-05
0.156249403953552 -0.000765800476074219
0.218749403953552 -0.00207853317260742
0.3125 -0.00539112091064453
0.343749403953552 -0.00657844543457031
0.4375 -0.00928163528442383
0.468749403953552 -0.0102658271789551
0.531249403953552 -0.0167346000671387
0.559430599212646 -0.0249533653259277
0.593749403953552 -0.0344219207763672
0.656249403953552 -0.0277347564697266
0.661793112754822 -0.0249533653259277
0.702293157577515 0.00629663467407227
0.718749403953552 0.0236401557922363
0.729268193244934 0.0375466346740723
0.744411945343018 0.0687966346740723
0.749143123626709 0.100046634674072
0.753130674362183 0.131296634674072
0.754268169403076 0.162546634674072
0.754530668258667 0.193796634674072
0.752843141555786 0.225046634674072
0.75493061542511 0.256296634674072
0.758736848831177 0.287546634674072
};
\addlegendentry{OM3, $D=32$}
\addplot [line width=2pt, sandybrown24516489, dash pattern=on 1pt off 3pt on 3pt off 3pt]
table {%
0 0
0.0468746423721313 -0.000117301940917969
0.09375 -0.000601768493652344
0.140624642372131 -0.00152349472045898
0.171874642372131 -0.00249195098876953
0.203124642372131 -0.00400781631469727
0.234527826309204 -0.00614833831787109
0.265624642372131 -0.00866413116455078
0.296874642372131 -0.0106797218322754
0.328124642372131 -0.0122890472412109
0.390624642372131 -0.0148200988769531
0.421874642372131 -0.0165390968322754
0.453124642372131 -0.0191950798034668
0.484374642372131 -0.0234918594360352
0.515624642372131 -0.0261015892028809
0.546874642372131 -0.0276012420654297
0.578124642372131 -0.0286483764648438
0.609374642372131 -0.0283985137939453
0.640624642372131 -0.025820255279541
0.663185834884644 -0.0193982124328613
0.671874642372131 -0.0170235633850098
0.696677803993225 -0.00614833831787109
0.703124642372131 -0.00174188613891602
0.715580940246582 0.00947666168212891
0.726268410682678 0.0251016616821289
0.734374642372131 0.0354299545288086
0.739009022712708 0.0407266616821289
0.74488091468811 0.0563516616821289
0.748527765274048 0.0719766616821289
0.749158978462219 0.0876016616821289
0.750321507453918 0.103226661682129
0.750721573829651 0.118851661682129
0.750584006309509 0.134476661682129
0.75109338760376 0.150101661682129
0.750221490859985 0.165726661682129
0.754687190055847 0.181351661682129
0.756608963012695 0.196976661682129
0.750715255737305 0.212601661682129
0.765624642372131 0.221929550170898
0.777530908584595 0.228226661682129
0.765624642372131 0.234304904937744
0.75026524066925 0.243851661682129
0.765624642372131 0.252336025238037
0.779655933380127 0.259476661682129
0.765624642372131 0.266554832458496
0.750130891799927 0.275101661682129
0.765624642372131 0.283960819244385
0.772327661514282 0.287343502044678
0.772327899932861 0.294335842132568
0.779030919075012 0.290726661682129
0.765624642372131 0.297945499420166
};
\addlegendentry{OM3, $D=64$}
\addplot [line width=2pt, darkorange24213334, dash pattern=on 1pt off 3pt on 3pt off 3pt]
table {%
0 0
0.03125 -0.000191211700439453
0.0546866655349731 -0.000562667846679688
0.0859366655349731 -0.00145292282104492
0.15625 -0.00392961502075195
0.195311665534973 -0.00499200820922852
0.234375 -0.00600385665893555
0.25 -0.00660181045532227
0.265625 -0.00746488571166992
0.273436665534973 -0.00797653198242188
0.304686665534973 -0.0109376907348633
0.328125 -0.0123085975646973
0.382811665534973 -0.0149297714233398
0.400661110877991 -0.016754150390625
0.414061665534973 -0.0184845924377441
0.429686665534973 -0.0197267532348633
0.460936665534973 -0.0212111473083496
0.484375 -0.0224299430847168
0.492186665534973 -0.0229296684265137
0.53125 -0.026425838470459
0.539061665534973 -0.026890754699707
0.570311665534973 -0.0277185440063477
0.59375 -0.0275077819824219
0.617186665534973 -0.0270156860351562
0.63304603099823 -0.0250000953674316
0.633280396461487 -0.0248827934265137
0.648436665534973 -0.0207657814025879
0.664061665534973 -0.020078182220459
0.674882531166077 -0.0148358345031738
0.679686665534973 -0.012601375579834
0.688702344894409 -0.00925779342651367
0.695311665534973 -0.00560951232910156
0.706827282905579 -0.00144529342651367
0.710936665534973 0.00305461883544922
0.713139772415161 0.00636720657348633
0.72763979434967 0.0141797065734863
0.729967951774597 0.0219922065734863
0.734483599662781 0.0298047065734863
0.735710144042969 0.0415239334106445
0.736046075820923 0.0454297065734863
0.741327404975891 0.0550742149353027
0.744108557701111 0.0610547065734863
0.747421026229858 0.0688672065734863
0.749171018600464 0.0766797065734863
0.749889731407166 0.0844922065734863
0.750108480453491 0.0962114334106445
0.750139832496643 0.104023933410645
0.750342965126038 0.127461433410645
0.750421047210693 0.139179706573486
0.757811665534973 0.145406246185303
0.760717868804932 0.146992206573486
0.757811665534973 0.148531436920166
0.750280380249023 0.154804706573486
0.757811665534973 0.160062313079834
0.762670993804932 0.162617206573486
0.757811665534973 0.165195465087891
0.750327348709106 0.170429706573486
0.764905452728271 0.178242206573486
0.750967860221863 0.186054706573486
0.761584758758545 0.191734313964844
0.765358567237854 0.193867206573486
0.757811665534973 0.198671817779541
0.75204598903656 0.201679706573486
0.757811665534973 0.204601764678955
0.765546083450317 0.209492206573486
0.757811665534973 0.214797019958496
0.752874135971069 0.217304706573486
0.757811665534973 0.219758033752441
0.765655398368835 0.225117206573486
0.757811665534973 0.232211112976074
0.756358504295349 0.232929706573486
0.757811665534973 0.233804702758789
0.764811635017395 0.240742206573486
0.758374214172363 0.248554706573486
0.765264749526978 0.256367206573486
0.75860857963562 0.264179706573486
0.765389800071716 0.271992206573486
0.760217905044556 0.279804706573486
0.764499187469482 0.287617206573486
0.762670993804932 0.295429706573486
0.763553977012634 0.299336433410645
};
\addlegendentry{OM3, $D=128$}
\end{axis}

\end{tikzpicture}

%% file: main.bbl
\begin{thebibliography}{10}
\providecommand \doibase [0]{http://dx.doi.org/}%

\bibitem{korner2005LatticeBoltzmannModel}
K{\"o}rner C, Thies M, Hofmann T, Th{\"u}rey N, R{\"u}de U. Lattice {{Boltzmann
  Model}} for {{Free Surface Flow}} for {{Modeling Foaming}}. {\it Journal of
  Statistical Physics} 2005\string; 121(1)\string: 179--196.
\newblock \href {\doibase 10.1007/s10955-005-8879-8} {doi:
  10.1007/s10955-005-8879-8}

\bibitem{scardovelli1999DirectNumericalSimulation}
Scardovelli R, Zaleski S. Direct Numerical Simulation of Free-Surface and
  Interfacial Flow. {\it Annual Review of Fluid Mechanics} 1999\string;
  31(1)\string: 567--603.
\newblock \href {\doibase 10.1146/annurev.fluid.31.1.567} {doi:
  10.1146/annurev.fluid.31.1.567}

\bibitem{hirt1981VolumeFluidVOF}
Hirt C, Nichols B. Volume of Fluid ({{VOF}}) Method for the Dynamics of Free
  Boundaries. {\it Journal of Computational Physics} 1981\string; 39(1).
\newblock \href {\doibase 10.1016/0021-9991(81)90145-5} {doi:
  10.1016/0021-9991(81)90145-5}

\bibitem{donath2011VerificationSurfaceTension}
Donath S, Mecke K, Rabha S, Buwa V, R{\"u}de U. Verification of Surface Tension
  in the Parallel Free Surface Lattice {{Boltzmann}} Method in {{waLBerla}}.
  {\it Computers \& Fluids} 2011\string; 45(1).
\newblock \href {\doibase 10.1016/j.compfluid.2010.12.027} {doi:
  10.1016/j.compfluid.2010.12.027}

\bibitem{zhao2013LatticeBoltzmannMethod}
Zhao Z, Huang P, Li Y, Li J. A Lattice {{Boltzmann}} Method for Viscous Free
  Surface Waves in Two Dimensions. {\it International Journal for Numerical
  Methods in Fluids} 2013\string; 71(2)\string: 223--248.
\newblock \href {\doibase 10.1002/fld.3660} {doi: 10.1002/fld.3660}

\bibitem{janssen2011FreeSurfaceFlow}
Jan{\ss}en C, Krafczyk M. Free Surface Flow Simulations on {{GPGPUs}} Using the
  {{LBM}}. {\it Computers \& Mathematics with Applications} 2011\string;
  61(12).
\newblock \href {\doibase 10.1016/j.camwa.2011.03.016} {doi:
  10.1016/j.camwa.2011.03.016}

\bibitem{lehmann2021EjectionMarineMicroplastics}
Lehmann M, Oehlschl{\"a}gel LM, H{\"a}usl FP, Held A, Gekle S. Ejection of
  Marine Microplastics by Raindrops: A Computational and Experimental Study.
  {\it Microplastics and Nanoplastics} 2021\string; 1(1).
\newblock \href {\doibase 10.1186/s43591-021-00018-8} {doi:
  10.1186/s43591-021-00018-8}

\bibitem{ammer2014SimulatingFastElectron}
Ammer R, Markl M, Ljungblad U, K{\"o}rner C, R{\"u}de U. Simulating Fast
  Electron Beam Melting with a Parallel Thermal Free Surface Lattice
  {{Boltzmann}} Method. {\it Computers \& Mathematics with Applications}
  2014\string; 67(2).
\newblock \href {\doibase 10.1016/j.camwa.2013.10.001} {doi:
  10.1016/j.camwa.2013.10.001}

\bibitem{becker2009CombinedLatticeBGK}
Becker J, Junk M, Kehrwald D, Th{\"o}mmes G, Yang Z. A Combined Lattice
  {{BGK}}/Level Set Method for Immiscible Two-Phase Flows. {\it Computers \&
  Mathematics with Applications} 2009\string; 58(5).
\newblock \href {\doibase 10.1016/j.camwa.2009.02.005} {doi:
  10.1016/j.camwa.2009.02.005}

\bibitem{lallemand2007LatticeBoltzmannFronttracking}
Lallemand P, Luo LS, Peng Y. A Lattice {{Boltzmann}} Front-Tracking Method for
  Interface Dynamics with Surface Tension in Two Dimensions. {\it Journal of
  Computational Physics} 2007\string; 226(2).
\newblock \href {\doibase 10.1016/j.jcp.2007.05.021} {doi:
  10.1016/j.jcp.2007.05.021}

\bibitem{gunstensen1991LatticeBoltzmannModel}
Gunstensen AK, Rothman DH, Zaleski S, Zanetti G. Lattice {{Boltzmann}} Model of
  Immiscible Fluids. {\it Physical Review A} 1991\string; 43(8).
\newblock \href {\doibase 10.1103/PhysRevA.43.4320} {doi:
  10.1103/PhysRevA.43.4320}

\bibitem{inamuro2004LatticeBoltzmannMethod}
Inamuro T, Ogata T, Tajima S, Konishi N. A Lattice {{Boltzmann}} Method for
  Incompressible Two-Phase Flows with Large Density Differences. {\it Journal
  of Computational Physics} 2004\string; 198(2).
\newblock \href {\doibase 10.1016/j.jcp.2004.01.019} {doi:
  10.1016/j.jcp.2004.01.019}

\bibitem{zheng2005LatticeBoltzmannInterface}
Zheng HW, Shu C, Chew YT. Lattice {{Boltzmann}} Interface Capturing Method for
  Incompressible Flows. {\it Physical Review E} 2005\string; 72(5).
\newblock \href {\doibase 10.1103/PhysRevE.72.056705} {doi:
  10.1103/PhysRevE.72.056705}

\bibitem{fakhari2017ImprovedLocalityPhasefield}
Fakhari A, Mitchell T, Leonardi C, Bolster D. Improved Locality of the
  Phase-Field Lattice-{{Boltzmann}} Model for Immiscible Fluids at High Density
  Ratios. {\it Physical Review E} 2017\string; 96(5).
\newblock \href {\doibase 10.1103/PhysRevE.96.053301} {doi:
  10.1103/PhysRevE.96.053301}

\bibitem{swift1995LatticeBoltzmannSimulation}
Swift MR, Osborn WR, Yeomans JM. Lattice {{Boltzmann Simulation}} of {{Nonideal
  Fluids}}. {\it Physical Review Letters} 1995\string; 75(5).
\newblock \href {\doibase 10.1103/PhysRevLett.75.830} {doi:
  10.1103/PhysRevLett.75.830}

\bibitem{shan1994SimulationNonidealGases}
Shan X, Chen H. Simulation of Nonideal Gases and Liquid-Gas Phase Transitions
  by the Lattice {{Boltzmann}} Equation. {\it Physical Review E} 1994\string;
  49(4).
\newblock \href {\doibase 10.1103/PhysRevE.49.2941} {doi:
  10.1103/PhysRevE.49.2941}

\bibitem{schwarzmeier2022ComparisonFreeSurface}
Schwarzmeier C, Holzer M, Mitchell T, Lehmann M, H{\"a}usl F, R{\"u}de U.
  Comparison of Free-Surface and Conservative {{Allen}}\textendash{{Cahn}}
  Phase-Field Lattice {{Boltzmann}} Method. {\it Journal of Computational
  Physics} 2023\string; 473\string: 111753.
\newblock \href {\doibase 10.1016/j.jcp.2022.111753} {doi:
  10.1016/j.jcp.2022.111753}

\bibitem{bogner2016CurvatureEstimationVolumeoffluid}
Bogner S, R{\"u}de U, Harting J. Curvature Estimation from a Volume-of-Fluid
  Indicator Function for the Simulation of Surface Tension and Wetting with a
  Free-Surface Lattice {{Boltzmann}} Method. {\it Physical Review E}
  2016\string; 93(4).
\newblock \href {\doibase 10.1103/PhysRevE.93.043302} {doi:
  10.1103/PhysRevE.93.043302}

\bibitem{bogner2017DirectNumericalSimulation}
Bogner S. {\it Direct {{Numerical Simulation}} of {{Liquid-Gas-Solid Flows
  Based}} on the {{Lattice Boltzmann Method}}}. PhD thesis.
  Friedrich-Alexander-Universit\"at Erlangen-N\"urnberg, {Erlangen};  2017.

\bibitem{thies2005LatticeBoltzmannModeling}
Thies M. {\it Lattice {{Boltzmann Modeling}} with {{Free Surfaces Applied}} to
  {{Formation}} of {{Metal Foams}}}. PhD thesis. Universit\"at
  Erlangen-N\"urnberg, {Erlangen};  2005.

\bibitem{bauer2021WaLBerlaBlockstructuredHighperformance}
Bauer M, Eibl S, Godenschwager C, et al. {{waLBerla}}: {{A}} Block-Structured
  High-Performance Framework for Multiphysics Simulations. {\it Computers \&
  Mathematics with Applications} 2021\string; 81.
\newblock \href {\doibase 10.1016/j.camwa.2020.01.007} {doi:
  10.1016/j.camwa.2020.01.007}

\bibitem{schwarzmeier2022ComparisonRefillingSchemes}
Schwarzmeier C, R{\"u}de U. Comparison of Refilling Schemes in the Free-Surface
  Lattice {{Boltzmann}} Method. {\it AIP Advances} 2022\string; 12(11)\string:
  23.
\newblock \href {\doibase 10.1063/5.0131159} {doi: 10.1063/5.0131159}

\bibitem{kruger2017LatticeBoltzmannMethod}
Kr{\"u}ger T, Kusumaatmaja H, Kuzmin A, Shardt O, Silva G, Viggen EM. {\it The
  Lattice {{Boltzmann}} Method: Principles and Practice}.
\newblock {Switzerland}: {Springer} .
\newblock 2017.

\bibitem{bauer2020TruncationErrorsD3Q19a}
Bauer M, Silva G, R{\"u}de U. Truncation Errors of the {{D3Q19}} Lattice Model
  for the Lattice {{Boltzmann}} Method. {\it Journal of Computational Physics}
  2020\string; 405(C).
\newblock \href {\doibase 10.1016/j.jcp.2019.109111} {doi:
  10.1016/j.jcp.2019.109111}

\bibitem{guo2002DiscreteLatticeEffects}
Guo Z, Zheng C, Shi B. Discrete Lattice Effects on the Forcing Term in the
  Lattice {{Boltzmann}} Method. {\it Physical Review E} 2002\string; 65(4).
\newblock \href {\doibase 10.1103/PhysRevE.65.046308} {doi:
  10.1103/PhysRevE.65.046308}

\bibitem{hou1996LatticeBoltzmannSubgrid}
Hou S, Sterling J, Chen S, Doolen GD. A {{Lattice Boltzmann Subgrid Model}} for
  {{High Reynolds Number Flows}}. In:  Lawniczak AT, Kapral R. \kern-2pt, eds.
  {\it Pattern {{Formation}} and {{Lattice Gas Automata}}}. 6 of {\it Fields
  {{Institute Communications}}}. {American Mathematical Society}. ; 1996

\bibitem{yu2005DNSDecayingIsotropic}
Yu H, Girimaji SS, Luo LS. {{DNS}} and {{LES}} of Decaying Isotropic Turbulence
  with and without Frame Rotation Using Lattice {{Boltzmann}} Method. {\it
  Journal of Computational Physics} 2005\string; 209(2).
\newblock \href {\doibase 10.1016/j.jcp.2005.03.022} {doi:
  10.1016/j.jcp.2005.03.022}

\bibitem{pohl2008HighPerformanceSimulation}
Pohl T. {\it High {{Performance Simulation}} of {{Free Surface Flows Using}}
  the {{Lattice Boltzmann Method}}}. PhD thesis. Universit\"at
  Erlangen-N\"urnberg, {Erlangen};  2008.

\bibitem{thurey2007PhysicallyBasedAnimation}
Th{\"u}rey N. {\it Physically Based {{Animation}} of {{Free Surface Flows}}
  with the {{Lattice Boltzmann Method}}}. PhD thesis. Universit\"at
  Erlangen-N\"urnberg, {Erlangen};  2007.

\bibitem{donath2011WettingModelsParallel}
Donath S. {\it Wetting {{Models}} for a {{Parallel High-Performance Free
  Surface Lattice Boltzmann Method}}}. PhD thesis. Universit\"at
  Erlangen-N\"urnberg, {Erlangen};  2011.

\bibitem{bogner2015BoundaryConditionsFree}
Bogner S, Ammer R, R{\"u}de U. Boundary Conditions for Free Interfaces with the
  Lattice {{Boltzmann}} Method. {\it Journal of Computational Physics}
  2015\string; 297.
\newblock \href {\doibase 10.1016/j.jcp.2015.04.055} {doi:
  10.1016/j.jcp.2015.04.055}

\bibitem{parker1992TwoThreeDimensional}
Parker BJ, Youngs DL. Two and Three Dimensional {{Eulerian}} Simulation and
  Fluid Flow with Material Interfaces. technical {{Report}} 01/92, {UK Atomic
  Weapons Establishment}; :   1992.

\bibitem{williams1999AccuracyConvergenceContinuum}
Williams MW, Kothe DB, Puckett EG. Accuracy and {{Convergence}} of {{Continuum
  Surface Tension Models}}. In:  Shyy W, Narayanan R. \kern-2pt, eds. {\it
  Fluid {{Dynamics}} at {{Interfaces}}}{Cambridge University Press}. first~ed.
  1999 (pp. 294--305).

\bibitem{anderl2014FreeSurfaceLattice}
Anderl D, Bogner S, Rauh C, R{\"u}de U, Delgado A. Free Surface Lattice
  {{Boltzmann}} with Enhanced Bubble Model. {\it Computers \& Mathematics with
  Applications} 2014\string; 67(2).
\newblock \href {\doibase 10.1016/j.camwa.2013.06.007} {doi:
  10.1016/j.camwa.2013.06.007}

\bibitem{ladd1994NumericalSimulationsParticulate}
Ladd AJC. Numerical Simulations of Particulate Suspensions via a Discretized
  {{Boltzmann}} Equation. {{Part}} 2. {{Numerical}} Results. {\it Journal of
  Fluid Mechanics} 1994\string; 271.
\newblock \href {\doibase 10.1017/S0022112094001783} {doi:
  10.1017/S0022112094001783}

\bibitem{ladd1994NumericalSimulationsParticulatea}
Ladd AJC. Numerical Simulations of Particulate Suspensions via a Discretized
  {{Boltzmann}} Equation. {{Part}} 1. {{Theoretical}} Foundation. {\it Journal
  of Fluid Mechanics} 1994\string; 271.
\newblock \href {\doibase 10.1017/S0022112094001771} {doi:
  10.1017/S0022112094001771}

\bibitem{thorimbert2021ImplementationLatticeBoltzmann}
Thorimbert Y, Chopard B, L{\"a}tt J. Implementation of Lattice {{Boltzmann}}
  Free-Surface and Shallow Water Models and Their Two-Way Coupling. {\it
  MethodsX} 2021\string; 8\string: 101338.
\newblock \href {\doibase 10.1016/j.mex.2021.101338} {doi:
  10.1016/j.mex.2021.101338}

\bibitem{biscarini2011ApplicationLatticeBoltzmann}
Biscarini C, Di~Francesco S, Mencattini M. Application of the Lattice
  {{Boltzmann}} Method for Large-scale Hydraulic Problems. {\it International
  Journal of Numerical Methods for Heat \& Fluid Flow} 2011\string;
  21(5)\string: 584--601.
\newblock \href {\doibase 10.1108/09615531111135846} {doi:
  10.1108/09615531111135846}

\bibitem{cubeddu2017SimulationsBubbleGrowth}
Cubeddu A, Rauh C, Ulrich V. Simulations of Bubble Growth and Interaction in
  High Viscous Fluids Using the Lattice {{Boltzmann}} Method. {\it
  International Journal of Multiphase Flow} 2017\string; 93\string: 108--114.
\newblock \href {\doibase 10.1016/j.ijmultiphaseflow.2017.04.001} {doi:
  10.1016/j.ijmultiphaseflow.2017.04.001}

\bibitem{zhao2017LBMLESSimulationTransient}
Zhao P, Li Q, Kuang SB, Zou Z. {{LBM-LES Simulation}} of the {{Transient
  Asymmetric Flow}} and {{Free Surface Fluctuations}} under {{Steady Operating
  Conditions}} of {{Slab Continuous Casting Process}}. {\it Metallurgical and
  Materials Transactions B} 2017\string; 48(1)\string: 456--470.
\newblock \href {\doibase 10.1007/s11663-016-0830-7} {doi:
  10.1007/s11663-016-0830-7}

\bibitem{chiappini2018AnalysisFluidMotion}
Chiappini D, Di~Ilio G, Bella G. Analysis of the {{Fluid Motion Induced}} by a
  {{Vibrating Lamina Through Free Surface-Lattice Boltzmann Coupled Method}}.
  In: {ASME}. {American Society of Mechanical Engineers}; 2018; {Pittsburgh,
  Pennsylvania, USA}\string: V009T12A003

\bibitem{bublik2021ExperimentalValidationNumerical}
Bubl{\'i}k O, Lobovsk{\'y} L, Heidler V, Mandys T, Vimmr J. Experimental
  Validation of Numerical Simulations of Free-Surface Flow within Casting Mould
  Cavities. {\it Engineering Computations} 2021\string; 38(10)\string:
  4024--4046.
\newblock \href {\doibase 10.1108/EC-08-2020-0458} {doi:
  10.1108/EC-08-2020-0458}

\bibitem{huang2021ThreedimensionalSimulationReservoir}
Huang Z, Diao W, Wu J, Cheng Y, Huai W. Three-Dimensional Simulation of
  Reservoir Temperature and Pollutant Transport by the Lattice {{Boltzmann}}
  Method. {\it Environmental Science and Pollution Research} 2021\string;
  28(1)\string: 459--472.
\newblock \href {\doibase 10.1007/s11356-020-10174-8} {doi:
  10.1007/s11356-020-10174-8}

\bibitem{dingemans1997WaterWavePropagation}
Dingemans MW. {\it Water {{Wave Propagation Over Uneven Bottoms}}: {{Part}} 1}.
  13 of {\it Advanced {{Series}} on {{Ocean Engineering}}}.
\newblock {World Scientific Publishing Company} .
\newblock 1997

\bibitem{lamb1975Hydrodynamics}
Lamb H. {\it Hydrodynamics}.
\newblock {Cambridge University Press}.
\newblock sixth~ed. 1975.

\bibitem{sato2022ComparativeStudyCumulant}
Sato K, Kawasaki K, Koshimura S. A Comparative Study of the Cumulant Lattice
  {{Boltzmann}} Method in a Single-Phase Free-Surface Model of Violent Flows.
  {\it Computers \& Fluids} 2022\string; 236.
\newblock \href {\doibase 10.1016/j.compfluid.2021.105303} {doi:
  10.1016/j.compfluid.2021.105303}

\bibitem{moraga2015VOFFVMPrediction}
Moraga NO, Lemus LA, Saavedra MA, {Lemus-Mondaca} RA. {{VOF}}/{{FVM}}
  Prediction and Experimental Validation for Shear-Thinning Fluid Column
  Collapse. {\it Computers \& Mathematics with Applications} 2015\string;
  69(2).
\newblock \href {\doibase 10.1016/j.camwa.2014.11.018} {doi:
  10.1016/j.camwa.2014.11.018}

\bibitem{martin1952PartIVExperimental}
Martin JC, Moyce WJ, Penney WG, Price AT, Thornhill CK. Part {{IV}}. {{An}}
  Experimental Study of the Collapse of Liquid Columns on a Rigid Horizontal
  Plane. {\it Philosophical Transactions of the Royal Society of London. Series
  A, Mathematical and Physical Sciences} 1952\string; 244(882).
\newblock \href {\doibase 10.1098/rsta.1952.0006} {doi: 10.1098/rsta.1952.0006}

\bibitem{rumble2021CRCHandbookChemistry}
Rumble J. \kern-2pt, ed.{\it {{CRC Handbook}} of {{Chemistry}} and
  {{Physics}}}.
\newblock {CRC Press}.
\newblock one hundred second~ed. 2021.

\bibitem{pavlidis1982AlgorithmsGraphicsImage}
Pavlidis T. {\it Algorithms for {{Graphics}} and {{Image Processing}}}.
\newblock {Berlin, Heidelberg}: {Springer Berlin Heidelberg} .
\newblock 1982

\bibitem{bugg2002VelocityFieldTaylor}
Bugg J, Saad G. The Velocity Field around a {{Taylor}} Bubble Rising in a
  Stagnant Viscous Fluid: Numerical and Experimental Results. {\it
  International Journal of Multiphase Flow} 2002\string; 28(5).
\newblock \href {\doibase 10.1016/S0301-9322(02)00002-2} {doi:
  10.1016/S0301-9322(02)00002-2}

\bibitem{wang2000SplashingImpactSingle}
Wang AB, Chen CC. Splashing Impact of a Single Drop onto Very Thin Liquid
  Films. {\it Physics of Fluids} 2000\string; 12(9).
\newblock \href {\doibase 10.1063/1.1287511} {doi: 10.1063/1.1287511}

\end{thebibliography}
